\documentclass[letterpaper,onecolumn,english,aps,rmp,floatfix]{revtex4}%
\usepackage{times}
\usepackage[T1]{fontenc}
\usepackage[latin1]{inputenc}
\usepackage{amsmath}
\usepackage{graphicx}
\usepackage{amssymb}
\usepackage{graphicx}
\usepackage{amscd}
\usepackage{bm}
\usepackage{amsfonts}
\usepackage{babel}%
\begin{document}
\title{Disorder-Driven Non-Fermi Liquid Behavior of Correlated Electrons}
\author{E.~Miranda}
\affiliation{Instituto de Física Gleb Wataghin, Unicamp, Caixa Postal 6165, 13083-970
Campinas, SP, Brazil}
\author{V.~Dobrosavljevi\'{c}}
\affiliation{Department of Physics and National High Magnetic Field Laboratory, Florida
State University, Tallahassee, FL 32306}
\date{\today{}}

\begin{abstract}
Systematic deviations from standard Fermi-liquid behavior have been widely
observed and documented in several classes of strongly correlated metals. For
many of these systems, mounting evidence is emerging that the anomalous
behavior is most likely triggered by the interplay of quenched disorder and
strong electronic correlations. In this review, we present a broad overview of
such disorder-driven non-Fermi-liquid behavior, and discuss various examples
where the anomalies have been studied in detail. We describe both their
phenomenological aspects as observed in experiment, and the current
theoretical scenarios that attempt to unravel their microscopic origin.

\end{abstract}
\maketitle
\tableofcontents

\section{Introduction}

\label{sec:Introduction}

Soon after the discovery of high temperature superconductivity in the cuprates
it was realized that the optimally doped compounds showed clear deviations
from Fermi liquid behavior. This was not the first time that a real system
seemed to violate Landau's paradigm. The heavy fermion superconductor
UBe$_{13}$ had previously been found to show anomalous normal state behavior
\cite{cox87}. Since then, the quest for such systems has been vigorous and
numerous puzzling compounds have been discovered. However, a coherent
theoretical framework with which to understand this behavior has not been
obtained. One of the few situations where there seems to be a general
agreement as to the inadequacy of Fermi liquid theory is the critical region
governed by a zero temperature quantum phase transition. Here, deviations from
Landau's predictions seem to be the norm, although a general theory of this
behavior is still lacking. However, the quantum critical non-Fermi liquid
(NFL) requires fine tuning to a special location in the phase diagram. In many
of the known compounds, NFL behavior seems to be better characterized as the
property of a whole phase instead of just a phase boundary.

Interestingly, NFL behavior in disordered systems rests on a much firmer
theoretical basis. Even an exactly solvable model is known where diverging
thermodynamical properties can be calculated \cite{mccoywu68,mccoy69}. In
addition, this anomalous behavior persists over an entire region of the phase
diagram and is not confined to a specific point. These so-called
Griffiths-McCoy phases had been originally proposed in disordered classical
systems, where the accompanying singularities are rather weak \cite{griffiths}%
. The quantum version of this phenomenon is, however, much stronger and has
now been established theoretically in many models. Furthermore, their
properties are not believed to be a peculiarity of the one-dimensional exactly
solvable model. Their relevance to real compounds has also been advocated in
many systems, from doped semi-conductors to disordered heavy fermion
compounds. The emergence of NFL behavior in a disordered context is thus seen
to be fairly natural.

The question of NFL behavior in various clean situations has been extensively
discussed in the literature. It is the purpose of this review to attempt to
bring together in a single article a number of theoretical models and analyses
in which disorder-induced NFL behavior has arisen. Although reference will be
made to the most important experimental findings, the focus will be on theory
rather than on experiments. For the latter, we will refer the reader to other reviews.

The issue of disorder-induced NFL behavior was discussed for the first time in
connection with the anomalous thermodynamic responses near the metal-insulator
transition in doped semiconductors \cite{paalanen91,sarachik95}. In this
context, the random singlet phase of Bhatt and Lee was an important pioneering
work \cite{bhattlee81,bhattlee82}. Since then, anomalous behavior has been
observed in various disordered heavy fermion compounds, where local moments
are well formed even in the clean limit \cite{stewartNFL}. More recently,
two-dimensional electron systems in semiconductor heterostructures have also
shown intriguing behavior \cite{abekravsarachik}. Finally, intrinsic and
extrinsic heterogeneities in high temperature superconductors have led to the
consideration of these disorder effects in that context as well. These
systems, therefore, have become the main focus of the theoretical efforts and
will take center stage in this review.

We will attempt to emphasize the general aspects of the various contexts in
which NFL behavior arises. As mentioned above, Griffiths phases play a central
role in many of these models. Although they come in different guises, all
Griffiths phases seem to share a few common generic features, which we will
describe. Once these features are present, most physical quantities can be
immediately obtained. The question of distinguishing among the different
microscopic mechanisms is therefore rather subtle. However, a few criteria can
be posed and perhaps used to make this distinction. We hope these will serve
as a guide to experimentalists.

This review is organized as follows. In Section~\ref{sec:fermiliquid}, we
briefly summarize the main ideas of Fermi liquid theory as applied to both
clean and disordered systems. Section~\ref{sec:Phenomenology} is devoted to a
general discussion of the phenomenology of the main non-Fermi liquid systems.
Subsection~\ref{sub:cleannfl} focuses on NFL behavior of clean systems,
whereas subsections~\ref{sub:dopedsemicond}, \ref{sub:nflkondoalloys} and
\ref{sub:metallic-glass} describe the anomalous behavior of doped
semiconductors, disordered heavy fermion systems, and the glassy regime close
to the metal-insulator transitions. The main theoretical approaches are
reviewed in Section~\ref{sec:Theoretical-approaches}.
Subsection~\ref{sub:Disordered-Hubbard-and-Kondo} is devoted to results on the
disordered Hubbard and Kondo/Anderson lattice models. We highlight the physics
of local moment formation in a disordered environment (\ref{sub:local-moments}%
), the random singlet phase (\ref{sub:randomsinglet}), the Kondo disorder
model (\ref{sub:kdm} and \ref{sub:dmft}), the electronic Griffiths phase close
to the disorder-induced metal-insulator transition (\ref{sub:statdmft},
\ref{sub:andersongriff} and \ref{sub:mottgriff}), and the incoherent transport
near the two-dimensional metal-insulator transition (\ref{sub:incoherent}).
The general topic of magnetic Griffiths phases is discussed in
subsection~\ref{sub:magneticgriff}. We first review their well-established
properties in the case of insulating magnets, where we emphasize the important
role played by the infinite randomness fixed point (\ref{sub:irfp}). The
generic mechanism behind magnetic Griffiths phases is described in
subsection~\ref{sub:griffgeneral}, the important effect of dissipation on the
dynamics of the Griffiths droplets is considered in
subsection~\ref{sub:metalgriff}, and further remarks on the applicability of
this scenario to real metallic systems are made in
subsection~\ref{sub:griffapplicab}. Section~\ref{sub:Itinerant-quantum-glass}
is devoted to the physics of glassy dynamics and ordering in metals. We wrap
up with conclusions and open questions in Section~\ref{sec:conclusions}.

\section{General predictions of Fermi liquid theory and its limitations}

\label{sec:fermiliquid}

\subsection{Fermi liquid theory for clean systems}

\label{sub:flt}

Interacting homogeneous systems of fermions in three dimensions are believed
to be described at low energies and long wave lengths by Landau's Fermi liquid
theory \cite{landauFL1,landauFL2}. Although there is no general rigorous
justification for it, Landau was able to use diagrammatic arguments to show
that his general framework is at least internally consistent \cite{landauFL3}.
More recently, these early demonstrations were put on a firmer basis in the
modern framework of the renormalization group, in which the asymptotic
correctness of the theory can be shown \cite{benfattogallavotti90,shankar94}.
In addition, our confidence in its correctness is boosted by simple model
systems where controlled calculations can be carried out, such as a dilute
Fermi gas with short-range interactions \cite{abrikkhalat58,galitskii58}, the
electron-phonon system \cite{migdal58}, and the very dense degenerate electron
plasma \cite{abrikosov62}. More importantly, its direct experimental
verification in the prototypical Fermi liquid system, the normal state of
liquid $^{3}$He between the onset of superfluidity at $T_{c}\approx3$ mK and
about 100 mK, can be taken as one of the most striking examples of Landau's
theory at work (see, e. g., \cite{leggett75}). Finally, Fermi liquid theory,
properly generalized to charged systems \cite{silin58a,silin58b}, forms the
foundation of our understanding of the behavior of electrons in metals.

There are excellent reviews of Fermi liquid theory to which the reader is
referred to \cite{pinesnozieres,agd,leggett75,baympethick78}. We will content
ourselves with a brief outline of its main assumptions and predictions. Fermi
liquid theory starts by recognizing that the low energy excitations of a Fermi
sea, say an added electron at a wave vector $\mathbf{k}$ with $\left|
\mathbf{k}\right|  =k>k_{F}$ and $k-k_{F}\ll k_{F}$, where $k_{F}$ is the
Fermi wave vector (we assume a rotationally invariant system, for simplicity),
have a very long lifetime. This is due to restrictions imposed by energy and
momentum conservation and the blocking of further occupation of the Fermi sea
by the Pauli exclusion principle. In this case, the lifetime can be shown to
be $\tau_{k}\sim\left(  k-k_{F}\right)  ^{2}$. The stability of these
quasi-particle excitations allows us to ignore their decay at first and treat
them (perturbatively) only at a later stage. The second ingredient of the
theory is sometimes called {}``adiabatic continuity'': the excitations of the
interacting fermions are in one-to-one correspondence to those of a
non-interacting Fermi gas, in other words, the slow {}``switching-on'' of the
interactions in a Fermi gas does not alter the nature of the excitation
spectrum. These two assumptions led Landau to write the total energy of weakly
excited states of the system as a functional of the various occupation numbers
of states labeled by momentum $\mathbf{k}$ and spin projection $\sigma=\pm$,
the quasi-particle states. More precisely, he used the deviations $\delta
n_{\mathbf{k},\sigma}$ from their occupations in the ground state (in this
review, we use units such that $\hbar=1$ and $k_{B}=1$)%
\begin{align}
E  &  =E_{0}+\sum_{\mathbf{k},\sigma}\epsilon\left(  \mathbf{k}\right)  \delta
n_{\mathbf{k},\sigma}\nonumber\\
&  +\sum_{\mathbf{k},\sigma;\mathbf{k}^{\prime},\sigma^{\prime}}f\left(
\mathbf{k},\sigma;\mathbf{k}^{\prime},\sigma^{\prime}\right)  \delta
n_{\mathbf{k},\sigma}\delta n_{\mathbf{k}^{\prime},\sigma^{\prime}}.
\label{eq:flenergy}%
\end{align}
Here, $E_{0}$ is the (usually unknown) ground state energy, $\epsilon\left(
\mathbf{k}\right)  \approx v_{F}\left(  k-k_{F}\right)  $ is the
quasi-particle dispersion, parametrized by $v_{F}=k_{F}/m^{\ast}$, where
$m^{\ast}$ is the effective mass of the quasi-particle, usually distinct from
its bare, free-electron value. Note that the Fermi wave vector $k_{F}$ is not
modified by interactions, a fact known as Luttinger's theorem
\cite{luttingertheorem}. The last term, quadratic in the occupations,
incorporates the self-consistent interactions among the quasi-particles and is
easily seen to be of the same order as the second term for weakly excited states.

Since only momenta close to the Fermi surface are needed for low-energy
excitations, we can take, to leading order, $\left|  \mathbf{k}\right|
=\left|  \mathbf{k}^{\prime}\right|  =k_{F}$ in the last term of Eq.
(\ref{eq:flenergy}) and then $f$ only depends on the angle $\theta$ between
$\mathbf{k}$ and $\mathbf{k}^{\prime}$. Moreover, in the absence of an applied
magnetic field, spin rotation invariance imposes additional constraints%
\begin{equation}
f\left(  \mathbf{k},\sigma;\mathbf{k}^{\prime},\sigma^{\prime}\right)
=f^{s}\left(  \theta\right)  +\sigma\sigma^{\prime}f^{a}\left(  \theta\right)
. \label{eq:flffunction}%
\end{equation}
Finally, it is convenient to decompose the angular dependence in Legendre
polynomials%
\begin{equation}
\nu_{F}f^{s,a}\left(  \theta\right)  =\sum_{l}\left(  2l+1\right)  F_{l}%
^{s,a}P_{l}\left(  \cos\theta\right)  , \label{eq:landauparameters}%
\end{equation}
where $\nu_{F}=Vm^{\ast}k_{F}/\pi²$ is the total density of states at the
Fermi level and $V$ is the volume of the system. The dimensionless constants
$F_{l}^{s,a}$ are known as Landau coefficients.

We thus see that the low-energy sector of the interacting system is completely
parametrized by the effective mass $m^{\ast}$ and the Landau coefficients, in
terms of which important physical properties can be written. For example, the
low temperature specific heat is linear in temperature with a coefficient
determined by the effective mass%
\begin{equation}
\frac{C_{V}\left(  T\right)  }{T}=\frac{1}{3}m^{\ast}k_{F}.
\label{eq:flspecificheat}%
\end{equation}
The magnetic susceptibility is a constant at low temperatures and related to
the Landau parameter $F_{0}^{a}$%
\begin{equation}
\chi\left(  T\right)  =\frac{m^{\ast}k_{F}\mu_{B}^{2}}{\pi^{2}\left(
1+F_{0}^{a}\right)  }, \label{eq:flsusceptibility}%
\end{equation}
where $\mu_{B}$ is the Bohr magneton. The compressibility $\kappa=-V\left(
\partial P/\partial V\right)  $, which can be accessed through the sound
velocity, is related to $F_{0}^{s}$%
\begin{equation}
\kappa=\frac{9\pi^{2}m^{\ast}}{k_{F}^{5}\left(  1+F_{0}^{s}\right)  }.
\label{eq:flcompressibility}%
\end{equation}
There are several other predictions which come out of the basic framework
outlined above, especially with regard to collective excitations (zero sound,
plasmons, spin waves) and transport properties, but we will not dwell on them,
referring the interested reader to the available reviews. We will only note
that the quadratic dependence on its excitation energy of the quasi-particle
lifetime leads to a quadratic temperature dependence of the resistivity at the
lowest temperatures. Moreover, in the $T\rightarrow0$ limit the resistivity
tends to a constant value determined by lattice imperfections and extrinsic
impurities, such that the Fermi liquid prediction is%
\begin{equation}
\rho\left(  T\right)  =\rho_{0}+AT^{2}. \label{eq:flresistivity}%
\end{equation}
The thermal conductivity of a Fermi liquid is proportional $T\sigma\left(
T\right)  $ (the Wiedemann-Franz law), where $\sigma\left(  T\right)
=1/\rho\left(  T\right)  $ is the electric conductivity.

The application of Fermi liquid theory to metals helps explain why a strongly
interacting electron gas in a clean crystal can behave almost like a free
electron gas and conduct heat and electricity so well at low temperatures.
Interestingly, very strong electronic correlations can put the theory to test
at the most extreme circumstances. In particular, in a certain class of
compounds containing rare-earth or actinide elements such as Ce or U (to be
reviewed later in Section~\ref{sub:qcp}), the effective mass is observed to be
2 to 3 orders of magnitude greater than the electron mass. These compounds
were therefore named {}``heavy fermions''. Nevertheless, even in these cases,
Fermi liquid theory provides a valid description in a number of cases,
although a growing number of exceptions have been discovered and investigated
in recent years. Furthermore, the idea of adiabatic continuation (the
one-to-one correspondence between the excitations of a simpler reference
system and another one of interest) has proven fruitful in contexts which go
far beyond Landau's original proposal. In particular, this philosophy has been
extended to the superfluid phases of $^{3}$He \cite{leggett75} and to nuclear
physics \cite{migdalbook67}. More importantly for the subject of this review,
Fermi liquid ideas form the basis of much of our understanding of interacting
electrons in disordered metals, as we will expand upon in the next Section.

\subsection{Fermi liquid theory for disordered systems}

\label{sub:fltdisorder}

While the original formulation and much of the later work on Fermi liquid
theory concentrated on clean metals, the relevant physical principles have a
much more general validity. This framework is flexible enough to be also
applicable not only in presence of arbitrary forms of randomness, but even for
finite size systems such as quantum dots, molecules, atoms, or atomic nuclei.
In electronic systems, the first systematic studies of the interplay of
interactions and disorder \cite{lr} emerged only in the last 25 years or so.
Most progress was achieved in the regime of weak disorder, where controlled
many-body calculations are possible using the disorder strength as a small parameter.

\subsubsection{Drude theory}

To lowest order in the disorder strength, one obtains the semi-classical
predictions of the Drude theory \cite{lr}, where the conductivity takes the
form
\begin{equation}
\sigma\approx\sigma_{o}=\frac{ne^{2}\tau_{tr}}{m}, \label{eq:drudeconduc}%
\end{equation}
where $n$ is the carrier concentration, $e$ the electron charge and $m$ its
band mass. According to Matthiessen's rule, the transport scattering rate
takes additive contributions from different scattering channels, viz.%
\begin{equation}
\tau_{tr}^{-1}=\tau_{el}^{-1}+\tau_{ee}^{-1}(T)+\tau_{ep}^{-1}(T)+\cdots.
\label{eq:mathiessen}%
\end{equation}
Here, $\tau_{el}^{-1}$ is the elastic scattering rate (describing impurity
scattering), and $\tau_{ee}^{-1}(T)$, $\tau_{ep}^{-1}(T)$,..., describe
inelastic scattering processes from electrons, phonons, etc. It is important
to note that in this picture the resistivity $\rho=\sigma^{-1}$ is a
monotonically increasing function of temperature
\begin{equation}
\rho(T)\approx\rho_{o}+AT^{n}, \label{eq:lowtresist}%
\end{equation}
since inelastic scattering is assumed to only increase at higher temperatures.
The residual resistivity $\rho_{o}=\sigma(T=0)^{-1}$ is thus viewed as a
measure of impurity (elastic) scattering. Within this formulation,
thermodynamic quantities are generally not expected to acquire any singular or
non-analytic corrections due to impurity scattering.

Drude theory encapsulates a very simple physical picture. It implicitly
assumes that the itinerant carriers undergo many collisions with unspecified
scattering centers, but that these scattering events remain independent and
uncorrelated, justifying Matthiessen's rule. At this level, inelastic
scattering processes are therefore assumed to be independent of impurity
scattering, and thus assume the same form as in standard Fermi liquid theory,
e.g. $\tau_{el}^{-1}\sim T^{2}$, etc. This simplifying assumption is better
justified at higher temperatures, where inelastic scattering processes erase
the phase memory of the electrons, and suppress quantum interference processes
arising from multiple scattering events.

\subsubsection{Perturbative quantum corrections}

At weak disorder, systematic corrections to the Drude theory were found
\cite{lr} to consist of several additive terms,
\begin{equation}
\sigma=\sigma_{o}+\delta\sigma_{wl}+\delta\sigma_{int}, \label{eq:conduccorr}%
\end{equation}
corresponding to the so-called {}``weak localization'' and {}``interaction''
corrections. These {}``hydrodynamic'' corrections are dominated by infrared
singularities, i.e., they acquire non-analytic contributions from small
momenta or equivalently large distances, and which generally lead to an
instability of the paramagnetic Fermi liquid in two dimensions. Specifically,
the weak localization corrections take the form%
\begin{equation}
\delta\sigma_{wl}=\frac{e^{2}}{\pi^{d}}\left[  l^{-(d-2)}-L_{Th}%
^{-(d-2)}\right]  , \label{eq:weakloccond}%
\end{equation}
where $l=v_{F}\tau$ is the mean free path, $d$ is the dimension of the system,
and $L_{Th}$ is the length scale over which the wave functions are coherent.
This effective system size is generally assumed to be a function of
temperature of the form $L_{Th}\sim T^{p/2}$, where the exponent $p$ depends
on the dominant source of decoherence through inelastic scattering. The
situation is simpler in the presence of a weak magnetic field where the weak
localization corrections are suppressed and the leading dependence comes from
the interaction corrections first discovered by Altshuler and Aronov
\cite{altshuler-79b}%
\begin{equation}
\delta\sigma_{int}=\frac{e^{2}}{\hslash}(c_{1}-c_{2}\widetilde{F}_{\sigma
})(T\tau)^{(d-2)/2}. \label{eq:intercorrcond}%
\end{equation}
Here, $c_{1}$ and $c_{2}$ are constants, and $\widetilde{F}_{\sigma}$ is an
interaction amplitude. In $d=3$, this leads to a square-root singularity
$\delta\sigma_{int}\sim\sqrt{T}$, and in $d=2$ to a logarithmic divergence
$\delta\sigma_{int}\sim\ln(T\tau)$. These corrections are generally more
singular than the temperature dependence of the Drude term, and thus they are
easily identified experimentally at the lowest temperatures. Indeed, the
$T^{1/2}$ law is commonly observed \cite{lr} in transport experiments in many
disordered metals at the lowest temperatures, typically below 500 mK.

Similar corrections have been predicted for other physical quantities, such as
the tunneling density of states and, more importantly, for thermodynamic
response functions. As in Drude theory, these quantities are not expected to
be appreciably affected by noninteracting localization processes, but singular
contributions are predicted from interaction corrections. In particular,
corrections to both the spin susceptibility $\chi$, and the specific heat
coefficient $\gamma=C_{V}/T$ were expected to take the general forms
$\delta\chi\sim\delta\gamma\sim T^{(d-2)/2}$, and thus in three dimensions%
\begin{align}
\chi &  =\chi_{o}+m_{\chi}\sqrt{T},\label{eq:intercorrsusc}\\
\gamma &  =\gamma_{o}+m_{\gamma}\sqrt{T}, \label{eq:intercorrgamma}%
\end{align}
where $m_{\chi}$ and $m_{\gamma}$ are constants.

As in conventional Fermi liquid theories, these corrections emerged already
when the interactions were treated at the lowest, Hartree-Fock level, as done
in the approach of Altshuler and Aronov \cite{altshuler-79b}. Higher order
corrections in the interaction amplitude were first incorporated by
Finkelshtein \cite{fink-jetp83,fink-jetp84}, demonstrating that the
predictions remained essentially unaltered, at least within the regime of weak
disorder. In this sense, Fermi liquid theory has been generalized to weakly
disordered metals, where its predictions have been confirmed in numerous
materials \cite{lr}.

\subsubsection{Scaling theories of disordered Fermi liquids}

In several systems, experimental tests were extended beyond the
regime of weak disorder, where at least at face value, the
perturbative predictions seem of questionable relevance.
Interestingly, a number of transport experiments
\cite{lr,paalanen91,sarachik95} seemed to indicate that some
predictions, such as the $T^{1/2}$ conductivity law, appear to
persist beyond the regime of weak disorder. At stronger disorder,
the system approaches a disorder-driven metal-insulator
transition. Since the ground-breaking experiment of Paalanen,
Rosenbaum, and Thomas in 1980 \cite{rosenbaum-prl80}, it became
clear that this is a continuous (second order) phase transition
\cite{paalanen82}, which bears many similarities to conventional
critical phenomena. This important observation has sparked a
veritable avalanche of experimental
\cite{lr,paalanen91,sarachik95} and theoretical
\cite{wegner76,wegner79,gang4,wegner80} works, most of which have
borrowed ideas from studies of second order phase transitions.
Indeed, many experimental results were interpreted using scaling
concepts \cite{lr}, culminating with the famous scaling theory of
localization \cite{gang4}.

The essential idea of these approaches focuses on the fact that a weak,
logarithmic instability of the clean Fermi liquid arises in two dimensions,
suggesting that $d=2$ corresponds to the lower critical dimension of the
problem. In conventional critical phenomena, such logarithmic corrections at
the lower critical dimension typically emerge due to long wavelength
fluctuations associated with spontaneously broken continuous symmetry. Indeed,
early work of Wegner \cite{wegner76,wegner79,wegner80} emphasized the analogy
between the localization transition and the critical behavior of Heisenberg
magnets. It mapped the problem onto a field theoretical nonlinear $\sigma
$-model and identified the hydrodynamic modes leading to singular corrections
in $d=2$. Since the ordered (metallic) phase is only marginally unstable in
two dimensions, the critical behavior in $d>2$ can be investigated by
expanding around two dimensions. Technically, this is facilitated by the fact
that in dimension $d=2+\varepsilon$ the critical value of disorder $W$ for the
metal-insulator transition is very small ($W_{c}\sim\varepsilon$), and thus
can be accessed using perturbative renormalization group (RG) approaches in
direct analogy to the procedures developed for Heisenberg magnets.

In this approach \cite{wegner76,wegner79,wegner80,gang4}, conductance is
identified as the fundamental scaling variable associated with the critical
point, which is an unstable fixed point of the RG flows. In this picture,
temperature scaling is obtained from examining the system at increasingly
longer length scales $L_{Th}\sim T^{p/2}$, which follows from the precise form
of the RG flows. In the metallic phase, under rescaling the conductance
$g\rightarrow\infty$, (corresponding to the reduction of effective disorder),
and the long distance behavior of all correlation functions is controlled by
the approach to the stable fixed point at $W=0$. In other words, the leading
low temperature behavior of all quantities should be \textit{identical} to
that calculated at infinitesimal disorder. This scaling argument therefore
provides strong justification for using the weak disorder predictions
throughout the entire metallic phase, \textit{provided that the temperature is
low enough. }

\subsubsection{Fermi liquid near localization transitions}

In the following years, these ideas were extended with a great deal of effort
in the formulation of a scaling theory of interacting disordered electrons by
Finkelshtein \cite{fink-jetp83} and many followers
\cite{cclm-prb84,kirkpatrick-rmp94}. While initially clad in an apparent veil
of quantum field theory jargon, these theories were later given a simple
physical interpretation in terms of Fermi liquid ideas \cite{ckl,kotliar-fl87}
for disordered electrons. Technical details of these theories are of
considerable complexity and the interested reader is referred to the original
literature \cite{fink-jetp83,fink-jetp84,cclm-prb84,kirkpatrick-rmp94}. Here
we just summarize the principal results, in order to clarify the constraints
imposed by these Fermi liquid approaches.

Within the Fermi liquid theory for disordered systems \cite{ckl,kotliar-fl87},
the low energy (low temperature) behavior of the system is characterized by a
small number of effective parameters, which include the diffusion constant
$D$, the frequency renormalization factor $Z$, and the interaction amplitudes
$\gamma_{s}$ and $\gamma_{t}$. These quantities can also be related to the
corresponding quasi-particle parameters which include the quasi-particle
density of states
\begin{equation}
\rho_{Q}=Z\rho_{o}, \label{eq:quasipartdos}%
\end{equation}
and the quasi-particle diffusion constant
\begin{equation}
D_{Q}=D/Z\sim D/\rho_{Q}. \label{eq:quasipartdiff}%
\end{equation}
Here, $\rho_{o}$ is the {}``bare'' density of states which describes the
noninteracting electrons. In the absence of interactions, the single-particle
density of states is only weakly modified by disorder and remains noncritical
(finite) at the transition \cite{dos0}.

Using these parameters, we can now express the thermodynamic response
functions as follows. We can write the compressibility
\begin{equation}
\chi_{c}=\frac{dn}{d\mu}=\rho_{Q}[1-2\gamma_{s}], \label{eq:compress}%
\end{equation}
the spin susceptibility
\begin{equation}
\chi_{s}=\mu_{B}^{2}\rho_{Q}[1-2\gamma_{t}], \label{eq:susceptibility}%
\end{equation}
and the specific heat
\begin{equation}
C_{V}=2\pi^{2}\rho_{Q}T/3. \label{eq:specificheat}%
\end{equation}
In addition, we can use the same parameters to express transport properties
such as the conductivity
\begin{equation}
\sigma=\frac{dn}{d\mu}D_{c}=\rho_{Q}D_{Q}, \label{eq:conductivity}%
\end{equation}
as well as the density-density and spin-spin correlation functions
\begin{equation}
\pi(q,\omega)=\frac{dn}{d\mu}\frac{D_{c}q^{2}}{D_{c}q^{2}-i\omega
}\;\;\;\;\;\;\;\;\chi_{s}(q,\omega)=\chi_{s}\frac{D_{s}q^{2}}{D_{s}%
q^{2}-i\omega}. \label{eq:correlations}%
\end{equation}
Here, we have expressed these properties in terms of the spin and charge
diffusion constants, which are defined as
\begin{equation}
D_{c}=\frac{D}{Z(1-2\gamma_{s})};\;\;\;\; D_{s}=\frac{D}{Z(1-2\gamma_{t})}.
\label{eq:diffusionconstants}%
\end{equation}

Note that the the quantity $D$ is \emph{not} the charge diffusion constant
$D_{c}$ that enters the Einstein relation, Eq. (\ref{eq:conductivity}). As we
can write $\sigma=\rho_{o}D$, and since $\rho_{o}$ is not critical at any type
of transition, the quantity $D$ (also called the {}``renormalized diffusion
constant'') has a critical behavior identical to that of the conductivity
$\sigma$. We also mention that the quasi-particle diffusion constant
$D_{Q}=D/Z$ has been physically interpreted as the heat diffusion constant.

\subsubsection{Critical scaling}

Finally, we discuss the scaling behavior of observables in the critical
region. In particular, if the scaling description
\cite{fink-jetp83,fink-jetp84,cclm-prb84,kirkpatrick-rmp94} is valid, the
conductivity can be written as
\begin{equation}
\sigma(t,T)=b^{-(d-2)}\; f_{\sigma}\;(b^{1/\nu}\; t,b^{z}T).
\label{eq:scalingcond}%
\end{equation}
Here, $b$ is the length rescaling factor, and $t=(n-n_{c})/n_{c}$ is the
dimensionless distance from the transition. We have also introduced the
correlation length exponent $\nu$ and the {}``dynamical exponent'' $z$. This
expression, first proposed by Wegner \cite{wegner76,wegner79,gang4,wegner80},
is expected to hold for {}``regular''types of transitions, where the critical
values of the interaction amplitudes remain finite. The conductivity exponent
$\mu$ in $\sigma\left(  T=0\right)  \sim t^{\mu}$ can be obtained by working
at low temperatures and choosing $b=t^{-\nu}$. We immediately see that
\begin{equation}
\sigma(T)\sim t^{\mu}\phi_{\sigma}(T/t^{\nu z}), \label{eq:scalingcond2}%
\end{equation}
where $\phi_{\sigma}(x)=f_{\sigma}(1,x)$ and
\begin{equation}
\mu=(d-2)\nu, \label{eq:wignerscaling}%
\end{equation}
a relation known as {}``Wegner scaling''. Finite temperature corrections in
the metallic phase are obtained by expanding%
\begin{equation}
\phi_{\sigma}(x)\approx1+ax^{\alpha}, \label{eq:condscalingfunction}%
\end{equation}
giving the low temperature conductivity of the form
\begin{equation}
\sigma(t,T)\approx\sigma_{o}(t)+m_{\sigma}(t)T^{\alpha}. \label{eq:lowtcond}%
\end{equation}
Here, $\sigma_{o}(t)\sim t^{\mu}$, and $m_{\sigma}(t)\sim t^{\mu-\alpha\nu z}
$. Since the scaling function $\phi_{\sigma}(x)$ is independent of $t$, the
exponent $\alpha$ must take an universal value in the entire metallic phase,
and thus it can be calculated at weak disorder, giving $\alpha=1/2$. This
scaling argument provides a formal justification for using the predictions
from perturbative quantum corrections as giving the leading low temperature
dependence in the entire metallic phase. Note, however, that according to this
result the prefactor $m_{\sigma}(t)$ is not correctly predicted by
perturbative calculations, since it undergoes Fermi liquid renormalizations
which can acquire a singular form in the critical region near the
metal-insulator transition.

The temperature dependence at the critical point (in the critical region) is
obtained if we put $t=0$ ($n=n_{c}$), and choose $b=T^{-1/z}$. We get
\begin{equation}
\sigma(t=0,T)\sim T^{(d-2)/z}. \label{eq:criticalcond}%
\end{equation}

An analogous argument can be used \cite{cdc-prb86} for the specific heat
coefficient $\gamma=C_{V}/T$ {[}it is important not to confuse this quantity
with the interaction amplitude $\gamma_{t}${]}%
\begin{equation}
\gamma(t,T)=b^{\kappa/\nu}\; f_{\gamma}\;(b^{1/\nu}\; t,b^{z}T).
\label{eq:gammascaling}%
\end{equation}
Choosing $b=t^{-\nu}$ and expanding in $T$ we find
\begin{equation}
\gamma(t,T)\approx\gamma_{o}(t)+m_{\gamma}(t)T^{1/2}, \label{eq:lowtgamma}%
\end{equation}
with $\gamma_{o}(t)\sim t^{-\kappa}$, $m_{\gamma}(t)\sim t^{-(\kappa+\nu
z/2)}$. Even though $\gamma_{o}=\gamma(t,T=0)$ can become singular at the
transition, it is expected to remain finite away from the transition ($t\neq
0$). We stress that the specific heat exponent $\kappa$ and the dynamical
exponent $z$ are \emph{not} independent quantities. In fact, Castellani
\emph{et al.} \cite{cdc-prb86} have proved that within Fermi liquid theory,
the following relation is obeyed
\begin{equation}
z=d+\frac{\kappa}{\nu}. \label{eq:exponentrelation}%
\end{equation}
Similar conclusions apply also to other quantities, such as the spin
susceptibility $\chi_{s}$, which should also remain bounded away from the
critical point and acquire universal $T^{1/2}$ corrections within the metallic
phase. Finally, the compressibility $\chi_{c}=dn/d\mu$ is generically expected
to remain nonsingular (finite) at the transition, since $n(\mu)$ is expected
to be a smooth function, except at special filling fractions. For example, if
one approaches \cite{mott-book90} a band or a Mott insulator, the
compressibility vanishes as a precursor of the gap opening at the Fermi surface.

All the above expressions are quite general, and can be considered to be a
phenomenological description \cite{ckl} of disordered Fermi liquids. On the
other hand, these relations tell us nothing about the specific \emph{values}
of the Landau parameters, or how they behave in the vicinity of the
metal-insulator transition. Perturbative renormalization group calculations
\cite{fink-jetp83,fink-jetp84,cclm-prb84,kirkpatrick-rmp94} based on the
$2+\varepsilon$ expansion have been used to make explicit predictions for the
values of the critical exponents and scaling functions for different
universality classes. Despite a great deal of effort invested in such
calculations, the predictions of these perturbative RG approaches have not met
almost any success in explaining the experimental data in the critical region
of the metal-insulator transition. We should emphasize, though, that
limitations associated with these weak-coupling theories do not invalidate the
potential applicability of Fermi liquid ideas \textit{per se}. On the other
hand, even in their most general form, these Fermi liquid considerations
predict that thermodynamic response functions such as $\chi$ and $\gamma$
remain finite at $T=0$ even within the disordered metallic phase. All
experiments that find a more singular behavior of these quantities should
therefore be regarded as violating the Fermi liquid theory - even when it is
properly generalized to disordered systems.

\subsection{Typical departures from Fermi liquid theory in real systems}

\label{sub:nfldepartures}

Fermi liquid theory describes the leading low energy excitations in a system
of fermions. These, as we have seen, are weakly interacting quasiparticles
characterized by several Landau parameters. Most remarkably, its validity is
by no means limited to systems with weak interactions. In several materials,
most notably heavy fermion systems, these many body renormalizations are
surprisingly large (e.g. $m^{\ast}/m\sim1000$), yet Fermi liquid theory is
known to apply at the lowest temperatures.

On the other hand, we should make it clear that precisely in such strongly
correlated systems the \textit{temperature range} where Fermi liquid theory
applies is often quite limited. The \char`\"{}coherence temperature\char`\"{}
$T^{\ast\text{ }}$below which it applies is typically inversely proportional
to the effective mass enhancement, and thus can be much smaller then the Fermi
temperature. Above $T^{\ast\text{ }}$ all physical quantities are dominated by
large incoherent electron-electron scattering, and Fermi liquid theory simply
ceases to be valid. In many heavy Fermion systems $T^{\ast\text{ }}$is found
to be comparable to the single-ion Kondo temperature, thus to typically be of
order $10-10^{2}K$.

\begin{figure}[ptb]
\begin{center}
\includegraphics[  width=3.4in,
keepaspectratio]{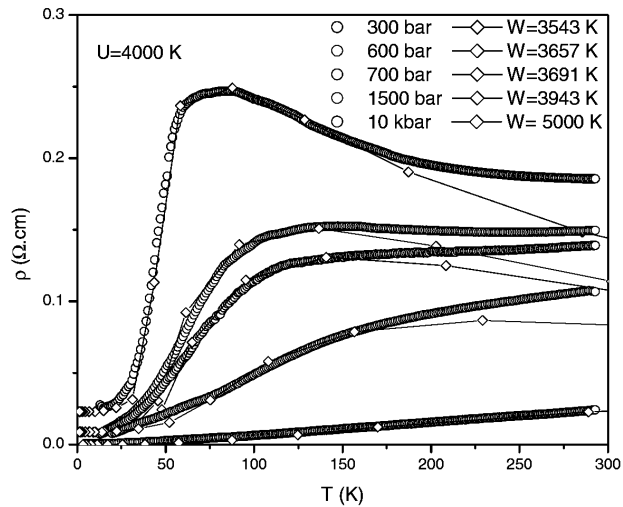}
\end{center}
\caption{Temperature dependence of the resistivity at different pressures for
a two-dimensional organic charge transfer $\kappa-(BEDT-TTF)_{2}%
Cu[N(CN)_{2}]Cl$, following Ref. \cite{limelette03prl}. The data (circles) are
compared to the DMFT predictions (diamonds), with a pressure dependence of the
bandwidth as indicated. }%
\label{cap:fig0}%
\end{figure}

Another interesting class of materials, where clear departures from Fermi
liquid theory have been observed at sufficiently high temperatures, are
systems close to the Mott transition. These include the transition metal
oxides such as $(V_{1-x}Cr_{x})_{2}O_{3}$ and chalcogenides such as
$NiS_{2-x}Se_{x}$ (for a review see Ref. \cite{imadareview}). In very recent
work, similar behavior was observed in another Mott system, a two-dimensional
organic charge transfer salt $\kappa-(BEDT-TTF)_{2}Cu[N(CN)_{2}]Cl$
\cite{limelette03prl}. In this material the system can be pressure-tuned
across the Mott transition, and on the metallic side the resistivity follows
the conventional $T^{2}$ law below a crossover temperature $T^{\ast}\sim50K$.
Above this temperature transport crosses over to an insulating-like form,
reflecting the destruction of heavy quasiparticles by strong inelastic
scattering (see Fig.~\ref{cap:fig0}). Clearly, Fermi liquid theory does not
apply above this coherence scale, but alternative DMFT approaches
\cite{limelette03prl} proved very successful in quantitatively fitting the
data over the entire temperature range.

The incoherent metallic behavior could be even more important in disordered
systems, where the coherence temperature $T^{\ast}$ can be viewed as a random,
position-dependent quantity $T^{\ast}(\mathbf{x})$. If this random quantity is
characterized by a sufficiently broad probability distribution function
$P(T^{\ast})$, then Fermi liquid theory may not be applicable in any
temperature regime. In the following, we examine several physical systems
where these phenomena are of potential importance, and then describe
theoretical efforts to address these issues and produce a consistent physical
picture of a disorder-driven non-Fermi liquid metallic state.

\section{Phenomenology of non-Fermi liquid behavior and experimental
realizations}

\label{sec:Phenomenology}

\subsection{NFL behavior in correlated systems with weak or no disorder}

\label{sub:cleannfl}

\subsubsection{Heavy fermion materials near magnetic quantum critical points}

\label{sub:qcp}

Heavy fermion systems have been the focus of intense study over the last three
decades as prime examples of strongly correlated electronic behavior. These
are metallic systems with an array of localized moments formed in the
incompletely-filled $f$-shells of rare-earth or actinide elements. An
important requirement for heavy fermion behavior is that the $f$-shell
electrons should be sufficiently close to a valence instability and therefore
should have a fairly low ionization energy in the metallic host. This
restricts the interesting behavior to a few elements, most often Ce and U, but
also Yb, Pr, Sm and some other less common cases. The proximity to a valence
instability promotes the enhancement of the hybridization between the
$f$-shell electrons and the conduction bands. This, in turn, leads to an
antiferromagnetic (AFM) interaction between the $f$-shell moment and the local
conduction electron spin density through essentially a super-exchange
mechanism. This interaction is otherwise much too weak in the lanthanide or
actinide series. The tendency towards local singlet formation through the
Kondo effect serves to strongly suppress the ubiquitous magnetic order seen in
most other intermetallics with localized $f$-moments. The hybridization of the
metallic carriers with the virtual bound states in the strongly correlated
$f$-shell leads to large effective mass renormalization factors (of order
$10^{2}-10^{3}$), whence the name {}``heavy fermions''. There are many
excellent reviews of heavy fermion physics to which the reader is referred to
\cite{stewartrev1,leeetal,steglichgrewe,hewson}.

A great part of the interesting physics of these compounds is a result of
competing tendencies: localization versus delocalization of the $f$-electrons
and magnetic order versus Kondo singlet formation. This complex interplay
provides the background where a wide variety of low temperature phases arise:
there are metals, insulators \cite{aepplifisk}, and superconductors
\cite{heffnernorman}. Furthermore, many exotic phenomena are also found, such
as small moment antiferromagnetism \cite{buyers96}, coexistence of
superconductivity and magnetism, and unconventional superconductivity with
more than one phase \cite{heffnernorman}.

Despite a bewildering zoo of correlated behavior, in recent years a common
theme has been the focus of attention. In many systems, the Néel temperature
is low enough that it can be tuned to zero by an external parameter, such as
pressure, chemical pressure (by alloying with an element with a different
ionic radius) or applied magnetic field. The zero temperature phase transition
between an antiferromagnet and a paramagnet as a function of the external
parameter is called a quantum phase transition. If this happens to be a second
order phase transition, the system will exhibit a diverging correlation length
and we expect the critical behavior to be classified in {}``universality
classes'', much like thermal second order phase transitions
\cite{mucioreview,sachdevbook,muciobook}. Indeed, many heavy fermion systems
can be tuned in just this way and a general theory of this so-called quantum
critical point (QCP) has been sought vigorously. The possibility of such a
quantum phase transition in heavy fermion systems is generally attributed to
Doniach, who discussed it early on in the context of a one-dimensional
effective model \cite{Doniach} (see also \cite{japiassuetal}). The associated
phase diagram is thus usually referred to as the Doniach phase diagram. It is
a natural consequence of the above mentioned competition between Kondo singlet
formation and AFM order.

\begin{figure}[ptb]
\begin{center}
\includegraphics[  width=3.4in,
keepaspectratio]{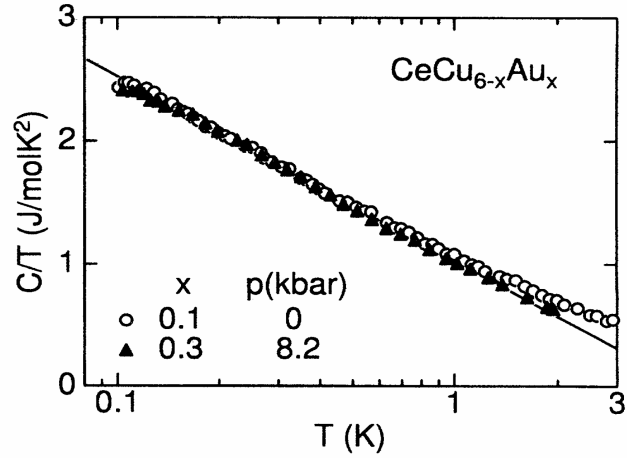}
\end{center}
\caption{Specific heat of CeCu$_{6-x}$Au$_{x}$ showing the logarithmic
divergence at criticality. Data are from Ref.~\cite{bogenlohn95}.}%
\label{cap:fig1}%
\end{figure}

Besides the natural interest in a catalogue of the possible universality
classes of such quantum phase transitions, the low temperature region in the
vicinity of the zero temperature quantum critical point is observed to be
characterized by strong deviations from Landau's Fermi liquid theory. A recent
exhaustive review of the experimental data on many heavy fermion systems
showing NFL behavior, including but not restricted to those where QCP physics
seems relevant, can be found in \cite{stewartNFL}. In many cases, the
following properties are observed:

\begin{itemize}
\item A diverging specific heat coefficient, often logarithmically, $C\left(
T\right)  /T\sim\gamma_{0}\log\left(  T_{0}/T\right)  $ (see
Fig.~\ref{cap:fig1}).

\item An anomalous temperature dependence of the resistivity, $\rho\left(
T\right)  \sim\rho_{0}+AT^{\alpha}$, where $\alpha<2$.

\item An anomalous Curie-Weiss law, $\chi^{-1}\left(  T\right)  \sim\chi
_{0}^{-1}+CT^{\beta}$, where $\beta<1$.
\end{itemize}

Despite an intensive theoretical onslaught, a coherent theoretical picture is
still lacking \cite{millissces99,pierspepinsirevaz}. The natural theoretical
description of the metallic QCP would seem to be one where the critical modes
are subject to the interactions generated in the presence of a Fermi sea, as
in the old paramagnon theory
\cite{izuyamaetal63,doniachengelsberg,berkschrieffer}. In particular, the
low-frequency dependence of the RPA susceptibility leads to the overdamping of
the critical modes, which can decay into particle-hole pairs (Landau damping).
As emphasized by Hertz, this increases the effective dimensionality of the
critical theory by the dynamical exponent $z$: $d_{eff}=d+z$ \cite{hertz} (see
also \cite{bealmonodmaki75}). The dynamical exponent $z$ determines the
relation between energy and length scales $E\sim L^{-z}$. Landau damping leads
to $\omega\sim-ik^{3}$($z=3$) in clean itinerant ferromagnets and $\omega
\sim-iq^{2}$ ($z=2$) in clean itinerant antiferromagnets ($q$ being the
deviation from the ordering vector), the latter being the case of interest in
heavy fermion materials. As a result, three-dimensional systems are
generically above their upper critical dimension, the dimension above which
mean field exponents become exact (equal to 4 in the usual thermal case for
Ising and Heisenberg symmetries). In the effective critical theory, the
critical modes are the more weakly interacting the longer the length scales,
rendering the theory asymptotically tractable
\cite{hertz,moryia,millis,mucioreview}. This is the weak coupling spin density
wave approach, sometimes named the Hertz-Millis scenario. In particular, at
the AFM QCP the specific heat is non-singular, $C\left(  T\right)
/T\sim\gamma_{0}-a\sqrt{T}$. Furthermore, scattering off the incipient AFM
order is singular only along \emph{lines} on the Fermi surface which are
connected by the AFM ordering wave vector $\mathbf{Q}$ ({}``hot spots'').
These are effectively short-circuited by the remainder of the Fermi surface
({}``cold spots''), which are hardly affected by criticality, leading in clean
samples to a Fermi liquid response $\rho\left(  T\right)  \sim\rho_{0}+AT^{2}$
\cite{hlubinarice95}. These results cannot explain the observed properties of
heavy fermion systems close to quantum criticality. Another consequence of the
higher effective dimensionality is the fact that the theory does not obey
hyperscaling. Hyperscaling is usually associated with the Josephson scaling
law: $2-\alpha=\nu d$, where $\alpha$ is the specific heat critical exponent
and $\nu$ is the correlation length exponent \cite{goldenfeldbook}. In a
quantum phase transition, $d$ should be replaced by the effective
dimensionality $d_{eff}$. However, the most important consequence of the
violation of hyperscaling is the absence of $\omega/T$ scaling in dynamical
responses which, in the AFM case, should involve the combination
$\omega/T^{3/2}$ instead \cite{pierspepinsirevaz}. However, $\omega/T$ scaling
is observed in a quantum critical heavy fermion system CeCu$_{5.9}$Au$_{0.1}$
\cite{schroederetal,schroedernature}. These results present major difficulties
for the description of the NFL behavior observed near a QCP in heavy fermion systems.

There have been proposals for amending the Hertz-Millis scenario in an effort
to account for the experimental results. In particular, we mention the effect
of disorder close to a quantum critical point \cite{rosch99,rosch00} and the
possible two-dimensional character of the spin fluctuation spectrum
\cite{roschetal97,mathuretal,stockertetal98,schroedernature}. Disorder acts by
relaxing the momentum conservation which restricts strong scattering to the
{}``hot lines'', effectively enhancing their influence on transport. We stress
that in this case disorder plays only an ancillary role and is not the driving
mechanism for NFL behavior. On the other hand, a two-dimensional spin
fluctuation spectrum places the system at the upper critical dimension and is
able to account for some of the observed critical exponents. None of these
approaches, however, is able to account for all the available data.
Alternatively, radical departures from the spin density wave critical theory
have been proposed. In one of these, the long time dynamics of the localized
moment acquires a non-trivial power-law dependence, whereas the spatial
fluctuations retain the usual mean-field form \cite{sietal}. This has been
dubbed `local quantum criticality' and is a natural scenario to incorporate
the $\omega/T$ scaling with non-trivial exponents observed over much of the
Brillouin zone. Its current form, however, still relies on a two-dimensional
spin fluctuation spectrum. In another approach, there is a phase transition
from the usual paramagnetic heavy Fermi liquid state to a non-trivial state
where the localized spins form a `spin liquid' weakly coupled to the
conduction sea \cite{senthiletal,senthiletal2}. An interesting consequence of
the latter proposals would be a drastic reduction of the Fermi surface volume
across the transition, presumably measurable through the Hall constant
\cite{pierspepinsirevaz}. This is still an open arena where perhaps new ideas
and new experiments will be necessary before further progress can be achieved.

\subsubsection{Marginal Fermi liquid behavior of high T$_{c}$ superconductors}

\label{sub:hightc}

\begin{figure}[ptb]
\begin{center}
\includegraphics[width=3.6in,
keepaspectratio]{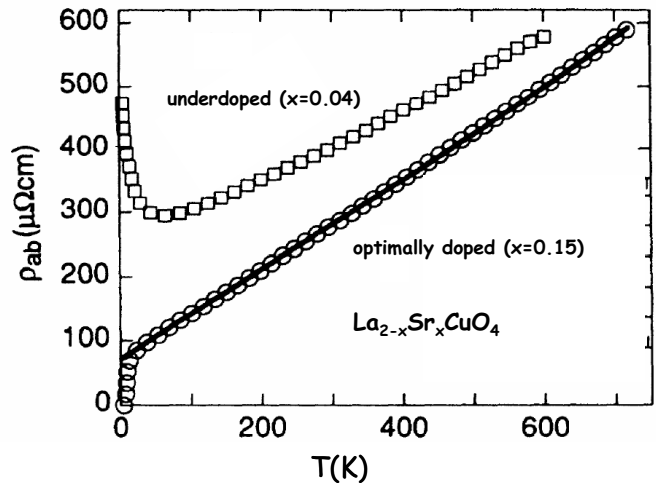}
\end{center}
\caption{Normal state in-plane resistivity of optimally doped
cuprate $La_{2-x} Sr_x CuO_4$ (circles). The striking linear
temperature dependence is observed around optimal doping,
persisting far above room temperature. Deviations from this
behavior are seen both below (squares) and above optimal doping.
Data are from
Ref.~\cite{takagi-prl92}.}%
\label{cap:fig2}%
\end{figure}

High temperature superconductors (HTS) have taken the center stage of modern
condensed matter theory ever since their discovery in 1987. Most of their
features continue to defy theoretical understanding, and even the origin of
superconducting pairing in these materials remains controversial. Most
remarkably, many features of the superconducting state seem to be less exotic
than the extremely unusual behavior observed in the normal phase. In fact, HTS
materials presented one of the first well documented examples featuring
deviations from Fermi liquid phenomenology, initiating much theoretical
activity and debate. On a phenomenological level, behavior close to
{}``optimal doping'' (the regime where the superconducting transition
temperature $T_{c}$ is the highest) seems to display {}``marginal Fermi
liquid'' (MFL) behavior. The most prominent feature observed in experiments is
the linear resistivity (see Fig.~\ref{cap:fig2}), which at optimal doping
persists in an enormous temperature range from a few Kelvin all the way to
much above room temperature \cite{takagi-prl92}. A phenomenological model
describing the MFL behavior of cuprates has been put forward a long time ago
\cite{mfl}, but its microscopic origin remains highly controversial. To our
knowledge no microscopic theory has so far been able to provide a microscopic
explanation for this puzzling behavior, despite years of effort and hundreds
of papers published on the subject.

Much theoretical and experimental effort over the years concentrated on
unraveling the microscopic origin of the superconducting pairing, a mechanism
that, one hopes, would also explain the puzzling features of the normal state.
Most proposed scenarios concentrated on exotic many-body mechanisms in an
assumed homogeneous conductor, thus ignoring the role of disorder or the
possible emergence of some form of random ordering. Many excellent reviews
exist detailing the current status of these research efforts and, given our
focus on the effects of disorder, will not be elaborated here. In the
following, we follow Ref.~\cite{christos-vlad04}, and briefly review those
experimental and theoretical efforts suggesting that the homogeneous picture
may be incomplete.

\begin{figure}[ptb]
\begin{center}
\includegraphics[  width=3.0in,
keepaspectratio]{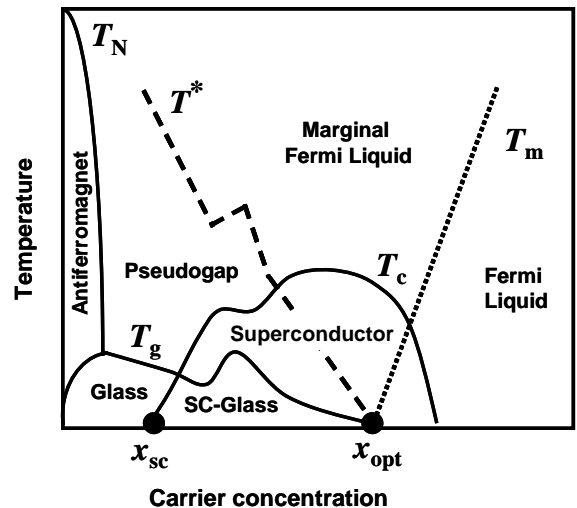}
\end{center}
\caption{Phase diagram of the archetypal HTS, following
Ref.~\cite{christos-vlad04}. $T_{N}$ is the Néel temperature, $T_{F}$ and
$T_{g}$ the onset of short range freezing to an electronic glass, and $T_{c}$
the superconducting transition temperature. At $x<x_{sc}$ the material is a
glassy insulator. At $x_{sc}<x<x_{opt}$ a microscopically inhomogeneous
conducting glassy state emerges, with intercalated superconducting and
magnetic regions. At $x=x_{opt}$ the system experiences a quantum glass
transition and at $x>x_{opt}$ the material transforms into a homogeneous metal
with BCS-like superconducting properties. The superfluid density is maximum at
$x=x_{opt}$. The crossover scales $T^{\ast}$ and $T_{m}$ characterizing
normal-state transport (see text for details), vanish at the quantum glass
transition.}%
\label{cap:fig3}%
\end{figure}

The first evidence of random ordering in HTS was obtained from the observation
of low temperature spin-glass ordering in the pseudo-gap phase. At doping
concentration $x$ larger than a few percent, antiferromagnetic ordering is
suppressed, but the short range magnetic order persists. The low-field
susceptibility displays a cusp at temperature $T=T_{g}$ and a thermal
hysteresis below it, characteristic of a spin glass transition. At $T<T_{g}$
the material displays memory effects like {}``traditional'' spin glasses and
is described by an Edwards-Anderson order parameter \cite{chou-prl95}. Such
magnetic measurements are not possible in the superconducting regime
($x>x_{sc}$), but muon spin relaxation ($\mu$SR) has been successful in
identifying the freezing of electronic moments under the superconducting dome
of various HTS systems
\cite{panagopoulos-ssc03,panagopoulos-prb02,kanigel-prl02,sanna-lanl04}. From
these studies one may conclude that in this regime glassiness coexists with
superconductivity on a microscopic scale throughout the bulk of the material.
Most remarkably, the observed spin-glass phase seems to end at a quantum
critical point precisely at optimal doping ($x=x_{opt}$), suggesting that
non-Fermi liquid behavior in the normal phase may be related to the emergence
of glassy ordering in the ground state (Fig.~\ref{cap:fig3}).

The correlation between spin order and charge transport is further emphasized
by experiments in high magnetic fields \cite{boebinger-prl96} where bulk
superconductivity was suppressed, revealing information about low-$T$ charge
transport in the normal phase. The resistivity data show a crossover in
$\rho_{ab}(T)$ from metallic to insulating-like behavior (resistivity minimum)
at a characteristic temperature $T^{\ast}$ which, similarly to $T_{g}$,
decreases upon doping, and seems to vanish at the putative quantum critical
point. In addition, a crossover temperature $T_{m}$ at $x>x_{opt}$ separating
marginal Fermi liquid transport at $T>T_{m}$ from more conventional metallic
behavior at $T<T_{m}$ also seems to drop \cite{naqib-physicaC03} to very small
values around optimal doping (see Fig.~\ref{cap:fig3}). At $x>x_{opt}$ the
ground state becomes metallic and homogeneous, with no evidence for glassiness
or other form of nano-scale heterogeneity
\cite{panagopoulos-ssc03,panagopoulos-prb02,boebinger-prl96,balakirev-nature03}%
. Remarkable independent evidence that a QCP is found precisely at $x=x_{opt}$
is provided by the observation of a sharp change in the superfluid density
$n_{s}(0)\sim1/\lambda_{ab}^{2}(0)$ (where $\lambda_{ab}(0)$ is the absolute
value of the in-plane penetration depth). At $x>x_{opt}$, $n_{s}(0)$ is mainly
doping independent (Fig.~\ref{cap:fig3}), while the $T$-dependence is in good
agreement with the BCS weak-coupling $d$-wave prediction
\cite{panagopoulos-ssc03}. At dopings below the quantum glass transition
$n_{s}(0) $ is rapidly suppressed (note the enhanced depletion near $x=1/8$
precisely where $T_{g}$ and $T^{\ast}$ are enhanced) and there is a marked
departure of $n_{s}(T)$ from the canonical weak coupling curve
\cite{panagopoulos-ssc03}. All these results provide strong evidence for a
sharp change in ground state properties at $x=x_{opt}$, and the emergence of
vanishing temperature scales as this point is approached - just as one expects
at a QCP.

The formation of an inhomogeneous state below optimal doping is consistent
with those theoretical scenarios that predict phase separation
\cite{gorkov-JETP87,kivelson-rmp03,dagotto-book} at low doping. Coulomb
interactions, however, enforce charge neutrality and prevent
\cite{kivelson-rmp03} global phase separation; instead, the carriers are
expected \cite{schmalian-prl00} to segregate into nano-scale domains - to form
a stripe/cluster glass \cite{schmalian-prl00}. As quantum fluctuations
increase upon doping \cite{pastor-prl99,mitglass-prl03}, such a glassy phase
should be eventually suppressed at a quantum critical point, around
$x=x_{opt}$. These ideas find striking support in very interesting STM studies
revealing nano-scale domains forming in the underdoped phase, as recently
reported by several research groups.

This scenario is consistent with another mysterious aspect of normal state
transport in the weakly underdoped regime ($x\lesssim x_{opt}$). Here, DC
transport has a much weaker \cite{boebinger-prl96} (although still
insulating-like) temperature dependence. However, the observed $\log T$
resistivity upturn in this region has been shown \cite{boebinger-prl96} to be
inconsistent with conventional localization/interaction corrections which
could indicate an insulating ground state. Instead, estimates
\cite{efetov-prl03} reveal this behavior to be consistent with that expected
for metallic droplet charging/tunnelling processes, as seen in quantum dots
and granular metals \cite{efetov-prl03}. These results suggest that in this
regime HTS are inhomogeneous metals, where conducting droplets connect
throughout the sample, and a metal-insulator transition in the normal phase
happens \textit{exactly} at $x=x_{sc}$. At lower densities the conducting
droplets remain isolated, and the material may be viewed as an insulating
cluster or stripe glass. As carrier concentration increases they connect and
the carriers are free to move throughout the sample, forming filaments or
{}``rivers''. This is, in fact, the point where free carriers emerge in Hall
effect data \cite{balakirev-nature03} and phase coherent bulk
superconductivity arises at $x>x_{sc}$. This observation suggests that it is
the \textit{inhomogeneous} nature of the underdoped glassy region which
controls and limits the extent of the superconducting phase at low doping.

At this time we still do not know if the formation of such inhomogeneous
states is indeed a fundamental property of HTS materials, or merely a
by-product of strong frustration and extrinsic disorder. Nevertheless, the
emerging evidence seems compelling enough by itself, as it opens the
possibility that inhomogeneities and glassy ordering may not be disregarded as
the possible origin of non-Fermi liquid behavior in the cuprates.

\subsection{Disorder-driven NFL behavior in correlated systems: doped
semiconductors}

\label{sub:dopedsemicond}

Doped semiconductors \cite{doped-book} have been studied for a long time, not
the least because of their enormous technological applications, but also
because of their relatively simple chemical composition allowing the
possibility for simple theoretical modelling \cite{doped-book}. Typically,
they are used to fabricate transistors and other devices, which essentially
can be used as electrical switches in logical integrated circuits. For this
reason, much attention has been devoted to understanding their behavior close
to the metal-insulator transition (MIT) \cite{mott-book90}, which typically
takes place when the Bohr radius $a_{o}$ of the donor atoms becomes comparable
to the typical carrier separation. Most applications are based on doped
silicon devices, where the critical concentration $n=n_{c}\sim10^{18}cm^{-3}$.
Since the Fermi energy of the carriers in this regime is fairly low
($T_{F}\approx100K$) \cite{paalanen91,sarachik95}, the behavior at room
temperatures is easy to understand using conventional solid-states theories
\cite{paalanen91,sarachik95}, and as such it has been qualitatively and even
quantitatively understood for many years. The evolution of the physical
properties as a function of $n$, however, is smooth at elevated temperatures
due to thermal activation.

\subsubsection{Metal-insulator transition}

A sharp metal-insulator transition is seen only at the lowest accessible
temperatures, and it is in this regime where most of the interesting physics
associated with quantum effects comes into play.

\begin{figure}[ptb]
\begin{center}
\includegraphics[  width=3.4in,
keepaspectratio]{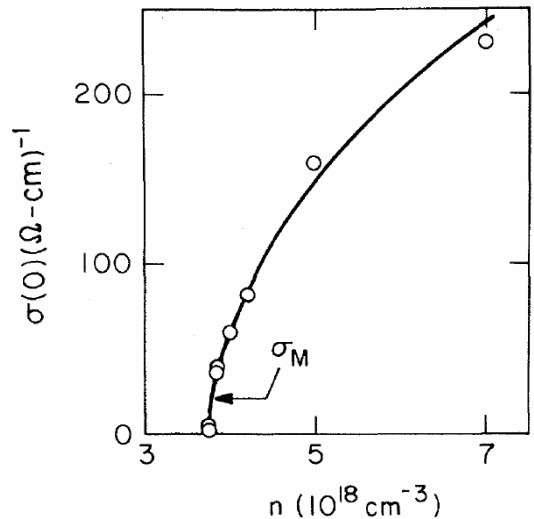}
\end{center}
\caption{Critical behavior of the conductivity extrapolated to $T\rightarrow0
$ for uncompensated Si:P. Note a sharp power-law behavior with exponent
$\mu\approx1/2$, extending over a surprisingly large concentration range. The
results reported are taken from the classic stress-tuned experiment of
Paalanen, Rosenbaum, and Thomas \cite{rosenbaum-prl80}. }%
\label{cap:fig4}%
\end{figure}

In the last 30 years, a great deal of effort has been devoted to the study of
detailed properties \cite{paalanen91,sarachik95} of doped semiconductors in
the critical region near the MIT. Early experiments
\cite{paalanen91,sarachik95} concentrated on transport measurements, and
reported a temperature dependence of the conductivity of the form%
\begin{equation}
\sigma(T)=\sigma_{o}+m\sqrt{T}, \label{eq:lowtcond2}%
\end{equation}
where $\sigma_{o}$ and $m$ are parameters that generally depended on the
carrier concentration $n$ and the magnetic field $B$. Using these
observations, one typically extrapolated these results to $T=0$, and examined
the critical behavior as the transition is approached. In agreement with
scaling predictions \cite{gang4,fink-jetp83,fink-jetp84,kirkpatrick-rmp94},
power-law behavior was reported \cite{rosenbaum-prl80,paalanen91,sarachik95}
of the form (see Fig.~\ref{cap:fig4})%
\begin{equation}
\sigma(T\rightarrow0)\sim(n-n_{c})^{\mu}. \label{eq:zerotcond}%
\end{equation}
The conductivity exponent was generally reported to take the value $\mu
\approx0.5$ for uncompensated materials (e.g. only donor dopants present as in
Si:P or only acceptors as in Si:B) in zero magnetic field. In the presence of
a magnetic field experiments \cite{paalanen91,sarachik95} showed $\mu\approx
1$, and the same behavior was reported in compensated materials (both donors
and acceptors present, such as Si:P,B). These findings had the general form
expected from Fermi liquid considerations and scaling approaches, although the
specific values for the conductivity exponent $\mu$ were not easy to
understand from the available microscopic theories \cite{kirkpatrick-rmp94}.
It was, however, generally felt \cite{lr} that the observed compensation
dependence reflects the enhanced role of the electronic correlations in the
uncompensated case, where the dopant impurity band is half-filled, and one
would expect a Mott metal-insulator transition in the absence of disorder.

\subsubsection{Thermodynamic anomalies}

If the proximity to the Mott transition \cite{mott-book90} is an important
consideration, then magnetism may be expected to have unusual features, since
the electrons turn into local magnetic moments \cite{mott-book90} in the Mott
insulator. Indeed, many strongly correlated metals near the Mott transition
(e.g. transition metal oxides \cite{mott-book90} such as $V_{2}O_{3}$) show
large spin susceptibility and specific heat enhancements, which are known to
be dominated by local magnetic moment \cite{marko,maekawa} physics.

\begin{figure}[ptb]
\begin{center}
\includegraphics[  width=3.4in,
keepaspectratio]{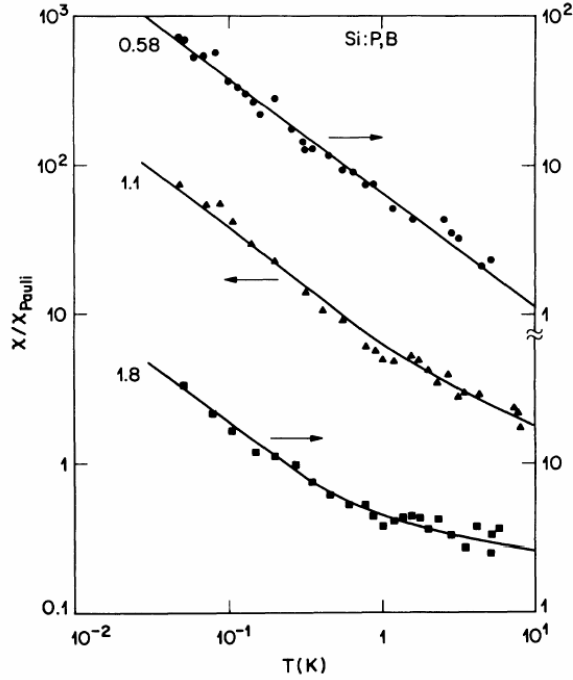}
\end{center}
\caption{Temperature dependence of the normalized susceptibility for three
different Si:P,B samples, with electron densities $n/n_{c}=0.58,1.1,1.8$, as
reported by Hirsch \emph{et al.} \cite{holcomb}. The low temperature behavior
looks qualitatively the same in a very broad concentration interval,
persisting well into the metallic side of the transition, in contrast to Fermi
liquid predictions. }%
\label{cap:fig5}%
\end{figure}

To test these ideas and also to explore the validity of generalized Fermi
liquid considerations \cite{ckl}, subsequent experiments turned to exploring
the thermodynamic behavior of the system. On the insulating side, one expects
\cite{doped-book} the donor electrons to be tightly bound to ionic centers,
thus behaving as spin 1/2 local magnetic moments. Because of the overlap of
the donor electron wave-functions, antiferromagnetic exchange is generated
between pairs of such magnetic moments, which takes the form \cite{doped-book}%
\begin{equation}
J(R)=J_{o}\exp\{-R/a\}. \label{eq:insulexchange}%
\end{equation}
The parameters $J_{o}$ and $a$ (the Bohr radius) can be calculated with
precision for all the known shallow impurity centers. Since the donor atoms
are randomly distributed throughout the host matrix, the insulator can
therefore be described as a random quantum antiferromagnet, which at low
temperature can be expected to exhibit random spin freezing, i.e. spin glass
behavior. Despite considerable effort, the search for such spin glass ordering
in the insulating phase has proven unsuccessful \cite{marko,maekawa}. Instead,
a very unusual thermodynamic response has been observed, with a diverging spin
susceptibility and sub-linear specific heat as a function of temperature, viz.%
\begin{align}
\chi(T)  &  \sim T^{\alpha-1},\label{eq:divergsusc}\\
\gamma(T)  &  =\frac{C}{T}\sim T^{\alpha-1}. \label{eq:diverggamma}%
\end{align}
The exponent $\alpha$ was found \cite{paalanen91,sarachik95} to weakly depend
on the details of specific system, generally taking the value $\alpha
\approx0.4$ for uncompensated and $\alpha\approx0.3$ for compensated systems.
Soon after the initial observations, this behavior on the insulating side was
qualitatively and even quantitatively explained by the {}``random singlet''
theory of Bhatt and Lee \cite{bhattlee81,bhattlee82} (see
Section~\ref{sub:randomsinglet}).

\subsubsection{Two-fluid model}

The real surprise, however, came from the observation that
essentially the same low temperature anomalies persisted
\cite{paalanen,schlager-epl97} well into the metallic side of the
transition. The data were fitted by a phenomenological
{}``two-fluid'' model \cite{marko,paalanen}, which assumed
coexistence of local magnetic moments and itinerant electrons, so
that (see
Fig.~\ref{cap:fig5})%
\begin{align}
\gamma/\gamma_{o}  &  =m^{\ast}/m_{o}^{\ast}+(T_{o}/T)^{\alpha-1}%
,\label{eq:twofluidgamma}\\
\chi/\chi_{o}  &  =m^{\ast}/m_{o}^{\ast}+\beta(T_{o}/T)^{\alpha-1}.
\label{eq:twofluidsusc}%
\end{align}
In these expressions, the first term is ascribed to itinerant
electrons which are assumed to form a Fermi liquid with band mass
$m_{o}^{\ast}$ and effective mass $m^{\ast}$. The second term
describes the contribution from localized magnetic moments, where
the deviation from the Curie law ($\alpha>1 $) is viewed as a
result of spin-spin interactions, similarly as in the Bhatt-Lee
theory \cite{bhattlee81,bhattlee82}. The exponent $\alpha$ was
found \cite{paalanen,schlager-epl97} to have a weak density
dependence, and essentially take the same value as on the
insulating side. The fitted values of the parameter $T_{o}$ can be
used to estimate the relative fraction of the electrons that form
the local moments. According to these estimates, between 10\% and
25\% of the electrons contribute to the formation of these
localized moments in the vicinity of the transition.

These experimental findings represented drastic violations of the Fermi liquid
predictions. Within the Fermi liquid picture \cite{ckl}, in any metal the
local magnetic moments should be screened by conduction electrons through the
Kondo effect or spin-spin interactions, so that $\gamma$ and $\chi$ should
remain finite as $T\rightarrow0$. Their precise value should be a function of
the corresponding Fermi liquid parameters $m^{\ast}$and $F_{o}^{a}$ (see
Eqs.~(\ref{eq:flspecificheat}) and (\ref{eq:flsusceptibility})), and as such
should have a singular behavior only as a phase transition is approached, not
within the metallic phase. In contrast, the singularities observed in these
materials display \textit{no observable anomaly} \cite{holcomb} as one crosses
from the metallic to the insulating side. In fact, the thermodynamic data on
their own have such a weak density dependence, that just on their basis one
could not even determine the critical concentration $n_{c}$. This behavior is
in sharp contrast to that observed in conventional Fermi liquids (e.g. clean
heavy fermion compounds), where sharp anomalies in both the transport and the
thermodynamic properties are clearly seen and well documented.

Since the physics associated with precursors of local moment magnetism is
typically associated with strong correlation effects, it seems very likely
\cite{lr} that the surprising behavior of doped semiconductors has the same
origin. If this is true, then the conventional weak-coupling approaches
\cite{kirkpatrick-rmp94} to electronic correlations seem ill suited to
describe this puzzling behavior. These ideas seem in agreement with the
original physical picture proposed by Mott \cite{mott-book90}, where the
Coulomb repulsion is viewed as a major driving force behind the
metal-insulator transition. Indeed, even early attempts to address the role of
correlation effects based on Hubbard models by Milovanovi\'{c}, Sachdev, and
Bhatt \cite{milovanovicetal89} were sufficient to indicate the emergence of
disorder-induced local moment formation around the transition region. This
theory was too simple to address more subtle questions such as the interaction
of such local moments and the conduction electrons
\cite{bhattfisher92,vladtedgabi,volfle}, but later work
\cite{dk-prl93,dk-prb94,motand,vladgabisdmft2} based on extended DMFT
approaches provided further support to the strong correlation picture.

\subsubsection{Asymptotic critical behavior or mean-field scaling?}

One more important aspect of the experimental data is worth
emphasizing. Several well defined features of these materials have
been clearly identified, both with respect to the thermodynamic
and the transport properties. These include a sharply defined
power-law behavior of the conductivity, with an exponent $\mu$
being a strong function of the magnetic field or the deviation
from half filling, and a smooth but singular thermodynamic
response. All of these features are clearly defined over a {\em
very broad} range of parameters covering, for example, in excess
of 100\% of the reduced concentration $\delta n=(n-n_{c})/n_{c}$.
Such behavior provides a strong indication that the puzzling
observations are not associated with complications arising within
the asymptotic critical region of a second-order phase transition.
In contrast, the behavior in the immediate vicinity of the
critical point is still a subject of some controversy
\cite{stupp-prl93,waffenschmit-prl99}, and may be dominated by
extrinsic (e.g. self-compensation) effects \cite{itoh-jpsj04}.

Therefore, theories designed to capture only long wavelength and
low energy excitations do not seem likely to be sufficient in
describing the most puzzling features of the experiments. To our
knowledge, none of the microscopic predictions of such {}``field
theoretical'' approaches \cite{kirkpatrick-rmp94} have been
experimentally verified; most observations seem drastically
different from what one expects based on these theories. Such
robust and systematic behavior observed in a broad parameter range
is more reminiscent of mean-field behavior, which typically
describes most of the experimental data around phase transitions,
except in extremely narrow critical regions. Typically, one has to
approach the critical point within a reduced temperature
$t=(T-T_{c})/T_{c}$ of the order of $10^{-3}$ or less in order to
observe the true asymptotic critical behavior. Unfortunately, an
appropriate mean-field description is not available yet for
metal-insulator transitions, despite years of effort. In our
opinion, what is needed to correctly describe the observed
non-Fermi liquid features are approaches that can explicitly
address the interplay of strong correlations and disorder.

\subsection{Disorder-driven NFL behavior in correlated systems: disordered
heavy fermion systems}

\label{sub:nflkondoalloys}

Anomalous NFL behavior has been observed in a large number of
non-stoichiometric heavy fermion compounds. An ample review of the
experimental situation until 2001 can be found in Stewart's review
\cite{stewartNFL}. Since the main focus of this review is on the theoretical
approaches to disordered NFL systems, we will not dwell much on the available
data on these systems. We will, however, give a bird's eye view of the subject.

We would first like to distinguish between the main systems of interest here
and those whose detailed description requires the consideration of disorder
effects but in which disorder is most likely \emph{not} the driving mechanism
of NFL behavior. This is especially important in those cases where the
proximity to a quantum critical point seems to be the origin of the anomalous
behavior. Indeed, the external tuning parameter in these systems is often
chemical pressure, namely, the substitution of small amounts of an
isoelectronic element with a slightly larger ionic radius with the aim of
expanding the lattice. Non-stoichiometry is then unavoidable but its effects
are considered secondary when compared with, say, the variation of the
couplings between neighboring atoms. Some studies of these effects, especially
on transport have been undertaken \cite{rosch99,rosch00}. A caveat is in
order, however. A well-known criterion due to Harris, establishes that if $\nu
d<2$, where $\nu$ is the correlation length critical exponent of the clean
system, then disorder is a \emph{relevant} perturbation, i.~e., the actual
critical behavior is modified in the dirty system \cite{harris74}. The clean
Hertz-Millis theory discussed in Section~\ref{sub:qcp} has a mean field
exponent $\nu=1/2$ and the Harris criterion indicates that disorder is
relevant. Therefore, even in this case it is not clear that disorder does not
play a more crucial role. We will discuss the interplay of disorder and
quantum critical behavior in both insulating and metallic systems in
Section~\ref{sub:magneticgriff}.

\begin{figure}[ptb]
\begin{center}
\includegraphics[  width=3.4in,
keepaspectratio]{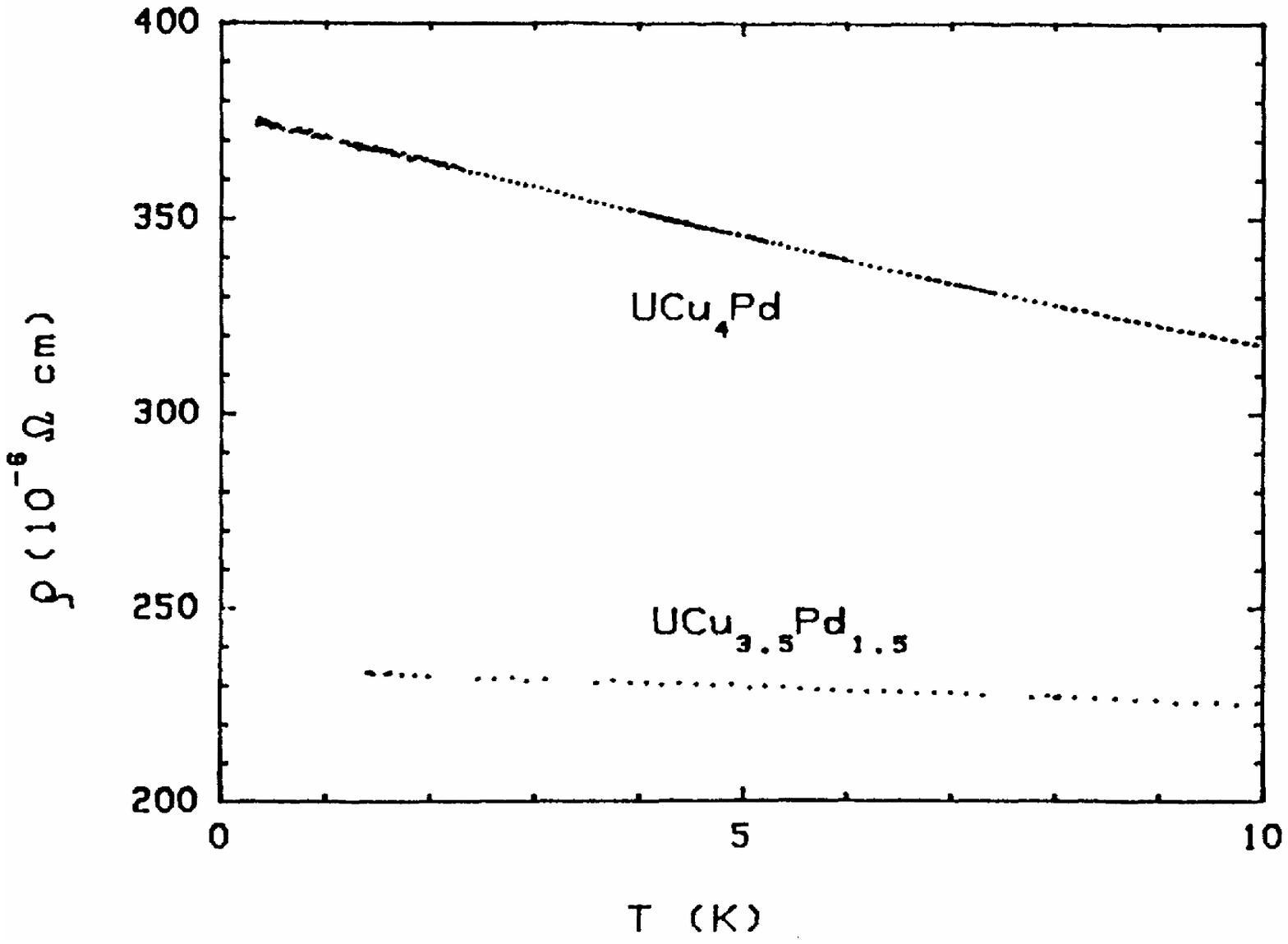}
\end{center}
\caption{Anomalous NFL resistivity of UCu$_{5-x}$Pd$_{x}$. Data are from
Ref.~\cite{andrakastewart}.}%
\label{cap:fig6}%
\end{figure}

There is of course no clear-cut way to separate systems whose behavior is
governed by the proximity to a QCP and those in which disorder is the driving
mechanism of NFL behavior. A practical rule of thumb is to check whether the
anomalous behavior exists in regions of the phase diagram distant from a clean
quantum phase transition, typically antiferromagnetism. This rule has been
adopted, for instance, by Stewart in his review \cite{stewartNFL}. Another
indication of the importance of disorder in heavy fermion alloys is a large
value of the residual resistivity $\rho_{0}=\rho\left(  T\rightarrow0\right)
$. Clean heavy fermion systems are characterized by large amounts of
scattering of the metallic carriers by the localized magnetic moments at high
temperatures, the so-called Kondo scattering \cite{kondo64}. As the
temperature is lowered, however, lattice translation invariance sets in and a
precipitous drop of the resistivity, by orders of magnitude, occurs below the
so-called coherence temperature $T_{coh}$, which is typically a few tens of
Kelvin. This delicate state is easily destroyed by disorder and large
incoherent Kondo scattering is then able to survive to the lowest temperatures
leading to values of $\rho_{0}$ on the order of hundreds of $\mu\Omega-cm$.
NFL behavior is observed in a number of heavy fermion alloys where no trace of
coherence remains. In these cases, besides a large value of $\rho_{0}$, the
leading low-temperature behavior is often \emph{linear} in temperature with a
negative coefficient: $\rho\left(  T\right)  \approx\rho_{0}-AT$, in sharp
contrast to the Fermi liquid behavior of dilute Kondo impurities $\rho\left(
T\right)  \approx\rho_{0}-BT^{2}$ \cite{Wilson2,nozieres74}. Prominent
examples of systems where this linear in temperature behavior is observed are
U$_{x}$ Y$_{1-x}$Pd$_{3}$ \cite{seamanetal91,andrakatsvelik91,ottetal93},
UCu$_{5-x}$Pd$_{x}$ ($x=1$ and $1.5$, see Fig.~\ref{cap:fig6})
\cite{andrakastewart,chaumaple96,weberetal01}, UCu$_{5-x}$Pt$_{x}$ ($x=0.5$,
$0.75 $ and $1$) \cite{chaumaple96,stewartNFL}, U$_{1-x}$Th$_{x}$Pd$_{2}%
$Al$_{3}$ \cite{mapleetal95}, Ce$_{0.1}$La$_{0.9}$Cu$_{2.2}$Si$_{2}$
\cite{andraka94}, UCu$_{4}$Ni \cite{delatorre98}, URh$_{2}$Ge$_{2}$ (as grown)
\cite{sullowetal00}, and U$_{2}$PdSi$_{2}$ \cite{lietal98}. The last two
systems and those of the form UCu$_{4}$M, although nominally stoichiometric
are actually crystallographically disordered, as evidenced by wide NMR lines
\cite{bernaletal}, $\mu$SR \cite{maclaughlinetal98} and EXAFS measurements
\cite{boothetal,baueretal,boothetal2}.

\begin{figure}[ptb]
\begin{center}
\includegraphics[  width=3.4in,
keepaspectratio]{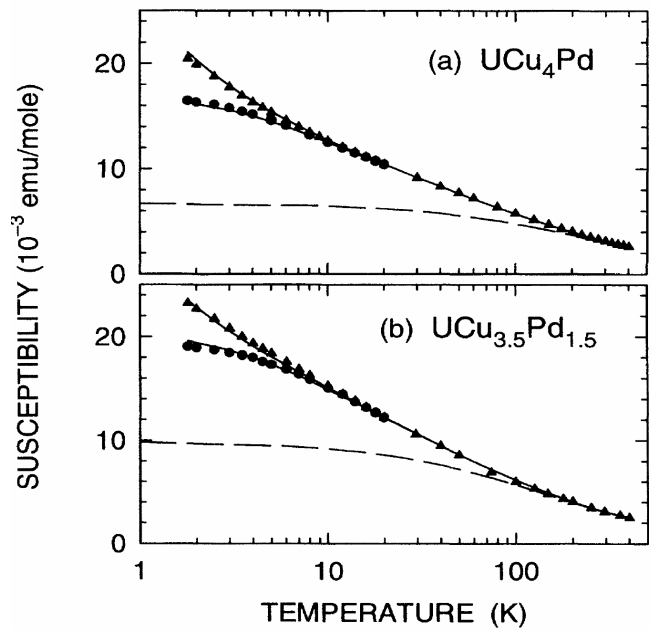}
\end{center}
\caption{Anomalous NFL magnetic susceptibility of UCu$_{5-x}$Pd$_{x}$ for
different applied fields: $H=5$ kOe (triangles) and $H=50$ kOe (circles). Data
are from Ref.~\cite{bernaletal}.}%
\label{cap:fig7}%
\end{figure}

Besides the anomalous transport, heavy fermion alloys also exhibit NFL
thermodynamic properties. Singular behavior is observed in both the specific
heat coefficient $\gamma\left(  T\right)  =C\left(  T\right)  /T$ and the
magnetic susceptibility $\chi\left(  T\right)  $ (see Fig.~\ref{cap:fig7}).
Ambiguities in the determination of the exact singularity are common since the
available temperature range is often not very broad. Therefore, there are
conflicting claims of logarithmic {[}$\sim\log\left(  T_{0}/T\right)  ${]} or
power law {[}$\sim T^{\lambda-1}${]} divergences, a small value of $\lambda$
in the latter being very difficult to discern from the former dependence. Once
again, we refer the reader to Stewart's review \cite{stewartNFL} for a summary
of the published results. In that article, the author also presents many data
previously published as being well described by a logarithmic dependence,
replotted and fitted to a weak power law with a small $\lambda$. These
discrepancies highlight the difficulty of extracting the nature of the
singular behavior from a limited range of temperatures. The only reliable way
to resolve these ambiguities is through the determination of error bars for
the exponents (as in classical critical phenomena). Unfortunately, this has
not, to our knowledge, been attempted.

Other experimental probes have also revealed NFL behavior in heavy fermion
alloys. Optical conductivity studies have shown that the frequency dependence
of the scattering rate at low temperatures is linear as opposed to the
quadratic dependence predicted by Fermi liquid theory. This behavior has been
seen in UCu$_{5-x}$Pd$_{x}$ \cite{degiorgiott96}, U$_{x}$ Y$_{1-x}$Pd$_{3}$
\cite{degiorgietal95}, and U$_{1-x}$Th$_{x}$Pd$_{2}$Al$_{3}$
\cite{degiorgietal96} (see also the review \cite{degiorgireview} for a more
detailed discussion). Furthermore, neutron scattering data on UCu$_{5-x}%
$Pd$_{x}$ have revealed $\omega/T$ scaling in the dynamical spin
susceptibility over a broad temperature and frequency range without any
significant wave vector dependence (other than the trivial Uranium atom form
factor) \cite{aronsonetal95,aronsonetal01}. Finally, we should mention
extensive NMR and $\mu$SR work on these systems
\cite{maclaughlinetal96,liuetal00,buttgenetal,dougetal3,dougetal,dougetal2,dougetal02,dougetal03,dougetal04}%
. These studies, together with the already cited EXAFS technique
\cite{boothetal,baueretal,boothetal2}, are invaluable tools to access spatial
fluctuations of local quantities, which are inevitably introduced by disorder.

\subsection{Disorder-driven NFL behavior in correlated systems: metallic glass
phases}

\label{sub:metallic-glass}

\subsubsection{Metallic glass phases}

Although usually not close to an ordered magnetic phase, disordered heavy
fermion systems very often show spin-glass freezing. This is perhaps not too
surprising given the interplay of disorder, local magnetic moments and
RKKY-induced frustration in these systems. However, freezing temperatures are
surprisingly low given the large concentrations of magnetic moments. A
possible explanation is the Kondo compensation by the conduction electrons,
which tends to favor a paramagnetic heavy-fermion state. Related to this is
the proposal that the quantum critical point separating the heavy-fermion
paramagnet and the spin glass phase is at the origin of the NFL behavior of
these alloys (see Section~\ref{sub:qcp-msg}). Spin-glass behavior is
usually detected by the difference between zero-field-cooled and field-cooled
magnetic susceptibility curves. Direct measurements of the spin dynamics in
$\mu$SR experiments has also proven to be a useful tool. Examples of systems
where spin-glass phases have been detected are Y$_{1-x}$U$_{x}$Pd$_{3}$
($x\agt0.2$) \cite{gajewskietal96}, UCu$_{5-x}$Pd$_{x}$ ($x\agt1$)
\cite{andrakastewart,vollmeretal,dougetal}, URh$_{2}$Ge$_{2}$
\cite{sullowetal00}, U$_{2}$PdSi$_{2}$ \cite{lietal98}, Ce$_{0.15}$La$_{0.85}%
$Cu$_{2}$Si$_{2}$ \cite{andraka94}, U$_{0.07}$Th$_{0.93}$Ru$_{2}$Si$_{2}$
\cite{stewartNFL}, U$_{1-x}$Y$_{x}$Al$_{2}$ ($0.3\leq x\leq0.7$)
\cite{mayretal97}, and U$_{2}$Cu$_{17-x}$Al$_{x}$ ($x=8$) \cite{pietrietal97}.
{}From these numerous examples, it becomes clear that spin-glass ordering is a
common low-temperature fate of heavy fermion alloys. It is likely that a
complete theoretical picture of the NFL behavior of these systems will have to
encompass the spin-glass state.

In many of these systems one can tune through a $T=0$ spin glass transition by
varying parameters such as doping pressure, etc. In such cases one expects new
physics associated with a spin-glass-paramagnet quantum critical point, but
very few experiments exist where systematic studies of this regime have been
reported in heavy fermion systems. On general grounds, one may expect such
disorder-driven QCPs to have a more complicated form then in the clean cases.
In recent work, theoretical scenarios have been proposed
\cite{tvojta03,vojtaschmalian04,noqcp}, suggesting that rare events and
dissipation may \char`\"{}smear\char`\"{} such phase transition, possibly
making it difficult even to detect the precise location of the critical point.
More experimental work is urgently needed to settle these interesting issues,
especially using the local probes (STM, NMR, $\mu$SR) which are better suited
to characterize such strongly inhomogeneous systems.

\subsubsection{Glassiness in the charge sector}

An interesting alternative to spin glass ordering is offered by the
possibility of glassy freezing of charge degrees of freedom. In many
disordered electronic systems \cite{lr}, electron-electron interactions and
disorder are equally important, and lead to a rich variety of behaviors which
remain difficult do understand. Their competition often leads to the emergence
of many metastable states and the resulting history-dependent glassy dynamics
of electrons. Theoretically, the possibility for glassy behavior in the charge
sector was anticipated a long time ago \cite{re:Efros75} in situations where
the electrons are strongly localized due to disorder. In the opposite limit,
for well delocalized electronic wave functions, one expects a single well
defined ground state and absence of glassiness. The behavior in the
intermediate region has proved more difficult to understand, and at present
little is known as to the precise role and stability of the glassy phase close
to the metal-insulator transition (MIT) \cite{mott-book90}.

\begin{figure}[ptb]
\begin{center}
\includegraphics[  width=3.4in,
keepaspectratio]{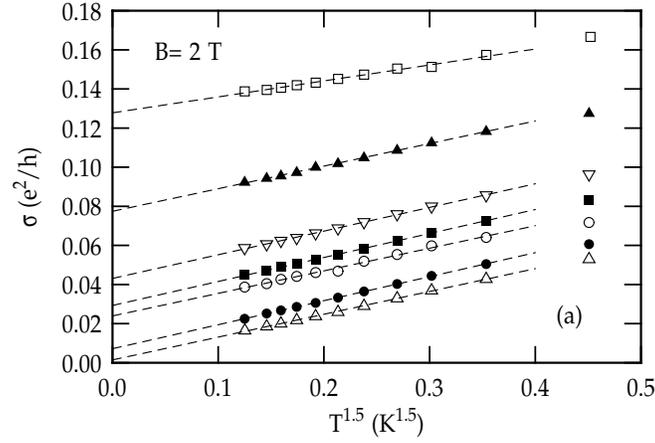}
\end{center}
\caption{Temperature dependence of the conductivity of two dimensional
electrons in silicon in the metallic glass phase, in the presence of a
parallel magnetic field ($B=2$~T). The data taken from Ref.~\cite{mag-glass}
show results for several electron densities {[}$n_{s}(10^{10}$cm$^{-2}%
)=11.9,11.6,11.3,11.2,11.0,10.9,10.7${]}, as the metal -insulator transition
is approached. }%
\label{cap:fig8}%
\end{figure}

{}From the experimental point of view, glassy behavior has often been observed
in sufficiently insulating materials \cite{films25,films24,films23}, but more
recent experiments \cite{bogdanovich-prl02,JJPRL02} have provided striking and
precise information on the regime closer to the metal-insulator transition.
These experiments on low density electrons in silicon MOSFET's have revealed
the existence of an intermediate metallic glass phase in low mobility (highly
disordered) samples \cite{bogdanovich-prl02}. Precise experimental studies of
low temperature transport in this regime have revealed an unusual low
temperature behavior of the resistivity of the form%
\[
\rho(T)=\rho(o)+AT^{3/2}.
\]
This behavior is observed throughout the metallic glass phase, allowing one to
systematically extrapolate the resistivity to $T=0$, and thus characterize the
approach to the metal-insulator transition. Such non-Fermi liquid behavior is
consistent with some theoretical predictions for metallic glasses
\cite{re:Sachdev96,Denis,arrachea}, but the same exponent $3/2$ is expected
for glassiness both in the spin and the charge sector. To determine the origin
of glassiness, the experiments were carried out \cite{mag-glass} in the
presence of large magnetic fields parallel to the 2D channel (such that the
field couples only to the electron spin), as shown in Fig~\ref{cap:fig8}.
Since the Fermi energy is relatively small in this density regime ($T_{F}%
\sim10K$), it is possible to completely spin polarize the electrons with
accessible fields, and thus completely eliminate any spin fluctuations.
Remarkably, all the signatures of the glassy behavior, as well as the
$T^{3/2}$ resistivity persisted in this regime, demonstrating that the
anomalies originate from the charge degrees of freedom.

These experimental results are certainly very compelling, as they indicate
that glassy ordering in the charge sector may very well play a crucial role in
most systems close to inhomogeneous insulating states. This possibility seems
even more reasonable having in mind that screening effects can be expected to
weaken as one approaches an insulator - thus strengthening the effects of the
Coulomb interactions. More experimental work is necessary to test these ideas,
especially using those techniques that directly couple to the charge degrees
of freedom. In this respect, global and local compressibility measurements, as
well as frequency-dependent dielectric constant experiments seem particularly
promising. Such experiments have recently been reported on some
transition-metal oxides, revealing glassy ordering, but similar work on other
materials are called for in order to determine how general these phenomena may be.

\section{Theoretical approaches}

\label{sec:Theoretical-approaches}

\subsection{Disordered Hubbard and Kondo lattice models}

\label{sub:Disordered-Hubbard-and-Kondo}

\subsubsection{Disordered-induced local moment formation}

\label{sub:local-moments}

The experiments of Paalanen \emph{et al.} were very successfully described
within a two-fluid model of itinerant carriers and localized magnetic moments
\cite{marko,paalanen}, as explained in Section~\ref{sub:dopedsemicond}. A
first attempt to directly test this phenomenology within a well-defined model
calculation was due to Milovanovi\'{c}, Sachdev and Bhatt
\cite{milovanovicetal89}. These authors investigated a disordered Hubbard
model with both diagonal (site energies) and off-diagonal (hopping) disorder.
The model parameters were chosen so as to faithfully describe the situation of
Si:P. In contrast to the weak-disorder approach of the scaling theory, their
work treats disorder exactly by numerical calculations while relying on the
mean-field Hartree-Fock treatment of interactions, as was done for the single
impurity Anderson model \cite{Anderson}. The latter is known to describe well
the phenomenon of local moment formation, here taken to signify a broad
temperature crossover with a Curie law magnetic susceptibility. However, it is
not capable of correctly accounting for the low-temperature quenching of these
moments by the conduction electrons, the Kondo effect. The results of
Milovanovi\'{c} \emph{et al.} show that a finite fraction of the electrons do
indeed exhibit a local moment Curie response even if the wave functions at the
Fermi level are extended, i.e. the system is still metallic. The fraction of
localized moments obtained is of order 10\% at 25\% above the critical
density, in reasonable order of magnitude agreement with the experiments
\cite{paalanen}. In other words, the results are compatible with the
observation of local moment responses on the metallic side of the
disorder-induced metal-insulator transition. A rather complete exploration of
the phase diagram of this model with the same method has also been carried out
\cite{tuschlogan93}, confirming this picture.

Extremal statistics arguments were used by Bhatt and Fisher to analyze local
moment formation and also to determine the effects of residual interactions
between these moments \cite{bhattfisher92}. They considered a disordered
Hubbard model very similar to the one studied by Milovanovi\'{c} \emph{et al.}
In such a system, rare disorder fluctuations can give rise to sites weakly
coupled to the rest of the lattice which, for sufficiently strong
interactions, can form localized moments. In the absence of the Kondo effect
or the RKKY interactions between these moments, they would give rise to a
Curie response as shown in reference~\cite{milovanovicetal89}. In the presence
of the itinerant electron fluid in the metallic phase, however, the Kondo
effect may quench these moments. The scale at which this quenching occurs is
the Kondo temperature $T_{K}\sim D\exp\left(  -1/\rho J\right)  $, where $D$
is the conduction electron half band width, $\rho$ is the density of states at
the Fermi level, and $J$ is the exchange coupling between the local moment and
the conduction electron fluid. Spatial fluctuations lead to a distribution of
couplings $J$ and consequently to a distribution of Kondo temperatures. Thus,
only those sites with $T_{K}<T$ will contribute significantly to the
thermodynamic properties. The overall behavior is still singular, though the
corrections to the Curie law are very mild%
\begin{equation}
\chi\left(  T\right)  \sim\frac{C\left(  T\right)  }{T}\sim\frac{1}{T}%
\exp\left\{  -A\ln^{d}\left[  \ln\left(  T_{0}/T\right)  \right]  \right\}  ,
\label{eq:bhattfisher}%
\end{equation}
where $A$ and $T_{0}$ are constants and $d$ is the dimensionality. The effects
of the RKKY interactions are stronger and lead, in the absence of ordering, to
a divergence that is slower than any power law%
\begin{equation}
\chi\left(  T\right)  \sim\frac{C\left(  T\right)  }{T}\sim\exp\left\{
B\ln^{1/d}\left(  T_{0}/T\right)  \right\}  , \label{eq:bhattfisher2}%
\end{equation}
where again $B$ is a constant. The picture that emerges is that of a random
singlet phase as will be explained below in Section~\ref{sub:randomsinglet}. A
phase transition into a spin-glass phase, however, may intervene at low temperatures.

This problem is further complicated by the fact that the Kondo temperature
depends exponentially both on the exchange coupling $J$ and the local
conduction electron density of states $\rho$. The distribution of $T_{K}$
calculated by Bhatt and Fisher did not take into account the fluctuations of
the latter. We emphasize that $\rho$ should be generalized, in a disordered
system, to the \emph{local} density of states $\rho\left(  \mathbf{R}\right)
$ at the position $\mathbf{R}$ of the magnetic moment. This is not a
self-averaging quantity and its fluctuations are expected to grow
exponentially as the disorder-induced metal-insulator transition is
approached. These effects were analyzed by Dobrosavljevi\'{c}, Kirkpatrick and
Kotliar \cite{vladtedgabi}. These authors emphasized the universal log-normal
form of the distribution of local density of states, which in turn leads to a
universal $T_{K}$ distribution. As usual for a broad distribution, the
magnetic susceptibility can be written as $\chi\left(  T\right)  \sim n\left(
T\right)  /T$, where $n\left(  T\right)  \sim T^{\alpha\left(  T\right)  }$ is
the temperature dependent number of unquenched spins at temperature $T$.
Although the exponent $\alpha\left(  T\right)  \rightarrow0$ as $T\rightarrow
0$, it does so in an extremely slow fashion and in practice, for reasonable
values of the parameters and for a strongly disordered but still metallic
system, $n\left(  T\right)  \sim50$\% for $T\sim10^{-4}-1$K.

\subsubsection{Random singlet phases}

\label{sub:randomsinglet}

Local moment formation is generically a crossover phenomenon which occurs at
intermediate temperatures whose signature is a Curie-Weiss magnetic
susceptibility. Possible low-temperature fates of these moments include some
form of magnetic ordering, such as antiferromagnetism and ferromagnetism or
spin-glass freezing. A novel kind of low temperature fixed point of disordered
localized magnetic moments, however, was proposed from studies of doped
semiconductors close to the localization transition, as was discussed in
Section~\ref{sub:dopedsemicond} \cite{bhattlee81,bhattlee82}. In these
systems, the magnetic susceptibility shows no signs of any type of ordering
but diverges with a non-trivial power law \cite{paalanen91,sarachik95}%
\begin{align}
\chi(T)  &  \sim T^{\alpha-1},\label{eq:powerlawsusc}\\
\gamma(T)  &  =\frac{C}{T}\sim T^{\alpha-1}. \label{eq:powerlawgamma}%
\end{align}
The theory of Bhatt and Lee \cite{bhattlee81,bhattlee82} was based upon a
generalization to three dimensions of a method introduced by Ma, Dasgupta and
Hu for one-dimensional disordered antiferromagnetic spin chains
\cite{madasguptahu,madasgupta}. The Hamiltonian of such a random spin system
is written as%
\begin{equation}
H=\sum_{i,j}J_{ij}\mathbf{S}_{i}\cdot\mathbf{S}_{j}, \label{eq:heisenberg}%
\end{equation}
where $J_{ij}>0$ is a random variable distributed according to some given
distribution function $P_{0}\left(  J_{ij}\right)  $. The method consists of
looking for the strongest coupling of a given realization of the random
lattice, say $\Omega$. At energy scales much smaller than $\Omega$, there is
only a small probability for this pair of spins to be in its excited triplet
state. We can then assume the pair is locked in its ground singlet state and
is magnetically inert at the scale considered. The pair is then effectively
{}``removed'' from the system. The remaining spins which had interactions with
these locked spins develop additional interactions induced by the removed
pair. If spins $i$ and $j$ form the removed pair and spins $k$ and $l $
interact with $i$ and $j$, respectively, the new interaction can be obtained
in second order of perturbation theory%
\begin{equation}
\Delta J_{kl}=\frac{J_{ik}J_{jl}}{2\Omega}. \label{eq:perttheory}%
\end{equation}
If initially $k$ and $l$ do not interact they will do so after the decimation
step. Their new coupling will be smaller than any of the removed ones.
Therefore, as more and more spins are decimated, the largest couplings are
progressively removed while new smaller ones are generated, thus changing the
shape of the {}``bare'' distribution into an effective {}``renormalized'' one
at lower energy scales. Formally, the renormalization group flow starts at
$P\left[  J,\Omega_{0}\right]  =P_{0}\left(  J\right)  $, where $\Omega_{0}$
is the initial largest value of the distribution. After many decimations,
$\Omega_{0}$ will have been decreased to $\Omega$ and the distribution will be
$P\left[  J,\Omega\right]  $. Bhatt and Lee implemented numerically the
decimation procedure we have just outlined using realistic initial
distributions of couplings appropriate for the insulating phase of doped
semiconductors. They showed that these initial distributions flowed towards
very broad distributions at small $J$ \cite{bhattlee81,bhattlee82}, leading to
diverging thermodynamic responses of the form given in
Eq.~(\ref{eq:powerlawgamma}), in remarkable agreement with the experiments.
The low energy state of the system showed no tendency towards ordering but was
characterized by the successive formation of singlet pairs between widely
separated spins at each decreasing energy scale. This novel disordered
magnetic state was dubbed a {}``random singlet phase''.

\subsubsection{Phenomenological Kondo disorder model}

\label{sub:kdm}

\begin{figure}[ptb]
\begin{center}
\includegraphics[  width=3.4in,
keepaspectratio]{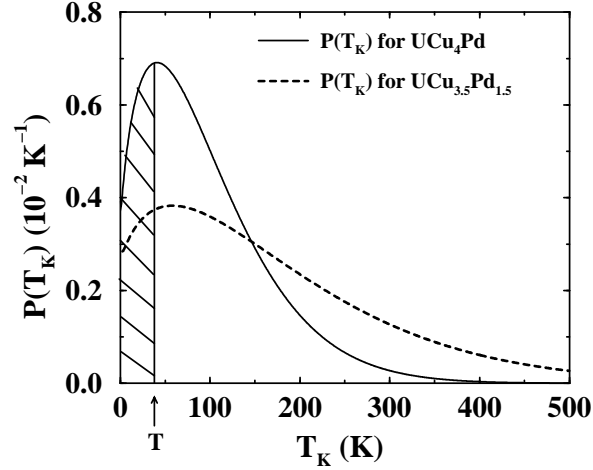}
\end{center}
\caption{Distribution of Kondo temperatures obtained within the Kondo disorder
model from an analysis of the disordered heavy fermion compounds UCu$_{5-x}%
$Pd$_{x}$ (Ref.~\cite{bernaletal}). Spins with $T_{K}<T$ (shaded area)
dominate the thermodynamic response.}%
\label{cap:fig9}%
\end{figure}

Although the importance of the distribution of Kondo temperatures was first
proposed with doped semiconductors in mind, the context of disordered heavy
fermion materials, where the Kondo effect is of primary importance, seemed
like a natural arena for its applications. Indeed, it was put to good use in
the attempts to understand both the temperature dependence of the Cu NMR
line-widths and the thermodynamic properties of the Kondo alloy UCu$_{5-x}%
$Pd$_{x}$ ($x=1,\,1.5$) by Bernal \emph{et al.} \cite{bernaletal}. As was
mentioned in Section~\ref{sub:nflkondoalloys}, this compound is known to
exhibit non-Fermi liquid behavior in many of its properties. Bernal \emph{et
al.} analyzed the broad Cu NMR lines in the following way. The Knight shift at
a particular Cu site positioned at $\mathbf{R}$, $K\left(  \mathbf{R}\right)
$, is proportional to the local spin susceptibility%
\begin{equation}
K\left(  \mathbf{R}\right)  =a\left(  \mathbf{R}\right)  \chi\left(
\mathbf{R}\right)  , \label{eq:knightshift}%
\end{equation}
where $a\left(  \mathbf{R}\right)  $ is the hyperfine coupling. If $a\left(
\mathbf{R}\right)  $ is assumed to have little variation in the sample, the
spatial fluctuations of the Knight shift can then be ascribed to the spatial
variations of the local susceptibility $\delta K=a\delta\chi$. If we use the
following fairly accurate form for the susceptibility of a single Kondo
impurity%
\begin{equation}
\chi\left(  T\right)  \sim\frac{1}{T+\alpha T_{K}}, \label{eq:kondosusc}%
\end{equation}
where as usual $T_{K}\approx D\exp\left[  -1/\left(  \rho J\right)  \right]
$, we can use the distribution of Kondo temperatures $P\left(  T_{K}\right)  $
to find the variance of the susceptibility and hence the line-width.
Furthermore, the bulk susceptibility and the specific heat can be obtained by
performing the same kind of average procedure over single-impurity results.
Bernal \emph{et al.} first fitted the susceptibility data by adjusting the
mean and the variance of a Gaussian distribution of the bare quantity $\rho
J$. Then, a fairly reasonable agreement with the NMR line-width and specific
heat data were obtained without further adjustments, explaining in particular
the strong anomalous temperature dependence of the line-widths. The
distribution of $\rho J$ used was not too broad. However, due to the
exponential form of $T_{K}\sim D\exp\left[  -1/\left(  \rho J\right)  \right]
$, even a fairly narrow distribution of $\rho J$ can lead to a very broad
$P\left(  T_{K}\right)  $. In the case of Bernal \emph{et al.}, $P\left(
T_{K}\right)  \rightarrow\mathrm{const.}$ as $T_{K}\rightarrow0$ (see
Fig.~\ref{cap:fig9}), so that%
\begin{align}
\overline{\chi}\left(  T\right)   &  \sim\int dT_{K}\frac{P\left(
T_{K}\right)  }{T+\alpha T_{K}}\nonumber\\
&  \sim\int_{0}^{\Lambda}dT_{K}\frac{P\left(  0\right)  }{T+\alpha T_{K}%
}+\mathrm{const.}\nonumber\\
&  \sim\ln\left(  \frac{T_{0}}{T}\right)  , \label{eq:kdmsusc}%
\end{align}
in good agreement with the experimental findings (see the solid lines of
Fig~\ref{cap:fig7}). The picture is again one where a few spins whose
$T_{K}<T$ (shaded area in Fig.~\ref{cap:fig9}) dominate the thermodynamic response.

\subsubsection{Dynamical mean field theory (DMFT)}

\label{sub:dmft}

The phenomenological model of Bernal \emph{et al.} made the seemingly
unjustified assumption of a collection of independent single Kondo impurities,
even though the alloys UCu$_{5-x}$Pd$_{x}$ contained a concentrated fcc
lattice of U ions. This assumption was put on a firmer theoretical basis by
Miranda, Dobrosavljevi\'{c}, and Kotliar
\cite{mirandavladgabi1,mirandavladgabi2,mirandavladgabi3}. These authors used
a dynamical mean field theory (DMFT) approach to the disordered Kondo (or
Anderson) lattice problem. The DMFT method has become a very useful tool in
the study of strongly correlated materials. It is the natural analogue of the
more familiar Weiss mean field theory of magnetic systems, which is here
generalized to describe quantum (especially fermionic) particles. Fairly
complete reviews of the subject are available
\cite{georgesrmp,pruschkeetalrev95}, and we will content ourselves with a
brief description, emphasizing the physical aspects.

For definiteness, \.{l}et us focus on a disordered Anderson lattice
Hamiltonian in usual notation%
\begin{align}
H_{AND}  &  =-t\sum_{\left\langle ij\right\rangle \sigma}\left(  c_{i\sigma
}^{\dagger}c_{j\sigma}^{\phantom{\dagger}}+\mathrm{H.}\,\mathrm{c.}\right)
+\sum_{j\sigma}\epsilon_{j}c_{j\sigma}^{\dagger}c_{j\sigma}%
^{\phantom{\dagger}}\nonumber\\
&  +\sum_{j\sigma}E_{fj}f_{j\sigma}^{\dagger}f_{j\sigma}^{\phantom{\dagger}}%
+U\sum_{j}f_{j\uparrow}^{\dagger}f_{j\uparrow}^{\phantom{\dagger}}%
f_{j\downarrow}^{\dagger}f_{j\downarrow}^{\phantom{\dagger}}\nonumber\\
&  +\sum_{j\sigma}\left(  V_{j}f_{j\sigma}^{\dagger}c_{j\sigma}%
^{\phantom{\dagger}}+\mathrm{H.}\,\mathrm{c.}\right)  , \label{eq:hamhybrid}%
\end{align}
where the conduction, $f$-electron, and hybridization energies $\epsilon_{j}$,
$E_{fj}$, and $V_{j}$ are, in principle, random variables distributed
according to $P_{\epsilon}\left(  \epsilon_{j}\right)  $, $P_{E_{f}}\left(
E_{fj}\right)  $, and $P_{V}\left(  V_{J}\right)  $, respectively. As in the
Weiss theory, one starts by singling out a particular lattice site, say $j$,
and writing out its effective action. A simplification is made here, by
neglecting all non-quadratic terms in the local fermionic operators, except
for the $U$-term. One gets%
\begin{align}
S_{eff}^{AND}\left(  j\right)   &  =S_{c}\left(  j\right)  +S_{f}\left(
j\right)  +S_{hyb}\left(  j\right)  ,\label{eq:siteaction}\\
S_{c}\left(  j\right)   &  =\sum_{\sigma}\int_{0}^{\beta}d\tau\int_{0}^{\beta
}d\tau^{\prime}c_{j\sigma}^{\dagger}\left(  \tau\right)  \left[  \delta\left(
\tau-\tau^{\prime}\right)  \right. \nonumber\\
&  \times\left.  \left(  \partial_{\tau}+\epsilon_{j}-\mu\right)  +\Delta
_{c}\left(  \tau-\tau^{\prime}\right)  \right]  c_{j\sigma}%
^{\phantom{\dagger}}\left(  \tau^{\prime}\right)  ,\label{eq:siteactionc}\\
S_{f}\left(  j\right)   &  =\int_{0}^{\beta}d\tau\left[  \sum_{\sigma
}f_{j\sigma}^{\dagger}\left(  \tau\right)  \left(  \partial_{\tau}+E_{fj}%
-\mu\right)  f_{j\sigma}^{\phantom{\dagger}}\left(  \tau\right)  \right.
\nonumber\\
&  +\left.  Uf_{j\uparrow}^{\dagger}\left(  \tau\right)  f_{j\uparrow
}^{\phantom{\dagger}}\left(  \tau\right)  f_{j\downarrow}^{\dagger}\left(
\tau\right)  f_{j\downarrow}^{\phantom{\dagger}}\left(  \tau\right)
\hspace{-2.7cm}%
\phantom{\sum_{\sigma}\left(\partial_{\tau}+E_{f}-\mu\right)}\right]
,\label{eq:siteactionf}\\
S_{hyb}\left(  j\right)   &  =\sum_{\sigma}\int_{0}^{\beta}d\tau\left[
V_{j}f_{j\sigma}^{\dagger}\left(  \tau\right)  c_{j\sigma}^{\phantom{\dagger}}%
\left(  \tau\right)  +\mathrm{H.}\,\mathrm{c.}\right]  .
\label{eq:siteactionhyb}%
\end{align}
The site $j$ {}``talks'' to the rest of the lattice only through the bath (or
{}``cavity'') function $\Delta_{cj}\left(  \tau\right)  $ in
Eq.~(\ref{eq:siteactionc}). For simplicity, we particularize the formulation
to the case of a Bethe lattice, in which the bath function is given by%
\begin{equation}
\Delta_{c}\left(  \tau\right)  =t^{2}\overline{G_{c}^{loc}\left(  \tau\right)
}. \label{eq:bath}%
\end{equation}
Here, the average local Green's function is obtained by first calculating the
local conduction electron Green's functions governed by the actions
(\ref{eq:siteaction})%
\begin{equation}
G_{c}\left(  j,\tau\right)  =-\left\langle T\left[  c_{j\sigma}%
^{\phantom{\dagger}}\left(  \tau\right)  c_{j\sigma}^{\dagger}\left(
0\right)  \right]  \right\rangle _{eff}, \label{eq:greenc}%
\end{equation}
and then averaging over all sites with the distributions of bare parameters
$P_{\epsilon}$, $P_{E_{f}}$, and $P_{V}$. Thus, all local correlations are
accounted for faithfully, while inter-site ones enter only through the bath
function. As in the Weiss theory, the whole procedure can be shown to be exact
in the limit of large coordination ({}``infinite dimensions'') if appropriate
scaling of the hopping is performed ($t\sim\widetilde{t}/\sqrt{z} $, with
$\widetilde{t}$ held constant as $z\rightarrow\infty$). Unlike the Weiss
theory, however, the {}``order parameter'' here, namely, $\Delta_{c}\left(
\tau\right)  $ is a \emph{function} not just a number (hence the name
{}``dynamical''). Correlation effects are incorporated in the step where the
conduction electron Green's function (\ref{eq:greenc}) is calculated. This is
equivalent to solving a single-impurity Anderson model \cite{georgeskotliar92}
and since the bare parameters are random, one has to solve a whole
\emph{ensemble} of these. The treatment of disorder within DMFT is equivalent
to the well-known coherent potential approximation (CPA) \cite{elliotetal74}.

The connection to the Kondo disorder model can now be made more apparent. The
\emph{ensemble} of single impurity problems described by the actions
(\ref{eq:siteaction}) is the equivalent of the collection of independent
single Kondo impurities of the phenomenological approach. However, the single
impurity problems are not really independent as each one {}``sees'' the same
bath function (\ref{eq:bath}), which in turn contains information from all the
other sites. The fully self-consistent calculation of the bath function,
however, shows that its precise form does not contain essential features being
as it is an average over many different sites. Each Anderson impurity has its
local Kondo temperature $T_{Kj}$, which in the Kondo limit can be written as
(taking $U\rightarrow\infty$)%
\begin{equation}
T_{Kj}=D\exp\left[  -\left|  E_{fj}\right|  /\left(  2\rho V_{j}^{2}\right)
\right]  . \label{eq:tkj}%
\end{equation}
The connection to the Kondo model is obtained from
$J_{j}=2V_{j}^{2}/\left| E_{fj}\right| $. A distribution of Kondo
temperatures then follows. The full solution is able to produce, with
a judicious choice of bare parameters, a $P\left( T_{K}\right) $ very
similar to the experimentally determined one for UCu$_{5-x}$Pd$_{x}$
based on the simple phenomenological Kondo disorder model (Section
\ref{sub:kdm}).

Within the DMFT framework, one can proceed to the calculation of
various physical properties. Bulk thermodynamic responses in the
particular case of the Anderson/Kondo lattice, can be very
accurately obtained through an \emph{ensemble} average of the
individual contributions from each site \cite{mirandavladgabi1},
thus justifying the procedure adopted by Bernal \emph{et al.}. The
reason for this is the dominance of the $f$-site contributions
over the conduction electron part \cite{mirandavladgabi1}.
Transport properties, however, cannot be so easily calculated and
require a special consideration of the coherence of the motion
through the lattice \cite{tesanovic-prb86}. This is particularly
striking in the clean limit, in which it is well known that the
$f$-ion contributions very accurately add up to give the
thermodynamic properties but the onset of heavy fermion coherence
in the resistivity is totally absent from a single impurity
description.

It is precisely with regard to transport properties that the DMFT approach is
able to go beyond the phenomenological Kondo disorder model. Assuming the same
$T_{K}$ distribution obtained experimentally by Bernal \emph{et al.}, one gets
a resistivity which has a large residual value and decreases linearly with
temperature with a negative slope
\cite{mirandavladgabi1,mirandavladgabi2,mirandavladgabi3}, in complete
agreement with the experiments on UCu$_{5-x}$Pd$_{x}$ \cite{andrakastewart}.
The large residual value is due to the total destruction of coherence by
disorder. The anomalous NFL linear dependence comes from the gradual
unquenching of local moments as the temperature is raised.

Other physical properties were calculated within DMFT, with good agreement
with experiments: the dynamical spin susceptibility
\cite{mirandavladgabi1,aronsonetal95}, the magneto-resistance \cite{chatto2},
and the optical conductivity \cite{chatto,degiorgiott96}. In all cases, NFL
behavior was tied to the finite weight of $P\left(  T_{K}\right)  $ as
$T_{K}\rightarrow0$.

One of the perceived deficiencies of this theory is its extreme sensitivity to
the choice of bare distributions. This is due to the exponential dependence of
the Kondo temperature on model parameters, which is hardly affected by the
self-consistency introduced by DMFT. In the particular case of UCu$_{5-x}%
$Pd$_{x}$, the finite intercept $P\left(  T_{K}=0\right)  $ is the most
important feature of the distribution and only fine tuning of the bare
distributions can provide the correct value. Besides, the thermodynamic
properties of many other compounds are more accurately described by power laws
\cite{stewartNFL,andradeetal}. These can be accommodated within the theory
through a power-law distribution of Kondo temperatures $P\left(  T_{K}\right)
\sim T_{K}^{\alpha-1}$. However, once again fine-tuning is necessary if one
wants to obtain such a distribution within DMFT.

\subsubsection{Statistical DMFT: localization effects and results of numerical
calculations}

\label{sub:statdmft}

The problem of fine tuning within DMFT was remedied in a more complete
approach. An important assumption of DMFT is the averaging procedure contained
in Eq.~(\ref{eq:bath}). It means that each site feels an \emph{average}
hybridization with its neighbors, which is why the procedure becomes
increasingly more accurate the larger the coordination. Fluctuations of this
quantity, usually associated with Anderson localization effects, can, however,
be incorporated in a so-called Statistical Dynamical Mean Field Theory
(statDMFT) \cite{motand,vladgabisdmft2}. A discussion as applied to the
Anderson lattice can be found in reference \cite{aguiaretal1}. The assumption
of retaining only on-site correlations, as in DMFT, is maintained in statDMFT.
Formally, this means that one still has to solve an \emph{ensemble} of single
impurity problems defined by the actions (\ref{eq:siteaction}). The only
difference comes from the prescription for the bath function, which is now
site-dependent: $\Delta_{c}\left(  \tau\right)  \rightarrow\Delta_{c}\left(
j,\tau\right)  $. If the lattice coordination is $z$, we can write%
\begin{equation}
\Delta_{c}\left(  j,\tau\right)  =t^{2}\sum_{l=1}^{z}G_{c}^{\left(  j\right)
}\left(  l,\tau\right)  , \label{eq:bath2}%
\end{equation}
where $G_{c}^{\left(  j\right)  }\left(  l,\tau\right)  $ is the local Green's
function of nearest-neighbor site $l$, \emph{after the removal of site} $j$.
Once again, we are particularizing to the Bethe lattice, which simplifies the
discussion considerably. The Green's function with a nearest-neighbor site
removed is defined as in Eq.~(\ref{eq:greenc}) but has to be calculated with a
different action. This action has the same form as (\ref{eq:siteaction}), but
now the bath function $\Delta_{c}\left(  \tau\right)  \rightarrow\Delta
_{c}^{\left(  j\right)  }\left(  l,\tau\right)  $, where%
\begin{equation}
\Delta_{c}^{\left(  j\right)  }\left(  l,\tau\right)  =t^{2}\sum_{m=1}%
^{z-1}G_{c}^{\left(  l\right)  }\left(  m,\tau\right)  . \label{eq:bath3}%
\end{equation}
Note that in (\ref{eq:bath3}) the summation has only $z-1$ terms, as opposed
to the $z$ terms of (\ref{eq:bath2}), since site $j$ has been removed. The
self-consistency loop is closed by noting that the Green's functions on the
right-hand side of Eq.~(\ref{eq:bath3}) are the same (i. e., obey the same
distributions) as the ones on the right-hand side of Eq.~(\ref{eq:bath2}). We
emphasize that the self-consistency problem is now no longer a simple
algebraic equation but rather a set of stochastic equations in the functions
$G_{c}\left(  j,\tau\right)  $ and $G_{c}^{\left(  j\right)  }\left(
l,\tau\right)  $, whose distributions have to be determined. If the
interactions are turned off, the treatment of disorder we have described is
equivalent to the self-consistent theory of localization \cite{abouetal}. The
latter is known to exhibit a disorder-induced Anderson metal-insulator
transition for coordination $z\geq3$. The formulation above can be generalized
to any lattice \cite{vladgabisdmft2}. However, the full set of stochastic
equations has to be solved numerically, in which case it is advantageous to
work in a Bethe lattice. In a realistic lattice, this approach offers the
advantage of treating the disorder exactly, albeit numerically, while
incorporating the local effects of correlations. In this sense, it is a
natural generalization of the work of Milovanovi\'{c} \emph{et al.}
\cite{milovanovicetal89}, to which it reduces if the single-impurity problems
are treated within Anderson's mean field theory \cite{Anderson}. More
sophisticated treatments of the single-impurity problems are, however,
possible, especially in the low temperature region.

The ability to access full distribution functions of physical quantities is a
especially appealing feature of statDMFT. In particular, the local density of
states%
\begin{equation}
\rho_{c}\left(  j,\omega\right)  =\frac{1}{\pi}\mathrm{Im}\left[  G_{c}\left(
j,\omega-i\delta\right)  \right]  \label{eq:localdos}%
\end{equation}
is intimately connected with the transport properties. It has the natural
interpretation of a \emph{escape rate} from site $j$ and is expected to be
finite if states at energy $\omega$ are extended and to vanish if they are
localized, as first pointed out by Anderson in his ground-breaking work
\cite{andersonloc}. Furthermore, the distribution of $\rho_{c}\left(
j,\omega\right)  $ can become very broad with increasing disorder. Although
the average density of states $\overline{\rho_{c}\left(  \omega\right)  }$
remains finite for any disorder strength, its typical value $\rho_{c}%
^{typ}\left(  \omega\right)  $ vanishes at the metal-insulator transition and
can serve as an order parameter for localization \cite{andersonloc}. A
convenient measure of $\rho_{c}^{typ}\left(  \omega\right)  $ is the geometric
average $\exp\left\{  \overline{\ln\left[  \rho\left(  \omega\right)  \right]
}\right\}  $.

Furthermore, other local quantities of interest can also be computed, such as
the distribution of Kondo temperatures $P\left(  T_{K}\right)  $ already
familiar from DMFT. Recalling that $T_{K}\approx D\exp\left[  -1/\left(  \rho
J\right)  \right]  $, we notice that the density of states $\rho$ {}``seen''
by any particular site $j$ is a fluctuating quantity in statDMFT and is thus a
new source of $T_{K}$ fluctuations. For the case of the $f$-electrons of the
disordered Anderson lattice, for example%
\begin{equation}
\rho_{j}=\frac{1}{\pi}\mathrm{Im}\left[  \frac{V_{j}^{2}}{\omega-\epsilon
_{j}+\mu-\Delta_{c}\left(  j,\omega-i\delta\right)  }\right]  .
\label{eq:fdos}%
\end{equation}
In particular, even if $J=2V^{2}/\left|  E_{f}\right|  $ is not random (e.~g.,
if only conduction electron diagonal disorder is present), there will be a
distribution of Kondo temperatures. Moreover, in DMFT, a discrete distribution
of $J$ leads necessarily to a discrete distribution of $T_{K}$, as $\rho$ is
fixed. This is no longer true in statDMFT, however. Even if $J$ (or,
equivalently, $V$ or $E_{f}$) is discrete, there will be a continuous
distribution of $\rho$ and consequently of $T_{K}$. This is because $\rho_{j}
$ at a particular site is influenced by fluctuations of other quantities at
sites which may be very distant from $j$, due to the extended nature of the
conduction electron wave function in the metallic phase.

\subsubsection{Electronic Griffiths phase in disordered Kondo lattice models
of dirty heavy-fermion materials}

\label{sub:andersongriff}

\begin{figure}[ptb]
\begin{center}
\includegraphics[  width=3.4in,
keepaspectratio]{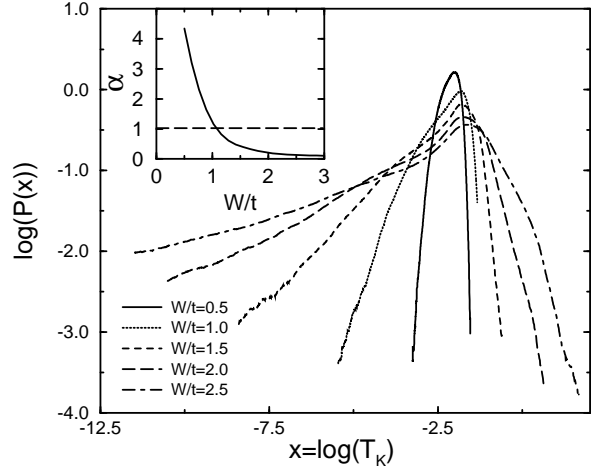}
\end{center}
\caption{Distribution of Kondo temperatures within statDFMT, showing $P\left(
T_{K}\right)  \sim T_{K}^{\alpha-1}$, with $\alpha\left(  W\right)  $
continuously varying with the strength of disorder. The onset of NFL behavior
occurs at $\alpha\le1$. From Ref.~\cite{mirandavlad1}.}%
\label{cap:fig10}%
\end{figure}

The application of statDMFT to the disordered Anderson lattice problem was
described in references
\cite{mirandavlad3,mirandavlad1,mirandavlad2,aguiaretal1}. The single-impurity
problems were solved at $T=0$ within the slave boson mean field field theory
\cite{barnes76,readnewns2,colemansb,colemanlong} or for $T\neq0$ with
second-order perturbation theory
\cite{kajuetergabi,meyernolting1,meyernolting2}. Several important features
deserve mention. The distribution of Kondo temperatures is generically
log-normal for weak disorder but slowly evolves towards a power-law at small
$T_{K}$,
\begin{equation}
P\left(  T_{K}\right)  \sim T_{K}^{\alpha-1}, \label{eq:ptkpower}%
\end{equation}
where the exponent $\alpha$ is a continuously varying function of the disorder
strength $W$, as shown in Fig.~\ref{cap:fig10}. This generic form is fairly
insensitive to the particular form of the bare distribution. This universality
reflects the mixing of many single-impurity problems which are connected by
the extended wave function of the conduction electrons within a correlation volume.

{}From the power-law distribution of energy scales, the usual phenomenology of a
Griffiths phase follows. Although our discussion in this Section is
self-contained, a complete description of a generic quantum Griffiths phase is
deferred to Section~\ref{sub:magneticgriff}. From the distribution
(\ref{eq:ptkpower}), the susceptibility and the specific heat coefficient, for
example, are given by%
\begin{align}
\chi\left(  T\right)   &  \sim\frac{C\left(  T\right)  }{T}\sim\int
dT_{K}\frac{P\left(  T_{K}\right)  }{T+\alpha T_{K}}\nonumber\\
&  \sim\frac{1}{T^{1-\alpha}}. \label{eq:grifsusc}%
\end{align}
For sufficient disorder, $\alpha<1$ and NFL behavior is observed. The marginal
case $\alpha=1$ corresponds to a log-divergence. Note that the point
$\alpha=1$ does not signal any phase transition. Other physical quantities
such as the non-linear susceptibility $\chi^{\left(  3\right)  }\left(
T\right)  $ start to diverge at a \emph{larger} value of $\alpha$
(corresponding to smaller disorder strength), since $\chi^{\left(  3\right)
}\left(  0\right)  \sim1/T_{K}^{3}$. Several other physical quantities can be
obtained as continuous functions of the exponent $\alpha$. Such behavior
appears consistent with many observations in disordered heavy fermion
materials which show NFL behavior \cite{andradeetal,stewartNFL}
(Section~\ref{sub:nflkondoalloys}).

Griffiths phases are common in the proximity of magnetic phase transitions, as
is explained in more detail in Section~\ref{sub:magneticgriff}. Here, however,
the Griffiths phase occurs in the absence of any form of magnetic ordering.
Rather, it is associated with the proximity to the disorder-induced
metal-insulator transition, which occurs for sufficiently large disorder
($W_{MIT}\approx12t$ for conduction electron diagonal disorder only
\cite{mirandavlad1}). This is somewhat reminiscent of the local moment phase
close to the metal-insulator transition in doped semiconductors, but is
different in two important respects: (i) in a disordered Kondo/Anderson
lattice, local moments are assumed stable even in the absence of disorder and
(ii) the NFL divergences begin to show up at fairly weak randomness and
persist for a wide range of disorder strength. Many other features of the
statDMFT solution of the disordered Anderson lattice, particularly concerning
transport properties, can be found in references
\cite{mirandavlad3,mirandavlad1,mirandavlad2,aguiaretal1}.

We should stress that the anomalous power laws obtained in this approach are a
direct consequence of the power law in the distribution of Kondo temperatures.
Interestingly, all that is needed is a power-law distribution of energy scales
for the spin fluctuators, \emph{not necessarily of Kondo origin or related to
localization effects}. Given this form of distribution the associated
Griffiths divergences follow immediately (see a discussion of generic
properties of Griffiths phases in Section~\ref{sub:griffgeneral}). As a
result, widely different microscopic mechanisms can give rise to the same
macroscopic behavior. This will be exemplified later when we discuss magnetic
Griffiths phases (Section~\ref{sub:magneticgriff}), which are usually tied to
a \emph{magnetic} phase transition. Whereas this highlights an interesting
universality of this phenomenology, it makes it hard to distinguish between
different microscopic mechanisms based solely upon macroscopic measurements.

\begin{figure}[ptb]
\begin{center}
\includegraphics[  width=3.4in,
keepaspectratio]{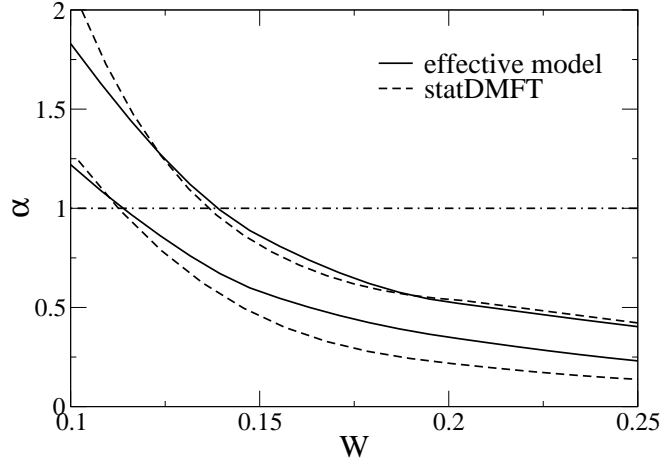}
\end{center}
\caption{Comparison between the $\alpha\left(  W\right)  $ exponent determined
within statDMFT and the one obtained in the effective model of
Ref.~\cite{tanaskovicetal04}.}%
\label{cap:fig11}%
\end{figure}

Some insight into the origin of the power law distribution of $T_{K}$ was
offered in reference~\cite{tanaskovicetal04}. In that work, it was shown how a
power-law distribution of Kondo temperatures can be easily obtained within a
DMFT treatment of the disordered Anderson lattice, if a Gaussian distribution
of conduction electron on-site energies%
\begin{equation}
P_{\epsilon}\left(  \epsilon_{j}\right)  =\frac{1}{\sqrt{2\pi}W}\exp\left(
-\frac{\epsilon_{j}^{2}}{2W^{2}}\right)  \label{eq:pofepsilon}%
\end{equation}
is used, with fixed values of $E_{f}$ and $V$. In this case, although the bath
function is fixed at its average value (Section~\ref{sub:dmft}), different
values of $\epsilon_{j}$ generate different Kondo temperatures (cf.
Eq.~(\ref{eq:siteaction})) according to%
\begin{equation}
T_{Kj}=T_{K}^{0}\exp\left(  -A\epsilon_{j}^{2}\right)  , \label{eq:toymodeltk}%
\end{equation}
where $A$ is a constant and $T_{K}^{0}$ is the Kondo temperature for
$\epsilon_{j}=0$. It is then straightforward to show that, up to log
corrections,%
\begin{equation}
P\left(  T_{K}\right)  \sim T_{K}^{\alpha-1}, \label{eq:ptkpower2}%
\end{equation}
where $\alpha^{-1}=2AW^{2}$. This type of argument, in which an energy scale
depends exponentially on a random variable, which in turn occurs with an
exponential probability, leading to a power law distribution of the energy
scale is very common in, and perhaps generic to Griffiths phases (see
Section~\ref{sub:griffgeneral}). Now, although the specific form of the
distribution (\ref{eq:pofepsilon}) is required within DMFT for the power law
to emerge, in statDMFT \emph{fluctuations of the bath function are generically
gaussian}, at least at weak disorder. By noting that the bath function enters
additively to $\epsilon_{j}$ in the single-site action (cf.
Eq.~(\ref{eq:siteactionc})), it is clear that the appropriate statistics is
naturally generated within this approach. Careful comparison between the two
approaches confirms that this is indeed the case (see Fig.~\ref{cap:fig11}),
thus clarifying the generic nature of the mechanism behind the Griffiths phase
in these systems.

\subsubsection{Electronic Griffiths phase near the Mott-Anderson
metal-insulator transition}

\label{sub:mottgriff}

The statDMFT approach can be usefully employed in the analysis of the
disordered Hubbard model, which serves as a prototypical model for both
disorder-induced (Anderson) as well as interaction-induced (Mott-Hubbard)
metal-insulator transitions. The Hamiltonian in this case is%
\begin{align}
H_{HUB}  &  =-t\sum_{\left\langle ij\right\rangle \sigma}\left(  c_{i\sigma
}^{\dagger}c_{j\sigma}^{\phantom{\dagger}}+\mathrm{H.}\,\mathrm{c.}\right)
+\sum_{j\sigma}\epsilon_{j}c_{j\sigma}^{\dagger}c_{j\sigma}%
^{\phantom{\dagger}}\nonumber\\
&  +U\sum_{j}c_{j\uparrow}^{\dagger}c_{j\uparrow}^{\phantom{\dagger}}%
c_{j\downarrow}^{\dagger}c_{j\downarrow}^{\phantom{\dagger}}.
\label{eq:disHubbard}%
\end{align}
The statDMFT equations in this case are very similar to the Anderson lattice
and we write them down here for completion. The auxiliary single-site actions
read \cite{motand,vladgabisdmft2}%
\begin{align}
S_{eff}^{HUB}\left(  j\right)   &  =\sum_{\sigma}\int_{0}^{\beta}d\tau\int
_{0}^{\beta}d\tau^{\prime}c_{j\sigma}^{\dagger}\left(  \tau\right)  \left[
\delta\left(  \tau-\tau^{\prime}\right)  \right. \nonumber\\
&  \times\left.  \left(  \partial_{\tau}+\epsilon_{j}-\mu\right)  +\Delta
_{c}\left(  j,\tau-\tau^{\prime}\right)  \right]  c_{j\sigma}%
^{\phantom{\dagger}}\left(  \tau^{\prime}\right) \nonumber\\
&  +U\int_{0}^{\beta}d\tau c_{j\uparrow}^{\dagger}\left(  \tau\right)
c_{j\uparrow}^{\phantom{\dagger}}\left(  \tau\right)  c_{j\downarrow}%
^{\dagger}\left(  \tau\right)  c_{j\downarrow}^{\phantom{\dagger}}\left(
\tau\right)  , \label{eq:siteactionhub}%
\end{align}
where the bath function is once again given by Eq.~(\ref{eq:bath2}). The
Green's function with a nearest-neighbor site removed is obtained from an
action of the form (\ref{eq:siteactionhub}), with a bath function given by
Eq.~(\ref{eq:bath3}), which closes the self-consistency loop (again, in a
Bethe lattice).

The analysis of the distributions of two local quantities which come out of
the single-site actions (\ref{eq:siteactionhub}) provide especially useful
insights into the physics of the disordered Hubbard model. One is the local
density of states at the Fermi level $\rho\left(  j,0\right)  $, already
defined in Eq.~(\ref{eq:localdos}). The other is the local Kondo temperature
of each single-site action $T_{Kj}$. The product of these two is the
quasi-particle weight $Z_{i}=\rho\left(  j,0\right)  T_{Kj}$. While the Kondo
temperature governs the thermodynamic response of the system, the local
density of states is related to the transport properties, as explained in
Section~\ref{sub:statdmft}.

\begin{figure}[ptb]
\begin{center}
\includegraphics[  width=3.4in,
keepaspectratio]{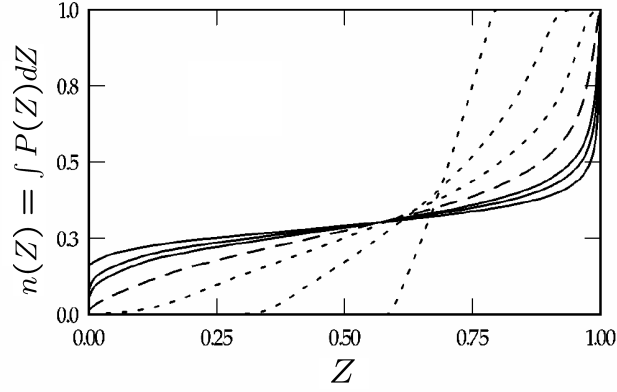}
\end{center}
\caption{Distribution of Kondo temperatures (here named $Z$) within statDMFT
for the disordered Hubbard model. From Ref.~\cite{motand}.}%
\label{cap:fig12}%
\end{figure}

The results of statDMFT as applied to the disordered Hubbard model show that,
similarly to the disordered Anderson lattice, a Griffiths phase emerges, with
NFL behavior, for intermediate values of disorder strength
\cite{motand,vladgabisdmft2}. This is signaled by a power-law distribution of
Kondo temperatures $P\left(  T_{K}\right)  \sim T_{K}^{\alpha-1}$, where
$\alpha$ varies continuously with disorder. The marginal value of $\alpha=1$,
at which the magnetic susceptibility diverges logarithmically with decreasing
temperature, occurs at $W\approx7t$, as shown in Fig.~\ref{cap:fig12}. This is
reminiscent of the behavior of doped semiconductors. In contrast to the
Anderson lattice though, the NFL behavior requires much stronger disorder.
This can be rationalized by noting that in contrast to the latter, where local
moments are stable in the clean limit, here they require quite a large amount
of randomness to appear.

Moreover, for low conduction electron filling ($n\approx0.3$), the typical
local density of states goes to zero at a large value of disorder strength
$W_{MIT}\approx11t$, signaling a metal-insulator transition
\cite{motand,vladgabisdmft2}. As in the non-interacting case, the average
value remains non-critical, actually diverging at $W_{MIT}$. The full
distribution of $\rho\left(  j,0\right)  $ is very close to a log-normal form,
with a width which increases as disorder increases. Closer to half-filling
($n\approx0.7$), however, the behavior is quite different. At these larger
densities the typical value of $\rho\left(  j,0\right)  $ remains finite for
very large values of disorder, suggesting that $W_{c}$ is pushed up
considerably \cite{vladgabisdmft2}. A similar {}``screening'' of the disorder
potential by strong interactions has been also found in a DMFT study of the
disordered Hubbard model \cite{tanaskovicetal03}.

\subsubsection{Incoherent metallic phase and the anomalous resistivity drop
near the 2D MIT}

\label{sub:incoherent}

The puzzle of the two-dimensional metal-insulator transition observed in
several materials \cite{abekravsarachik} remains a largely unsolved problem.
One of the most intriguing features of these systems is the large resistivity
drop observed in a temperature range comparable to the Fermi energy. This
suggests the importance of inelastic scattering processes, a feature usually
absent in Fermi liquid based approaches, or at most included only in a
perturbative fashion. An initial attempt at the incorporation of such
inelastic effects was made within a DMFT framework \cite{aguiaretal2}. A
half-filled disordered Hubbard model was considered in that work. The
associated single-impurity problems were solved within second-order
perturbation theory \cite{kajuetergabi} and checked against Quantum Monte
Carlo calculations. The results were also compared with a Hartree-Fock
impurity solver, which does not include inelastic scattering.

The most important result of that work is the demonstration of the large
contribution of inelastic processes to transport properties in a region of
parameter space where kinetic energy, interaction and disorder strength are of
comparable size. This is especially striking when one compares the results of
DMFT with Hartree-Fock. Whereas in the former, a drop of order 10 in the total
scattering rate is obtained, only a very weak drop is observed in the latter
for the same values of disorder and interaction. This large difference is
clarified when one separates the inelastic from the elastic parts. The
Hartree-Fock scattering rate, which is completely elastic, is comparable to
the elastic contribution in DMFT. However, for comparable values of disorder,
Fermi energy and interaction strengths, the contribution of inelastic
scattering is seen to completely overwhelm the weak elastic processes, in
effect determining the size and temperature dependence of the total rate.

Furthermore, keeping the interaction and Fermi energy of equal
size and decreasing randomness, one is able to enhance even more
the size of the drop in the scattering rate. The more gradual drop
for larger disorder is attributed to a wide distribution of
coherence scales. These scales set the boundary between coherent
transport at low temperatures and incoherent
inelastic-process-dominated transport at higher temperatures. Once
a distribution of these scales is generated, different spatial
regions of the system become incoherent at different temperatures
as the temperatures is raised, leading to a gradual rise in the
overall scattering rate. This rise is approximately linear in the
range where the various coherence scales are equally frequent. An
approximately linear rise of the scattering rate is observed in Si
MOSFET's \cite{abekravsarachik}. Finally, the coherence scale is
inversely proportional to the local effective mass, which in turn
is tied to the thermodynamic response of the system. Thus, when
the disorder is comparable to the interaction strength, a
sufficiently broad distribution of coherence scales is generated
such that very small scales have finite weight, leading to an
enhanced overall effective mass and thermodynamic properties.
Results similar to those DMFT findings were also obtained from
exact numerical studies of finite size lattices
\cite{scalettar-prl99,scalettar-prl01,scalettar-prl03}, but more
work is needed to gain a more definitive understanding of this
regime.

\subsection{Magnetic Griffiths phases}

\label{sub:magneticgriff}

\subsubsection{Quantum Griffiths phases in insulating magnets with disorder:
random singlet formation and the infinite-randomness fixed point}

\label{sub:irfp}

We have seen in Section~\ref{sub:randomsinglet}, how the Bhatt and Lee
numerical study of random antiferromagnetic spin systems
\cite{bhattlee81,bhattlee82} led to the concept of a random singlet phase.
Further insight into this kind of behavior came from a rather complete
analytical treatment of the nearest-neighbor random antiferromagnetic chain
(Eq.~(\ref{eq:heisenberg})) by D. S. Fisher \cite{fisherrandomchain} using the
renormalization group procedure invented by Ma, Dasgupta and Hu and described
in Section~\ref{sub:randomsinglet}. Fisher showed that an essentially exact
solution of the problem could be obtained asymptotically at low energies. He
showed that the relevant fixed point distribution of couplings is infinitely
broad and the system slowly approaches it as it is probed at lower and lower
energies. The system is governed by an {}``infinite randomness fixed point''
(IRFP) which reflects the true physical significance of the random singlet
phase. Note that, if the distribution is very broad, the side couplings of the
strongest bond $\Omega$, $J_{ik}$ and $J_{jl}$ in Eq.~(\ref{eq:perttheory}),
are almost certainly much smaller than $\Omega$, rendering the perturbation
theory result essentially exact. It is in this sense that the Fisher solution
is asymptotically exact. Initial decimations will be inaccurate but if the
system flows towards broader and broader distributions, then the results will
become increasingly more accurate and asymptotically exact. Interestingly,
results at weak disorder \cite{dotyfisher} showed that the clean system is
unstable with respect to infinitesimal disorder and numerical results seem to
confirm that practically any initial randomness is in the basin of attraction
of the IRFP found by Fisher. We will highlight the some of the most important
physical properties of this type of phase. This is a vast subject and we refer
the reader to some recent reviews of experimental and theoretical results of
disordered spin chains and ladders \cite{continentinoborate04,igoimonthus}.

When the system is governed by the IRFP it exhibits many unusual properties.
In particular, the relation between energy ($\Omega$) and length ($L$) scales
is {}``activated''%
\begin{equation}
\Omega\sim\Omega_{0}\exp\left(  -L^{\psi}\right)  , \label{eq:activdynscaling}%
\end{equation}
where $\psi=1/2$ in the Heisenberg chain case. This should be contrasted with
the more common dynamical scaling observed in quantum critical systems
$\Omega\sim L^{-z}$, where $z$ is the dynamical exponent \cite{hertz} (see
Section~\ref{sub:qcp}). Formally, we can say that the {}``activated dynamical
scaling'' of Eq.~(\ref{eq:activdynscaling}) corresponds to a divergent
dynamical exponent $z\rightarrow\infty$. This has many important consequences.
The magnetic susceptibility of the system at temperature $T$ can be obtained
by scaling down to $\Omega=T$. It can then be very accurately calculated
through%
\begin{equation}
\chi\left(  T\right)  \sim\frac{n\left(  \Omega=T\right)  }{T},
\label{eq:irfpsusc}%
\end{equation}
where it is assumed that the spins that were decimated at larger scales form
magnetically inert singlets while the remaining ones, whose density is
$n\left(  \Omega=T\right)  $ are almost free and contribute a Curie-like term.
This estimate is better, the broader the distribution of couplings. If we now
write%
\begin{equation}
n^{-1}\left(  \Omega=T\right)  \sim L\left(  \Omega=T\right)  \sim\left[
\ln\left(  \Omega_{0}/T\right)  \right]  ^{1/\psi}, \label{eq:irfpnumber}%
\end{equation}
where use was made of Eq.~(\ref{eq:activdynscaling}), we can write for the
random Heisenberg chain%
\begin{equation}
\chi\left(  T\right)  \sim\frac{1}{T\ln^{1/\psi}\left(  \frac{\Omega_{0}}%
{T}\right)  }\sim\frac{1}{T\ln^{2}\left(  \frac{\Omega_{0}}{T}\right)  }.
\label{eq:irfpsusc2}%
\end{equation}
Similarly, the entropy density $s\left(  T\right)  \sim n\left(
\Omega=T\right)  \ln2$, from which follows%
\begin{equation}
C\left(  T\right)  /T\sim\frac{1}{T\left[  \ln\left(  \Omega_{0}/T\right)
\right]  ^{\left(  \psi+1\right)  /\psi}}\sim\frac{1}{T\ln^{3}\left(
\Omega_{0}/T\right)  }. \label{eq:irfpgamma}%
\end{equation}
Furthermore, the spin-spin correlation functions are very broadly distributed.
Fisher showed that while the \emph{average} correlation function decays as a
power law ($C_{ij}\equiv\left\langle \mathbf{S}_{i}\cdot\mathbf{S}%
_{j}\right\rangle $)%
\begin{equation}
\overline{C_{ij}}\sim\frac{\left(  -1\right)  ^{i-j}}{\left|  i-j\right|
^{2}}, \label{eq:meancorr}%
\end{equation}
the \emph{typical} one, given by the geometric average, decays as a stretched
exponential%
\begin{equation}
\exp\left(  \overline{\ln\left|  C_{ij}\right|  }\right)  \sim\exp\left(
-\left|  i-j\right|  ^{\psi}\right)  \sim\exp\left(  -\sqrt{\left|
i-j\right|  }\right)  . \label{eq:typicalcorr}%
\end{equation}
This large difference between average and typical correlations is a result of
the fact that a typical pair of spins separated by a distance $L\sim\left|
i-j\right|  $ does \emph{not} form a random singlet and is therefore very
weakly correlated. Those rare pairs that do form a random singlet, however,
are strongly correlated and dominate the average.

Further extensions of these studies deserve mention. Random chains with higher
spins have been investigated. Integer spin chains are especially interesting
because the clean system is gapped (Haldane gap
\cite{haldaneconj1,haldaneconj2}). It has been shown that the $S=1$ chain (and
presumably other integer $S$ ones) are stable with respect to weak enough
disorder, retaining a (pseudo-)gap
\cite{boechatetal96,yangetal,hymanyang97,monthusetal97,monthusetal98,hida99,saguiaetal}%
For larger disorder, the pseudogap is destroyed and a Griffiths phase is
realized, with a diverging power-law susceptibility with a varying
non-universal exponent. At sufficiently large disorder, a random singlet phase
is obtained governed by an IRFP. Similar results were also obtained in other
systems for which the clean analogue is gapped: random chains in which the gap
is induced by a dimerization of the couplings
\cite{hymanetal,yangetal,heneliusgirvin98} and the two-leg ladder
\cite{melinetal,yusufyang1,hoyosmiranda1}. A Griffiths phase was also
identified in a disordered Kondo necklace model \cite{Doniach}, in which the
spin sector of the one-dimensional Kondo lattice is mimicked by a double spin
chain system \cite{rappoportetal03}.

It has been proposed that the $S=3/2$ antiferromagnetic random chain realizes
two kinds of random singlet phases separated by a quantum phase transition
\cite{refaeletal02}. At weaker randomness, the \emph{excitations} of the
random singlet phase have $S=1/2$ whereas in the stronger disorder phase they
have $S=3/2$. The quantum critical point between these two phases can be
viewed as a multicritical point in a larger parameter space which includes the
effects of an added dimerization of the couplings \cite{damlehuse}. The
physics of this and other higher spin multicritical points has been shown to
be governed by infinite randomness fixed points with different exponents: for
example, $\psi=1/4$ for the $S=3/2$ chain. The phase diagram of the $S=3/2$
chain proposed in Ref.~\cite{refaeletal02} has been recently disputed, however
\cite{saguiaetal03,carlonetal04}.

Another important universality class of disordered spin chains is found in
systems with random ferromagnetic and antiferromagnetic interactions. In this
case, it is possible to use a generalization of the decimation procedure
described above for any spin size and both signs of couplings
\cite{westerbergetal95,westerbergetal,hida97,frischmuthsigrist97,frischmuthetal99,hikiharaetal}%
. In general, the renormalization group flow generates spins whose average
size grows without limit as a power law. In fact, these large spin clusters
grow in a fashion which can be described as a {}``random walk in spin space'',
so that the total spin and the cluster size scale as%
\begin{equation}
\overline{S^{2}}\sim L. \label{eq:squaredspin}%
\end{equation}
Moreover, the energy-length scale relation has a conventional power-law form
$\Omega\sim L^{-z}$. For disorder distributions which are not too singular,
the dynamical exponent has a universal value $z\approx4.5$ (for stronger
disorder, the value of $z$ is non-universal). Employing similar arguments as
outlined above for the random singlet phase one finds%
\begin{align}
\chi\left(  T\right)   &  \sim\frac{1}{T},\label{eq:largespinsusc}\\
C\left(  T\right)  /T  &  \sim\frac{\ln\left(  1/T\right)  }{T^{1-1/z}}.
\label{eq:largespingamma}%
\end{align}
Interestingly, the $z$ exponent does not show up in the susceptibility to
leading order. This is because there is a cancellation of the temperature
dependences of the number of spin clusters and squared spin of the clusters
(Eq.~(\ref{eq:squaredspin}))%
\begin{equation}
\chi\left(  T\right)  \sim\frac{\overline{S^{2}}\left(  \Omega=T\right)
n\left(  \Omega=T\right)  }{T}\sim\frac{1}{T}. \label{eq:largespinsusc2}%
\end{equation}

This phase has been dubbed a {}``large spin phase'' and has been also
identified in some spin ladder systems
\cite{melinetal,yusufyang03,yusufyang2,hoyosmiranda1}. In fact, the larger
connectivity of spin ladders and higher dimensional systems favors the
formation of ferromagnetic couplings and large spins in the decimation
procedure. This is easily seen if we consider the decimation of an
antiferromagnetically coupled spin pair such that two other spins are
initially antiferromagnetically coupled to the \emph{same spin}. It is
physically evident and easy to show that the effective coupling generated
between the latter spins is \emph{ferromagnetic}. Thus, unless the lattice
geometry is fine tuned (an example of which is the two-leg spin ladder
\cite{melinetal,yusufyang1,hoyosmiranda1}), there is a strong tendency towards
a large spin phase formation. The singular behavior first discovered by Bhatt
and Lee in higher dimensions is thus only a crossover and gives way to a large
spin phase behavior at lower temperatures \cite{motrunichetal01,linetal03}.

Generically, Griffiths phases seem to be a ubiquitous phenomenon in these
random insulating magnets, usually in the vicinity of an infinite randomness
fixed point. Systems with an Ising symmetry are especially interesting. Once
again, a great deal of insight has been gained from an essentially exact
analytical solution in the vicinity and at the quantum critical point of the
random transverse field Ising chain due to D. S. Fisher
\cite{fishertransising,fishertransising2} (see also \cite{shankarmurthy87}).
The clean version of this system has a quantum phase transition between an
Ising ferromagnet and a disordered paramagnet
\cite{liebschultzmattis61,pfeuty70}. The disordered version of the model is
related to the McCoy-Wu model \cite{mccoywu68,mccoywu69,mccoy69}, in which
context quantum Griffiths singularities were first studied (hence the name
Griffiths-McCoy singularities sometimes used). It can be written as%
\begin{equation}
H_{TFIM}=-\sum_{i}J_{i}\sigma_{i}^{z}\sigma_{i+1}^{z}-\sum_{i}h_{i}\sigma
_{i}^{x}. \label{eq:rtfimham}%
\end{equation}
It also has a quantum phase transition tuned by the difference between the
average exchange ($\Delta_{J}=\overline{\ln J}$) and the average transverse
field ($\Delta_{h}=\overline{\ln h}$). Through a generalization of the
Ma-Dasgupta-Hu strong disorder renormalization group procedure, Fisher was
able to solve for the effective distributions of exchange and transverse field
couplings in the quantum critical region. The quantum critical point itself
was shown to have all the features of an IRFP, including activated dynamical
scaling (with $\psi=1/2$), diverging thermodynamic responses%
\begin{align}
\chi\left(  T\right)   &  \sim\frac{\left[  \ln\left(  1/T\right)  \right]
^{2\phi-2}}{T},\label{eq:rtfimsusc}\\
C\left(  T\right)  /T  &  \sim\frac{1}{T\ln^{3}\left(  1/T\right)  },
\label{eq:rtfimgamma}%
\end{align}
and widely different average and typical correlation functions ($C^{a}\left(
x\right)  \equiv\left\langle \sigma_{i}^{a}\cdot\sigma_{i+x}^{a}\right\rangle
$, $a=x,z$)%
\begin{align}
\overline{C^{a}}\left(  x\right)   &  \sim\frac{1}{x^{2-\phi}}%
,\label{eq:rtfimmeancorr}\\
\overline{\ln C^{a}}\left(  x\right)   &  \sim-\sqrt{x},
\label{eq:rtfimtypcorr}%
\end{align}
where $\phi=\left(  1+\sqrt{5}\right)  /2$, the golden mean. Note the
similarity with the result for the Heisenberg case if $\phi$ is taken to be
zero. The singular behavior is attributed to the presence of large spin
clusters at low energies which can tunnel between two different configurations
with reversed magnetizations. The behavior off but near criticality, however,
is characteristic of a Griffiths phase. Defining
\begin{equation}
\delta=\frac{\Delta_{h}-\Delta_{J}}{\mathrm{var}\left(  \ln h\right)
+\mathrm{var}\left(  \ln J\right)  }, \label{eq:rtfimdelta}%
\end{equation}
where $\mathrm{var}\left(  O\right)  \equiv\overline{O^{2}}-\overline{O}^{2}$
is the variance of the variable $O$, it is found that, in the slightly
disordered region ($\Delta_{h}\agt\Delta_{J}$)%
\begin{align}
\chi\left(  T\right)   &  \sim\delta^{4-2\phi}\frac{\left[  \ln\left(
1/T\right)  \right]  ^{2}}{T^{1-1/z}},\label{eq:rtfimsusc2}\\
C\left(  T\right)  /T  &  \sim\delta^{3}\frac{1}{T^{1-1/z}},
\label{eq:rtfimgamma2}%
\end{align}
whereas in the slightly ordered region ($\Delta_{h}\alt\Delta_{J}$)%
\begin{align}
\chi\left(  T\right)   &  \sim\left|  \delta\right|  ^{2-2\phi}\frac
{1}{T^{1+1/z}},\label{eq:rtfimsusc3a}\\
C\left(  T\right)  /T  &  \sim\left|  \delta\right|  ^{3}\frac{1}{T^{1-1/z}},
\label{eq:rtfimgamma3a}%
\end{align}
where $z\sim1/\left(  2\left|  \delta\right|  \right)  $ is a non-universal
disorder-dependent exponent, typical of Griffiths phases. An exact expression
for $z$ in terms of the bare distributions was obtained in
Ref.~\cite{igloietal01}. The typical and average correlations both decay
exponentially at large distances but with different correlation lengths%
\begin{align}
\xi_{typ}  &  \sim\frac{1}{\left|  \delta\right|  },\label{eq:typcorrlength}\\
\xi_{av}  &  \sim\frac{1}{\delta^{2}}, \label{eq:avcorrlength}%
\end{align}
such that $\xi_{av}\gg\xi_{typ}$ and $\overline{C^{z}}\left(  x\right)
\gg\exp\overline{\ln C^{z}}\left(  x\right)  $. The exponent of the
correlation length of the average correlation function is the one that
satisfies (in fact, saturates) the bound $\nu\geq2/d=2$ of
Ref.~\cite{chayesetal86}, which is the generalization of the Harris criterion
\cite{harris74} to a disordered critical point.

There has been independent (i.e., not based on the approximate RG scheme)
numerical confirmation of Fisher's results on the 1D random transverse field
Ising model through the mapping to free fermions \cite{youngrieger96}.
Furthermore, similar behavior was also found in higher dimensions by Quantum
Monte Carlo \cite{pichetal98} and by a numerical implementation of the above
decimation procedure \cite{motrunichetal01}. In addition, a random
\emph{dilute} Ising model in a transverse field has been shown to have these
same general properties close to the percolation transition in any dimension
$d>1$ \cite{senthilsachdev}.

A similar scenario has also been discussed phenomenologically for a quantum
Ising spin glass transition in higher dimensions \cite{thillhuse95}. The
presence of a Griffiths phase in a model of a quantum Ising spin glass in a
transverse field has been confirmed numerically in $d=2$ and 3
\cite{guoetal94,riegeryoung94,riegeryoung96,guoetal96}, although the IRFP was
neither confirmed nor ruled out at the quantum phase transition point due to
the small lattice sizes used in those studies. However, it is believed that
this quantum critical point is also governed by an IRFP. The argument is based
on the IRFP nature of the random transverse field Ising model quantum critical
point in higher dimensions \cite{pichetal98,motrunichetal01} and the
realization that frustration is \emph{irrelevant} at the IRFP
\cite{motrunichetal01}. Indeed, at low energies the weakest link of a given
loop of spins is almost certainly infinitely weaker than the other links.
Therefore, one can neglect it and the loop is always unfrustrated.

\subsubsection{General properties of quantum Griffiths phases}

\label{sub:griffgeneral}

The general nature of Griffiths phases can be elucidated in a simple and
general fashion as we now show. We have seen in
Section~\ref{sub:andersongriff} how it arises in the context of the electronic
Griffiths phase. In that case, the fluctuators are local moments coupled by a
Kondo interaction $J_{i}$ to the fluid of conduction electrons. The energy
scale governing the local moment dynamics is the Kondo temperature, which
depends exponentially on the combination $\rho_{i}J_{i}$, $T_{Ki}\approx
D\exp\left(  -1/\rho_{i}J_{i}\right)  $, where $\rho_{i}$ is the local
conduction electron density of states. We showed how the exponential tail of
the distribution of the quantity $1/\rho_{i}J_{i}$ leads to a power law
distribution of Kondo temperatures, $P\left(  T_{K}\right)  \sim T_{K}%
^{\alpha-1}$. This exponential tail is naturally obtained from localization
effects. The final result is a collection of essentially independent Kondo
spins whose characteristic energy scales are distributed in a power law fashion.

The magnetic Griffiths phase arises in a wholly analogous fashion. In
the vicinity of a quantum phase transition spatial fluctuations due to
disorder generate droplets of the ordered phase inside a system which
is on the whole in the disordered phase and vice-versa. Let us for
simplicity focus on the disordered side of the transition. For the
simplest case of uncorrelated disorder, the statistics of these rare
regions is Poissonian and the probability of a region of the ordered
phase with volume $L^{d}$ is%
\begin{equation}
P\left(  L^{d}\right)  \sim\exp\left(  -cL^{d}\right)  , \label{eq:poisson}%
\end{equation}
where $c$ is a constant tuned by the disorder strength. This droplet of
ordered phase has a net magnetic moment. In the case of a ferromagnetic
transition, the moment is proportional to the droplet size, whereas in the
case of an antiferromagnet the net droplet moment is $\propto L^{\left(
d-1\right)  /2}$ due to random incomplete spin cancellations on the surface of
the droplet. These net moments are subject to quantum tunneling between states
with reversed magnetizations. Since the driving mechanism is single-spin
tunneling the total tunneling rate of a droplet is exponential on the number
of spins%
\begin{equation}
\Delta\sim\omega_{0}\exp\left(  -bL^{d}\right)  , \label{eq:tunnelingrate}%
\end{equation}
where $b$ is another constant related to the microscopic tunneling mechanism
and $\omega_{0}$ is some frequency cutoff. The tunneling rate gives the energy
splitting between the lowest and first excited states of the droplet, higher
excited states lying far above in energy. Now, the interesting thing is that
the exponentially small probability of a large cluster is {}``compensated'' by
the exponentially large tunneling times between states, such that the
distribution of energy splittings (gaps) is given by a \emph{power law}%
\begin{equation}
P\left(  \Delta\right)  \sim\int dL^{d}P\left(  L^{d}\right)  \delta\left[
\Delta-\omega_{0}\exp\left(  -bL^{d}\right)  \right]  \sim\Delta^{\alpha-1},
\label{eq:pdelta}%
\end{equation}
where $\alpha=c/b$. We are then left with a distribution of independent
droplets, whose frequencies are distributed according to a power law. Thus,
although the microscopic origin of the magnetic and the electronic Griffiths
phases are very different, the end result, namely, a collection of fluctuators
whose energy scales are distributed in a power-law fashion, is common to both mechanisms.

Given the power-law distribution of energy scales, thermodynamic and dynamical
properties follow immediately. For example, the spin susceptibility at
temperature $T$ can be calculated by assuming that all the droplets (Kondo
spins) with $\Delta>T$ ($T_{K}>T$) are frozen in the non-magnetic ground
state, whereas all the others are essentially free and give a Curie response.
The error introduced by the borderline moments is small if the distribution is
broad as in Eq.~(\ref{eq:pdelta}). Thus, the susceptibility is given by the
number of free spins at temperature $T$, $n\left(  T\right)  =\int_{0}%
^{T}d\Delta P\left(  \Delta\right)  $ times $1/T$%
\begin{equation}
\chi\left(  T\right)  \sim\frac{n\left(  T\right)  }{T}\sim\frac
{1}{T^{1-\alpha}}. \label{eq:griffsusc}%
\end{equation}
Note that, in the magnetic droplet case, we have neglected the droplet moment
$\mu\propto L^{\phi}\propto\ln^{\phi/d}\Delta$ (where $\phi$ depends on the
nature of the magnetic order), because it only gives rise to a negligible
logarithmic correction to the power law (see, however,
Section~\ref{sub:griffapplicab}). Analogously, the total entropy $S\left(
T\right)  \propto n\left(  T\right)  $, such that the specific heat is given
by%
\begin{equation}
\frac{C\left(  T\right)  }{T}=\frac{dS\left(  T\right)  }{dT}\sim P\left(
T\right)  \sim\frac{1}{T^{1-\alpha}}. \label{eq:griffgamma}%
\end{equation}
Similarly, the droplet dynamical susceptibility is%
\begin{equation}
\chi_{drop}^{\prime\prime}\left(  \omega\right)  \sim\tanh\left(  \frac
{\Delta}{2T}\right)  \delta\left(  \omega-\Delta\right)  .
\label{eq:dropdynsusc}%
\end{equation}
Averaging over the droplet distribution%
\begin{equation}
\chi^{\prime\prime}\left(  \omega\right)  \sim\omega^{\alpha-1}\tanh\left(
\frac{\omega}{2T}\right)  . \label{eq:griffdynsusc}%
\end{equation}
Interestingly, if the fluctuator has relaxational dynamics (as in the case of
a Kondo spin)%
\begin{equation}
\chi_{rel}^{\prime\prime}\left(  \omega\right)  \sim\frac{\omega\chi\left(
T\right)  \Gamma\left(  T\right)  }{\omega^{2}+\Gamma^{2}\left(  T\right)  },
\label{eq:reldynsusc}%
\end{equation}
where $\Gamma\left(  T\right)  $ is the relaxation rate and $\chi\left(
T\right)  $ is the static susceptibility. Typically at high temperatures,
$\Gamma\left(  T\right)  \sim T$, whereas as $T\rightarrow0$, $\Gamma\left(
T\right)  \sim\Delta$. Since $\chi\left(  T\right)  \Gamma\left(  T\right)  $
is an almost constant smooth function of $T$, we can ignore it in
Eq.~(\ref{eq:reldynsusc}). We thus obtain at low temperatures after averaging%
\begin{equation}
\chi^{\prime\prime}\left(  \omega\right)  \sim\overline{\frac{\omega}%
{\omega^{2}+\Delta^{2}}}\sim\omega^{\alpha-1}, \label{eq:griffdynsusc2}%
\end{equation}
which has the same form as Eq.~(\ref{eq:griffdynsusc}).

Similar considerations apply to the magnetization as a function of
magnetic field $H$. The fluctuator magnetization has the following
limiting behaviors\begin{equation}
M_{fluc}\left(H,T,\Delta\right)\sim\left\{ \begin{array}{cc}
\mu & H\gg T,\Delta;\\
H/\Delta & H\ll T\ll\Delta;\\
H/T & H\ll\Delta\ll T,\end{array}\right.\label{eq:fluctuatormagn}\end{equation}
 where $\mu$ is the fluctuator moment. This applies both to a magnetic
droplet in a transverse field~\cite{castronetojones} and to a Kondo
spin ($\Delta=T_{K}$)~\cite{hewson}. Averaging over $\Delta$ with
$P\left(\Delta\right)$ from Eq.~(\ref{eq:pdelta}), it is clear
that the large field behavior is dominated by the fluctuators with
$\Delta\ll H$, such that\begin{eqnarray}
M\left(H,T\right) & = & \overline{M_{fluc}\left(H,T,\Delta\right)}=\overline{\mu}\int_{0}^{H}d\Delta P\left(\Delta\right)\\
 & = & \overline{\mu}n\left(H\right)\propto H^{\alpha},\label{eq:griffmagnetization}\end{eqnarray}
where $\overline{\mu}$ is an average moment (we have again neglected
the possible logarithmic contribution coming from the moment size).
At lower fields, this crosses over to a linear behavior, with a coefficient
given by the (singular) susceptibility of Eq.~(\ref{eq:griffsusc}).

The physical properties of Griffiths phases are therefore rather generic and
are determined almost exclusively by the exponent of the power law
distribution of energy scales, quite independent of the underlying mechanism.
This makes it very difficult to distinguish microscopic models solely on the
basis of this generic power law behavior of macroscopic properties. We will,
however, point out in Section~\ref{sub:griffapplicab} some possible
measurements which can distinguish between different microscopic mechanisms.

\subsubsection{Possibility of quantum Griffiths phases in itinerant random
magnets}

\label{sub:metalgriff}

The discovery of a plethora of quantum Griffiths phases in the vicinity of
phase transitions of several models of disordered insulating magnets
immediately suggests the possibility of similar phenomena in metallic
compounds. As we have seen in Section~\ref{sub:nflkondoalloys}, many of the
systems of interest have the required ingredients, namely, the presence of
extrinsic disorder and/or some kind of magnetic order whose critical
temperature can be tuned to zero by some external parameter such as external
or chemical pressure. Furthermore, the diverging susceptibility with tunable
exponents observed close to the metal-insulator transition in doped
semiconductors \emph{on the metallic side} was found to have many similarities
to Griffiths singularities
\cite{bhattlee81,bhattlee82,paalanen,bhattfisher92,vladtedgabi}.

\begin{itemize}
\item \emph{Quantum Griffiths phase in the theory of Castro Neto, Castilla and
Jones }
\end{itemize}

The possibility of a magnetic Griffiths phase as the origin of NFL behavior in
Kondo alloys was first put forward by Castro Neto, Castilla and Jones
\cite{castronetoetal1}. In that work, a disordered anisotropic Kondo lattice
system was analyzed, since heavy fermion systems are known to exhibit large
spin-orbit interactions which tend to strongly break Heisenberg spin SU(2)
symmetry. Due to the competition between the RKKY interaction and the Kondo
effect in a disordered environment, it was argued that large spatial
fluctuations would give rise to regions where either of the two competing
tendencies would dominate. The local moments that belong to a region where the
local Kondo temperature is especially large are efficiently quenched at low
temperatures and contribute to the formation of a disordered heavy Fermi
liquid. On the other hand, anomalously small Kondo temperature fluctuations
give rise to local spins which, though still subject to weak Kondo spin-flip
processes, are left to interact with other similar partners through the RKKY
interaction. Assuming strong anisotropy, the magnetic interactions lead to a
random Ising model for the unquenched spins. Furthermore, the Kondo spin-flip
terms were shown to give rise to an effective transverse field, i.e., if the
Ising model variables are $\sigma_{i}^{z}$ , anisotropic Kondo scattering
generates a local field in the $x$-direction. One is then left with an
effective random Ising model in a transverse field, which was discussed in
Section~\ref{sub:irfp}. As we have seen, there is numerical evidence that the
order-disorder quantum phase transition of this model is governed by an IRFP
flanked on both sides by Griffiths phases with non-universal tunable
exponents. In the picture proposed by Castro Neto \emph{et al.}, alloying
tends to enhance quantum (Kondo) correlations, which act as a destabilizing
mechanism on long-range magnetic order. Thus, a power law distribution of
energy scales is immediately obtained as is generically expected within
Griffiths phases, as explained in Section~\ref{sub:griffgeneral}. The
calculation of physical quantities then follows from the arguments of that
Section and they obtained, in terms of a non-universal tunable parameter
$\lambda$ (which is the same as the $\alpha$ of Section~\ref{sub:griffgeneral}%
)%
\begin{equation}
\chi\left(  T\right)  \sim\frac{C\left(  T\right)  }{T}\sim T^{\lambda-1},
\label{eq:suscandgamma}%
\end{equation}
\begin{equation}
\chi^{\left(  3\right)  }\left(  T\right)  \sim T^{\lambda-3}, \label{eq:chi3}%
\end{equation}
\begin{equation}
\chi_{loc}^{\prime\prime}\left(  \omega\right)  \sim\omega^{\lambda-1}%
\tanh\left(  \omega/T\right)  , \label{eq:chidoubleprime}%
\end{equation}
\begin{equation}
T_{1}^{-1}\left(  T\right)  \sim\omega^{\lambda-2}T\tanh\left(  \omega
/T\right)  , \label{eq:t1}%
\end{equation}
\begin{equation}
\delta\chi\left(  T\right)  /\chi\left(  T\right)  \sim T^{-\lambda/2},
\label{eq:delchi}%
\end{equation}
where $\chi^{\left(  3\right)  }\left(  T\right)  $ is the non-linear magnetic
susceptibility, $\chi_{loc}^{\prime\prime}\left(  \omega\right)  $ is the
imaginary part of the local frequency-dependent susceptibility, $T_{1}\left(
T\right)  $ is the NMR spin relaxation time, and $\delta\chi\left(  T\right)
$ is the root mean square susceptibility due to the disorder fluctuations. As
we have seen, $\lambda<1$ implies NFL behavior. Some of these predictions have
found experimental support in the NFL Kondo alloys Th$_{1-x}$U$_{x}$Pd$_{2}%
$Al$_{3}$, Y$_{1-x}$U$_{x}$Pd$_{3}$, and UCu$_{5-x}$M$_{x}$ (M = Pd, Pt ,
where the specific heat and susceptibility values could be well fitted to the
forms in Eq.~(\ref{eq:suscandgamma}) with $\lambda$ ranging from about 0.6 to
about 1 \cite{andradeetal}. Some other systems have been reanalyzed by Stewart
in his review\ \cite{stewartNFL} in light of the above results and found to be
describable by such power laws with appropriately chosen values of $\lambda$.

Here again, the anomalous power laws are a result of a power-law distribution
of energy scales for the spin fluctuators (see the discussion in
Section~\ref{sub:griffgeneral}). In the specific case here, in which the
system teeters at the onset of magnetic order, these are large clusters of one
phase (say, the ordered one) inside the other phase (the disordered one).
Sample averaging then leads to the results in Eqs.~(\ref{eq:suscandgamma}%
-\ref{eq:delchi}). Although these are different from the low-$T_{K}$ spins of
the electronic Griffiths phase, the same phenomenology is obtained in either case.

\begin{itemize}
\item \emph{Effects of dissipation by the conduction electrons }
\end{itemize}

These early results of the magnetic Griffiths phase route towards NFL behavior
did not, however, take into account the effect of dissipation on the tunneling
of the Ising spins caused by the excitation of conduction electron
particle-hole pairs of the metallic host. These were later incorporated in a
more complete treatment by Castro Neto and Jones \cite{castronetojones}. In
that work, the authors considered essentially the same anisotropic Kondo
lattice model. The same competition between the RKKY interaction and the Kondo
effect in a disordered environment was argued to lead, in general, to the
formation of clusters of spins locked together by their mutual interactions
and with widely distributed sizes. These clusters are able to tunnel at low
temperatures between two different configurations with reversed signs of
magnetization (in the ferromagnetic case) or staggered magnetization (in the
antiferromagnetic case). The tunneling is caused either by residual
anisotropic RKKY interactions or by the Kondo spin-flip scattering with the
conduction electrons. The latter was called a \emph{cluster Kondo effect}. The
picture is thus similar to the previous work but now there is a distribution
of cluster sizes $P\left(  N\right)  $, which is argued to be given by
percolation theory as%
\begin{equation}
P\left(  N\right)  =\frac{N^{1-\theta}e^{-N/N_{\xi}}}{\Gamma\left(
2-\theta\right)  N_{\xi}^{2-\theta}}. \label{eq:percolationpofn}%
\end{equation}
Here, $\theta$ is a critical exponent and $N_{\xi}$ is a correlation volume
(given in terms of the number of spins enclosed by it) that diverges as
$N\sim\left|  p-p_{c}\right|  ^{-\nu}$ close to the percolation threshold
$p_{c} $. The origin of percolation here is the alloying-induced growth of the
quenched spin fluid, which acts as an inert pervasive heavy Fermi liquid.

For a general cluster with $N$ spins, the \emph{bare} tunneling splitting
between the two configurations scales like
\begin{equation}
\Delta_{0}\left(  N\right)  =\omega_{0}e^{-\gamma N}, \label{eq:tunnelsplit}%
\end{equation}
where $\gamma$ is related to the single-spin-flip mechanism and $\omega_{0}$
is an attempt frequency. Moreover, the authors also included the
renormalization generated by dissipation due to the conduction electrons in
the cluster tunneling processes. The dissipation constant of a cluster was
shown to scale like $\alpha\left(  N\right)  =\left(  N/N_{c}\right)
^{\varphi}$, where $\varphi$ is an exponent dependent on the specific ordering
wave vector or tunneling mechanism. This dissipative two-level system is known
to be governed by the scale \cite{leggettetal87,lesagesaleur98}%
\begin{equation}
T_{K}\left(  N\right)  \sim\frac{\Lambda}{\alpha\left(  N\right)  }\left[
\frac{\Delta_{0}\left(  N\right)  }{\Lambda}\right]  ^{1/\left[
1-\alpha\left(  N\right)  \right]  }, \label{eq:clustertk}%
\end{equation}
where $\Lambda$ is a non-universal high-energy cutoff. The renormalization
acts to suppress the tunneling rate, which should be viewed as an effective
\emph{cluster} Kondo temperature. When $\alpha\left(  N\right)  <1$, the
cluster exhibits damped oscillations between the two configurations, whereas
for $\alpha\left(  N\right)  >1$ dissipation completely suppresses tunneling
and the cluster is frozen, behaving like a superparamagnetic particle. From
the form of $\alpha\left(  N\right)  $ it can be seen that tunneling will
cease to happen for sufficiently large clusters, i.e., for $N>N_{c}$, where
$\alpha\left(  N_{c}\right)  =1$. The authors then proceeded to treat this
collection of clusters of different sizes by a generalization of the quantum
droplet model of Thill and Huse \cite{thillhuse95} to the dissipative case.

The $T_{K}$ distribution for small values of $T_{K}$ is a result of the
exponentially rare large clusters imposed by Eq.~(\ref{eq:percolationpofn})
and the exponentially large tunneling times given by
Eqs.~(\ref{eq:tunnelsplit}) and (\ref{eq:clustertk}) for large values of
$N<N_{c}$. As usual (see the discussion in Section~\ref{sub:griffgeneral}),
this yields the familiar Griffiths phase power law dependence $P\left(
T_{K}\right)  \sim T_{K}^{1-\lambda}$, with a tunable $\lambda,$ which in turn
is at the origin of the singular thermodynamic properties. The new element
here is the cutoff imposed by dissipation at $N=N_{c}$. It replaces the above
power law, at very low $T_{K}$, by an extremely singular form
\begin{equation}
P\left(  T_{K}\right)  \sim\frac{1}{T_{K}\ln\left(  T_{K}/\omega_{0}\right)
}. \label{eq:singularptk}%
\end{equation}
The power law behavior is valid above a crossover scale $T_{K}\agt T_{K}%
^{\ast}$, whereas Eq.~(\ref{eq:singularptk}) is obtained for $T_{K}%
\alt T_{K}^{\ast}$, where $T_{K}^{\ast}$ can be related to $N_{c}$
\cite{castronetojones}. Thus, for scales above $T_{K}^{\ast}$ Griffiths
effects should be observable. However, for temperatures below $T_{K}^{\ast}$,
more singular responses are obtained, such as $\chi\left(  T\right)
\sim1/\left[  T\ln\left(  \Lambda/T\right)  \right]  $ (superparamagnetism).
The conclusion then is that the dissipation caused by a metallic environment
has a dramatic effect on Griffiths phase behavior, effectively suppressing it
at low enough temperatures and confining it to an intermediate temperature
crossover regime.

\begin{itemize}
\item \emph{Theory of Millis, Morr, and Schmalian }
\end{itemize}

Another important work on the possibility of Griffiths singularities in
metallic disordered systems is the one by Millis, Morr, and Schmalian
\cite{millisetal01,millisetal02}. These authors considered the effects of
disorder fluctuations on an almost critical system with Ising symmetry both
with and without dissipation due to conduction electrons. The approach is
based on a coarse-grained Landau action description, possibly modified by an
ohmic dissipation term, in which disorder fluctuations couple to the quadratic
term, locally changing the value of $T_{c}$. The authors confined themselves
to situations in which the effective dimensionality in the Hertz sense
\cite{hertz} $d_{eff}=d+z$ is above the upper critical dimension 4 and the
analysis can be carried out at the mean field level (see Section~\ref{sub:qcp}%
). They then considered the behavior of rare disorder fluctuations which are
able to nucleate a locally ordered region ({}``droplet'') inside the
paramagnetic phase. In the first work \cite{millisetal01}, the behavior of a
single droplet was analyzed. Although they considered point, line and planar
defects, we will focus here on the case of point droplets. It was shown that
the order parameter decays very slowly ($\sim1/r$) in the region right outside
the droplet core (defined approximately as the region where the local
$T_{c}\left(  r\right)  >T$) and not further from it than the clean
correlation length $\xi$. This slow decay region, which can be very large
close to criticality, is essential when it comes to dissipation. They show
that these droplets will tunnel between two reversed-magnetization
configurations at a rate given by the usual expression for a dissipative
two-level system \cite{leggettetal87}, analogous to Eq.~(\ref{eq:clustertk}).
However, the dissipation constant is far greater than the critical value and
actually \emph{diverges} at criticality. Therefore, tunneling will be totally
suppressed sufficiently close to the quantum critical point. Millis \emph{et
al.} clarify this result by noting that the very large effective droplet size
implies that many conduction electron angular momentum channels will be
scattered by it, as opposed to the simple s-wave case of a point scatterer.
Similar considerations were also made concerning the effect of magnetic
impurities in a system close to a ferromagnetic quantum critical point
\cite{larkinmelnikov72,maebashietal02}.

In a subsequent paper, Millis \emph{et al.} extended this analysis and
considered the collective effect of a distribution of droplets with different
sizes and strengths \cite{millisetal02}. By relating the energetics and the
size of a single droplet and using standard statistical methods, they
determined the form of the droplet distribution function. Using the results
for the tunneling rate of a single droplet the overall response of the system
could be calculated. It was shown how the usual Griffiths singularities of an
insulating (i.~e., non-dissipative) nearly critical magnet (as expounded in
Section~\ref{sub:irfp}) can be re-obtained within this macroscopic
Landau-functional scheme. More importantly, in the metallic case, the
suppression of tunneling by dissipation was shown to completely destroy
Griffiths behavior. Indeed, since dissipation effectively freezes \emph{most}
droplets, their contribution is that of essentially classical degrees of
freedom, giving rise to simple superparamagnetism $\chi\left(  T\right)
\sim1/T$. Close to criticality, the largest contribution to the thermodynamic
properties comes from such frozen superparamagnetic droplets. Moreover, in the
contribution of the unfrozen droplets (i.~e., those that do tunnel) only those
droplets \emph{on the verge} of classicality, and consequently with very low
tunneling rates, contribute significantly, leading again to $\chi\left(
T\right)  \sim1/T$. The results of Millis \emph{et al.} thus cast some doubts
on the relevance of a magnetic Griffiths phase picture to the physics of NFL
disordered heavy fermion compounds.

Even though differing in their detailed approach, the results obtained by
Millis \emph{et al.} are in general broad agreement with those of Castro Neto
and Jones. In both cases, disorder fluctuations lead to the formation of large
clusters of ordered material inside the disordered phase, which are able to
tunnel between two magnetic configurations. Very large clusters are
exponentially rare. Their tunneling rates, however, are exponentially small,
leading to a power-law distribution of small energy scales (tunneling energy
splittings) and quantum Griffiths singularities at intermediate temperatures.
Finally, dissipation imposes an upper cutoff $N_{c}$ on the fluctuating
cluster sizes: above this threshold, the clusters are frozen and behave like
superparamagnetic particles with a classical Curie response. The frozen
cluster contribution eventually swamps the quantum droplets and kills the
Griffiths singularities at the lowest temperatures. Their discrepant
conclusions with regard to the applicability of their theories to Kondo alloys
boils down to the assumptions made in each case about the strength of
dissipation (see the recent exchange
\cite{castrojones04,millisetal04,castrojones04b}). Millis \emph{et al.} argue
that in heavy fermion systems the local moments are strongly coupled to the
conduction electrons such that the most natural assumptions lead to a
dissipation constant which is of the same order as the other energy scales. As
a result, the largest fluctuating cluster $N_{c}$ would be fairly small (of
the order of a few sites) and the energy scale separating Griffiths behavior
from superparamagnetic response would be of order of the energy cutoff (e.g.,
the Kondo temperature of the clean system). This would leave only a rather
small window of temperatures in which quantum Griffiths effects might be
observable, effectively ruling out this mechanism as the source of NFL
behavior in disordered heavy fermion systems. Castro Neto and Jones, on the
other hand, argue that the conduction electron density of states in the region
of the ordered cluster is not renormalized by the Kondo effect and is small.
The dissipation is therefore considerably reduced and $N_{c}$ is
correspondingly enhanced, leaving a sufficiently wide range of temperatures
for the Griffiths singularities to be detectable in heavy fermion systems.
Further progress in determining which of these pictures, if any, applies to
heavy fermion alloys will likely come from a direct determination of the
strength of dissipation in real systems, either theoretically or from experiments.

\begin{itemize}
\item \emph{Vojta theory of QCP {}``rounding''}
\end{itemize}

The essential gist of the results of Millis \emph{et al.} can be put in a
simpler, more general form and its consequences extended, as shown by T. Vojta
\cite{tvojta03}. The argument is based on the same coarse-grained
Ising-symmetry Landau approach adopted by Millis \emph{et al}. Vojta pointed
out that droplet regions at $T=0$ are bounded in the spatial directions but
infinite along the imaginary time dimension. Since the ohmic dissipation
($\sim\left|  \omega_{n}\right|  $) term of the Hertz action is equivalent to
a $\sim1/\left(  \tau-\tau^{\prime}\right)  ^{2}$ kernel in the imaginary time
direction, each droplet at $T=0$ is equivalent to one-dimensional Ising models
with $\sim1/r^{2}$ interactions. These are known to undergo a phase transition
at finite temperature (equivalent to a finite coupling constant in the quantum
case, which means a large enough droplet), i.e., they are above their lower
critical dimension \cite{thouless69}. Therefore, sufficiently large droplets
will order (freeze) close enough to the critical point. Interestingly, this is
in complete agreement with results of Millis \emph{et al} and lends a new,
simpler perspective to the rather involved calculations of that work.
Considering the statistics of the droplet sizes for different types of
disorder distributions, it was shown then that the clean quantum phase
transition is {}``rounded'' by disorder. For a Gaussian disorder distribution
for example, the magnetization in $d$ dimensions at $T=0$ is \emph{always
finite}%
\begin{equation}
M\sim\exp\left(  -Bg^{2-d/\phi}\right)  , \label{eq:vojtamag}%
\end{equation}
where $g$ measures deviations from the clean critical point, $B$ is a
constant, and $\phi$ is the finite-size scaling shift exponent of the clean
$d$-dimensional system. The term {}``rounding'' here relates to the fact that
the onset of order is \emph{not} a collective effect as in conventional phase
transitions but is rather the result of a sum of many finite-size frozen
droplet contributions. At finite temperatures, droplet-droplet interactions
induce a conventional thermal phase transition. If these interactions are
assumed to be exponential in the droplet separation, the critical temperature
is a double exponential of $g$%
\begin{equation}
\ln T_{c}\sim-\exp\left(  Ag^{2-d/\phi}\right)  , \label{eq:vojtatc}%
\end{equation}
where $A$ is a constant. Finally, for non-dissipative dynamics $\left|
\omega_{n}\right|  $ is replaced by $\omega_{n}^{2}$. The equivalent
one-dimensional Ising model is short-ranged and thus \emph{at the lower
critical dimension}. This leads naturally to the Griffiths phenomena already
identified by other methods. The analysis of Vojta, therefore, puts a number
of different results on a common setting by means of some simple and
transparent arguments.

More recently, the analysis of the Ising symmetry has been extended to the
Heisenberg (in fact, continuous, $O\left(  N\right)  $) case by T. Vojta and
J. Schmalian \cite{vojtaschmalian04}. Whereas the dissipative Ising droplet is
above its lower critical dimension (when viewed as a one-dimensional system in
the time direction with $1/r^{2}$ interactions), the Heisenberg analogue is
\emph{at its lower critical dimension} \cite{joyce69,bruno01}. Therefore, in
the Vojta language, the itinerant, disordered, nearly critical Heisenberg
magnet is analogous to the insulating, disordered, nearly critical Ising
system. It is no surprise then that Griffiths singularities should arise in
this system as well. Indeed, this is the main result of
Ref.~\cite{vojtaschmalian04}, which shows that a power-law distribution of
cluster tunneling scales is obtained in the by now familiar fashion: $P\left(
\Delta\right)  \sim\Delta^{d/z-1}$. In particular, the non-universal tunable
dynamical exponent $z$ could be calculated in a controlled $1/N$ expansion as
a function of disorder strength. Similar calculations in an XY magnet were
also performed in Ref~\cite{lohetal05} with similar conclusions. The presence
of Griffiths singularities in nearly critical metals with continuous order
parameter symmetry suggests the exciting possibility that the associated
critical point, by analogy with the insulating Ising case, is also governed by
an IRFP.

\begin{itemize}
\item \emph{Long range RKKY interactions and the cluster glass phase }
\end{itemize}

\begin{figure}[ptb]
\begin{center}
\includegraphics[  width=3.4in,
keepaspectratio]{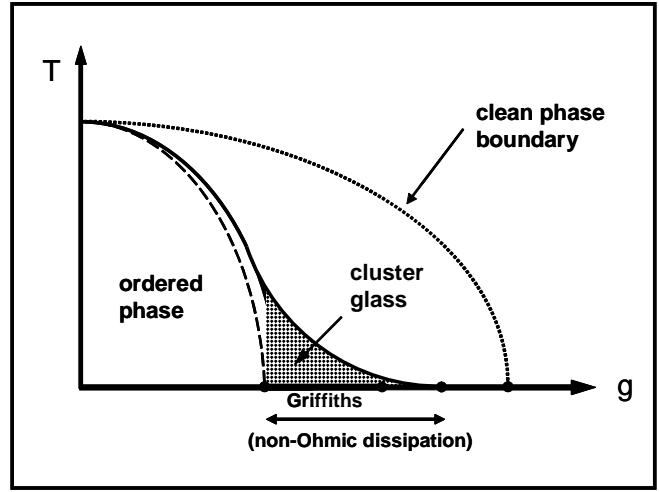}
\end{center}
\caption{Phase diagram of an itinerant system with disorder, close to the
clean quantum critical point (QCP), following Ref.~\cite{noqcp}. Shown are the
clean critical line (dotted) and that in the presence of disorder if non-Ohmic
dissipation is ignored (dashed). In this case, a Griffiths phase emerges close
to the magnetically ordered phase. When dissipation is present, sufficiently
large droplets {}``freeze'', leading to the formation of the {}``cluster
glass'' phase (shaded region), preceding the uniform ordering. }%
\label{cap:fig13}%
\end{figure}

The Griffiths phase of itinerant systems with continuous symmetry was obtained
by neglecting the effects of droplet-droplet interactions. These were
incorporated recently in an extended dynamical mean field theory fashion
\cite{noqcp}. These interactions are generically long-ranged in metallic
systems due to the RKKY interactions. The presence of disorder is known to
lead to an exponential suppression of the \emph{average} RKKY interaction,
because it introduces a random phase in the oscillating part. However, the
fluctuations around the average retain their long-ranged $\sim1/r^{d}$
amplitude and are random in sign \cite{jagannathanetal88,narozhnyetal00}. A
particular droplet will therefore be coupled to a large number of other
droplets and subject to their Weiss dynamical molecular field. A
self-consistent treatment leads to a Weiss field dynamics set by the
\emph{average local} susceptibility of all the other droplets $\overline{\chi
}_{loc}\left(  \omega_{n}\right)  $. In the very dilute limit far from the
critical point, the results of Vojta and Schmalian, which neglect
droplet-droplet interactions, can be taken as a zeroth order approximation and
an instability analysis can be performed. This has the usual power law form%
\begin{align}
\overline{\chi}_{loc}\left(  \omega_{n}\right)   &  \sim\int d\Delta
\Delta^{d/z-1}\frac{1}{\Delta+\left|  \omega_{n}\right|  }\nonumber\\
&  \sim\overline{\chi}_{loc}\left(  0\right)  -\widetilde{\gamma}\left|
\omega_{n}\right|  ^{d/z-1}+\mathcal{O}\left(  \left|  \omega_{n}\right|
\right)  . \label{eq:heisgriffdynsusc}%
\end{align}
The Weiss dynamical field adds to the usual ohmic term
($\sim\left| \omega_{n}\right|  $) and \emph{dominates at low
frequencies} if $d/z<2$, which occurs even before the Griffiths
singularity in the susceptibility (which requires $d/z<1$). This
sub-ohmic dissipation term places a single droplet \emph{above}
its lower critical dimension \cite{joyce69,bruno01}, leading again
to freezing of sufficiently large droplets and eventually to
ordering, in analogy with the Ising system. The conclusion is
that, within this picture, long-ranged droplet-droplet
interactions generate \emph{additional dissipation}, over and
above the one induced by the itinerant carriers, which is able to
trigger magnetic order of a cluster-glass type. Interestingly, the
existence of a very similar cluster glass phase was proposed in
early work, over thirty years ago \cite{sherrington74}. The
Griffiths phase will be {}``hidden'' inside this ordered phase
(Fig.~\ref{cap:fig13}). It should be stressed that this argument
cannot be generalized to insulating systems, where the presence of
short-ranged interactions casts doubts on the validity of the mean
field treatment, at least in low dimensions. Likewise, for
itinerant systems, this approach is likely to break down in low
dimensions and the Griffiths phase scenario may survive. An
interesting unsolved question is the value of the lower critical
dimension for this class of disordered systems.

\begin{itemize}
\item \emph{Internal quantum dynamics of droplets }
\end{itemize}

Finally, we should like to mention work which takes into account the internal
quantum dynamics of droplets with a special geometry. Shah and Millis
considered a {}``necklace'' of spins in a ring geometry coupled as a finite
XXZ chain \cite{shahmillis03}. They showed that internal quantum fluctuations
of the necklace facilitate the reversal of the droplet magnetization, whose
amplitude now decays as a power of the number of droplet spins, instead of an
exponential. As a result, the nature of the droplet dynamics is closer to the
multi-channel Kondo effect and the spin flips actually tend to proliferate at
lower temperatures. The special geometry of a circular ring enforces a
residual chiral symmetry of the necklace which protects a two-channel fixed
point, with its well-known anomalous properties
\cite{nozieresblandin80,affleckludwig91,affleckludwig91b,affleckludwig91c,ludwigaffleck91}%
. This symmetry is not expected to be present in the case of disorder-induced
droplets. Kondo-type dynamics at low temperatures was also proposed for
magnetic impurities with XY symmetry in a nearly ferromagnetic metal
\cite{lohetal05}.

\subsubsection{On the applicability of the magnetic Griffiths phase
phenomenology to metallic disordered systems}

\label{sub:griffapplicab}

An important unsolved question that remains is whether a magnetic Griffiths
phase (i.e., a Griffiths phase associated with a magnetic phase transition)
can be realized in an observable temperature/frequency range in metallic
systems. As already discussed in Section~\ref{sub:metalgriff}, a great deal of
controversy surrounds the question of whether dissipation due to the
conduction electrons is weak enough to allow for Griffiths singularities to be
observable in an appreciable temperature range. However, a number of
additional points, which are independent of the answer to this question, are
also worth mentioning regarding the applicability of this scenario to real
metallic systems.

\begin{itemize}
\item \emph{Less singular thermodynamic response of heavy fermion systems:}
\end{itemize}

We have seen that power law anomalies are observed in both doped
semiconductors and heavy fermion materials which conform, in principle, with
the Griffiths phase phenomenology (Sections~\ref{sub:dopedsemicond} and
\ref{sub:nflkondoalloys}). There are some trends, however, which seem to
distinguish these two classes of systems. Griffiths phases are characterized
by power-law divergent thermodynamic properties such as%
\begin{align}
\chi\left(  T\right)   &  \sim\frac{1}{T^{1-\alpha}},\\
\frac{C\left(  T\right)  }{T}  &  \sim\frac{1}{T^{1-\alpha}},
\end{align}
where the power-law exponent $\alpha$ is non-universal and tunable by the
disorder strength. NFL behavior is signaled by $\alpha<1$. A significant
feature of the phenomenology of the above compounds is the fact that $\alpha$
typically lies in the range $0.7-1.0$ ($\alpha=1$ implying a logarithmic
divergence) in the case of heavy fermion compounds (with a few cases below 0.7
but always above 0.5, see \cite{stewartNFL} and \cite{andradeetal}), whereas
doped semiconductors show a more singular response $\alpha\approx0.3-0.4$
\cite{paalanen91,sarachik95}. A possible explanation to this interesting trend
is the fact that localized magnetic moments are well formed and stable in
heavy fermion systems, whereas they are induced by disorder fluctuations in
doped semiconductors. Therefore, interactions among local moments are likely
to be \emph{stronger} in the former than in the latter. Such interactions,
usually neglected in Griffiths phase theories, where the droplets are assumed
dilute and independent, contribute to quench the local moments and may be at
the origin of the less singular spin entropy available at low temperatures in
heavy fermion materials.

\begin{itemize}
\item \emph{Wilson ratio:}
\end{itemize}

As shown in Section~\ref{sub:griffgeneral}, the Griffiths phase thermodynamic
responses can be obtained from the scaling of a few physical quantities with
temperature. The results on the random quantum Ising model are very
instructive in this respect \cite{fishertransising,fishertransising2}. The
spin susceptibility, for example, is given by%
\begin{equation}
\chi\left(  T\right)  \sim\frac{\mu^{2}\left(  T\right)  n\left(  T\right)
}{T},
\end{equation}
where $\mu\left(  T\right)  $ is the average value of the magnetic moment per
cluster and $n\left(  T\right)  $ is the number of active clusters, both at
temperature $T$. The specific heat is obtained similarly from the entropy%
\begin{align}
S\left(  T\right)   &  \sim n\left(  T\right)  \ln2,\\
C\left(  T\right)  /T  &  \sim\frac{dS}{dT}\,.
\end{align}
As mentioned in Section~\ref{sub:irfp}, in the disordered Griffiths phase of
the one-dimensional random Ising chain in a transverse field,%
\begin{align}
\chi\left(  T\right)   &  \sim\delta^{4-2\phi}\frac{\left[  \ln\left(
1/T\right)  \right]  ^{2}}{T^{1-1/z}},\label{eq:rtfimsusc3}\\
C\left(  T\right)  /T  &  \sim\delta^{3}\frac{1}{T^{1-1/z}}.
\label{eq:rtfimgamma3}%
\end{align}
It can be seen that the Wilson ratio%
\begin{equation}
R_{W}\propto\frac{T\chi\left(  T\right)  }{C\left(  T\right)  }\sim\left[
\ln\left(  1/T\right)  \right]  ^{2}, \label{eq:wilsonratio}%
\end{equation}
which is logarithmically divergent as $T\rightarrow0$. The origin of this
divergence can be traced back to the scaling of the magnetic moment of the
ordered (ferromagnetic) clusters with size or energy scales, which leads to
$\mu\sim\ln\left(  \Omega_{o}/T\right)  $. Thus, the Wilson ratio is a useful
quantity to estimate the cluster moment at low temperatures, as the cluster
number density which appears both in the susceptibility and the specific heat
cancels out. These results are expected to be valid even in higher dimensions.
Since we expect the cluster moments to grow with size, both in ferromagnetic
and in antiferromagnetic Griffiths phases (in the latter, $\mu\sim L^{\left(
d-1\right)  /2}$), the Wilson ratio should diverge, albeit slowly, as
$T\rightarrow0$. Although perhaps hard to determine, a careful examination of
the Wilson ratio may be a useful guide to test the applicability of the
magnetic Griffiths phase scenario.

\begin{itemize}
\item \emph{Observed entropy and the size of Griffiths droplets:}
\end{itemize}

The generic quantum magnetic Griffiths phase is built upon large clusters
which tunnel between two reversed (staggered or uniform) magnetization states.
Therefore, the available entropy per cluster is $S_{cl}=k_{B}\ln2$, the other
microscopic degrees of freedom being effectively frozen. On the other hand,
the relevant clusters are generically assumed to be large, with a typical
number of $N_{\xi}$spins per cluster. If the total number of clusters is of
order $N_{cl}$, the total entropy can be estimated as $S\approx N_{cl}S_{cl}$,
and the entropy per spin is%
\begin{equation}
\frac{S}{N_{spins}}\approx\frac{N_{cl}S_{cl}}{N_{spins}}\approx\frac{S_{cl}%
}{N_{\xi}}\sim\frac{k_{B}\ln2}{N_{\xi}}, \label{eq:entropyperspin}%
\end{equation}
where $N_{spins}$ is the total number of spins. We conclude that the entropy
per mole of spins is \emph{decreased from the typical value of} $R\ln2$
\emph{by a number of the order of the typical cluster size.} We thus expect
that conventional magnetic Griffiths behavior should be characterized by
divergent power laws with \emph{rather small amplitudes}. Conversely, if
Griffiths phase behavior with a sizable ($\sim R\ln2$) molar spin entropy is
observed, it necessarily follows that the relevant fluctuators involve a
small, of order 1, number of spins. All the candidate heavy fermion compounds
analyzed if Refs.~\cite{stewartNFL} and \cite{andradeetal} have molar spin
entropies of the order of $R$, which seems to leave little room for large
clusters \cite{aguiaretal1}. Therefore, the electronic Griffiths phase
(Section~\ref{sub:andersongriff}), whose fluctuators are individual Kondo
spins seems a much more viable explanation. Of course, if dissipation freezes
clusters larger than a critical size $N_{c}$, these clusters will contribute a
Curie-like term, which should be taken into account. However, it is clear that
the above argument also limits the low temperature entropy of the frozen
clusters, even though their contribution is more singular. The conclusion that
$S_{molar}\sim R$ implies small fluctuators seems therefore inescapable.

\begin{itemize}
\item \emph{Temperature range of Griffiths phase behavior:}
\end{itemize}

Another question relates to the range of temperatures where the Griffiths
singularities may be observable. We have seen that the strength of dissipation
generated by the conduction electrons is a very important input in the
determination of this range (see Section~\ref{sub:metalgriff}). Only very
small dissipation rates are compatible with a wide range of Griffiths
singularities. However, \emph{even if the dissipation is negligibly weak}, the
range of temperatures available for Griffiths anomalies is still fairly
restricted. Let us consider the phenomenology of insulating disordered magnets
discussed in Section~\ref{sub:irfp}. As we have seen, all the known Griffiths
phases in these systems occur in the vicinity of a phase or a point which is
governed by an infinite randomness fixed point. Formally, this is usually seen
through the divergence of the dynamical exponent $z$ as the IRFP is
approached. Now, as the system approaches the IRFP, there is a line of
temperature crossover that approaches zero, as in any quantum critical point.
Above the crossover line, the system is still governed by the IRFP, even off
criticality. Only below the crossover line can the non-critical behavior, in
particular the quantum Griffiths singularities, be observed \cite{sachdevbook}%
. The functional form of this crossover line can be easily obtained for
systems governed by an IRFP \cite{fishertransising2}. The primary feature of
the IRFP is the {}``activated dynamical scaling'' which relates energy and
length scales (see Section~\ref{sub:irfp})%
\begin{equation}
\Omega\sim\Omega_{0}\exp\left(  -L^{\psi}\right)  .
\label{eq:activdynscaling2}%
\end{equation}
Close to criticality, there is a correlation length $\xi$ which sets the scale
beyond which Eq.~(\ref{eq:activdynscaling2}) no longer holds and conventional
dynamical scaling is recovered. As shown by Fisher \cite{fishertransising2},
this is the {}``true'' correlation length in the sense of the criterion of
Ref.~\cite{chayesetal86}: beyond this scale, the system {}``knows'' it is
non-critical and most magnetic clusters can be treated as effectively
independent, leading to typical Griffiths singularities. This {}``true''
correlation length diverges as a power law as the IRFP is approached
($\nu\rightarrow0$)%
\begin{equation}
\xi\sim\frac{1}{\delta^{\nu}}, \label{eq:truecorrlength}%
\end{equation}
where $\nu$ satisfies the Harris criterion $\nu d\geq2$
\cite{harris74,chayesetal86}. This is the correlation length of the average
correlation function. By plugging this characteristic length into
Eq.~(\ref{eq:activdynscaling2}) and setting $\Omega=T_{cross}$ we can
determine the crossover line%
\begin{equation}
T_{cross}\sim\Omega_{0}\exp\left(  -\frac{1}{\delta^{\nu\psi}}\right)  .
\label{eq:crossover}%
\end{equation}
In the particular case of the 1D Ising model in a transverse field $\nu\psi
=1$. We thus see that, quite generically, quantum Griffiths behavior is
expected to occur below an energy scale that is \emph{exponentially smaller}
than the natural energy scales of the problem. Above this scale, the more
singular universal behavior characteristic of the IRFP should be observed.
This represents a severe restriction on the range of temperatures where
Griffiths effects could be observed. Apparently, the diluted Ising model
analyzed in Ref.~\cite{senthilsachdev} is an exception to this kind of
behavior. This is probably due to the peculiar percolating character of the
transition in that case. Since metallic systems are characterized by
long-ranged RKKY interactions, it is reasonable to expect that percolation is
less likely to play a significant role and the more generic behavior of
Eq.~(\ref{eq:crossover}) should apply.

\subsection{Itinerant quantum glass phases and their precursors}

\label{sub:Itinerant-quantum-glass}

\subsubsection{Inherent instability of the electronic Griffiths phases to
spin-glass ordering}

\label{sub:instabilitygriff}

So far, we have discussed the electronic and the magnetic Griffiths phase
scenarios for disorder-induced non-Fermi liquid behavior. Both pictures
envision the formation of rare regions with anomalously slow dynamics, which
under certain conditions dominate the low temperature properties. Neither
picture, however, seems satisfactory for the following key reason: in both
cases the resulting NFL behavior is characterized by power law anomalies, with
non-universal, rapidly varying exponents. In contrast, most experimental data
seem to show reasonably weak anomalies, close to marginal Fermi liquid
behavior \cite{stewartNFL}.

Physically, it is clear what is missing from the theory. Similarly as magnetic
Griffiths phases, the electronic Griffiths phase is characterized
\cite{mirandavlad1,tanaskovicetal04} by a broad distribution $P\left(
T_{K}\right)  \sim\left(  T_{K}\right)  ^{\alpha-1}$ of local energy scales
(Kondo temperatures), with the exponent $\alpha\sim W^{-2}$ rapidly decreasing
with disorder $W$. At any given temperature, the local moments with
$T_{K}(i)<T$ remain unscreened. As disorder increases, the number of such
unscreened spins rapidly proliferates. Within the existing theory
\cite{mirandavladgabi1,mirandavladgabi2,mirandavlad1,tanaskovicetal04} these
unscreened spins act essentially as free local moments and provide a very
large contribution to the thermodynamic response. In a more realistic
description, however, even the Kondo-unscreened spins are \emph{not}
completely free, since the metallic host generates long-ranged
Ruderman-Kittel-Kasuya-Yosida (RKKY) interactions even between relatively
distant spins.

In a disordered metal, impurity scattering introduces random phase
fluctuations in the usual periodic oscillations of the RKKY interaction,
which, however, retains its power law form (although its \textit{average}
value decays exponentially \cite{jagannathanetal88,narozhnyetal00}). Hence,
such an interaction acquires a random amplitude $J_{ij}$ of zero mean but
finite variance $<J_{ij}^{2}(R)>\sim R^{-2d}$
\cite{jagannathanetal88,narozhnyetal00}. As a result, in a disordered metallic
host, a given spin is effectively coupled with random but long range
interactions to many other spins, often leading to spin-glass freezing at the
lowest temperatures. How this effect is particularly important in Griffiths
phases can also be seen from the mean-field stability criterion
\cite{braymoore80} for spin glass ordering, which takes the form%
\begin{equation}
J\chi_{loc}(T)=1. \label{eq:sg}%
\end{equation}
Here, $J$ is a characteristic interaction scale for the RKKY interactions, and
$\chi_{loc}(T)$ is the disorder average of the local spin susceptibility. As
we generally expect $\chi_{loc}(T)$ to diverge within a Griffiths phase, this
arguments strongly suggests that in the presence of RKKY interactions such
systems should have an inherent instability to finite (even if very low)
temperature spin glass ordering.

Similarly as other forms of magnetic order, the spin glass ordering is
typically reduced by quantum fluctuations (e.g. the Kondo effect) which are
enhanced by coupling of the local moments to itinerant electrons. Sufficiently
strong quantum fluctuations can completely suppress spin-glass ordering even
at $T=0$, leading to a quantum critical point separating metallic spin glass
from the conventional Fermi liquid ground state. As in other QCP's, one
expects the precursors to magnetic ordering to emerge even before the
transition is reached and produce non-Fermi liquid behavior within the
corresponding quantum critical region. Since many systems where
disorder-driven NFL behavior is observed are not too far from incipient
spin-glass ordering, it is likely that these effects play an important role,
and should be theoretically examined in detail.

{}From the theoretical point of view, a number of recent works have examined the
general role of quantum fluctuations in glassy systems, and the associated
quantum critical behavior. Most of the results obtained so far have
concentrated on the behavior within the mean-field picture (i.e., in the limit
of large coordination), where a consistent description of the QCP behavior has
been obtained for several models. In a few cases \cite{re:Read95}, corrections
to mean-field theory have been examined, but the results appear inconclusive
and controversial at this time. In the following, we briefly review the most
important results obtained within the mean-field approaches.

\subsubsection{Quantum critical behavior in insulating and metallic spin
glasses}

\label{sub:qcp-msg}

\begin{itemize}
\item \emph{Ising spin glass in a transverse field}
\end{itemize}

\begin{figure}[ptb]
\begin{center}
\includegraphics[  width=3.4in,
keepaspectratio]{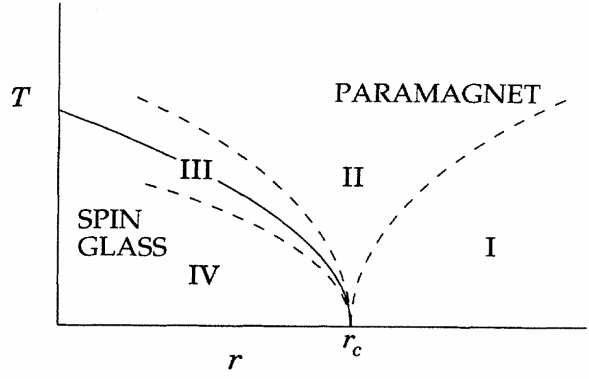}
\end{center}
\caption{Generic phase diagram (following \cite{re:Read95}) of the quantum
critical behavior for spin glasses. The parameter $r$, which measures the
quantum fluctuations, can represent the transverse field for localized spin
models or the Fermi energy in metallic spin glasses.}%
\label{cap:fig14}%
\end{figure}

The simplest framework to study the quantum critical behavior of spin glasses
is provided by localized spin models such as the infinite-range Ising model in
a transverse field (IMTF) with random exchange interactions $J_{ij}$ of zero
mean and variance $J^{2}/N$ ($N\longrightarrow\infty$ is the number of lattice
sites).%
\begin{equation}
H_{TFIM}=-\sum_{ij}J_{ij}\sigma_{i}^{z}\sigma_{j}^{z}-\Gamma\sum_{i}\sigma
_{i}^{x}. \label{eqisingsg}%
\end{equation}
In the classical limit ($\Gamma=0$), this model reduces to the well-studied
Sherrington-Kirkpatrick model \cite{re:Mezard86}, where spins freeze with
random orientations below a critical temperature $T_{SG}(\Gamma=0)=J$. Quantum
fluctuations are introduced by turning on the transverse field, which induces
up-down spin flips with tunneling rate $\sim\Gamma$. As $\Gamma$ grows, the
critical temperature $T_{SG}(\Gamma)$ decreases, until the quantum critical
point is reached at $\Gamma=\Gamma_{c}\approx0.731J$, signaling the $T=0$
transition from a spin-glass to a quantum-disordered paramagnetic state
(Fig.~\ref{cap:fig14}).

Similarly as in DMFT theories for electronic systems, such infinite range
models can be formally reduced to a self-consistent solution of an appropriate
quantum impurity problem, as first discussed in the context of quantum spin
glasses by Bray and Moore \cite{braymoore80}. Early work quickly established
the phase diagram \cite{dobrosavljevic87} of this model, but the dynamics near
the quantum critical point proved more difficult to unravel, even when the
critical point is approached form the quantum-disordered side. Here, the
problem reduces to solving for the dynamics of a single Ising spin in a
transverse field, described by an effective Hamiltonian \cite{re:Miller93} of
the form%
\[
H=-\frac{1}{2}J^{2}\int\int d\tau d\tau^{\prime}\sigma^{z}(\tau)\chi(\tau
-\tau^{\prime})\sigma^{z}(\tau^{\prime})+\Gamma\int d\tau\sigma^{x}(\tau).
\]
Physically, the interaction of the considered spin with the spin fluctuations
of its environment generates the retarded interaction described by the
{}``memory kernel'' $\chi(\tau-\tau^{\prime})$. An appropriate
self-consistency condition relates the memory kernel to the disorder-averaged
local dynamical susceptibility of the quantum spin
\[
\chi(\tau-\tau^{\prime})=\overline{\left\langle T\sigma^{z}(\tau)\sigma
^{z}(\tau^{\prime})\right\rangle }.
\]
A complete solution of the quantum critical behavior can be obtained, as first
established in a pioneering work by Miller and Huse \cite{re:Miller93}. These
authors have set up a diagrammatic perturbation theory for the dynamic
susceptibility, showing that the leading loop approximation already captures
the exact quantum critical behavior, as the higher order corrections provide
only quantitative renormalizations. The dynamical susceptibility takes the
general form%
\[
\chi(\omega_{n})=\chi_{o}+\left(  \omega_{n}^{2}+\Delta^{2}\right)  ^{1/2},
\]
where the local static susceptibility $\chi_{o}$ remains finite throughout the
critical regime, and the spin excitations exist above a gap
\[
\Delta\sim\left(  r/\left|  \ln r\right|  \right)  ^{1/2}%
\]
which vanishes as the transition is approached from the paramagnetic side
(here $r=\left(  \Gamma-\Gamma_{c}\right)  /\Gamma_{c}$ measures the distance
from the critical point).

\begin{itemize}
\item \emph{The quantum spin-glass phase and the replicon mode }
\end{itemize}

The validity of this solution was confirmed by a generalization \cite{ye93} to
the $M$-component rotor model (the Ising model belongs to the same
universality class as the $M=2$ rotor model), which can be solved in closed
form in the large $M$ limit. This result, which proves to be exact to all
orders in the $1/M$ expansion, could be extended even to the spin glass phase,
where a full replica symmetric solution was obtained. Most remarkably, the
spin excitation spectrum remains gapless ($\Delta=0$) throughout the ordered
phase. Such gapless excitations commonly occur in ordered states with broken
continuous symmetry, but are generally not expected in classical or quantum
models with a discrete symmetry of the order parameter. In glassy phases (at
least within mean-field solutions), however, gapless excitations generically
arise for both classical and quantum models. Here, they reflect the marginal
stability \cite{re:Mezard86} found in the presence of replica symmetry
breaking, a phenomenon which reflects the high degree of frustration in these
systems. The role of the Goldstone mode in this case is played by the
so-called {}``replicon'' mode, which describes the collective low energy
excitations characterizing the glassy state.

A proper treatment of the low energy excitations in this regime requires
special attention to the role of replica symmetry breaking (RSB) in the
$T\rightarrow0$ limit. The original work \cite{ye93} suggested that RSB is
suppressed at $T=0$, so that the simpler replica symmetric solution can be
used at low temperatures. Later work \cite{georgesetal01}, however,
established that the full RSB solution must be considered before taking the
$T\rightarrow0$ limit, and only then can the correct form of the leading low
temperature corrections (e.g., the linear $T$-dependence of the specific heat)
be obtained.

\begin{itemize}
\item \emph{Physical content of the mean-field solution }
\end{itemize}

In appropriate path-integral language
\cite{dobrosavljevic87,ye93}, the problem can be shown to reduce
to solving a one dimensional classical Ising model with long-range
interactions, the form of which must be self-consistently
determined. Such classical spin chains with long range
interactions in general can be highly nontrivial. Some important
examples are the Kondo problem
\cite{Yuval-Anderson1,Yuval-Anderson2,Yuval-Anderson3}, and the
dissipative two-level system \cite{leggettetal87}, both of which
map to an Ising chain with $1/\tau^{2}$ interactions. Quantum
phase transitions in these problems correspond to the
Kosterlitz-Thouless transition found in the Ising chain
\cite{kosterlitz76prl}, the description of which required a
sophisticated renormalization-group analysis. Why then is the
solution of the quantum Ising spin glass model so simple? The
answer was provided in the paper by Ye, Sachdev, and Read
\cite{ye93}, which emphasized that the critical state (and the RSB
spin-glass state) does not correspond to the critical point, but
rather to the high-temperature phase of the equivalent Ising
chain, where a perturbative solution is sufficient. In Kondo
language, this state corresponds to the Fermi-liquid solution
characterized by a finite Kondo scale, as demonstrated by a
quantum Monte-Carlo calculation of Rozenberg and Grempel
\cite{rozenberg98prl}, which also confirmed other predictions of
the analytical theory.

{}From a more general perspective, the possibility of obtaining a simple
analytical solution for quantum critical dynamics has a simple origin. It
follows from the fact that all corrections to Gaussian (i.e. Landau) theory
are irrelevant above the upper critical dimension, as first established by the
Hertz-Millis theory \cite{hertz,millis} for conventional quantum criticality.
The mean-field models become exact in the limit of infinite dimensions, hence
the Gaussian solution of Refs. \cite{re:Miller93,ye93} becomes exact. Leading
corrections to mean-field theory for rotor models were examined by an
$\varepsilon$-expansion below the upper critical dimension $d_{c}=8$ for the
rotor models by Read, Sachdev, and Ye, but these studies found run-away flows,
presumably indicating non-perturbative effects that require more sophisticated
theoretical tools. Most likely these include Griffiths phase phenomena
controlled by the infinite randomness fixed point, as already discussed in
Section~\ref{sub:irfp}.

\begin{itemize}
\item \emph{Metallic spin glasses }
\end{itemize}


A particularly interesting role of the low-lying excitations associated with
the spin-glass phase is found in metallic spin glasses. Here the quantum
fluctuations are provided by the Kondo coupling between the conduction
electrons and local moments, and therefore can be tuned by controlling the
Fermi energy in the system. The situation is again the simplest for Ising
spins where an itinerant version of the rotor model of Sengupta and Georges
\cite{senguptageorges} can be considered, and similar results have obtained
for the {}``spin-density glass'' model of Sachdev, Read, and Oppermann
\cite{sachdevreadopper}. The essential new feature in these models is the
presence of itinerant electrons which, as in the Hertz-Millis approach
\cite{hertz,millis}, have to be formally integrated out before an effective
order-parameter theory can be obtained. This is justified \textit{provided}
that the quasiparticles remain well defined at the quantum critical point,
i.e. that the quasiparticle weight $Z\sim T_{K}$ remains finite and the Kondo
effect remains operative. The validity of these assumptions is by no means
obvious, and led to considerable controversy before a detailed quantum Monte
Carlo solution of the model became available \cite{rozenberg99prb}, confirming
the proposed scenario.

Under these assumptions, the theory can again be solved in closed form, and we
only quote the principal results. Physically, the essential modification is
that the presence of itinerant electron now induces Landau damping, which
creates dissipation for the collective mode. As a result, the dynamics is
modified, and the local dynamic susceptibility now takes the following form%
\begin{equation}
\chi(\omega_{n})=\chi_{o}+\left(  \left|  \omega_{n}\right|  +\omega^{\ast
}\right)  ^{1/2}.
\end{equation}
The dynamics is characterized by the crossover scale $\omega^{\ast}\sim r$
($r$ measures the distance to the transition) which defines a crossover
temperature $T^{\ast}\sim\omega^{\ast}$ separating the Fermi liquid regime (at
$T\ll T^{\ast}$) from the quantum critical regime (at $T\gg T^{\ast}$). At the
critical point $\chi(\omega_{n})=\chi_{o}+\left|  \omega_{n}\right|  ^{1/2}$,
leading to non-Fermi liquid behavior of all physical quantities, which acquire
a leading low-temperature correction of the $T^{3/2}$ form. This is a rather
mild violation of Fermi liquid theory, since both the static spin
susceptibility and the specific heat coefficient remain finite at the QCP. A
more interesting feature, which is specific to glassy systems, is the
persistence of such quantum critical NFL behavior \textit{throughout} the
metallic glass phase, reflecting the role of the replicon mode.

\subsubsection{Spin-liquid behavior, destruction of the Kondo effect by
bosonic dissipation, and fractionalization}

\label{sub:spinliquid}

\begin{itemize}
\item \emph{Quantum Heisenberg spin glass and the spin-liquid solution }
\end{itemize}

Quantum spin glass behavior proves to be much more interesting in the case of
Heisenberg spins, where the Berry phase term \cite{Fradkin} plays a highly
nontrivial role, completely changing the dynamics even within the paramagnetic
phase. While the existence of a finite temperature spin-glass transition was
established even in early work \cite{braymoore80}, solving for the details of
the dynamics proved difficult until the remarkable work of Sachdev and Ye
\cite{sachdevye}. By a clever use of large-$N$ methods, these authors
identified a striking \textit{spin-liquid} solution within the paramagnetic
phase. In contrast to the nonsingular behavior of Ising or rotor quantum spin
glasses, the dynamical susceptibility now displays a logarithmic singularity
at low frequency. On the real axis it takes the form
\[
\chi(\omega)\sim\ln(1/\left|  \omega\right|  )+i\frac{\pi}{2}sign(\omega).
\]
A notable feature of this solution is that it is precisely of the form
postulated for {}``marginal Fermi liquid'' phenomenology \cite{mfl} of doped
cuprates. The specific heat is also found to assume a singular form
$C\sim\sqrt{T}$, which was shown \cite{georgesetal01} to reflect a nonzero
extensive entropy if the spin liquid solution is extrapolated to $T=0$. Of
course, the spin-liquid solution becomes unstable at a finite ordering
temperature, and the broken symmetry state has to be examined to discuss the
low temperature properties of the model.

Subsequent work \cite{georgesetal01} demonstrated that this mean-field
solution remains valid for all finite $N$ and generalized the solution to the
spin-glass (ordered) phase. A closed set of equations describing the low
temperature thermodynamics in the spin glass phase was obtained, which was
very recently re-examined in detail \cite{rozenberg04prl} to reveal fairly
complicated behavior.

\begin{itemize}
\item \emph{Metallic Heisenberg spin glasses and fractionalization }
\end{itemize}

Even more interesting is the fate of this spin liquid solution in itinerant
systems, where an additional Kondo coupling is added between the local moments
and the conduction electrons. The mean-field approach can be extended to this
interesting situation by examining a Kondo-Heisenberg spin glass model
\cite{burdinetal} with the Hamiltonian
\begin{equation}
H_{KH}=-t\sum_{\langle ij\rangle\sigma}(c_{i\sigma}^{\dagger}c_{j\sigma
}^{\phantom{\dagger}}+\mbox{H. c.})+J_{K}\sum_{i}\mathbf{S}_{i}\cdot
\mathbf{s}_{i}+\sum_{\langle ij\rangle}J_{ij}\mathbf{S}_{i}\cdot\mathbf{S}%
_{j}. \label{eq:spinliqham}%
\end{equation}

In the regime where the scale of the RKKY interaction $J=\left\langle
J_{ij}^{2}\right\rangle ^{1/2}$ is small compared to the Kondo coupling
$J_{K}$, one expects Kondo screening to result in standard Fermi liquid
behavior. In the opposite limit, however, the spin fluctuations associated
with the retarded RKKY interactions may be able to adversely affect the Kondo
screening, and novel metallic behavior could emerge. This intriguing
possibility can be precisely investigated in the mean-field (infinite range)
limit, where the problem reduces to a single-impurity action of the form
\cite{burdinetal}%
\begin{align}
S_{eff}^{KH}  &  =\sum_{\sigma}\int_{0}^{\beta}d\tau c_{\sigma}^{\dagger
}\left(  \tau\right)  \left(  \partial_{\tau}-\mu+v_{j}\right)  c_{\sigma
}^{\phantom{\dagger}}\left(  \tau\right) \nonumber\\
&  -t^{2}\sum_{\sigma}\int_{0}^{\beta}d\tau\int_{0}^{\beta}d\tau^{\prime
}c_{\sigma}^{\dagger}(\tau)G_{c}(\tau-\tau^{\prime})c_{\sigma}%
^{\phantom{\dagger}}(\tau^{\prime})\nonumber\\
&  +J_{K}\int_{0}^{\beta}d\tau\mathbf{S}(\tau)\cdot\mathbf{s}(\tau)\nonumber\\
&  -\frac{J^{2}}{2}\int_{0}^{\beta}d\tau\int_{0}^{\beta}d\tau^{\prime}%
\chi(\tau-\tau^{\prime})\mathbf{S}(\tau)\cdot\mathbf{S}(\tau^{\prime}).
\label{eq:spinliqaction}%
\end{align}

Such a single-impurity action (\ref{eq:spinliqaction}) describes the so-called
Bose-Fermi Kondo (BFK) impurity model
\cite{qimiao96prl,sengupta,zhusi02,zaranddemler02} where, in addition to the
coupling to the fermionic bath of conduction electrons, the Kondo spin also
interacts with a bosonic bath of spin fluctuations, with local spectral
density $\chi\left(  \omega_{n}\right)  $. Because the same BFK model also
appears in {}``extended'' DMFT theories \cite{qimiao00prb} of quantum
criticality in clean systems
\cite{qimiao96prl,sengupta,sietal,zhusi02,zaranddemler02}, its properties have
been studied in detail and are by now well understood.

\begin{figure}[ptb]
\begin{center}
\includegraphics[  width=3.4in,
keepaspectratio]{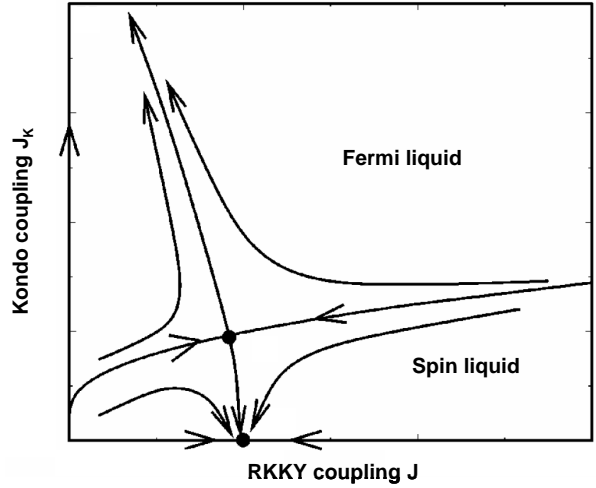}
\end{center}
\caption{Phase diagram of the Bose-Fermi Kondo model in the presence of a
sub-Ohmic bosonic bath \cite{qimiao96prl,sengupta,zhusi02,zaranddemler02}.
Kondo screening is destroyed for sufficiently large dissipation (RKKY coupling
to spin fluctuations). }%
\label{cap:fig15}%
\end{figure}

In the absence of the RKKY coupling ($J=0$), the ground state of the impurity
is a Kondo singlet for any value of $J_{K}\neq0$. By contrast, when $J>0$, the
dissipation induced by the bosonic bath tends to destabilize the Kondo effect.
For a bosonic bath of {}``Ohmic'' form ($\chi\left(  \omega_{n}\right)
=\chi_{o}-C\left|  \omega_{n}\right|  $), this effect only leads to a finite
decrease of the Kondo temperature, but the Fermi liquid behavior persists. In
contrast, for {}``sub-Ohmic'' dissipation ($\chi\left(  \omega_{n}\right)
=\chi_{o}-C\left|  \omega_{n}\right|  ^{1-\epsilon}$ with $\epsilon>0)$ two
different phases exist, and for sufficiently large RKKY coupling the Kondo
effect is destroyed. The two regimes are separated by a quantum phase
transition (see Fig.~\ref{cap:fig15}).

Of course, in the considered Kondo lattice model with additional RKKY
interactions, the form of the bosonic bath $\chi\left(  \omega_{n}\right)  $
is self-consistently determined, and can take different forms as the RKKY
coupling $J$ is increased. The model was analytically solved within a large
$N$ approach by Burdin \emph{et al.} \cite{burdinetal}, who calculated the
evolution of the Fermi liquid coherence scale $T^{\ast}$, and the
corresponding quasiparticle weight $Z$ in the presence of RKKY interactions.
Within the paramagnetic phase both $T^{\ast}(J)$ and $Z(J)$ are found to
decrease with $J$ until the Kondo effect (and thus the Fermi liquid) is
destroyed at $J=J_{c}\approx10T_{K}^{o}$ (here $T_{K}^{o}$ is the $J=0$ Kondo
temperature), where both scales vanish \cite{burdinetal,tanaskovicetal05}. At
$T>T^{\ast}(J)$ (and of course at any temperature for $J>J_{c}$) the spins
effectively decouple from conduction electrons, and spin liquid behavior,
essentially identical to that of the insulating model, is established. Thus,
sufficiently strong and frustrating RKKY interactions are able to suppress
Fermi liquid behavior, and marginal Fermi liquid behavior emerges in a
metallic system. This phenomenon, corresponding to spin-charge separation
resulting from the destruction of the Kondo effect, is sometimes called
{}``fractionalization''
\cite{colemanandrei,kaganetal92,demleretal02,senthiletal,senthiletal2}. Such
behavior has often been advocated as an appealing scenario for exotic phases
of strongly correlated electrons, but with the exception of the described
model, there are very few well established results and model calculations to
support its validity. Finally, we should mention related work
\cite{parcolletgeorges} on doped Mott insulators with random exchanges, with
many similarities with the above picture.

We should note, however, that this exotic solution is valid only within the
paramagnetic phase, which is generally expected to become unstable to magnetic
(spin glass) ordering at sufficiently low temperatures. Since
fractionalization emerges only for sufficiently large RKKY coupling (in the
large $N$ model $J_{c}\approx10T_{K}^{o}$), while in general one expects
magnetic ordering to take place already at $J\sim T_{K}^{o}$ (according to the
famous Doniach criterion \cite{Doniach}), one expects \cite{burdinetal} the
system to magnetically order much before the Kondo temperature vanishes. If
this is true, then one expects the quantum critical behavior to be very
similar to metallic Ising spin glasses, i.e., to assume the conventional
Hertz-Millis form, at least the for mean-field spin glass models we discussed
here. The precise relevance of this paramagnetic spin liquid solution thus
remains unclear, at least for systems with weak or no disorder in the
conduction band.

\begin{itemize}
\item \emph{Fractionalized two-fluid behavior of electronic Griffiths phases }
\end{itemize}

The situation seems more promising in the presence of sufficient amounts of
disorder, where the electronic Griffiths phase forms. As we have seen in Sec.
\ref{sub:andersongriff}, here the disorder generates a very broad distribution
of local Kondo temperatures, making the system much more sensitive to RKKY
interactions. This mechanism has recently been studied within an extended DMFT
approach \cite{tanaskovicetal05}, which is able to incorporate both the
formation of the Griffiths phase, and the effects of frustrating RKKY
interactions leading to spin-glass dynamics. At the local impurity level, the
problem still reduced to the Bose-Fermi Kondo model, but the presence of
conduction electron disorder qualitatively modifies the self-consistency
conditions determining the form of $\chi\left(  \omega_{n}\right)  $.

To obtain a sufficient condition for decoupling, we examine the stability of
the Fermi liquid solution, by considering the limit of infinitesimal RKKY
interactions. To leading order we replace $\chi(\omega_{n})\longrightarrow
\chi_{o}\left(  \omega_{n}\right)  \equiv\chi(\omega_{n} ;J=0)$, \linebreak
and the calculation reduces to the {}\textquotedblleft bare
model\textquotedblright\ of Ref.~\cite{tanaskovicetal05}. In this case, from
Eq.~(\ref{eq:ptkpower}), $P\left(  T_{K}\right)  \sim T_{K}^{\alpha-1}$, where
$\alpha\sim1/W^{2}$, and%
\begin{equation}
\chi_{0}\left(  \omega_{n}\right)  \sim\int dT_{K}P\left(  T_{K}\right)
\chi\left(  \omega_{n},T_{K}\right)  \sim\chi_{0}\left(  0\right)
-C_{0}\left|  \omega_{n}\right|  ^{1-\epsilon}, \label{eq:chizero}%
\end{equation}
where $\epsilon=2-\alpha$. Thus, for sufficiently strong disorder (i.e. within
the electronic Griffiths phase), even the {}``bare'' bosonic bath is
sufficiently singular to generate decoupling. The critical value of $W$ will
be modified by self-consistency, but it is clear that decoupling will occur
for sufficiently large disorder.

Once decoupling is present, the system is best viewed as composed of two
fluids, one made up of a fraction $n$ of decoupled spins, and the other of a
fraction $(1-n)$ of Kondo screened spins. The self-consistent $\chi\left(
\omega_{n}\right)  $ acquires contributions from both fluids
\begin{equation}
\chi\left(  \omega_{n}\right)  =n\chi_{dc}\left(  \omega_{n}\right)  +\left(
1-n\right)  \chi_{s}\left(  \omega_{n}\right)  . \label{eq:twofluidchi}%
\end{equation}
A careful analysis \cite{tanaskovicetal05} shows that, for a bath
characterized by an exponent $\epsilon$%
\begin{align}
\chi_{dc}\left(  \omega_{n}\right)   &  \sim\chi_{dc}\left(  0\right)
-C\left|  \omega_{n}\right|  ^{1-\left(  2-\epsilon\right)  }%
;\label{eq:decoupledchi}\\
\chi_{s}\left(  \omega_{n}\right)   &  \sim\chi_{s}\left(  0\right)
-C^{\prime}\left|  \omega_{n}\right|  ^{1-\left(  2-\epsilon-1/\nu\right)  },
\label{eq:screenedchi}%
\end{align}
where $\nu=\nu\left(  \epsilon\right)  $ is a critical exponent governing how
the Kondo scale vanishes at the quantum critical point of the Bose-Fermi
model. Since $\nu>0$, the contribution of the decoupled fluid is more singular
and dominates at lower frequencies. Self-consistency then yields $\epsilon=1$,
as in the familiar spin liquid state of Sachdev and Ye \cite{sachdevye}. For
$\epsilon=1$, the local susceptibility is logarithmically divergent (both in
$\omega_{n}$ and $T$). This does not necessarily mean that the bulk
susceptibility, which is the experimentally relevant quantity, behaves in the
same manner \cite{parcolletgeorges}. More work remains to be done to determine
the precise low temperature form of this and other physical quantities, and to
assert the relevance of this mechanism for specific materials.

\begin{figure}[ptb]
\begin{center}
\includegraphics[  width=3.4in,
keepaspectratio]{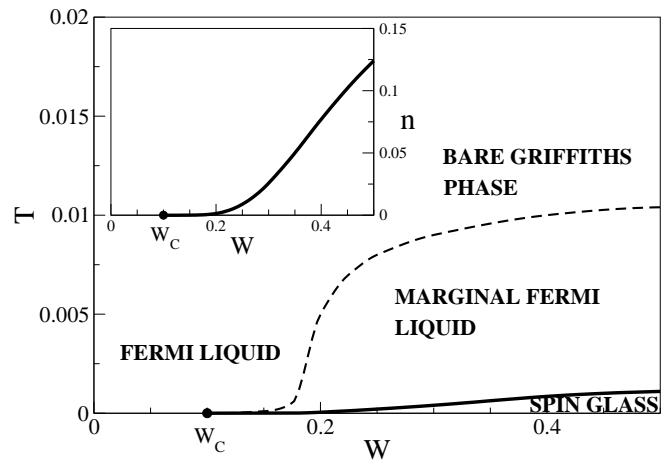}
\end{center}
\caption{Phase diagram of the electronic Griffiths phase model with RKKY
interactions \cite{tanaskovicetal05}. The inset shows the fraction of
decoupled spins as a function of the disorder strength $W$.}%
\label{cap:fig16}%
\end{figure}

As in the case where conduction electron disorder is absent, the spin liquid
state is unstable towards spin-glass ordering at sufficiently low
temperatures. However, numerical estimates for the Griffiths phase model
\cite{tanaskovicetal05} suggest a surprisingly wide temperature window where
the marginal behavior should persist above the ordering temperature.
Fig.~\ref{cap:fig16} represents the predicted phase diagram of this model. For
weak disorder the system is in the Fermi liquid phase, while for $W>W_{c}$ the
marginal Fermi liquid phase emerges. The crossover temperature (dashed line)
delimiting this regime can be estimated from the frequency up to which the
logarithmic behavior of the local dynamical susceptibility $\chi(i\omega) $ is
observed. The spin glass phase, obtained from Eq.~(\ref{eq:sg}), appears only
at the lowest temperatures, well below the marginal Fermi liquid boundary.
Interestingly, recent experiments \cite{dougetal} have indeed found evidence
of dynamical spin freezing in the milliKelvin temperature range for the same
Kondo alloys that display normal phase NFL behavior in a much broader
temperature window. 

The two-fluid phenomenology of the disordered Kondo lattice we have described
above is very reminiscent of earlier work on the clean Kondo lattice, where
the conduction electrons effectively decouple from the local moments, the
latter forming a spin liquid state
\cite{colemanandrei,kaganetal92,demleretal02,senthiletal,senthiletal2}. The
major difference between the results presented in this Section and these other
cases is that here local spatial disorder fluctuations lead to an
\emph{inhomogeneous} coexistence of the two fluids, as each site decouples or
not from the conduction electrons depending on its local properties. The
discussed mean-field models should be considered as merely the first examples
of this fascinating physics. The specific features of the spin liquid behavior
that was obtained from these models may very well prove to be too restrictive
and perhaps even inaccurate. For example, the specific heat enhancements may
well be overestimated, reflecting the residual $T=0$ entropy of the mean-field
models. Nevertheless, the physics of Kondo screening being destroyed by the
interplay of disorder and RKKY interactions will almost certainly play a
central role in determining the properties of many NFL systems, and clearly
needs to be investigated in more detail in the future.

\subsubsection{Electron glasses, freezing in the charge sector, and the
quantum AT line}

\label{sub:electronglass}

Glassy behavior in disordered electronic systems is not limited to phenomena
associated with the freezing of spins. In fact, glassy physics in the charge
sector was already envisioned a long time ago in the pioneering works of Efros
and Shklovskii \cite{re:Efros75,doped-book} and Pollak \cite{re:Pollak84} on
disordered insulators. It is expected to result from the competition of the
long-ranged Coulomb interaction and disorder. The physical picture that has
emerged from these works is easy to understand. In absence of disorder, the
Coulomb repulsion tends to keep the electrons as far from each other as
possible. If quantum or thermal fluctuations are sufficiently small, the
electrons tend to assume a periodic pattern, leading to the formation of a
charge density wave (e.g. a Wigner crystal). In contrast, disorder tends to
arrange the electrons in a random fashion, opposing the periodic pattern
favored by the Coulomb interaction. When both effects are comparable the
system is frustrated: there now exist many different low energy configurations
separated by potential barriers, leading to glassy dynamics.

Of course, these glassy effects are most pronounced in disordered insulators,
where quantum fluctuations are minimized by electron localization. Over the
last thirty years a large number of theoretical studies have concentrated on
the physics of the Coulomb glass. In this review we will not attempt to
describe in detail this large body of work, since many of these results do not
directly touch upon the non-Fermi liquid physics observed in metals. We will
limit our attention to those works that concentrated on the manifestations of
electronic glass behavior on the metallic side of the metal-insulator
transition. The original works of Efros and Shklovskii
\cite{re:Efros75,doped-book} and Pollak \cite{re:Pollak84} concentrated on
classical models for the electron glass, as did the work of many followers.
Because most workers used numerical approaches to investigate the problem, it
proved difficult to incorporate the effects of quantum fluctuations into this
picture, which cannot be avoided close to the metal-insulator transition.

\begin{figure}[ptb]
\begin{center}
\includegraphics[  width=3.4in,
keepaspectratio]{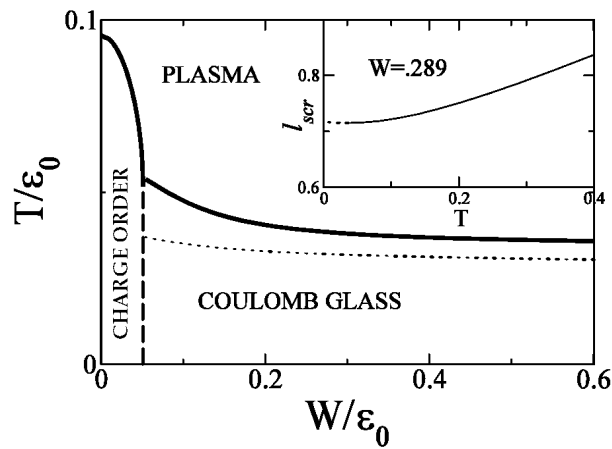}
\end{center}
\caption{Phase diagram of the three-dimensional classical Coulomb glass, as
obtained from the EDMFT approach, following Ref. \cite{pankov05prl}. The full
horizontal line indicates the glass transition temperature, and the dotted
line shows where the entropy of the fluid solution turns negative. The inset
shows the temperature dependence of the screening length as a function of
temperature.}%
\label{cap:fig17}%
\end{figure}

More recently, extended DMFT approaches (EDMFT) \cite{chitra00prl,qimiao00prb}
were shown to capture many relevant aspect of the classical Coulomb glasses,
such as the formation of the Coulomb gap \cite{pastor-prl99,pankov05prl}
(Fig.~\ref{cap:fig17}). In addition, these theories were able to discuss the
effects of quantum fluctuations \cite{pastor-prl99,mitglass-prl03} in the
vicinity to the metal-insulator transition. In the following we briefly
describe the physical picture of the Coulomb glass emerging from these theories.

\begin{itemize}
\item \emph{Coulomb gap, the replicon mode, and self-organized criticality}
\end{itemize}

The classical Coulomb glass model of Efros and Shklovskii (ES)
\cite{re:Efros75,doped-book} is given by the Hamiltonian%
\begin{equation}
H=\sum_{i}v_{i}n_{i}+\frac{1}{2}\sum_{ij}\frac{e^{2}}{r_{ij}}(n_{i}%
-K)(n_{j}-K). \label{cgham}%
\end{equation}
Here $n_{i}=0,1$ is the electron occupation number, and $v_{i}$ is a Gaussian
distributed random potential of variance $W^{2}$. In their classic work on
this model ES presented convincing evidence that at $T=0$ a soft {}``Coulomb
gap'' emerges in the single-particle density of states (DOS) which, in
arbitrary dimension $d$, takes the form%
\begin{equation}
g(\varepsilon)\sim\varepsilon^{d-1}.
\end{equation}
>From a general point of view this result is quite surprising. It indicates a
power-law distribution of excitation energies, i.e. the absence of a
characteristic energy scale for excitations above the ground state. Such
behavior is common in models with broken continuous symmetry, where it
reflects the corresponding Goldstone modes, but is generally not expected in
discrete symmetry models, such as the one used by ES. Within the EDMFT theory
of the Coulomb glass \cite{pastor-prl99,pankov05prl,muller04prl} this puzzling
feature finds a natural explanation: it reflects the emergence of the soft
{}``replicon'' mode, similarly to the classical and quantum spin-glass models
we have already discussed. As a result, again in direct analogy with spin
glass models, the ordered state of the Coulomb glass is expected to display
self-organized criticality \cite{re:Pazmandi99} and hysteresis behavior
\cite{horbach02} characterized by avalanches on all scales.

Another interesting feature of the Coulomb glass is worth mentioning. As
explicitly demonstrated by examining the broken replica symmetric solution of
the mean-field equations \cite{muller04prl}, the screening length is found to
diverge as $\ell_{scr}\sim1/T$ as a result of the vanishing zero-field cooled
compressibility \cite{pastor-prl99} within the glassy phase. This result is
significant because it explains the absence of screening in the ground state
of the Coulomb glass, in agreement with the assumptions of the ES theory. It
also indicates the divergence of the effective coordination number in the
$T\longrightarrow0$ limit, giving further credence to the predictions of the
mean-field approach.

\begin{itemize}
\item \emph{Quantum Coulomb glass and the metallic glass phase }
\end{itemize}

\begin{figure}[ptb]
\begin{center}
\includegraphics[  width=3.2in,
keepaspectratio]{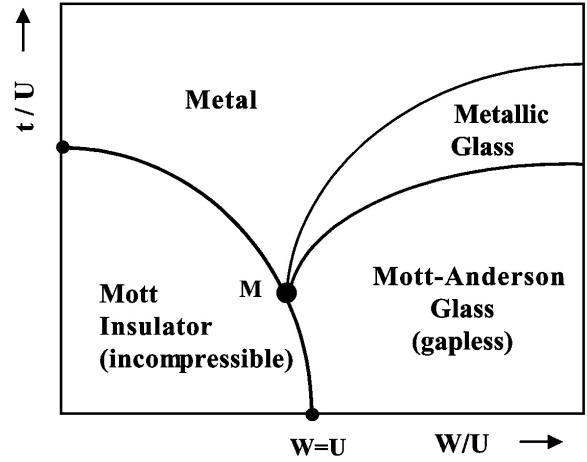}
\end{center}
\caption{Global phase diagram for the disordered Hubbard model
\cite{mitglass-prl03}, as a function of the hopping element $t$ and the
disordered strength $W$, both expressed in units of the on-site interaction
$U$. The size of the metallic glass phase is determined by the strength of the
inter-site Coulomb interaction $V$.}%
\label{cap:fig18}%
\end{figure}

When quantum fluctuations are introduced in the model by including hopping
amplitudes $t_{ij}$ between lattice sites , one must examine the $T=0$
transition where glassy freezing of electrons is suppressed. Such quantum
melting of the electron glass was examined in Refs.
\cite{pastor-prl99,mitglass-prl03}. Interestingly, the theory shows
\cite{mitglass-prl03} that the approach to an Anderson-like insulator is a
singular perturbation to the stability of the glassy phase. Physically, this
reflects the fact that the quantum fluctuations in this model arise from the
mobility of itinerant electrons - which is expected to vanish precisely at the
metal-insulator transition. As a result, the glass transition is predicted to
generically \textit{precede} the metal-insulator transition
(Fig.~\ref{cap:fig18}), giving rise to an intermediate metallic glass phase,
consistent with some recent experiments \cite{bogdanovich-prl02}.

\begin{itemize}
\item \emph{Critical behavior along the quantum AT line }
\end{itemize}

As we can see from Eq.~(\ref{cgham}), the Coulomb glass Hamiltonian is
essentially identical to that of an antiferromagnetic random field Ising model
(RFIM) with long range interactions. The random potential $v_{i}$ plays the
role of a random symmetry-breaking field, and as a result, the finite
temperature glass transition in our model assumes the character of a
de-Almeida-Thouless (AT) line \cite{re:AT,re:Mezard86}. The $T=0$ transition
separating the Fermi liquid from a NFL metallic glass phase thus assumes the
character of a {}``quantum AT line''. A complete description of this quantum
critical point can be obtained within the mean-field formulation, which has
been studied in detail in Ref. \cite{Denis}. Similarly as in metallic spin
glasses (see Section \ref{sub:qcp-msg}), within the quantum critical regime and
the entire metallic glass phase, the resistivity acquires a non-Fermi liquid
$T^{3/2}$ form, as seen in some experiments \cite{bogdanovich-prl02}. In
contrast to the metallic spin glass scenario, the approach to the quantum AT
line is characterized \cite{Denis} by a crossover scale
\begin{equation}
T^{\ast}\sim\delta t^{2},
\end{equation}
where $\delta t=(t-t_{c})/t_{c}$ measures the distance to the critical point.
This indicates a much broader quantum critical region in this case, which may
explain the apparent suppression of weak localization (Fermi liquid)
corrections in two dimensional electron systems near the metal-insulator
transition \cite{abekravsarachik}.

\section{Conclusions and open problems}

\label{sec:conclusions}

We have seen throughout this review that many routes can lead to
disorder-induced NFL behavior. In this concluding section, we would like to
pause to analyze general trends of what has been done so far, what is still
missing, and what are the most promising directions for future work.

We have seen how the phenomenology of quantum Griffiths singularities seems to
afford a natural description for many of the anomalies seen in several
systems, from doped semiconductors to disordered heavy fermion systems. On the
other hand, we have listed several outstanding points which still need to (and
should) be clarified. More generally, however, we would like to stress a
different question. Griffiths phases rely on the existence of rare disorder
configurations which generate essentially non-interacting fluctuators. It is
the slow quantum dynamics of these fluctuators which eventually leads to the
known singularities. This situation is often described as being one in which
quantum mechanics (tunneling) is a {}``dangerously irrelevant operator''
responsible for changing the scaling dimension of otherwise classical (and
non-interacting) objects. On the other hand, we have described how
droplet-droplet interactions can dramatically change this picture and be one
possible destabilizing factor of Griffiths physics
(Section~\ref{sub:metalgriff}). This effect represents essentially the
\emph{delocalization} of the local (droplet) spin modes which are at the heart
of the Griffiths phases. An important outstanding issue in this context is the
determination of the lower critical dimension for spin mode localization. This
is especially important in the case of metallic systems where the spin
interactions are long-ranged.

At or below this lower critical dimension, there appears the question of the
existence of an infinite randomness fixed point in metallic systems with
continuous symmetry. These represent the best candidates so far for a quantum
Griffiths phase which can survive down to zero temperature, as discussed in
Section~\ref{sub:metalgriff}. Below the (still unknown) lower critical
dimension for spin mode localization, and based on the well understood case of
insulating systems, we would expect that this Griffiths phase would be a
precursor of a phase transition governed by an infinite randomness fixed
point. In fact, this is probably the only known route for such a point that
does not require fine tuning in systems with continuous symmetry. Indeed, as
explained in Section~\ref{sub:irfp}, except for a few special model systems
(the nearest-neighbor random antiferromagnetic Heisenberg and XX chains being
the most prominent examples), almost all other perturbations, such as
increased connectivity and/or ferromagnetic interactions, seem to lead to a
large spin phase. These include the important case studied by Bhatt and Lee in
the context of doped semiconductors. Therefore, the rigorous establishment of
the stability of a Griffiths phase and an associated infinite randomness fixed
point in systems with continuous symmetry and dissipation would be an
important achievement.

Above the lower critical dimension, we expect glassy dynamics to take center
stage. Here, very interesting possibilities exist for non-trivial physics even
in the paramagnetic phase, as hinted by the results and references quoted in
Section~\ref{sub:spinliquid}. Other questions which arise are: (i) Is there a
Griffiths phase close to the spin glass quantum critical point? (ii) Is it
governed by an infinite randomness fixed point or is it characterized by
finite disorder? (iii) How is the charge transport affected by the spin
dynamics? (iv) What is the role of dissipation?

The exploration of glassiness in the charge sector also deserves further
attention. Here again long range Coulomb interactions play a crucial role. The
interplay with the spin sector is still largely unexplored. This seems
especially important in the case of the disorder-induced metal-insulator
transition both in 3 and in 2 dimensions.

We hope that addressing and answering some of these questions will provide
important clues as to the origin of non-Fermi liquid behavior in disordered systems.

\begin{acknowledgments}
We have benefited over the years from discussions with many colleagues and
collaborators. Some deserve special mention: E. Abrahams, M. C. O. Aguiar, M.
Aronson, R. N. Bhatt, A. H. Castro Neto, P. Coleman, M. A. Continentino, D.
Dalidovich, A. Georges, L. P. Gor'kov, D. R. Grempel, J. A. Hoyos, G. Kotliar,
D. E. MacLaughlin, A. J. Millis, D. K. Morr, J. A. Mydosh, C. Panagopoulos, D.
Popovi\'{c}, M. J. Rozenberg, S. Sachdev, J. Schmalian, Q. Si, G. Stewart, S.
Sülow, D. Tanaskovi\'{c}, C. M. Varma, A. P. Vieira, T. Vojta, K. Yang, and G. Zaránd.

This work was supported by FAPESP through grant 01/00719-8 (E.M.),
by CNPq through grant 302535/02-0 (E.M.), by the NSF through grant
NSF-0234215 (V.D.), and the National High Magnetic Field
Laboratory (V.D. and E.M.).

\end{acknowledgments}


\begin{thebibliography}{310}
\expandafter\ifx\csname
natexlab\endcsname\relax\def\natexlab#1{#1}\fi
\expandafter\ifx\csname bibnamefont\endcsname\relax
  \def\bibnamefont#1{#1}\fi
\expandafter\ifx\csname bibfnamefont\endcsname\relax
  \def\bibfnamefont#1{#1}\fi
\expandafter\ifx\csname citenamefont\endcsname\relax
  \def\citenamefont#1{#1}\fi
\expandafter\ifx\csname url\endcsname\relax
  \def\url#1{\texttt{#1}}\fi
\expandafter\ifx\csname
urlprefix\endcsname\relax\def\urlprefix{URL }\fi
\providecommand{\bibinfo}[2]{#2}
\providecommand{\eprint}[2][]{\url{#2}}

\bibitem[{\citenamefont{Abou-Chacra}
  \emph{et~al.}(1973)\citenamefont{Abou-Chacra, Anderson, and
  Thouless}}]{abouetal}
\bibinfo{author}{\bibnamefont{Abou-Chacra}, \bibfnamefont{R.}},
  \bibinfo{author}{\bibfnamefont{P.~W.} \bibnamefont{Anderson}}, and
  \bibinfo{author}{\bibfnamefont{D.~J.} \bibnamefont{Thouless}},
  \bibinfo{year}{1973}, \bibinfo{journal}{J. Phys. C}
  \textbf{\bibinfo{volume}{6}}, \bibinfo{pages}{1734}.

\bibitem[{\citenamefont{Abrahams} \emph{et~al.}(1979)\citenamefont{Abrahams,
  Anderson, Licciardello, and Ramakrishnan}}]{gang4}
\bibinfo{author}{\bibnamefont{Abrahams}, \bibfnamefont{E.}},
  \bibinfo{author}{\bibfnamefont{P.~W.} \bibnamefont{Anderson}},
  \bibinfo{author}{\bibfnamefont{D.~C.} \bibnamefont{Licciardello}}, and
  \bibinfo{author}{\bibfnamefont{T.~V.} \bibnamefont{Ramakrishnan}},
  \bibinfo{year}{1979}, \bibinfo{journal}{Phys. Rev. Lett.}
  \textbf{\bibinfo{volume}{42}}, \bibinfo{pages}{673}.

\bibitem[{\citenamefont{Abrahams} \emph{et~al.}(2001)\citenamefont{Abrahams,
  Kravchenko, and Sarachik}}]{abekravsarachik}
\bibinfo{author}{\bibnamefont{Abrahams}, \bibfnamefont{E.}},
  \bibinfo{author}{\bibfnamefont{S.~V.} \bibnamefont{Kravchenko}}, and
  \bibinfo{author}{\bibfnamefont{M.~P.} \bibnamefont{Sarachik}},
  \bibinfo{year}{2001}, \bibinfo{journal}{Rev. Mod. Phys.}
  \textbf{\bibinfo{volume}{73}}, \bibinfo{pages}{251}.

\bibitem[{\citenamefont{Abrikosov}(1962)}]{abrikosov62}
\bibinfo{author}{\bibnamefont{Abrikosov}, \bibfnamefont{A.~A.}},
  \bibinfo{year}{1962}, \bibinfo{journal}{Sov. Phys. JETP}
  \textbf{\bibinfo{volume}{14}}, \bibinfo{pages}{408}.

\bibitem[{\citenamefont{Abrikosov} \emph{et~al.}(1975)\citenamefont{Abrikosov,
  Gor'kov, and Dzyaloshinskii}}]{agd}
\bibinfo{author}{\bibnamefont{Abrikosov}, \bibfnamefont{A.~A.}},
  \bibinfo{author}{\bibfnamefont{L.~P.} \bibnamefont{Gor'kov}}, and
  \bibinfo{author}{\bibfnamefont{I.~E.} \bibnamefont{Dzyaloshinskii}},
  \bibinfo{year}{1975}, \emph{\bibinfo{title}{Methods of {Q}uantum {F}ield
  {T}heory in {S}tatistical {P}hysics}} (\bibinfo{publisher}{Dover},
  \bibinfo{address}{New York}).

\bibitem[{\citenamefont{Abrikosov and Khalatnikov}(1958)}]{abrikkhalat58}
\bibinfo{author}{\bibnamefont{Abrikosov}, \bibfnamefont{A.~A.}}, and
  \bibinfo{author}{\bibfnamefont{I.~M.} \bibnamefont{Khalatnikov}},
  \bibinfo{year}{1958}, \bibinfo{journal}{Sov. Phys. JETP}
  \textbf{\bibinfo{volume}{6}}, \bibinfo{pages}{888}.

\bibitem[{\citenamefont{Aeppli and Fisk}(1992)}]{aepplifisk}
\bibinfo{author}{\bibnamefont{Aeppli}, \bibfnamefont{G.}}, and
  \bibinfo{author}{\bibfnamefont{Z.}~\bibnamefont{Fisk}}, \bibinfo{year}{1992},
  \bibinfo{journal}{Comments Condens. Matter Phys.}
  \textbf{\bibinfo{volume}{16}}, \bibinfo{pages}{155}.

\bibitem[{\citenamefont{Affleck and
  Ludwig}(1991{\natexlab{a}})}]{affleckludwig91b}
\bibinfo{author}{\bibnamefont{Affleck}, \bibfnamefont{I.}}, and
  \bibinfo{author}{\bibfnamefont{A.~W.~W.} \bibnamefont{Ludwig}},
  \bibinfo{year}{1991}{\natexlab{a}}, \bibinfo{journal}{Nuc. Phys. B}
  \textbf{\bibinfo{volume}{360}}, \bibinfo{pages}{641}.

\bibitem[{\citenamefont{Affleck and
  Ludwig}(1991{\natexlab{b}})}]{affleckludwig91}
\bibinfo{author}{\bibnamefont{Affleck}, \bibfnamefont{I.}}, and
  \bibinfo{author}{\bibfnamefont{A.~W.~W.} \bibnamefont{Ludwig}},
  \bibinfo{year}{1991}{\natexlab{b}}, \bibinfo{journal}{Nuc. Phys. B}
  \textbf{\bibinfo{volume}{352}}, \bibinfo{pages}{849}.

\bibitem[{\citenamefont{Affleck and
  Ludwig}(1991{\natexlab{c}})}]{affleckludwig91c}
\bibinfo{author}{\bibnamefont{Affleck}, \bibfnamefont{I.}}, and
  \bibinfo{author}{\bibfnamefont{A.~W.~W.} \bibnamefont{Ludwig}},
  \bibinfo{year}{1991}{\natexlab{c}}, \bibinfo{journal}{Phys. Rev. Lett.}
  \textbf{\bibinfo{volume}{67}}, \bibinfo{pages}{161}.

\bibitem[{\citenamefont{Aguiar} \emph{et~al.}(2003)\citenamefont{Aguiar,
  Miranda, and Dobrosavljevi\'{c}}}]{aguiaretal1}
\bibinfo{author}{\bibnamefont{Aguiar}, \bibfnamefont{M.~C.~O.}},
  \bibinfo{author}{\bibfnamefont{E.}~\bibnamefont{Miranda}}, and
  \bibinfo{author}{\bibfnamefont{V.}~\bibnamefont{Dobrosavljevi\'{c}}},
  \bibinfo{year}{2003}, \bibinfo{journal}{Phys. Rev. B}
  \textbf{\bibinfo{volume}{68}}, \bibinfo{pages}{125104}.

\bibitem[{\citenamefont{Aguiar} \emph{et~al.}(2004)\citenamefont{Aguiar,
  Miranda, Dobrosavljevi\'{c}, Abrahams, and Kotliar}}]{aguiaretal2}
\bibinfo{author}{\bibnamefont{Aguiar}, \bibfnamefont{M.~C.~O.}},
  \bibinfo{author}{\bibfnamefont{E.}~\bibnamefont{Miranda}},
  \bibinfo{author}{\bibfnamefont{V.}~\bibnamefont{Dobrosavljevi\'{c}}},
  \bibinfo{author}{\bibfnamefont{E.}~\bibnamefont{Abrahams}}, and
  \bibinfo{author}{\bibfnamefont{G.}~\bibnamefont{Kotliar}},
  \bibinfo{year}{2004}, \bibinfo{journal}{Europhys. Lett.}
  \textbf{\bibinfo{volume}{67}}, \bibinfo{pages}{226}.

\bibitem[{\citenamefont{de~Almeida and Thouless}(1978)}]{re:AT}
\bibinfo{author}{\bibnamefont{de~Almeida}, \bibfnamefont{J.~R.~L.}}, and
  \bibinfo{author}{\bibfnamefont{D.~J.} \bibnamefont{Thouless}},
  \bibinfo{year}{1978}, \bibinfo{journal}{J. Phys. A}
  \textbf{\bibinfo{volume}{11}}, \bibinfo{pages}{983}.

\bibitem[{\citenamefont{Altshuler and Aronov}(1979)}]{altshuler-79b}
\bibinfo{author}{\bibnamefont{Altshuler}, \bibfnamefont{B.~L.}}, and
  \bibinfo{author}{\bibfnamefont{A.~B.} \bibnamefont{Aronov}},
  \bibinfo{year}{1979}, \bibinfo{journal}{JETP Lett.}
  \textbf{\bibinfo{volume}{50}}, \bibinfo{pages}{968}.

\bibitem[{\citenamefont{Anderson and Yuval}(1969)}]{Yuval-Anderson1}
\bibinfo{author}{\bibnamefont{Anderson}, \bibfnamefont{P.}}, and
  \bibinfo{author}{\bibfnamefont{G.}~\bibnamefont{Yuval}},
  \bibinfo{year}{1969}, \bibinfo{journal}{Phys. Rev. Lett.}
  \textbf{\bibinfo{volume}{23}}, \bibinfo{pages}{89}.

\bibitem[{\citenamefont{Anderson} \emph{et~al.}(1970)\citenamefont{Anderson,
  Yuval, and Hamman}}]{Yuval-Anderson3}
\bibinfo{author}{\bibnamefont{Anderson}, \bibfnamefont{P.}},
  \bibinfo{author}{\bibfnamefont{G.}~\bibnamefont{Yuval}}, and
  \bibinfo{author}{\bibfnamefont{D.}~\bibnamefont{Hamman}},
  \bibinfo{year}{1970}, \bibinfo{journal}{Phys. Rev. B}
  \textbf{\bibinfo{volume}{1}}, \bibinfo{pages}{4464}.

\bibitem[{\citenamefont{Anderson}(1958)}]{andersonloc}
\bibinfo{author}{\bibnamefont{Anderson}, \bibfnamefont{P.~W.}},
  \bibinfo{year}{1958}, \bibinfo{journal}{Phys. Rev.}
  \textbf{\bibinfo{volume}{109}}, \bibinfo{pages}{1492}.

\bibitem[{\citenamefont{Anderson}(1961)}]{Anderson}
\bibinfo{author}{\bibnamefont{Anderson}, \bibfnamefont{P.~W.}},
  \bibinfo{year}{1961}, \bibinfo{journal}{Phys. Rev.}
  \textbf{\bibinfo{volume}{124}}, \bibinfo{pages}{41}.

\bibitem[{\citenamefont{de~Andrade}
  \emph{et~al.}(1998)\citenamefont{de~Andrade, Chau, Dickey, Dilley, Freeman,
  Gajewski, Maple, Movshovich, Neto, Castilla, and Jones}}]{andradeetal}
\bibinfo{author}{\bibnamefont{de~Andrade}, \bibfnamefont{M.~C.}},
  \bibinfo{author}{\bibfnamefont{R.}~\bibnamefont{Chau}},
  \bibinfo{author}{\bibfnamefont{R.~P.} \bibnamefont{Dickey}},
  \bibinfo{author}{\bibfnamefont{N.~R.} \bibnamefont{Dilley}},
  \bibinfo{author}{\bibfnamefont{E.~J.} \bibnamefont{Freeman}},
  \bibinfo{author}{\bibfnamefont{D.~A.} \bibnamefont{Gajewski}},
  \bibinfo{author}{\bibfnamefont{M.~B.} \bibnamefont{Maple}},
  \bibinfo{author}{\bibfnamefont{R.}~\bibnamefont{Movshovich}},
  \bibinfo{author}{\bibfnamefont{A.~H.~C.} \bibnamefont{Neto}},
  \bibinfo{author}{\bibfnamefont{G.}~\bibnamefont{Castilla}}, and
  \bibinfo{author}{\bibfnamefont{B.~A.} \bibnamefont{Jones}},
  \bibinfo{year}{1998}, \bibinfo{journal}{Phys. Rev. Lett.}
  \textbf{\bibinfo{volume}{81}}, \bibinfo{pages}{5620}.

\bibitem[{\citenamefont{Andraka}(1994)}]{andraka94}
\bibinfo{author}{\bibnamefont{Andraka}, \bibfnamefont{B.}},
  \bibinfo{year}{1994}, \bibinfo{journal}{Phys. Rev. B}
  \textbf{\bibinfo{volume}{49}}, \bibinfo{pages}{3589}.

\bibitem[{\citenamefont{Andraka and Stewart}(1993)}]{andrakastewart}
\bibinfo{author}{\bibnamefont{Andraka}, \bibfnamefont{B.}}, and
  \bibinfo{author}{\bibfnamefont{G.~R.} \bibnamefont{Stewart}},
  \bibinfo{year}{1993}, \bibinfo{journal}{Phys. Rev. B}
  \textbf{\bibinfo{volume}{47}}, \bibinfo{pages}{3208}.

\bibitem[{\citenamefont{Andraka and Tsvelik}(1991)}]{andrakatsvelik91}
\bibinfo{author}{\bibnamefont{Andraka}, \bibfnamefont{B.}}, and
  \bibinfo{author}{\bibfnamefont{A.~M.} \bibnamefont{Tsvelik}},
  \bibinfo{year}{1991}, \bibinfo{journal}{Phys. Rev. Lett.}
  \textbf{\bibinfo{volume}{67}}, \bibinfo{pages}{2886}.

\bibitem[{\citenamefont{Aronson} \emph{et~al.}(1995)\citenamefont{Aronson,
  Obsorn, Robinson, Lynn, Chau, Seaman, and Maple}}]{aronsonetal95}
\bibinfo{author}{\bibnamefont{Aronson}, \bibfnamefont{M.~C.}},
  \bibinfo{author}{\bibfnamefont{R.}~\bibnamefont{Obsorn}},
  \bibinfo{author}{\bibfnamefont{R.~A.} \bibnamefont{Robinson}},
  \bibinfo{author}{\bibfnamefont{J.~W.} \bibnamefont{Lynn}},
  \bibinfo{author}{\bibfnamefont{R.}~\bibnamefont{Chau}},
  \bibinfo{author}{\bibfnamefont{C.~L.} \bibnamefont{Seaman}}, and
  \bibinfo{author}{\bibfnamefont{M.~B.} \bibnamefont{Maple}},
  \bibinfo{year}{1995}, \bibinfo{journal}{Phys. Rev. Lett.}
  \textbf{\bibinfo{volume}{75}}, \bibinfo{pages}{725}.

\bibitem[{\citenamefont{Aronson} \emph{et~al.}(2001)\citenamefont{Aronson,
  Osborn, Chau, Maple, Rainford, and Murani}}]{aronsonetal01}
\bibinfo{author}{\bibnamefont{Aronson}, \bibfnamefont{M.~C.}},
  \bibinfo{author}{\bibfnamefont{R.}~\bibnamefont{Osborn}},
  \bibinfo{author}{\bibfnamefont{R.}~\bibnamefont{Chau}},
  \bibinfo{author}{\bibfnamefont{M.~B.} \bibnamefont{Maple}},
  \bibinfo{author}{\bibfnamefont{B.~D.} \bibnamefont{Rainford}}, and
  \bibinfo{author}{\bibfnamefont{A.~P.} \bibnamefont{Murani}},
  \bibinfo{year}{2001}, \bibinfo{journal}{Phys. Rev. Lett.}
  \textbf{\bibinfo{volume}{87}}, \bibinfo{pages}{197205}.

\bibitem[{\citenamefont{Arrachea} \emph{et~al.}(2004)\citenamefont{Arrachea,
  Dalidovich, Dobrosavljevi\'c, and Rozenberg}}]{arrachea}
\bibinfo{author}{\bibnamefont{Arrachea}, \bibfnamefont{L.}},
  \bibinfo{author}{\bibfnamefont{D.}~\bibnamefont{Dalidovich}},
  \bibinfo{author}{\bibfnamefont{V.}~\bibnamefont{Dobrosavljevi\'c}}, and
  \bibinfo{author}{\bibfnamefont{M.~J.} \bibnamefont{Rozenberg}},
  \bibinfo{year}{2004}, \bibinfo{journal}{Phys. Rev. B}
  \textbf{\bibinfo{volume}{69}}, \bibinfo{pages}{064419}.

\bibitem[{\citenamefont{Balakirev} \emph{et~al.}(2003)\citenamefont{Balakirev,
  Betts, Migliori, Ono, Ando, and Boebinger}}]{balakirev-nature03}
\bibinfo{author}{\bibnamefont{Balakirev}, \bibfnamefont{F.~F.}},
  \bibinfo{author}{\bibfnamefont{J.~B.} \bibnamefont{Betts}},
  \bibinfo{author}{\bibfnamefont{A.}~\bibnamefont{Migliori}},
  \bibinfo{author}{\bibfnamefont{S.}~\bibnamefont{Ono}},
  \bibinfo{author}{\bibfnamefont{Y.}~\bibnamefont{Ando}}, and
  \bibinfo{author}{\bibfnamefont{G.~S.} \bibnamefont{Boebinger}},
  \bibinfo{year}{2003}, \bibinfo{journal}{Nature}
  \textbf{\bibinfo{volume}{424}}, \bibinfo{pages}{912}.

\bibitem[{\citenamefont{Barnes}(1976)}]{barnes76}
\bibinfo{author}{\bibnamefont{Barnes}, \bibfnamefont{S.~E.}},
  \bibinfo{year}{1976}, \bibinfo{journal}{J. Phys. F}
  \textbf{\bibinfo{volume}{7}}, \bibinfo{pages}{1375}.

\bibitem[{\citenamefont{Bauer} \emph{et~al.}(2002)\citenamefont{Bauer, Booth,
  Kwei, Chau, and Maple}}]{baueretal}
\bibinfo{author}{\bibnamefont{Bauer}, \bibfnamefont{E.~D.}},
  \bibinfo{author}{\bibfnamefont{C.~H.} \bibnamefont{Booth}},
  \bibinfo{author}{\bibfnamefont{G.~H.} \bibnamefont{Kwei}},
  \bibinfo{author}{\bibfnamefont{R.}~\bibnamefont{Chau}}, and
  \bibinfo{author}{\bibfnamefont{M.~B.} \bibnamefont{Maple}},
  \bibinfo{year}{2002}, \bibinfo{journal}{Phys. Rev. B}
  \textbf{\bibinfo{volume}{65}}, \bibinfo{pages}{245114}.

\bibitem[{\citenamefont{Baym and Pethick}(1991)}]{baympethick78}
\bibinfo{author}{\bibnamefont{Baym}, \bibfnamefont{G.}}, and
  \bibinfo{author}{\bibfnamefont{C.}~\bibnamefont{Pethick}},
  \bibinfo{year}{1991}, \emph{\bibinfo{title}{Landau {F}ermi {L}iquid {T}heory:
  {C}oncepts and {A}pplications}} (\bibinfo{publisher}{Wiley},
  \bibinfo{address}{New York}).

\bibitem[{\citenamefont{B{\'e}al-Monod and Maki}(1975)}]{bealmonodmaki75}
\bibinfo{author}{\bibnamefont{B{\'e}al-Monod}, \bibfnamefont{M.~T.}}, and
  \bibinfo{author}{\bibfnamefont{K.}~\bibnamefont{Maki}}, \bibinfo{year}{1975},
  \bibinfo{journal}{Phys. Rev. Lett.} \textbf{\bibinfo{volume}{34}},
  \bibinfo{pages}{1461}.

\bibitem[{\citenamefont{Belitz and Kirkpatrick}(1994)}]{kirkpatrick-rmp94}
\bibinfo{author}{\bibnamefont{Belitz}, \bibfnamefont{D.}}, and
  \bibinfo{author}{\bibfnamefont{T.~R.} \bibnamefont{Kirkpatrick}},
  \bibinfo{year}{1994}, \bibinfo{journal}{Rev. Mod. Phys.}
  \textbf{\bibinfo{volume}{66}}, \bibinfo{pages}{261}.

\bibitem[{\citenamefont{Beloborodov}
  \emph{et~al.}(2003)\citenamefont{Beloborodov, Efetov, Lopatin, and
  Vinokur}}]{efetov-prl03}
\bibinfo{author}{\bibnamefont{Beloborodov}, \bibfnamefont{I.~S.}},
  \bibinfo{author}{\bibfnamefont{K.~B.} \bibnamefont{Efetov}},
  \bibinfo{author}{\bibfnamefont{A.}~\bibnamefont{Lopatin}}, and
  \bibinfo{author}{\bibfnamefont{V.}~\bibnamefont{Vinokur}},
  \bibinfo{year}{2003}, \bibinfo{journal}{Phys. Rev. Lett.}
  \textbf{\bibinfo{volume}{91}}, \bibinfo{pages}{246801}.

\bibitem[{\citenamefont{Benfatto and Gallavotti}(1990)}]{benfattogallavotti90}
\bibinfo{author}{\bibnamefont{Benfatto}, \bibfnamefont{G.}}, and
  \bibinfo{author}{\bibfnamefont{G.}~\bibnamefont{Gallavotti}},
  \bibinfo{year}{1990}, \bibinfo{journal}{Phys. Rev. B}
  \textbf{\bibinfo{volume}{42}}, \bibinfo{pages}{9967}.

\bibitem[{\citenamefont{Berk and Schrieffer}(1966)}]{berkschrieffer}
\bibinfo{author}{\bibnamefont{Berk}, \bibfnamefont{N.~F.}}, and
  \bibinfo{author}{\bibfnamefont{J.~R.} \bibnamefont{Schrieffer}},
  \bibinfo{year}{1966}, \bibinfo{journal}{Phys. Rev. Lett.}
  \textbf{\bibinfo{volume}{17}}, \bibinfo{pages}{433}.

\bibitem[{\citenamefont{Bernal} \emph{et~al.}(1995)\citenamefont{Bernal,
  MacLaughlin, Lukefahr, and Andraka}}]{bernaletal}
\bibinfo{author}{\bibnamefont{Bernal}, \bibfnamefont{O.~O.}},
  \bibinfo{author}{\bibfnamefont{D.~E.} \bibnamefont{MacLaughlin}},
  \bibinfo{author}{\bibfnamefont{H.~G.} \bibnamefont{Lukefahr}}, and
  \bibinfo{author}{\bibfnamefont{B.}~\bibnamefont{Andraka}},
  \bibinfo{year}{1995}, \bibinfo{journal}{Phys. Rev. Lett.}
  \textbf{\bibinfo{volume}{75}}, \bibinfo{pages}{2023}.

\bibitem[{\citenamefont{Bhatt and Fisher}(1992)}]{bhattfisher92}
\bibinfo{author}{\bibnamefont{Bhatt}, \bibfnamefont{R.~N.}}, and
  \bibinfo{author}{\bibfnamefont{D.~S.} \bibnamefont{Fisher}},
  \bibinfo{year}{1992}, \bibinfo{journal}{Phys. Rev. Lett.}
  \textbf{\bibinfo{volume}{68}}, \bibinfo{pages}{3072}.

\bibitem[{\citenamefont{Bhatt and Lee}(1981)}]{bhattlee81}
\bibinfo{author}{\bibnamefont{Bhatt}, \bibfnamefont{R.~N.}}, and
  \bibinfo{author}{\bibfnamefont{P.~A.} \bibnamefont{Lee}},
  \bibinfo{year}{1981}, \bibinfo{journal}{J. Appl. Phys.}
  \textbf{\bibinfo{volume}{52}}, \bibinfo{pages}{1703}.

\bibitem[{\citenamefont{Bhatt and Lee}(1982)}]{bhattlee82}
\bibinfo{author}{\bibnamefont{Bhatt}, \bibfnamefont{R.~N.}}, and
  \bibinfo{author}{\bibfnamefont{P.~A.} \bibnamefont{Lee}},
  \bibinfo{year}{1982}, \bibinfo{journal}{Phys. Rev. Lett.}
  \textbf{\bibinfo{volume}{48}}, \bibinfo{pages}{344}.

\bibitem[{\citenamefont{Boebinger} \emph{et~al.}(1996)\citenamefont{Boebinger,
  Ando, Passner, Kimura, Okuya, Shimoyama, Kishio, Tamasaku, Ichikawa, and
  Uchida}}]{boebinger-prl96}
\bibinfo{author}{\bibnamefont{Boebinger}, \bibfnamefont{G.~S.}},
  \bibinfo{author}{\bibfnamefont{Y.}~\bibnamefont{Ando}},
  \bibinfo{author}{\bibfnamefont{A.}~\bibnamefont{Passner}},
  \bibinfo{author}{\bibfnamefont{T.}~\bibnamefont{Kimura}},
  \bibinfo{author}{\bibfnamefont{M.}~\bibnamefont{Okuya}},
  \bibinfo{author}{\bibfnamefont{J.}~\bibnamefont{Shimoyama}},
  \bibinfo{author}{\bibfnamefont{K.}~\bibnamefont{Kishio}},
  \bibinfo{author}{\bibfnamefont{K.}~\bibnamefont{Tamasaku}},
  \bibinfo{author}{\bibfnamefont{N.}~\bibnamefont{Ichikawa}}, and
  \bibinfo{author}{\bibfnamefont{S.}~\bibnamefont{Uchida}},
  \bibinfo{year}{1996}, \bibinfo{journal}{Phys. Rev. Lett.}
  \textbf{\bibinfo{volume}{77}}, \bibinfo{pages}{5417}.

\bibitem[{\citenamefont{Boechat} \emph{et~al.}(1996)\citenamefont{Boechat,
  Saguia, and Continentino}}]{boechatetal96}
\bibinfo{author}{\bibnamefont{Boechat}, \bibfnamefont{B.}},
  \bibinfo{author}{\bibfnamefont{A.}~\bibnamefont{Saguia}}, and
  \bibinfo{author}{\bibfnamefont{M.~A.} \bibnamefont{Continentino}},
  \bibinfo{year}{1996}, \bibinfo{journal}{Solid State Commun.}
  \textbf{\bibinfo{volume}{98}}, \bibinfo{pages}{411}.

\bibitem[{\citenamefont{Bogdanovich and Popovi\'c}(2002)}]{bogdanovich-prl02}
\bibinfo{author}{\bibnamefont{Bogdanovich}, \bibfnamefont{S.}}, and
  \bibinfo{author}{\bibfnamefont{D.}~\bibnamefont{Popovi\'c}},
  \bibinfo{year}{2002}, \bibinfo{journal}{Phys. Rev. Lett.}
  \textbf{\bibinfo{volume}{88}}, \bibinfo{pages}{236401}.

\bibitem[{\citenamefont{Bogenberger and v.~L{\"o}hneysen}(1995)}]{bogenlohn95}
\bibinfo{author}{\bibnamefont{Bogenberger}, \bibfnamefont{B.}}, and
  \bibinfo{author}{\bibfnamefont{H.}~\bibnamefont{v.~L{\"o}hneysen}},
  \bibinfo{year}{1995}, \bibinfo{journal}{Phys. Rev. Lett.}
  \textbf{\bibinfo{volume}{74}}, \bibinfo{pages}{1016}.

\bibitem[{\citenamefont{Booth} \emph{et~al.}(1998)\citenamefont{Booth,
  MacLaughlin, Heffner, Chau, Maple, and Kwei}}]{boothetal}
\bibinfo{author}{\bibnamefont{Booth}, \bibfnamefont{C.~H.}},
  \bibinfo{author}{\bibfnamefont{D.~E.} \bibnamefont{MacLaughlin}},
  \bibinfo{author}{\bibfnamefont{R.~H.} \bibnamefont{Heffner}},
  \bibinfo{author}{\bibfnamefont{R.}~\bibnamefont{Chau}},
  \bibinfo{author}{\bibfnamefont{M.~B.} \bibnamefont{Maple}}, and
  \bibinfo{author}{\bibfnamefont{G.~H.} \bibnamefont{Kwei}},
  \bibinfo{year}{1998}, \bibinfo{journal}{Phys. Rev. Lett.}
  \textbf{\bibinfo{volume}{81}}, \bibinfo{pages}{3960}.

\bibitem[{\citenamefont{Booth} \emph{et~al.}(2002)\citenamefont{Booth, Scheidt,
  Killer, Weber, and Kehrein}}]{boothetal2}
\bibinfo{author}{\bibnamefont{Booth}, \bibfnamefont{C.~H.}},
  \bibinfo{author}{\bibfnamefont{E.-W.} \bibnamefont{Scheidt}},
  \bibinfo{author}{\bibfnamefont{U.}~\bibnamefont{Killer}},
  \bibinfo{author}{\bibfnamefont{A.}~\bibnamefont{Weber}}, and
  \bibinfo{author}{\bibfnamefont{S.}~\bibnamefont{Kehrein}},
  \bibinfo{year}{2002}, \bibinfo{journal}{Phys. Rev. B}
  \textbf{\bibinfo{volume}{66}}, \bibinfo{pages}{140402}.

\bibitem[{\citenamefont{Bray and Moore}(1980)}]{braymoore80}
\bibinfo{author}{\bibnamefont{Bray}, \bibfnamefont{A.~J.}}, and
  \bibinfo{author}{\bibfnamefont{M.~A.} \bibnamefont{Moore}},
  \bibinfo{year}{1980}, \bibinfo{journal}{J. Phys. C}
  \textbf{\bibinfo{volume}{13}}, \bibinfo{pages}{L655}.

\bibitem[{\citenamefont{Bruno}(2001)}]{bruno01}
\bibinfo{author}{\bibnamefont{Bruno}, \bibfnamefont{P.}}, \bibinfo{year}{2001},
  \bibinfo{journal}{Phys. Rev. Lett.} \textbf{\bibinfo{volume}{87}},
  \bibinfo{pages}{137203}.

\bibitem[{\citenamefont{Burdin} \emph{et~al.}(2002)\citenamefont{Burdin,
  Grempel, and Georges}}]{burdinetal}
\bibinfo{author}{\bibnamefont{Burdin}, \bibfnamefont{S.}},
  \bibinfo{author}{\bibfnamefont{D.~R.} \bibnamefont{Grempel}}, and
  \bibinfo{author}{\bibfnamefont{A.}~\bibnamefont{Georges}},
  \bibinfo{year}{2002}, \bibinfo{journal}{Phys. Rev. B}
  \textbf{\bibinfo{volume}{66}}, \bibinfo{pages}{045111}.

\bibitem[{\citenamefont{B{\"u}ttgen}
  \emph{et~al.}(2000)\citenamefont{B{\"u}ttgen, Trinkl, Weber, Hemberger,
  Loidl, and Kehrein}}]{buttgenetal}
\bibinfo{author}{\bibnamefont{B{\"u}ttgen}, \bibfnamefont{N.}},
  \bibinfo{author}{\bibfnamefont{W.}~\bibnamefont{Trinkl}},
  \bibinfo{author}{\bibfnamefont{J.-E.} \bibnamefont{Weber}},
  \bibinfo{author}{\bibfnamefont{J.}~\bibnamefont{Hemberger}},
  \bibinfo{author}{\bibfnamefont{A.}~\bibnamefont{Loidl}}, and
  \bibinfo{author}{\bibfnamefont{S.}~\bibnamefont{Kehrein}},
  \bibinfo{year}{2000}, \bibinfo{journal}{Phys. Rev. B}
  \textbf{\bibinfo{volume}{62}}, \bibinfo{pages}{11545}.

\bibitem[{\citenamefont{Buyers}(1996)}]{buyers96}
\bibinfo{author}{\bibnamefont{Buyers}, \bibfnamefont{W.~J.~L.}},
  \bibinfo{year}{1996}, \bibinfo{journal}{Physica B}
  \textbf{\bibinfo{volume}{223-224}}, \bibinfo{pages}{9}.

\bibitem[{\citenamefont{Camjayi and Rozenberg}(2003)}]{rozenberg04prl}
\bibinfo{author}{\bibnamefont{Camjayi}, \bibfnamefont{A.}}, and
  \bibinfo{author}{\bibfnamefont{M.~J.} \bibnamefont{Rozenberg}},
  \bibinfo{year}{2003}, \bibinfo{journal}{Phys. Rev. Lett.}
  \textbf{\bibinfo{volume}{90}}, \bibinfo{pages}{217202}.

\bibitem[{\citenamefont{Carlon} \emph{et~al.}(2004)\citenamefont{Carlon,
  Lajk{\'o}, Rieger, and Igl{\'o}i}}]{carlonetal04}
\bibinfo{author}{\bibnamefont{Carlon}, \bibfnamefont{E.}},
  \bibinfo{author}{\bibfnamefont{P.}~\bibnamefont{Lajk{\'o}}},
  \bibinfo{author}{\bibfnamefont{H.}~\bibnamefont{Rieger}}, and
  \bibinfo{author}{\bibfnamefont{F.}~\bibnamefont{Igl{\'o}i}},
  \bibinfo{year}{2004}, \bibinfo{journal}{Phys. Rev. B}
  \textbf{\bibinfo{volume}{69}}, \bibinfo{pages}{144416}.

\bibitem[{\citenamefont{Castellani and Castro}(1986)}]{cdc-prb86}
\bibinfo{author}{\bibnamefont{Castellani}, \bibfnamefont{C.}}, and
  \bibinfo{author}{\bibfnamefont{C.~D.} \bibnamefont{Castro}},
  \bibinfo{year}{1986}, \bibinfo{journal}{Phys. Rev. B}
  \textbf{\bibinfo{volume}{34}}, \bibinfo{pages}{5935}.

\bibitem[{\citenamefont{Castellani}
  \emph{et~al.}(1984)\citenamefont{Castellani, Castro, Lee, and
  Ma}}]{cclm-prb84}
\bibinfo{author}{\bibnamefont{Castellani}, \bibfnamefont{C.}},
  \bibinfo{author}{\bibfnamefont{C.~D.} \bibnamefont{Castro}},
  \bibinfo{author}{\bibfnamefont{P.~A.} \bibnamefont{Lee}}, and
  \bibinfo{author}{\bibfnamefont{M.}~\bibnamefont{Ma}}, \bibinfo{year}{1984},
  \bibinfo{journal}{Phys. Rev. B} \textbf{\bibinfo{volume}{30}},
  \bibinfo{pages}{527}.

\bibitem[{\citenamefont{Castellani}
  \emph{et~al.}(1987)\citenamefont{Castellani, Kotliar, and Lee}}]{ckl}
\bibinfo{author}{\bibnamefont{Castellani}, \bibfnamefont{C.}},
  \bibinfo{author}{\bibfnamefont{B.~G.} \bibnamefont{Kotliar}}, and
  \bibinfo{author}{\bibfnamefont{P.~A.} \bibnamefont{Lee}},
  \bibinfo{year}{1987}, \bibinfo{journal}{Phys. Rev. Lett.}
  \textbf{\bibinfo{volume}{56}}, \bibinfo{pages}{1179}.

\bibitem[{\citenamefont{Castro~Neto}
  \emph{et~al.}(1998)\citenamefont{Castro~Neto, Castilla, and {B. A.
  Jones}}}]{castronetoetal1}
\bibinfo{author}{\bibnamefont{Castro~Neto}, \bibfnamefont{A.~H.}},
  \bibinfo{author}{\bibfnamefont{G.}~\bibnamefont{Castilla}}, and
  \bibinfo{author}{\bibnamefont{{B. A. Jones}}}, \bibinfo{year}{1998},
  \bibinfo{journal}{Phys. Rev. Lett.} \textbf{\bibinfo{volume}{81}},
  \bibinfo{pages}{3531}.

\bibitem[{\citenamefont{Castro~Neto and Jones}(2000)}]{castronetojones}
\bibinfo{author}{\bibnamefont{Castro~Neto}, \bibfnamefont{A.~H.}}, and
  \bibinfo{author}{\bibfnamefont{B.~A.} \bibnamefont{Jones}},
  \bibinfo{year}{2000}, \bibinfo{journal}{Phys. Rev. B}
  \textbf{\bibinfo{volume}{62}}(\bibinfo{number}{22}), \bibinfo{pages}{14975}.

\bibitem[{\citenamefont{Castro~Neto and
  Jones}(2004{\natexlab{a}})}]{castrojones04b}
\bibinfo{author}{\bibnamefont{Castro~Neto}, \bibfnamefont{A.~H.}}, and
  \bibinfo{author}{\bibfnamefont{B.~A.} \bibnamefont{Jones}},
  \bibinfo{year}{2004}{\natexlab{a}}, \bibinfo{journal}{cond-mat/0412020} .

\bibitem[{\citenamefont{Castro~Neto and
  Jones}(2004{\natexlab{b}})}]{castrojones04}
\bibinfo{author}{\bibnamefont{Castro~Neto}, \bibfnamefont{A.~H.}}, and
  \bibinfo{author}{\bibfnamefont{B.~A.} \bibnamefont{Jones}},
  \bibinfo{year}{2004}{\natexlab{b}}, \bibinfo{journal}{cond-mat/0411197} .

\bibitem[{\citenamefont{Chattopadhyay and Jarrell}(1997)}]{chatto}
\bibinfo{author}{\bibnamefont{Chattopadhyay}, \bibfnamefont{A.}}, and
  \bibinfo{author}{\bibfnamefont{M.}~\bibnamefont{Jarrell}},
  \bibinfo{year}{1997}, \bibinfo{journal}{Phys. Rev. B}
  \textbf{\bibinfo{volume}{56}}, \bibinfo{pages}{2920}.

\bibitem[{\citenamefont{Chattopadhyay}
  \emph{et~al.}(1998)\citenamefont{Chattopadhyay, Jarrell, Krishnamurthy, Ng,
  Sarrao, and Fisk}}]{chatto2}
\bibinfo{author}{\bibnamefont{Chattopadhyay}, \bibfnamefont{A.}},
  \bibinfo{author}{\bibfnamefont{M.}~\bibnamefont{Jarrell}},
  \bibinfo{author}{\bibfnamefont{H.~R.} \bibnamefont{Krishnamurthy}},
  \bibinfo{author}{\bibfnamefont{H.~K.} \bibnamefont{Ng}},
  \bibinfo{author}{\bibfnamefont{J.}~\bibnamefont{Sarrao}}, and
  \bibinfo{author}{\bibfnamefont{Z.}~\bibnamefont{Fisk}}, \bibinfo{year}{1998},
  \bibinfo{journal}{cond-mat/9805127} .

\bibitem[{\citenamefont{Chau and Maple}(1996)}]{chaumaple96}
\bibinfo{author}{\bibnamefont{Chau}, \bibfnamefont{R.}}, and
  \bibinfo{author}{\bibfnamefont{M.~B.} \bibnamefont{Maple}},
  \bibinfo{year}{1996}, \bibinfo{journal}{J. Phys.: Condens. Matter}
  \textbf{\bibinfo{volume}{8}}, \bibinfo{pages}{9939}.

\bibitem[{\citenamefont{Chayes} \emph{et~al.}(1986)\citenamefont{Chayes,
  Chayes, Fisher, and Spencer}}]{chayesetal86}
\bibinfo{author}{\bibnamefont{Chayes}, \bibfnamefont{J.~T.}},
  \bibinfo{author}{\bibfnamefont{L.}~\bibnamefont{Chayes}},
  \bibinfo{author}{\bibfnamefont{D.~S.} \bibnamefont{Fisher}}, and
  \bibinfo{author}{\bibfnamefont{T.}~\bibnamefont{Spencer}},
  \bibinfo{year}{1986}, \bibinfo{journal}{Phys. Rev. Lett.}
  \textbf{\bibinfo{volume}{57}}, \bibinfo{pages}{2999}.

\bibitem[{\citenamefont{Chitra and Kotliar}(2000)}]{chitra00prl}
\bibinfo{author}{\bibnamefont{Chitra}, \bibfnamefont{R.}}, and
  \bibinfo{author}{\bibfnamefont{G.}~\bibnamefont{Kotliar}},
  \bibinfo{year}{2000}, \bibinfo{journal}{Phys. Rev. Lett.}
  \textbf{\bibinfo{volume}{84}}, \bibinfo{pages}{3678}.

\bibitem[{\citenamefont{Chou} \emph{et~al.}(1995)\citenamefont{Chou, Belk,
  Kastner, Birgeneau, and Aharony}}]{chou-prl95}
\bibinfo{author}{\bibnamefont{Chou}, \bibfnamefont{F.~C.}},
  \bibinfo{author}{\bibfnamefont{N.}~\bibnamefont{Belk}},
  \bibinfo{author}{\bibfnamefont{M.}~\bibnamefont{Kastner}},
  \bibinfo{author}{\bibfnamefont{R.}~\bibnamefont{Birgeneau}}, and
  \bibinfo{author}{\bibfnamefont{A.}~\bibnamefont{Aharony}},
  \bibinfo{year}{1995}, \bibinfo{journal}{Phys. Rev. Lett.}
  \textbf{\bibinfo{volume}{75}}, \bibinfo{pages}{2204}.

\bibitem[{\citenamefont{Coleman}(1984)}]{colemansb}
\bibinfo{author}{\bibnamefont{Coleman}, \bibfnamefont{P.}},
  \bibinfo{year}{1984}, \bibinfo{journal}{Phys. Rev. B}
  \textbf{\bibinfo{volume}{29}}, \bibinfo{pages}{3035}.

\bibitem[{\citenamefont{Coleman}(1987)}]{colemanlong}
\bibinfo{author}{\bibnamefont{Coleman}, \bibfnamefont{P.}},
  \bibinfo{year}{1987}, \bibinfo{journal}{Phys. Rev. B}
  \textbf{\bibinfo{volume}{35}}, \bibinfo{pages}{5072}.

\bibitem[{\citenamefont{Coleman and Andrei}(1987)}]{colemanandrei}
\bibinfo{author}{\bibnamefont{Coleman}, \bibfnamefont{P.}}, and
  \bibinfo{author}{\bibfnamefont{N.}~\bibnamefont{Andrei}},
  \bibinfo{year}{1987}, \bibinfo{journal}{J. Phys.: Condens. Matter}
  \textbf{\bibinfo{volume}{1}}, \bibinfo{pages}{4057}.

\bibitem[{\citenamefont{Coleman} \emph{et~al.}(2001)\citenamefont{Coleman,
  P{\'e}pin, Si, and Ramazashvili}}]{pierspepinsirevaz}
\bibinfo{author}{\bibnamefont{Coleman}, \bibfnamefont{P.}},
  \bibinfo{author}{\bibfnamefont{C.}~\bibnamefont{P{\'e}pin}},
  \bibinfo{author}{\bibfnamefont{Q.}~\bibnamefont{Si}}, and
  \bibinfo{author}{\bibfnamefont{R.}~\bibnamefont{Ramazashvili}},
  \bibinfo{year}{2001}, \bibinfo{journal}{J. Phys.: Condens. Matter}
  \textbf{\bibinfo{volume}{13}}, \bibinfo{pages}{R723}.

\bibitem[{\citenamefont{Continentino}(1994)}]{mucioreview}
\bibinfo{author}{\bibnamefont{Continentino}, \bibfnamefont{M.~A.}},
  \bibinfo{year}{1994}, \bibinfo{journal}{Phys. Rep.}
  \textbf{\bibinfo{volume}{239}}, \bibinfo{pages}{179}.

\bibitem[{\citenamefont{Continentino}(2001)}]{muciobook}
\bibinfo{author}{\bibnamefont{Continentino}, \bibfnamefont{M.~A.}},
  \bibinfo{year}{2001}, \emph{\bibinfo{title}{Quantum scaling in many-body
  systems}} (\bibinfo{publisher}{World Scientific},
  \bibinfo{address}{Singapore}).

\bibitem[{\citenamefont{Continentino}
  \emph{et~al.}(2004)\citenamefont{Continentino, Fernandes, Guimaraes, Boechat,
  and Saguia}}]{continentinoborate04}
\bibinfo{author}{\bibnamefont{Continentino}, \bibfnamefont{M.~A.}},
  \bibinfo{author}{\bibfnamefont{J.~C.} \bibnamefont{Fernandes}},
  \bibinfo{author}{\bibfnamefont{R.~B.} \bibnamefont{Guimaraes}},
  \bibinfo{author}{\bibfnamefont{B.}~\bibnamefont{Boechat}}, and
  \bibinfo{author}{\bibfnamefont{A.}~\bibnamefont{Saguia}},
  \bibinfo{year}{2004}, in \emph{\bibinfo{booktitle}{Magnetic {M}aterials}},
  edited by \bibinfo{editor}{\bibnamefont{{A. Narlikar}}}
  (\bibinfo{publisher}{Springer}, \bibinfo{address}{Germany}).

\bibitem[{\citenamefont{Continentino}
  \emph{et~al.}(1989)\citenamefont{Continentino, Japiassu, and
  Troper}}]{japiassuetal}
\bibinfo{author}{\bibnamefont{Continentino}, \bibfnamefont{M.~A.}},
  \bibinfo{author}{\bibfnamefont{G.~M.} \bibnamefont{Japiassu}}, and
  \bibinfo{author}{\bibfnamefont{A.}~\bibnamefont{Troper}},
  \bibinfo{year}{1989}, \bibinfo{journal}{Phys. Rev. B}
  \textbf{\bibinfo{volume}{39}}, \bibinfo{pages}{9734}.

\bibitem[{\citenamefont{Cox}(1987)}]{cox87}
\bibinfo{author}{\bibnamefont{Cox}, \bibfnamefont{D.~L.}},
  \bibinfo{year}{1987}, \bibinfo{journal}{Phys. Rev. Lett.}
  \textbf{\bibinfo{volume}{59}}, \bibinfo{pages}{1240}.

\bibitem[{\citenamefont{Dagotto}(2002)}]{dagotto-book}
\bibinfo{author}{\bibnamefont{Dagotto}, \bibfnamefont{E.}},
  \bibinfo{year}{2002}, \emph{\bibinfo{title}{Nanoscale Phase Separation and
  Colossal Magnetoresistance}} (\bibinfo{publisher}{Springer-Verlag, Berlin}).

\bibitem[{\citenamefont{Dalidovich and Dobrosavljevi\'c}(2002)}]{Denis}
\bibinfo{author}{\bibnamefont{Dalidovich}, \bibfnamefont{D.}}, and
  \bibinfo{author}{\bibfnamefont{V.}~\bibnamefont{Dobrosavljevi\'c}},
  \bibinfo{year}{2002}, \bibinfo{journal}{Phys. Rev. B}
  \textbf{\bibinfo{volume}{66}}, \bibinfo{pages}{081107}.

\bibitem[{\citenamefont{Damle and Huse}(2002)}]{damlehuse}
\bibinfo{author}{\bibnamefont{Damle}, \bibfnamefont{K.}}, and
  \bibinfo{author}{\bibfnamefont{D.~A.} \bibnamefont{Huse}},
  \bibinfo{year}{2002}, \bibinfo{journal}{Phys. Rev. Lett.}
  \textbf{\bibinfo{volume}{89}}, \bibinfo{pages}{277203}.

\bibitem[{\citenamefont{Dasgupta and k.~Ma}(1980)}]{madasgupta}
\bibinfo{author}{\bibnamefont{Dasgupta}, \bibfnamefont{C.}}, and
  \bibinfo{author}{\bibfnamefont{S.}~\bibnamefont{k.~Ma}},
  \bibinfo{year}{1980}, \bibinfo{journal}{Phys. Rev. B}
  \textbf{\bibinfo{volume}{22}}, \bibinfo{pages}{1305}.

\bibitem[{\citenamefont{Degiorgi}(1999)}]{degiorgireview}
\bibinfo{author}{\bibnamefont{Degiorgi}, \bibfnamefont{L.}},
  \bibinfo{year}{1999}, \bibinfo{journal}{Rev. Mod. Phys.}
  \textbf{\bibinfo{volume}{71}}, \bibinfo{pages}{687}.

\bibitem[{\citenamefont{Degiorgi and Ott}(1996)}]{degiorgiott96}
\bibinfo{author}{\bibnamefont{Degiorgi}, \bibfnamefont{L.}}, and
  \bibinfo{author}{\bibfnamefont{H.~R.} \bibnamefont{Ott}},
  \bibinfo{year}{1996}, \bibinfo{journal}{J. Phys.: Condens. Matter}
  \textbf{\bibinfo{volume}{8}}, \bibinfo{pages}{9901}.

\bibitem[{\citenamefont{Degiorgi} \emph{et~al.}(1995)\citenamefont{Degiorgi,
  Ott, and Hulliger}}]{degiorgietal95}
\bibinfo{author}{\bibnamefont{Degiorgi}, \bibfnamefont{L.}},
  \bibinfo{author}{\bibfnamefont{H.~R.} \bibnamefont{Ott}}, and
  \bibinfo{author}{\bibfnamefont{F.}~\bibnamefont{Hulliger}},
  \bibinfo{year}{1995}, \bibinfo{journal}{Phys. Rev. B}
  \textbf{\bibinfo{volume}{52}}, \bibinfo{pages}{42}.

\bibitem[{\citenamefont{Degiorgi} \emph{et~al.}(1996)\citenamefont{Degiorgi,
  Wachter, Maple, de~Andrade, and Herrmann}}]{degiorgietal96}
\bibinfo{author}{\bibnamefont{Degiorgi}, \bibfnamefont{L.}},
  \bibinfo{author}{\bibfnamefont{P.}~\bibnamefont{Wachter}},
  \bibinfo{author}{\bibfnamefont{M.~B.} \bibnamefont{Maple}},
  \bibinfo{author}{\bibfnamefont{M.~C.} \bibnamefont{de~Andrade}}, and
  \bibinfo{author}{\bibfnamefont{J.}~\bibnamefont{Herrmann}},
  \bibinfo{year}{1996}, \bibinfo{journal}{Phys. Rev. B}
  \textbf{\bibinfo{volume}{54}}, \bibinfo{pages}{6065}.

\bibitem[{\citenamefont{Demler} \emph{et~al.}(2002)\citenamefont{Demler, Nayak,
  Kee, Kim, and Senthil}}]{demleretal02}
\bibinfo{author}{\bibnamefont{Demler}, \bibfnamefont{E.}},
  \bibinfo{author}{\bibfnamefont{C.}~\bibnamefont{Nayak}},
  \bibinfo{author}{\bibfnamefont{H.-Y.} \bibnamefont{Kee}},
  \bibinfo{author}{\bibfnamefont{Y.-B.} \bibnamefont{Kim}}, and
  \bibinfo{author}{\bibfnamefont{T.}~\bibnamefont{Senthil}},
  \bibinfo{year}{2002}, \bibinfo{journal}{Phys. Rev. B}
  \textbf{\bibinfo{volume}{65}}, \bibinfo{pages}{155103}.

\bibitem[{\citenamefont{Denteneer and Scalettar}(2003)}]{scalettar-prl03}
\bibinfo{author}{\bibnamefont{Denteneer}, \bibfnamefont{P.~J.~H.}}, and
  \bibinfo{author}{\bibfnamefont{R.~T.} \bibnamefont{Scalettar}},
  \bibinfo{year}{2003}, \bibinfo{journal}{Phys. Rev. Lett.}
  \textbf{\bibinfo{volume}{90}}, \bibinfo{pages}{246401}.

\bibitem[{\citenamefont{Denteneer} \emph{et~al.}(1999)\citenamefont{Denteneer,
  Scalettar, and Trivedi}}]{scalettar-prl99}
\bibinfo{author}{\bibnamefont{Denteneer}, \bibfnamefont{P.~J.~H.}},
  \bibinfo{author}{\bibfnamefont{R.~T.} \bibnamefont{Scalettar}}, and
  \bibinfo{author}{\bibfnamefont{N.}~\bibnamefont{Trivedi}},
  \bibinfo{year}{1999}, \bibinfo{journal}{Phys. Rev. Lett.}
  \textbf{\bibinfo{volume}{83}}, \bibinfo{pages}{4610}.

\bibitem[{\citenamefont{Denteneer} \emph{et~al.}(2001)\citenamefont{Denteneer,
  Scalettar, and Trivedi}}]{scalettar-prl01}
\bibinfo{author}{\bibnamefont{Denteneer}, \bibfnamefont{P.~J.~H.}},
  \bibinfo{author}{\bibfnamefont{R.~T.} \bibnamefont{Scalettar}}, and
  \bibinfo{author}{\bibfnamefont{N.}~\bibnamefont{Trivedi}},
  \bibinfo{year}{2001}, \bibinfo{journal}{Phys. Rev. Lett.}
  \textbf{\bibinfo{volume}{87}}, \bibinfo{pages}{146401}.

\bibitem[{\citenamefont{Dobrosavljevi\'c and G.Kotliar}(1993)}]{dk-prl93}
\bibinfo{author}{\bibnamefont{Dobrosavljevi\'c}, \bibfnamefont{V.}}, and
  \bibinfo{author}{\bibnamefont{G.Kotliar}}, \bibinfo{year}{1993},
  \bibinfo{journal}{Phys. Rev. Lett.} \textbf{\bibinfo{volume}{71}},
  \bibinfo{pages}{3218}.

\bibitem[{\citenamefont{Dobrosavljevi\'c and G.Kotliar}(1994)}]{dk-prb94}
\bibinfo{author}{\bibnamefont{Dobrosavljevi\'c}, \bibfnamefont{V.}}, and
  \bibinfo{author}{\bibnamefont{G.Kotliar}}, \bibinfo{year}{1994},
  \bibinfo{journal}{Phys. Rev. B} \textbf{\bibinfo{volume}{50}},
  \bibinfo{pages}{1430}.

\bibitem[{\citenamefont{Dobrosavljevi\'c and G.Kotliar}(1997)}]{motand}
\bibinfo{author}{\bibnamefont{Dobrosavljevi\'c}, \bibfnamefont{V.}}, and
  \bibinfo{author}{\bibnamefont{G.Kotliar}}, \bibinfo{year}{1997},
  \bibinfo{journal}{Phys. Rev. Lett.} \textbf{\bibinfo{volume}{78}},
  \bibinfo{pages}{3943}.

\bibitem[{\citenamefont{Dobrosavljevi\'{c}}
  \emph{et~al.}(1992)\citenamefont{Dobrosavljevi\'{c}, Kirkpatrick, and
  Kotliar}}]{vladtedgabi}
\bibinfo{author}{\bibnamefont{Dobrosavljevi\'{c}}, \bibfnamefont{V.}},
  \bibinfo{author}{\bibfnamefont{T.~R.} \bibnamefont{Kirkpatrick}}, and
  \bibinfo{author}{\bibfnamefont{B.~G.} \bibnamefont{Kotliar}},
  \bibinfo{year}{1992}, \bibinfo{journal}{Phys. Rev. Lett.}
  \textbf{\bibinfo{volume}{69}}, \bibinfo{pages}{1113}.

\bibitem[{\citenamefont{Dobrosavljevi\'{c} and Kotliar}(1998)}]{vladgabisdmft2}
\bibinfo{author}{\bibnamefont{Dobrosavljevi\'{c}}, \bibfnamefont{V.}}, and
  \bibinfo{author}{\bibfnamefont{G.}~\bibnamefont{Kotliar}},
  \bibinfo{year}{1998}, \bibinfo{journal}{Philos. Trans. R. Soc. London A}
  \textbf{\bibinfo{volume}{356}}, \bibinfo{pages}{57}.

\bibitem[{\citenamefont{Dobrosavljevi\'{c} and Miranda}(2005)}]{noqcp}
\bibinfo{author}{\bibnamefont{Dobrosavljevi\'{c}}, \bibfnamefont{V.}}, and
  \bibinfo{author}{\bibfnamefont{E.}~\bibnamefont{Miranda}},
  \bibinfo{year}{2005}, \bibinfo{journal}{Phys. Rev. Lett.}
  \textbf{\bibinfo{volume}{94}}, \bibinfo{pages}{187203}.

\bibitem[{\citenamefont{Dobrosavljevi\'c and Stratt}(1987)}]{dobrosavljevic87}
\bibinfo{author}{\bibnamefont{Dobrosavljevi\'c}, \bibfnamefont{V.}}, and
  \bibinfo{author}{\bibfnamefont{R.~M.} \bibnamefont{Stratt}},
  \bibinfo{year}{1987}, \bibinfo{journal}{Phys. Rev. B}
  \textbf{\bibinfo{volume}{36}}, \bibinfo{pages}{8484}.

\bibitem[{\citenamefont{Dobrosavljevi\'c}
  \emph{et~al.}(2003)\citenamefont{Dobrosavljevi\'c, Tanaskovi\'c, and
  Pastor}}]{mitglass-prl03}
\bibinfo{author}{\bibnamefont{Dobrosavljevi\'c}, \bibfnamefont{V.}},
  \bibinfo{author}{\bibfnamefont{D.}~\bibnamefont{Tanaskovi\'c}}, and
  \bibinfo{author}{\bibfnamefont{A.~A.} \bibnamefont{Pastor}},
  \bibinfo{year}{2003}, \bibinfo{journal}{Phys. Rev. Lett.}
  \textbf{\bibinfo{volume}{90}}, \bibinfo{pages}{016402}.

\bibitem[{\citenamefont{Doniach}(1977)}]{Doniach}
\bibinfo{author}{\bibnamefont{Doniach}, \bibfnamefont{S.}},
  \bibinfo{year}{1977}, \bibinfo{journal}{Physica B}
  \textbf{\bibinfo{volume}{91}}, \bibinfo{pages}{231}.

\bibitem[{\citenamefont{Doniach and Engelsberg}(1966)}]{doniachengelsberg}
\bibinfo{author}{\bibnamefont{Doniach}, \bibfnamefont{S.}}, and
  \bibinfo{author}{\bibfnamefont{S.}~\bibnamefont{Engelsberg}},
  \bibinfo{year}{1966}, \bibinfo{journal}{Phys. Rev. Lett.}
  \textbf{\bibinfo{volume}{17}}, \bibinfo{pages}{750}.

\bibitem[{\citenamefont{Doty and Fisher}(1992)}]{dotyfisher}
\bibinfo{author}{\bibnamefont{Doty}, \bibfnamefont{C.~A.}}, and
  \bibinfo{author}{\bibfnamefont{D.~S.} \bibnamefont{Fisher}},
  \bibinfo{year}{1992}, \bibinfo{journal}{Phys. Rev. B}
  \textbf{\bibinfo{volume}{45}}, \bibinfo{pages}{2167}.

\bibitem[{\citenamefont{Efros and Shklovskii}(1975)}]{re:Efros75}
\bibinfo{author}{\bibnamefont{Efros}, \bibfnamefont{A.~L.}}, and
  \bibinfo{author}{\bibfnamefont{B.~I.} \bibnamefont{Shklovskii}},
  \bibinfo{year}{1975}, \bibinfo{journal}{J. Phys. C}
  \textbf{\bibinfo{volume}{8}}, \bibinfo{pages}{L49}.

\bibitem[{\citenamefont{Elliott} \emph{et~al.}(1974)\citenamefont{Elliott,
  Krumhansl, and Leath}}]{elliotetal74}
\bibinfo{author}{\bibnamefont{Elliott}, \bibfnamefont{R.~J.}},
  \bibinfo{author}{\bibfnamefont{J.~A.} \bibnamefont{Krumhansl}}, and
  \bibinfo{author}{\bibfnamefont{P.~L.} \bibnamefont{Leath}},
  \bibinfo{year}{1974}, \bibinfo{journal}{Rev. Mod. Phys.}
  \textbf{\bibinfo{volume}{46}}, \bibinfo{pages}{465}.

\bibitem[{\citenamefont{Finkel'stein}(1983)}]{fink-jetp83}
\bibinfo{author}{\bibnamefont{Finkel'stein}, \bibfnamefont{A.~M.}},
  \bibinfo{year}{1983}, \bibinfo{journal}{Zh. Eksp. Teor. Fiz.}
  \textbf{\bibinfo{volume}{84}}, \bibinfo{pages}{168}, \bibinfo{note}{[Sov.
  Phys. JETP {\bf 57}, 97 (1983)]}.

\bibitem[{\citenamefont{Finkel'stein}(1984)}]{fink-jetp84}
\bibinfo{author}{\bibnamefont{Finkel'stein}, \bibfnamefont{A.~M.}},
  \bibinfo{year}{1984}, \bibinfo{journal}{Zh. Eksp. Teor. Fiz.}
  \textbf{\bibinfo{volume}{86}}, \bibinfo{pages}{367}, \bibinfo{note}{[Sov.
  Phys. JETP {\bf 59}, 212 (1983)]}.

\bibitem[{\citenamefont{Fisher}(1992)}]{fishertransising}
\bibinfo{author}{\bibnamefont{Fisher}, \bibfnamefont{D.~S.}},
  \bibinfo{year}{1992}, \bibinfo{journal}{Phys. Rev. Lett.}
  \textbf{\bibinfo{volume}{69}}, \bibinfo{pages}{534}.

\bibitem[{\citenamefont{Fisher}(1994)}]{fisherrandomchain}
\bibinfo{author}{\bibnamefont{Fisher}, \bibfnamefont{D.~S.}},
  \bibinfo{year}{1994}, \bibinfo{journal}{Phys. Rev. B}
  \textbf{\bibinfo{volume}{50}}, \bibinfo{pages}{3799}.

\bibitem[{\citenamefont{Fisher}(1995)}]{fishertransising2}
\bibinfo{author}{\bibnamefont{Fisher}, \bibfnamefont{D.~S.}},
  \bibinfo{year}{1995}, \bibinfo{journal}{Phys. Rev. B}
  \textbf{\bibinfo{volume}{51}}, \bibinfo{pages}{6411}.

\bibitem[{\citenamefont{Fradkin}(1991)}]{Fradkin}
\bibinfo{author}{\bibnamefont{Fradkin}, \bibfnamefont{E.}},
  \bibinfo{year}{1991}, \emph{\bibinfo{title}{Field Theories of Condensed
  Matter Systems}} (\bibinfo{publisher}{Adisson-Wesley},
  \bibinfo{address}{Redwood City, California}).

\bibitem[{\citenamefont{Frischmuth and Sigrist}(1997)}]{frischmuthsigrist97}
\bibinfo{author}{\bibnamefont{Frischmuth}, \bibfnamefont{B.}}, and
  \bibinfo{author}{\bibfnamefont{M.}~\bibnamefont{Sigrist}},
  \bibinfo{year}{1997}, \bibinfo{journal}{Phys. Rev. Lett.}
  \textbf{\bibinfo{volume}{79}}, \bibinfo{pages}{147}.

\bibitem[{\citenamefont{Frischmuth}
  \emph{et~al.}(1999)\citenamefont{Frischmuth, Sigrist, Ammon, and
  Troyer}}]{frischmuthetal99}
\bibinfo{author}{\bibnamefont{Frischmuth}, \bibfnamefont{B.}},
  \bibinfo{author}{\bibfnamefont{M.}~\bibnamefont{Sigrist}},
  \bibinfo{author}{\bibfnamefont{B.}~\bibnamefont{Ammon}}, and
  \bibinfo{author}{\bibfnamefont{M.}~\bibnamefont{Troyer}},
  \bibinfo{year}{1999}, \bibinfo{journal}{Phys. Rev. B}
  \textbf{\bibinfo{volume}{60}}, \bibinfo{pages}{3388}.

\bibitem[{\citenamefont{Gajewski} \emph{et~al.}(1996)\citenamefont{Gajewski,
  Dilley, Chau, and Maple}}]{gajewskietal96}
\bibinfo{author}{\bibnamefont{Gajewski}, \bibfnamefont{D.~A.}},
  \bibinfo{author}{\bibfnamefont{N.~R.} \bibnamefont{Dilley}},
  \bibinfo{author}{\bibfnamefont{R.}~\bibnamefont{Chau}}, and
  \bibinfo{author}{\bibfnamefont{M.~B.} \bibnamefont{Maple}},
  \bibinfo{year}{1996}, \bibinfo{journal}{J. Phys.: Condens. Matter}
  \textbf{\bibinfo{volume}{8}}, \bibinfo{pages}{9793}.

\bibitem[{\citenamefont{Galitskii}(1958)}]{galitskii58}
\bibinfo{author}{\bibnamefont{Galitskii}, \bibfnamefont{V.~M.}},
  \bibinfo{year}{1958}, \bibinfo{journal}{Sov. Phys. JETP}
  \textbf{\bibinfo{volume}{7}}, \bibinfo{pages}{104}.

\bibitem[{\citenamefont{Georges and Kotliar}(1992)}]{georgeskotliar92}
\bibinfo{author}{\bibnamefont{Georges}, \bibfnamefont{A.}}, and
  \bibinfo{author}{\bibfnamefont{G.}~\bibnamefont{Kotliar}},
  \bibinfo{year}{1992}, \bibinfo{journal}{Phys. Rev. B}
  \textbf{\bibinfo{volume}{45}}, \bibinfo{pages}{6479}.

\bibitem[{\citenamefont{Georges} \emph{et~al.}(1996)\citenamefont{Georges,
  Kotliar, Krauth, and Rozenberg}}]{georgesrmp}
\bibinfo{author}{\bibnamefont{Georges}, \bibfnamefont{A.}},
  \bibinfo{author}{\bibfnamefont{G.}~\bibnamefont{Kotliar}},
  \bibinfo{author}{\bibfnamefont{W.}~\bibnamefont{Krauth}}, and
  \bibinfo{author}{\bibfnamefont{M.~J.} \bibnamefont{Rozenberg}},
  \bibinfo{year}{1996}, \bibinfo{journal}{Rev. Mod. Phys.}
  \textbf{\bibinfo{volume}{68}}, \bibinfo{pages}{13}.

\bibitem[{\citenamefont{Georges} \emph{et~al.}(2001)\citenamefont{Georges,
  Parcollet, and Sachdev}}]{georgesetal01}
\bibinfo{author}{\bibnamefont{Georges}, \bibfnamefont{A.}},
  \bibinfo{author}{\bibfnamefont{O.}~\bibnamefont{Parcollet}}, and
  \bibinfo{author}{\bibfnamefont{S.}~\bibnamefont{Sachdev}},
  \bibinfo{year}{2001}, \bibinfo{journal}{Phys. Rev. B}
  \textbf{\bibinfo{volume}{63}}, \bibinfo{pages}{134406}.

\bibitem[{\citenamefont{Goldenfeld}(1992)}]{goldenfeldbook}
\bibinfo{author}{\bibnamefont{Goldenfeld}, \bibfnamefont{N.}},
  \bibinfo{year}{1992}, \emph{\bibinfo{title}{Lectures on phase transitions and
  the renormalization group}} (\bibinfo{publisher}{Addison-Wesley},
  \bibinfo{address}{Reading}).

\bibitem[{\citenamefont{Gor'kov and Sokol}(1987)}]{gorkov-JETP87}
\bibinfo{author}{\bibnamefont{Gor'kov}, \bibfnamefont{L.~P.}}, and
  \bibinfo{author}{\bibfnamefont{A.~V.} \bibnamefont{Sokol}},
  \bibinfo{year}{1987}, \bibinfo{journal}{JETP Lett.}
  \textbf{\bibinfo{volume}{46}}, \bibinfo{pages}{420}.

\bibitem[{\citenamefont{Grewe and Steglich}(1991)}]{steglichgrewe}
\bibinfo{author}{\bibnamefont{Grewe}, \bibfnamefont{N.}}, and
  \bibinfo{author}{\bibfnamefont{F.}~\bibnamefont{Steglich}},
  \bibinfo{year}{1991}, in \emph{\bibinfo{booktitle}{Handbook on the {P}hysics
  and {C}hemistry of {R}are {E}arths}}, edited by
  \bibinfo{editor}{\bibnamefont{{K. A. Geschneider Jr. and L. Eyring}}}
  (\bibinfo{publisher}{Elsevier}, \bibinfo{address}{Amsterdam}),
  volume~\bibinfo{volume}{14}, p. \bibinfo{pages}{343}.

\bibitem[{\citenamefont{Griffiths}(1969)}]{griffiths}
\bibinfo{author}{\bibnamefont{Griffiths}, \bibfnamefont{R.~B.}},
  \bibinfo{year}{1969}, \bibinfo{journal}{Phys. Rev. Lett.}
  \textbf{\bibinfo{volume}{23}}, \bibinfo{pages}{17}.

\bibitem[{\citenamefont{Guo} \emph{et~al.}(1994)\citenamefont{Guo, Bhatt, and
  Huse}}]{guoetal94}
\bibinfo{author}{\bibnamefont{Guo}, \bibfnamefont{M.}},
  \bibinfo{author}{\bibfnamefont{R.~N.} \bibnamefont{Bhatt}}, and
  \bibinfo{author}{\bibfnamefont{D.~A.} \bibnamefont{Huse}},
  \bibinfo{year}{1994}, \bibinfo{journal}{Phys. Rev. Lett.}
  \textbf{\bibinfo{volume}{72}}, \bibinfo{pages}{4137}.

\bibitem[{\citenamefont{Guo} \emph{et~al.}(1996)\citenamefont{Guo, Bhatt, and
  Huse}}]{guoetal96}
\bibinfo{author}{\bibnamefont{Guo}, \bibfnamefont{M.}},
  \bibinfo{author}{\bibfnamefont{R.~N.} \bibnamefont{Bhatt}}, and
  \bibinfo{author}{\bibfnamefont{D.~A.} \bibnamefont{Huse}},
  \bibinfo{year}{1996}, \bibinfo{journal}{Phys. Rev. B}
  \textbf{\bibinfo{volume}{54}}, \bibinfo{pages}{3336}.

\bibitem[{\citenamefont{Haldane}(1983{\natexlab{a}})}]{haldaneconj1}
\bibinfo{author}{\bibnamefont{Haldane}, \bibfnamefont{F.~D.~M.}},
  \bibinfo{year}{1983}{\natexlab{a}}, \bibinfo{journal}{Phys. Lett. A}
  \textbf{\bibinfo{volume}{93}}, \bibinfo{pages}{464}.

\bibitem[{\citenamefont{Haldane}(1983{\natexlab{b}})}]{haldaneconj2}
\bibinfo{author}{\bibnamefont{Haldane}, \bibfnamefont{F.~D.~M.}},
  \bibinfo{year}{1983}{\natexlab{b}}, \bibinfo{journal}{Phys. Rev. Lett.}
  \textbf{\bibinfo{volume}{50}}, \bibinfo{pages}{1153}.

\bibitem[{\citenamefont{Harris}(1974)}]{harris74}
\bibinfo{author}{\bibnamefont{Harris}, \bibfnamefont{A.~B.}},
  \bibinfo{year}{1974}, \bibinfo{journal}{J. Phys. C}
  \textbf{\bibinfo{volume}{7}}, \bibinfo{pages}{1671}.

\bibitem[{\citenamefont{Heffner and Norman}(1996)}]{heffnernorman}
\bibinfo{author}{\bibnamefont{Heffner}, \bibfnamefont{R.~H.}}, and
  \bibinfo{author}{\bibfnamefont{M.~R.} \bibnamefont{Norman}},
  \bibinfo{year}{1996}, \bibinfo{journal}{Comments Condens. Matter Phys.}
  \textbf{\bibinfo{volume}{17}}, \bibinfo{pages}{361}.

\bibitem[{\citenamefont{Henelius and Girvin}(1998)}]{heneliusgirvin98}
\bibinfo{author}{\bibnamefont{Henelius}, \bibfnamefont{P.}}, and
  \bibinfo{author}{\bibfnamefont{S.~M.} \bibnamefont{Girvin}},
  \bibinfo{year}{1998}, \bibinfo{journal}{Phys. Rev. B}
  \textbf{\bibinfo{volume}{57}}, \bibinfo{pages}{11457}.

\bibitem[{\citenamefont{Hertz}(1976)}]{hertz}
\bibinfo{author}{\bibnamefont{Hertz}, \bibfnamefont{J.~A.}},
  \bibinfo{year}{1976}, \bibinfo{journal}{Phys. Rev. B}
  \textbf{\bibinfo{volume}{14}}, \bibinfo{pages}{1165}.

\bibitem[{\citenamefont{Hewson}(1993)}]{hewson}
\bibinfo{author}{\bibnamefont{Hewson}, \bibfnamefont{A.~C.}},
  \bibinfo{year}{1993}, \emph{\bibinfo{title}{The {K}ondo {P}roblem to {H}eavy
  {F}ermions}} (\bibinfo{publisher}{Cambrige University Press},
  \bibinfo{address}{Cambridge}).

\bibitem[{\citenamefont{Hida}(1997)}]{hida97}
\bibinfo{author}{\bibnamefont{Hida}, \bibfnamefont{K.}}, \bibinfo{year}{1997},
  \bibinfo{journal}{J. Phys. Soc. Jpn.} \textbf{\bibinfo{volume}{62}},
  \bibinfo{pages}{330}.

\bibitem[{\citenamefont{Hida}(1999)}]{hida99}
\bibinfo{author}{\bibnamefont{Hida}, \bibfnamefont{K.}}, \bibinfo{year}{1999},
  \bibinfo{journal}{Phys. Rev. Lett.} \textbf{\bibinfo{volume}{83}},
  \bibinfo{pages}{3297}.

\bibitem[{\citenamefont{Hikihara} \emph{et~al.}(1999)\citenamefont{Hikihara,
  Furusaki, and Sigrist}}]{hikiharaetal}
\bibinfo{author}{\bibnamefont{Hikihara}, \bibfnamefont{T.}},
  \bibinfo{author}{\bibfnamefont{A.}~\bibnamefont{Furusaki}}, and
  \bibinfo{author}{\bibfnamefont{M.}~\bibnamefont{Sigrist}},
  \bibinfo{year}{1999}, \bibinfo{journal}{Phys. Rev. B}
  \textbf{\bibinfo{volume}{60}}, \bibinfo{pages}{12116}.

\bibitem[{\citenamefont{Hirsch} \emph{et~al.}(1992)\citenamefont{Hirsch,
  Holcomb, Bhatt, and Paalanen}}]{holcomb}
\bibinfo{author}{\bibnamefont{Hirsch}, \bibfnamefont{M.~J.}},
  \bibinfo{author}{\bibfnamefont{D.~F.} \bibnamefont{Holcomb}},
  \bibinfo{author}{\bibfnamefont{R.~N.} \bibnamefont{Bhatt}}, and
  \bibinfo{author}{\bibfnamefont{M.~A.} \bibnamefont{Paalanen}},
  \bibinfo{year}{1992}, \bibinfo{journal}{Phys. Rev. Lett.}
  \textbf{\bibinfo{volume}{68}}, \bibinfo{pages}{1418}.

\bibitem[{\citenamefont{Hlubina and Rice}(1995)}]{hlubinarice95}
\bibinfo{author}{\bibnamefont{Hlubina}, \bibfnamefont{R.}}, and
  \bibinfo{author}{\bibfnamefont{T.~M.} \bibnamefont{Rice}},
  \bibinfo{year}{1995}, \bibinfo{journal}{Phys. Rev. B}
  \textbf{\bibinfo{volume}{51}}, \bibinfo{pages}{9253}.

\bibitem[{\citenamefont{Hoyos and Miranda}(2004)}]{hoyosmiranda1}
\bibinfo{author}{\bibnamefont{Hoyos}, \bibfnamefont{J.~A.}}, and
  \bibinfo{author}{\bibfnamefont{E.}~\bibnamefont{Miranda}},
  \bibinfo{year}{2004}, \bibinfo{journal}{Phys. Rev. B}
  \textbf{\bibinfo{volume}{69}}, \bibinfo{pages}{214411}.

\bibitem[{\citenamefont{Hyman and Yang}(1997)}]{hymanyang97}
\bibinfo{author}{\bibnamefont{Hyman}, \bibfnamefont{R.~A.}}, and
  \bibinfo{author}{\bibfnamefont{K.}~\bibnamefont{Yang}}, \bibinfo{year}{1997},
  \bibinfo{journal}{Phys. Rev. Lett.} \textbf{\bibinfo{volume}{78}},
  \bibinfo{pages}{1783}.

\bibitem[{\citenamefont{Hyman} \emph{et~al.}(1996)\citenamefont{Hyman, Yang,
  Bhatt, and Girvin}}]{hymanetal}
\bibinfo{author}{\bibnamefont{Hyman}, \bibfnamefont{R.~A.}},
  \bibinfo{author}{\bibfnamefont{K.}~\bibnamefont{Yang}},
  \bibinfo{author}{\bibfnamefont{R.~N.} \bibnamefont{Bhatt}}, and
  \bibinfo{author}{\bibfnamefont{S.~M.} \bibnamefont{Girvin}},
  \bibinfo{year}{1996}, \bibinfo{journal}{Phys. Rev. Lett.}
  \textbf{\bibinfo{volume}{76}}, \bibinfo{pages}{839}.

\bibitem[{\citenamefont{Igl{\'o}i} \emph{et~al.}(2001)\citenamefont{Igl{\'o}i,
  Juh{\'a}sz, and Lajk{\'o}}}]{igloietal01}
\bibinfo{author}{\bibnamefont{Igl{\'o}i}, \bibfnamefont{F.}},
  \bibinfo{author}{\bibfnamefont{R.}~\bibnamefont{Juh{\'a}sz}}, and
  \bibinfo{author}{\bibfnamefont{P.}~\bibnamefont{Lajk{\'o}}},
  \bibinfo{year}{2001}, \bibinfo{journal}{Phys. Rev. Lett.}
  \textbf{\bibinfo{volume}{86}}, \bibinfo{pages}{1343}.

\bibitem[{\citenamefont{Igl{\'o}i and Monthus}(2005)}]{igoimonthus}
\bibinfo{author}{\bibnamefont{Igl{\'o}i}, \bibfnamefont{F.}}, and
  \bibinfo{author}{\bibfnamefont{C.}~\bibnamefont{Monthus}},
  \bibinfo{year}{2005}, \bibinfo{journal}{cond-mat/0502448} .

\bibitem[{\citenamefont{Imada} \emph{et~al.}(1998)\citenamefont{Imada,
  Fujimori, and Tokura}}]{imadareview}
\bibinfo{author}{\bibnamefont{Imada}, \bibfnamefont{M.}},
  \bibinfo{author}{\bibfnamefont{A.}~\bibnamefont{Fujimori}}, and
  \bibinfo{author}{\bibfnamefont{Y.}~\bibnamefont{Tokura}},
  \bibinfo{year}{1998}, \bibinfo{journal}{Rev. Mod. Phys.}
  \textbf{\bibinfo{volume}{70}}, \bibinfo{pages}{1039}.

\bibitem[{\citenamefont{Itoh} \emph{et~al.}(2004)\citenamefont{Itoh, Watanabe,
  Ootuka, Haller, and T.Ohtsuki}}]{itoh-jpsj04}
\bibinfo{author}{\bibnamefont{Itoh}, \bibfnamefont{K.~M.}},
  \bibinfo{author}{\bibfnamefont{M.}~\bibnamefont{Watanabe}},
  \bibinfo{author}{\bibfnamefont{Y.}~\bibnamefont{Ootuka}},
  \bibinfo{author}{\bibfnamefont{E.~E.} \bibnamefont{Haller}}, and
  \bibinfo{author}{\bibnamefont{T.Ohtsuki}}, \bibinfo{year}{2004},
  \bibinfo{journal}{Jour. Phys. Soc. Japan} \textbf{\bibinfo{volume}{73}},
  \bibinfo{pages}{173}.

\bibitem[{\citenamefont{Izuyama} \emph{et~al.}(1963)\citenamefont{Izuyama, Kim,
  and Kubo}}]{izuyamaetal63}
\bibinfo{author}{\bibnamefont{Izuyama}, \bibfnamefont{T.}},
  \bibinfo{author}{\bibfnamefont{D.~J.} \bibnamefont{Kim}}, and
  \bibinfo{author}{\bibfnamefont{R.}~\bibnamefont{Kubo}}, \bibinfo{year}{1963},
  \bibinfo{journal}{J. Phys. Soc. Jpn.} \textbf{\bibinfo{volume}{18}},
  \bibinfo{pages}{1925}.

\bibitem[{\citenamefont{Jagannathan}
  \emph{et~al.}(1988)\citenamefont{Jagannathan, Abrahams, and
  Stephen}}]{jagannathanetal88}
\bibinfo{author}{\bibnamefont{Jagannathan}, \bibfnamefont{A.}},
  \bibinfo{author}{\bibfnamefont{E.}~\bibnamefont{Abrahams}}, and
  \bibinfo{author}{\bibfnamefont{M.~J.} \bibnamefont{Stephen}},
  \bibinfo{year}{1988}, \bibinfo{journal}{Phys. Rev. B}
  \textbf{\bibinfo{volume}{37}}, \bibinfo{pages}{436}.

\bibitem[{\citenamefont{Jaroszy\'nski}
  \emph{et~al.}(2002)\citenamefont{Jaroszy\'nski, Popovi\'c, and
  Klapwijk}}]{JJPRL02}
\bibinfo{author}{\bibnamefont{Jaroszy\'nski}, \bibfnamefont{J.}},
  \bibinfo{author}{\bibfnamefont{D.}~\bibnamefont{Popovi\'c}}, and
  \bibinfo{author}{\bibfnamefont{T.~M.} \bibnamefont{Klapwijk}},
  \bibinfo{year}{2002}, \bibinfo{journal}{Phys. Rev. Lett.}
  \textbf{\bibinfo{volume}{89}}, \bibinfo{pages}{276401}.

\bibitem[{\citenamefont{Jaroszynski}
  \emph{et~al.}(2004)\citenamefont{Jaroszynski, Popovic, and
  Klapwijk}}]{mag-glass}
\bibinfo{author}{\bibnamefont{Jaroszynski}, \bibfnamefont{J.}},
  \bibinfo{author}{\bibfnamefont{D.}~\bibnamefont{Popovic}}, and
  \bibinfo{author}{\bibfnamefont{T.~M.} \bibnamefont{Klapwijk}},
  \bibinfo{year}{2004}, \bibinfo{journal}{Phys. Rev. Lett.}
  \textbf{\bibinfo{volume}{92}}, \bibinfo{pages}{226403}.

\bibitem[{\citenamefont{Joyce}(1969)}]{joyce69}
\bibinfo{author}{\bibnamefont{Joyce}, \bibfnamefont{G.~S.}},
  \bibinfo{year}{1969}, \bibinfo{journal}{J. Phys. C}
  \textbf{\bibinfo{volume}{2}}, \bibinfo{pages}{1531}.

\bibitem[{\citenamefont{Kagan} \emph{et~al.}(1992)\citenamefont{Kagan, Kikoin,
  and Prokof'ev}}]{kaganetal92}
\bibinfo{author}{\bibnamefont{Kagan}, \bibfnamefont{Y.}},
  \bibinfo{author}{\bibfnamefont{K.~A.} \bibnamefont{Kikoin}}, and
  \bibinfo{author}{\bibfnamefont{N.~V.} \bibnamefont{Prokof'ev}},
  \bibinfo{year}{1992}, \bibinfo{journal}{Physica B}
  \textbf{\bibinfo{volume}{182}}, \bibinfo{pages}{201}.

\bibitem[{\citenamefont{Kajueter and Kotliar}(1996)}]{kajuetergabi}
\bibinfo{author}{\bibnamefont{Kajueter}, \bibfnamefont{H.}}, and
  \bibinfo{author}{\bibfnamefont{G.}~\bibnamefont{Kotliar}},
  \bibinfo{year}{1996}, \bibinfo{journal}{Phys. Rev. Lett.}
  \textbf{\bibinfo{volume}{77}}, \bibinfo{pages}{131}.

\bibitem[{\citenamefont{Kanigel} \emph{et~al.}(2002)\citenamefont{Kanigel,
  Keren, Eckstein, Knizhnik, Lord, and Amato}}]{kanigel-prl02}
\bibinfo{author}{\bibnamefont{Kanigel}, \bibfnamefont{A.}},
  \bibinfo{author}{\bibfnamefont{A.}~\bibnamefont{Keren}},
  \bibinfo{author}{\bibfnamefont{Y.}~\bibnamefont{Eckstein}},
  \bibinfo{author}{\bibfnamefont{A.}~\bibnamefont{Knizhnik}},
  \bibinfo{author}{\bibfnamefont{J.}~\bibnamefont{Lord}}, and
  \bibinfo{author}{\bibfnamefont{A.}~\bibnamefont{Amato}},
  \bibinfo{year}{2002}, \bibinfo{journal}{Phys. Rev. Lett.}
  \textbf{\bibinfo{volume}{88}}, \bibinfo{pages}{137003}.

\bibitem[{\citenamefont{Kivelson} \emph{et~al.}(2003)\citenamefont{Kivelson,
  Bindloss, Fradkin, Oganesyan, Tranquada, Kapitulnik, and
  Howald}}]{kivelson-rmp03}
\bibinfo{author}{\bibnamefont{Kivelson}, \bibfnamefont{S.~A.}},
  \bibinfo{author}{\bibfnamefont{I.~P.} \bibnamefont{Bindloss}},
  \bibinfo{author}{\bibfnamefont{E.}~\bibnamefont{Fradkin}},
  \bibinfo{author}{\bibfnamefont{V.}~\bibnamefont{Oganesyan}},
  \bibinfo{author}{\bibfnamefont{J.~M.} \bibnamefont{Tranquada}},
  \bibinfo{author}{\bibfnamefont{A.}~\bibnamefont{Kapitulnik}}, and
  \bibinfo{author}{\bibfnamefont{C.}~\bibnamefont{Howald}},
  \bibinfo{year}{2003}, \bibinfo{journal}{Rev. Mod. Phys.}
  \textbf{\bibinfo{volume}{75}}, \bibinfo{pages}{1201}.

\bibitem[{\citenamefont{Kondo}(1964)}]{kondo64}
\bibinfo{author}{\bibnamefont{Kondo}, \bibfnamefont{J.}}, \bibinfo{year}{1964},
  \bibinfo{journal}{Prog. Theor. Phys.} \textbf{\bibinfo{volume}{32}},
  \bibinfo{pages}{37}.

\bibitem[{\citenamefont{Kosterlitz}(1976)}]{kosterlitz76prl}
\bibinfo{author}{\bibnamefont{Kosterlitz}, \bibfnamefont{J.~M.}},
  \bibinfo{year}{1976}, \bibinfo{journal}{Phys. Rev. Lett.}
  \textbf{\bibinfo{volume}{37}}, \bibinfo{pages}{1577}.

\bibitem[{\citenamefont{Kotliar}(1987)}]{kotliar-fl87}
\bibinfo{author}{\bibnamefont{Kotliar}, \bibfnamefont{G.}},
  \bibinfo{year}{1987}, \emph{\bibinfo{title}{Anderson Localization}}
  (\bibinfo{publisher}{Springer, Berlin}).

\bibitem[{\citenamefont{Lakner} \emph{et~al.}(1994)\citenamefont{Lakner,
  L{\"o}hneysen, Langenfeld, and W{\"o}lfle}}]{volfle}
\bibinfo{author}{\bibnamefont{Lakner}, \bibfnamefont{M.}},
  \bibinfo{author}{\bibfnamefont{H.~V.} \bibnamefont{L{\"o}hneysen}},
  \bibinfo{author}{\bibfnamefont{A.}~\bibnamefont{Langenfeld}}, and
  \bibinfo{author}{\bibfnamefont{P.}~\bibnamefont{W{\"o}lfle}},
  \bibinfo{year}{1994}, \bibinfo{journal}{Phys. Rev. B}
  \textbf{\bibinfo{volume}{50}}, \bibinfo{pages}{17064}.

\bibitem[{\citenamefont{Landau}(1957{\natexlab{a}})}]{landauFL2}
\bibinfo{author}{\bibnamefont{Landau}, \bibfnamefont{L.~D.}},
  \bibinfo{year}{1957}{\natexlab{a}}, \bibinfo{journal}{Sov. Phys. JETP}
  \textbf{\bibinfo{volume}{5}}, \bibinfo{pages}{101}.

\bibitem[{\citenamefont{Landau}(1957{\natexlab{b}})}]{landauFL1}
\bibinfo{author}{\bibnamefont{Landau}, \bibfnamefont{L.~D.}},
  \bibinfo{year}{1957}{\natexlab{b}}, \bibinfo{journal}{Sov. Phys. JETP}
  \textbf{\bibinfo{volume}{3}}, \bibinfo{pages}{920}.

\bibitem[{\citenamefont{Landau}(1959)}]{landauFL3}
\bibinfo{author}{\bibnamefont{Landau}, \bibfnamefont{L.~D.}},
  \bibinfo{year}{1959}, \bibinfo{journal}{Sov. Phys. JETP}
  \textbf{\bibinfo{volume}{8}}, \bibinfo{pages}{70}.

\bibitem[{\citenamefont{Larkin and Melnikov}(1972)}]{larkinmelnikov72}
\bibinfo{author}{\bibnamefont{Larkin}, \bibfnamefont{A.~I.}}, and
  \bibinfo{author}{\bibfnamefont{V.~I.} \bibnamefont{Melnikov}},
  \bibinfo{year}{1972}, \bibinfo{journal}{Sov. Phys. JETP}
  \textbf{\bibinfo{volume}{34}}, \bibinfo{pages}{656}.

\bibitem[{\citenamefont{Lee and Ramakrishnan}(1985)}]{lr}
\bibinfo{author}{\bibnamefont{Lee}, \bibfnamefont{P.~A.}}, and
  \bibinfo{author}{\bibfnamefont{T.~V.} \bibnamefont{Ramakrishnan}},
  \bibinfo{year}{1985}, \bibinfo{journal}{Rev. Mod. Phys.}
  \textbf{\bibinfo{volume}{57}}, \bibinfo{pages}{287}.

\bibitem[{\citenamefont{Lee} \emph{et~al.}(1986)\citenamefont{Lee, Rice,
  Serene, Sham, and Wilkins}}]{leeetal}
\bibinfo{author}{\bibnamefont{Lee}, \bibfnamefont{P.~A.}},
  \bibinfo{author}{\bibfnamefont{T.~M.} \bibnamefont{Rice}},
  \bibinfo{author}{\bibfnamefont{J.~W.} \bibnamefont{Serene}},
  \bibinfo{author}{\bibfnamefont{L.~J.} \bibnamefont{Sham}}, and
  \bibinfo{author}{\bibfnamefont{J.~W.} \bibnamefont{Wilkins}},
  \bibinfo{year}{1986}, \bibinfo{journal}{Comments Condens. Matter Phys.}
  \textbf{\bibinfo{volume}{12}}, \bibinfo{pages}{99}.

\bibitem[{\citenamefont{Leggett}(1975)}]{leggett75}
\bibinfo{author}{\bibnamefont{Leggett}, \bibfnamefont{A.~J.}},
  \bibinfo{year}{1975}, \bibinfo{journal}{Rev. Mod. Phys.}
  \textbf{\bibinfo{volume}{47}}, \bibinfo{pages}{331}.

\bibitem[{\citenamefont{Leggett} \emph{et~al.}(1987)\citenamefont{Leggett,
  Chakravarty, Dorsey, Fisher, Garg, and Zwerger}}]{leggettetal87}
\bibinfo{author}{\bibnamefont{Leggett}, \bibfnamefont{A.~J.}},
  \bibinfo{author}{\bibfnamefont{S.}~\bibnamefont{Chakravarty}},
  \bibinfo{author}{\bibfnamefont{A.~T.} \bibnamefont{Dorsey}},
  \bibinfo{author}{\bibfnamefont{M.~P.~A.} \bibnamefont{Fisher}},
  \bibinfo{author}{\bibfnamefont{A.}~\bibnamefont{Garg}}, and
  \bibinfo{author}{\bibfnamefont{W.}~\bibnamefont{Zwerger}},
  \bibinfo{year}{1987}, \bibinfo{journal}{Rev. Mod. Phys.}
  \textbf{\bibinfo{volume}{59}}, \bibinfo{pages}{1}.

\bibitem[{\citenamefont{Lesage and Saleur}(1998)}]{lesagesaleur98}
\bibinfo{author}{\bibnamefont{Lesage}, \bibfnamefont{F.}}, and
  \bibinfo{author}{\bibfnamefont{H.}~\bibnamefont{Saleur}},
  \bibinfo{year}{1998}, \bibinfo{journal}{Phys. Rev. Lett.}
  \textbf{\bibinfo{volume}{80}}, \bibinfo{pages}{4370}.

\bibitem[{\citenamefont{Li} \emph{et~al.}(1998)\citenamefont{Li, Shiokawa,
  Homma, Uesawa, D{\"o}nni, Suzuki, Haga, Yamamoto, Honma, and
  Onuki}}]{lietal98}
\bibinfo{author}{\bibnamefont{Li}, \bibfnamefont{D.~X.}},
  \bibinfo{author}{\bibfnamefont{Y.}~\bibnamefont{Shiokawa}},
  \bibinfo{author}{\bibfnamefont{Y.}~\bibnamefont{Homma}},
  \bibinfo{author}{\bibfnamefont{A.}~\bibnamefont{Uesawa}},
  \bibinfo{author}{\bibfnamefont{A.}~\bibnamefont{D{\"o}nni}},
  \bibinfo{author}{\bibfnamefont{T.}~\bibnamefont{Suzuki}},
  \bibinfo{author}{\bibfnamefont{Y.}~\bibnamefont{Haga}},
  \bibinfo{author}{\bibfnamefont{E.}~\bibnamefont{Yamamoto}},
  \bibinfo{author}{\bibfnamefont{T.}~\bibnamefont{Honma}}, and
  \bibinfo{author}{\bibfnamefont{Y.}~\bibnamefont{Onuki}},
  \bibinfo{year}{1998}, \bibinfo{journal}{Phys. Rev. B}
  \textbf{\bibinfo{volume}{57}}, \bibinfo{pages}{7434}.

\bibitem[{\citenamefont{Lieb} \emph{et~al.}(1961)\citenamefont{Lieb, Schultz,
  and Mattis}}]{liebschultzmattis61}
\bibinfo{author}{\bibnamefont{Lieb}, \bibfnamefont{E.}},
  \bibinfo{author}{\bibfnamefont{T.}~\bibnamefont{Schultz}}, and
  \bibinfo{author}{\bibfnamefont{D.}~\bibnamefont{Mattis}},
  \bibinfo{year}{1961}, \bibinfo{journal}{Ann. Phys. (N. Y.)}
  \textbf{\bibinfo{volume}{16}}, \bibinfo{pages}{407}.

\bibitem[{\citenamefont{Limelette} \emph{et~al.}(2003)\citenamefont{Limelette,
  Wzietek, Florens, Georges, Costi, Pasquier, Jerome, Meziere, and
  Batail}}]{limelette03prl}
\bibinfo{author}{\bibnamefont{Limelette}, \bibfnamefont{P.}},
  \bibinfo{author}{\bibfnamefont{P.}~\bibnamefont{Wzietek}},
  \bibinfo{author}{\bibfnamefont{S.}~\bibnamefont{Florens}},
  \bibinfo{author}{\bibfnamefont{A.}~\bibnamefont{Georges}},
  \bibinfo{author}{\bibfnamefont{T.~A.} \bibnamefont{Costi}},
  \bibinfo{author}{\bibfnamefont{C.}~\bibnamefont{Pasquier}},
  \bibinfo{author}{\bibfnamefont{D.}~\bibnamefont{Jerome}},
  \bibinfo{author}{\bibfnamefont{C.}~\bibnamefont{Meziere}}, and
  \bibinfo{author}{\bibfnamefont{P.}~\bibnamefont{Batail}},
  \bibinfo{year}{2003}, \bibinfo{journal}{Phys. Rev. Lett.}
  \textbf{\bibinfo{volume}{91}}, \bibinfo{pages}{016401}.

\bibitem[{\citenamefont{Lin} \emph{et~al.}(2003)\citenamefont{Lin, M{\'e}lin,
  Rieger, and Igl{\'o}i}}]{linetal03}
\bibinfo{author}{\bibnamefont{Lin}, \bibfnamefont{Y.-C.}},
  \bibinfo{author}{\bibfnamefont{R.}~\bibnamefont{M{\'e}lin}},
  \bibinfo{author}{\bibfnamefont{H.}~\bibnamefont{Rieger}}, and
  \bibinfo{author}{\bibfnamefont{F.}~\bibnamefont{Igl{\'o}i}},
  \bibinfo{year}{2003}, \bibinfo{journal}{Phys. Rev. B}
  \textbf{\bibinfo{volume}{68}}, \bibinfo{pages}{024424}.

\bibitem[{\citenamefont{Liu} \emph{et~al.}(2000)\citenamefont{Liu, MacLaughlin,
  Neto, Lukefahr, Thompson, Sarrao, and Fisk}}]{liuetal00}
\bibinfo{author}{\bibnamefont{Liu}, \bibfnamefont{C.-Y.}},
  \bibinfo{author}{\bibfnamefont{D.~E.} \bibnamefont{MacLaughlin}},
  \bibinfo{author}{\bibfnamefont{A.~H.~C.} \bibnamefont{Neto}},
  \bibinfo{author}{\bibfnamefont{H.~G.} \bibnamefont{Lukefahr}},
  \bibinfo{author}{\bibfnamefont{J.~D.} \bibnamefont{Thompson}},
  \bibinfo{author}{\bibfnamefont{J.~L.} \bibnamefont{Sarrao}}, and
  \bibinfo{author}{\bibfnamefont{Z.}~\bibnamefont{Fisk}}, \bibinfo{year}{2000},
  \bibinfo{journal}{Phys. Rev. B} \textbf{\bibinfo{volume}{61}},
  \bibinfo{pages}{432}.

\bibitem[{\citenamefont{Loh} \emph{et~al.}(2005)\citenamefont{Loh, Tripathi,
  and Turkalov}}]{lohetal05}
\bibinfo{author}{\bibnamefont{Loh}, \bibfnamefont{Y.~L.}},
  \bibinfo{author}{\bibfnamefont{V.}~\bibnamefont{Tripathi}}, and
  \bibinfo{author}{\bibfnamefont{M.}~\bibnamefont{Turkalov}},
  \bibinfo{year}{2005}, \bibinfo{journal}{Phys. Rev. B}
  \textbf{\bibinfo{volume}{71}}, \bibinfo{pages}{024429}.

\bibitem[{\citenamefont{Ludwig and Affleck}(1991)}]{ludwigaffleck91}
\bibinfo{author}{\bibnamefont{Ludwig}, \bibfnamefont{A.~W.~W.}}, and
  \bibinfo{author}{\bibfnamefont{I.}~\bibnamefont{Affleck}},
  \bibinfo{year}{1991}, \bibinfo{journal}{Phys. Rev. Lett}
  \textbf{\bibinfo{volume}{67}}, \bibinfo{pages}{3160}.

\bibitem[{\citenamefont{Luttinger}(1960)}]{luttingertheorem}
\bibinfo{author}{\bibnamefont{Luttinger}, \bibfnamefont{J.~M.}},
  \bibinfo{year}{1960}, \bibinfo{journal}{Phys. Rev.}
  \textbf{\bibinfo{volume}{119}}, \bibinfo{pages}{1153}.

\bibitem[{\citenamefont{Ma} \emph{et~al.}(1979)\citenamefont{Ma, Dasgupta, and
  Hu}}]{madasguptahu}
\bibinfo{author}{\bibnamefont{Ma}, \bibfnamefont{S.~k.}},
  \bibinfo{author}{\bibfnamefont{C.}~\bibnamefont{Dasgupta}}, and
  \bibinfo{author}{\bibfnamefont{C.~k.} \bibnamefont{Hu}},
  \bibinfo{year}{1979}, \bibinfo{journal}{Phys. Rev. Lett.}
  \textbf{\bibinfo{volume}{43}}, \bibinfo{pages}{1434}.

\bibitem[{\citenamefont{MacLaughlin}
  \emph{et~al.}(2001)\citenamefont{MacLaughlin, Bernal, Heffner, Nieuwenhuys,
  Rose, Sonier, Andraka, Chau, and Maple}}]{dougetal}
\bibinfo{author}{\bibnamefont{MacLaughlin}, \bibfnamefont{D.~E.}},
  \bibinfo{author}{\bibfnamefont{O.~O.} \bibnamefont{Bernal}},
  \bibinfo{author}{\bibfnamefont{R.~H.} \bibnamefont{Heffner}},
  \bibinfo{author}{\bibfnamefont{G.~J.} \bibnamefont{Nieuwenhuys}},
  \bibinfo{author}{\bibfnamefont{M.~S.} \bibnamefont{Rose}},
  \bibinfo{author}{\bibfnamefont{J.~E.} \bibnamefont{Sonier}},
  \bibinfo{author}{\bibfnamefont{B.}~\bibnamefont{Andraka}},
  \bibinfo{author}{\bibfnamefont{R.}~\bibnamefont{Chau}}, and
  \bibinfo{author}{\bibfnamefont{M.~B.} \bibnamefont{Maple}},
  \bibinfo{year}{2001}, \bibinfo{journal}{Phys. Rev. Lett.}
  \textbf{\bibinfo{volume}{87}}, \bibinfo{pages}{066402}.

\bibitem[{\citenamefont{MacLaughlin}
  \emph{et~al.}(1996)\citenamefont{MacLaughlin, Bernal, and
  Lukefahr}}]{maclaughlinetal96}
\bibinfo{author}{\bibnamefont{MacLaughlin}, \bibfnamefont{D.~E.}},
  \bibinfo{author}{\bibfnamefont{O.~O.} \bibnamefont{Bernal}}, and
  \bibinfo{author}{\bibfnamefont{H.~G.} \bibnamefont{Lukefahr}},
  \bibinfo{year}{1996}, \bibinfo{journal}{J. Phys.: Condens. Matter}
  \textbf{\bibinfo{volume}{8}}, \bibinfo{pages}{9855}.

\bibitem[{\citenamefont{MacLaughlin}
  \emph{et~al.}(2002{\natexlab{a}})\citenamefont{MacLaughlin, Bernal, Sonier,
  Heffner, Taniguchi, and Miyako}}]{dougetal2}
\bibinfo{author}{\bibnamefont{MacLaughlin}, \bibfnamefont{D.~E.}},
  \bibinfo{author}{\bibfnamefont{O.~O.} \bibnamefont{Bernal}},
  \bibinfo{author}{\bibfnamefont{J.~E.} \bibnamefont{Sonier}},
  \bibinfo{author}{\bibfnamefont{R.~H.} \bibnamefont{Heffner}},
  \bibinfo{author}{\bibfnamefont{T.}~\bibnamefont{Taniguchi}}, and
  \bibinfo{author}{\bibfnamefont{Y.}~\bibnamefont{Miyako}},
  \bibinfo{year}{2002}{\natexlab{a}}, \bibinfo{journal}{Phys. Rev. B}
  \textbf{\bibinfo{volume}{65}}, \bibinfo{pages}{184401}.

\bibitem[{\citenamefont{MacLaughlin}
  \emph{et~al.}(2004)\citenamefont{MacLaughlin, Heffner, Bernal, Ishida,
  Sonier, Nieuwenhuys, Maple, and Stewart}}]{dougetal04}
\bibinfo{author}{\bibnamefont{MacLaughlin}, \bibfnamefont{D.~E.}},
  \bibinfo{author}{\bibfnamefont{R.~H.} \bibnamefont{Heffner}},
  \bibinfo{author}{\bibfnamefont{O.~O.} \bibnamefont{Bernal}},
  \bibinfo{author}{\bibfnamefont{K.}~\bibnamefont{Ishida}},
  \bibinfo{author}{\bibfnamefont{J.~E.} \bibnamefont{Sonier}},
  \bibinfo{author}{\bibfnamefont{G.~J.} \bibnamefont{Nieuwenhuys}},
  \bibinfo{author}{\bibfnamefont{M.~B.} \bibnamefont{Maple}}, and
  \bibinfo{author}{\bibfnamefont{G.~R.} \bibnamefont{Stewart}},
  \bibinfo{year}{2004}, \bibinfo{journal}{J. Phys.: Condens. Matter}
  \textbf{\bibinfo{volume}{16}}, \bibinfo{pages}{S4479}.

\bibitem[{\citenamefont{MacLaughlin}
  \emph{et~al.}(2002{\natexlab{b}})\citenamefont{MacLaughlin, Heffner, Bernal,
  Nieuwenhuys, Sonier, Rose, Chau, Maple, and Andraka}}]{dougetal02}
\bibinfo{author}{\bibnamefont{MacLaughlin}, \bibfnamefont{D.~E.}},
  \bibinfo{author}{\bibfnamefont{R.~H.} \bibnamefont{Heffner}},
  \bibinfo{author}{\bibfnamefont{O.~O.} \bibnamefont{Bernal}},
  \bibinfo{author}{\bibfnamefont{G.~J.} \bibnamefont{Nieuwenhuys}},
  \bibinfo{author}{\bibfnamefont{J.~E.} \bibnamefont{Sonier}},
  \bibinfo{author}{\bibfnamefont{M.~S.} \bibnamefont{Rose}},
  \bibinfo{author}{\bibfnamefont{R.}~\bibnamefont{Chau}},
  \bibinfo{author}{\bibfnamefont{M.~B.} \bibnamefont{Maple}}, and
  \bibinfo{author}{\bibfnamefont{B.}~\bibnamefont{Andraka}},
  \bibinfo{year}{2002}{\natexlab{b}}, \bibinfo{journal}{Physica B}
  \textbf{\bibinfo{volume}{312-313}}, \bibinfo{pages}{453}.

\bibitem[{\citenamefont{MacLaughlin}
  \emph{et~al.}(1998)\citenamefont{MacLaughlin, Heffner, Nieuwenhuys, Luke,
  Fudamoto, Uemura, Chau, Maple, and Andraka}}]{maclaughlinetal98}
\bibinfo{author}{\bibnamefont{MacLaughlin}, \bibfnamefont{D.~E.}},
  \bibinfo{author}{\bibfnamefont{R.~H.} \bibnamefont{Heffner}},
  \bibinfo{author}{\bibfnamefont{G.~J.} \bibnamefont{Nieuwenhuys}},
  \bibinfo{author}{\bibfnamefont{G.~M.} \bibnamefont{Luke}},
  \bibinfo{author}{\bibfnamefont{Y.}~\bibnamefont{Fudamoto}},
  \bibinfo{author}{\bibfnamefont{Y.~J.} \bibnamefont{Uemura}},
  \bibinfo{author}{\bibfnamefont{R.}~\bibnamefont{Chau}},
  \bibinfo{author}{\bibfnamefont{M.~B.} \bibnamefont{Maple}}, and
  \bibinfo{author}{\bibfnamefont{B.}~\bibnamefont{Andraka}},
  \bibinfo{year}{1998}, \bibinfo{journal}{Phys. Rev. B}
  \textbf{\bibinfo{volume}{58}}, \bibinfo{pages}{11849}.

\bibitem[{\citenamefont{MacLaughlin}
  \emph{et~al.}(2000)\citenamefont{MacLaughlin, Heffner, Sonier, Nieuwenhuys,
  Chau, Maple, Andraka, Luke, Fudamoto, Uemura, Amato, and Baines}}]{dougetal3}
\bibinfo{author}{\bibnamefont{MacLaughlin}, \bibfnamefont{D.~E.}},
  \bibinfo{author}{\bibfnamefont{R.~H.} \bibnamefont{Heffner}},
  \bibinfo{author}{\bibfnamefont{J.~E.} \bibnamefont{Sonier}},
  \bibinfo{author}{\bibfnamefont{G.~J.} \bibnamefont{Nieuwenhuys}},
  \bibinfo{author}{\bibfnamefont{R.}~\bibnamefont{Chau}},
  \bibinfo{author}{\bibfnamefont{M.~B.} \bibnamefont{Maple}},
  \bibinfo{author}{\bibfnamefont{B.}~\bibnamefont{Andraka}},
  \bibinfo{author}{\bibfnamefont{G.~M.} \bibnamefont{Luke}},
  \bibinfo{author}{\bibfnamefont{Y.}~\bibnamefont{Fudamoto}},
  \bibinfo{author}{\bibfnamefont{Y.~J.} \bibnamefont{Uemura}},
  \bibinfo{author}{\bibfnamefont{A.}~\bibnamefont{Amato}}, and
  \bibinfo{author}{\bibfnamefont{C.}~\bibnamefont{Baines}},
  \bibinfo{year}{2000}, \bibinfo{journal}{Physica B}
  \textbf{\bibinfo{volume}{289-290}}, \bibinfo{pages}{15}.

\bibitem[{\citenamefont{MacLaughlin}
  \emph{et~al.}(2003)\citenamefont{MacLaughlin, Rose, Young, Bernal, Heffner,
  Morris, Ishida, Nieuwenhuys, and Sonier}}]{dougetal03}
\bibinfo{author}{\bibnamefont{MacLaughlin}, \bibfnamefont{D.~E.}},
  \bibinfo{author}{\bibfnamefont{M.~S.} \bibnamefont{Rose}},
  \bibinfo{author}{\bibfnamefont{B.-L.} \bibnamefont{Young}},
  \bibinfo{author}{\bibfnamefont{O.~O.} \bibnamefont{Bernal}},
  \bibinfo{author}{\bibfnamefont{R.~H.} \bibnamefont{Heffner}},
  \bibinfo{author}{\bibfnamefont{G.~D.} \bibnamefont{Morris}},
  \bibinfo{author}{\bibfnamefont{K.}~\bibnamefont{Ishida}},
  \bibinfo{author}{\bibfnamefont{G.~J.} \bibnamefont{Nieuwenhuys}}, and
  \bibinfo{author}{\bibfnamefont{J.~E.} \bibnamefont{Sonier}},
  \bibinfo{year}{2003}, \bibinfo{journal}{Physica B}
  \textbf{\bibinfo{volume}{326}}, \bibinfo{pages}{381}.

\bibitem[{\citenamefont{Maebashi} \emph{et~al.}(2002)\citenamefont{Maebashi,
  Miyake, and Varma}}]{maebashietal02}
\bibinfo{author}{\bibnamefont{Maebashi}, \bibfnamefont{H.}},
  \bibinfo{author}{\bibfnamefont{K.}~\bibnamefont{Miyake}}, and
  \bibinfo{author}{\bibfnamefont{C.~M.} \bibnamefont{Varma}},
  \bibinfo{year}{2002}, \bibinfo{journal}{Phys. Rev. Lett.}
  \textbf{\bibinfo{volume}{88}}, \bibinfo{pages}{226403}.

\bibitem[{\citenamefont{Maple} \emph{et~al.}(1995)\citenamefont{Maple,
  de~Andrade, Herrmann, Dalichaouch, Gajewski, Seaman, Chau, Movshovich,
  Aronson, and Osborn}}]{mapleetal95}
\bibinfo{author}{\bibnamefont{Maple}, \bibfnamefont{M.~B.}},
  \bibinfo{author}{\bibfnamefont{M.~C.} \bibnamefont{de~Andrade}},
  \bibinfo{author}{\bibfnamefont{J.}~\bibnamefont{Herrmann}},
  \bibinfo{author}{\bibfnamefont{Y.}~\bibnamefont{Dalichaouch}},
  \bibinfo{author}{\bibfnamefont{D.~A.} \bibnamefont{Gajewski}},
  \bibinfo{author}{\bibfnamefont{C.~L.} \bibnamefont{Seaman}},
  \bibinfo{author}{\bibfnamefont{R.}~\bibnamefont{Chau}},
  \bibinfo{author}{\bibfnamefont{R.}~\bibnamefont{Movshovich}},
  \bibinfo{author}{\bibfnamefont{M.~C.} \bibnamefont{Aronson}}, and
  \bibinfo{author}{\bibfnamefont{R.}~\bibnamefont{Osborn}},
  \bibinfo{year}{1995}, \bibinfo{journal}{J. Low. Temp. Phys.}
  \textbf{\bibinfo{volume}{99}}, \bibinfo{pages}{223}.

\bibitem[{\citenamefont{Mathur} \emph{et~al.}(1998)\citenamefont{Mathur,
  Grosche, Julian, Walker, Freye, Haselwimmer, and Lonzarich}}]{mathuretal}
\bibinfo{author}{\bibnamefont{Mathur}, \bibfnamefont{N.~D.}},
  \bibinfo{author}{\bibfnamefont{F.~M.} \bibnamefont{Grosche}},
  \bibinfo{author}{\bibfnamefont{S.~R.} \bibnamefont{Julian}},
  \bibinfo{author}{\bibfnamefont{I.~R.} \bibnamefont{Walker}},
  \bibinfo{author}{\bibfnamefont{D.~M.} \bibnamefont{Freye}},
  \bibinfo{author}{\bibfnamefont{R.~K.~W.} \bibnamefont{Haselwimmer}}, and
  \bibinfo{author}{\bibfnamefont{G.~G.} \bibnamefont{Lonzarich}},
  \bibinfo{year}{1998}, \bibinfo{journal}{Nature}
  \textbf{\bibinfo{volume}{394}}, \bibinfo{pages}{39}.

\bibitem[{\citenamefont{Mayr} \emph{et~al.}(1997)\citenamefont{Mayr,
  v.~Blanckenhagen, and Stewart}}]{mayretal97}
\bibinfo{author}{\bibnamefont{Mayr}, \bibfnamefont{F.}},
  \bibinfo{author}{\bibfnamefont{G.-F.} \bibnamefont{v.~Blanckenhagen}}, and
  \bibinfo{author}{\bibfnamefont{G.~R.} \bibnamefont{Stewart}},
  \bibinfo{year}{1997}, \bibinfo{journal}{Phys. Rev. B}
  \textbf{\bibinfo{volume}{55}}, \bibinfo{pages}{947}.

\bibitem[{\citenamefont{McCoy}(1969)}]{mccoy69}
\bibinfo{author}{\bibnamefont{McCoy}, \bibfnamefont{B.~M.}},
  \bibinfo{year}{1969}, \bibinfo{journal}{Phys. Rev. B}
  \textbf{\bibinfo{volume}{188}}, \bibinfo{pages}{1014}.

\bibitem[{\citenamefont{McCoy and Wu}(1968)}]{mccoywu68}
\bibinfo{author}{\bibnamefont{McCoy}, \bibfnamefont{B.~M.}}, and
  \bibinfo{author}{\bibfnamefont{T.~T.} \bibnamefont{Wu}},
  \bibinfo{year}{1968}, \bibinfo{journal}{Phys. Rev. B}
  \textbf{\bibinfo{volume}{176}}, \bibinfo{pages}{631}.

\bibitem[{\citenamefont{McCoy and Wu}(1969)}]{mccoywu69}
\bibinfo{author}{\bibnamefont{McCoy}, \bibfnamefont{B.~M.}}, and
  \bibinfo{author}{\bibfnamefont{T.~T.} \bibnamefont{Wu}},
  \bibinfo{year}{1969}, \bibinfo{journal}{Phys. Rev. B}
  \textbf{\bibinfo{volume}{188}}, \bibinfo{pages}{982}.

\bibitem[{\citenamefont{M{\'e}lin} \emph{et~al.}(2002)\citenamefont{M{\'e}lin,
  Lin, Lajk{\'o}, Rieger, and Igl{\'o}i}}]{melinetal}
\bibinfo{author}{\bibnamefont{M{\'e}lin}, \bibfnamefont{R.}},
  \bibinfo{author}{\bibfnamefont{Y.-C.} \bibnamefont{Lin}},
  \bibinfo{author}{\bibfnamefont{P.}~\bibnamefont{Lajk{\'o}}},
  \bibinfo{author}{\bibfnamefont{H.}~\bibnamefont{Rieger}}, and
  \bibinfo{author}{\bibfnamefont{F.}~\bibnamefont{Igl{\'o}i}},
  \bibinfo{year}{2002}, \bibinfo{journal}{Phys. Rev. B}
  \textbf{\bibinfo{volume}{65}}, \bibinfo{pages}{104415}.

\bibitem[{\citenamefont{Meyer and Nolting}(2000{\natexlab{a}})}]{meyernolting2}
\bibinfo{author}{\bibnamefont{Meyer}, \bibfnamefont{D.}}, and
  \bibinfo{author}{\bibfnamefont{W.}~\bibnamefont{Nolting}},
  \bibinfo{year}{2000}{\natexlab{a}}, \bibinfo{journal}{Phys. Rev. B}
  \textbf{\bibinfo{volume}{62}}, \bibinfo{pages}{5657}.

\bibitem[{\citenamefont{Meyer and Nolting}(2000{\natexlab{b}})}]{meyernolting1}
\bibinfo{author}{\bibnamefont{Meyer}, \bibfnamefont{D.}}, and
  \bibinfo{author}{\bibfnamefont{W.}~\bibnamefont{Nolting}},
  \bibinfo{year}{2000}{\natexlab{b}}, \bibinfo{journal}{Phys. Rev. B}
  \textbf{\bibinfo{volume}{61}}, \bibinfo{pages}{13465}.

\bibitem[{\citenamefont{M{\'e}zard}
  \emph{et~al.}(1986)\citenamefont{M{\'e}zard, Parisi, and
  Virasoro}}]{re:Mezard86}
\bibinfo{author}{\bibnamefont{M{\'e}zard}, \bibfnamefont{M.}},
  \bibinfo{author}{\bibfnamefont{G.}~\bibnamefont{Parisi}}, and
  \bibinfo{author}{\bibfnamefont{M.~A.} \bibnamefont{Virasoro}},
  \bibinfo{year}{1986}, \emph{\bibinfo{title}{Spin Glass theory and beyond}}
  (\bibinfo{publisher}{World Scientific}, \bibinfo{address}{Singapore}).

\bibitem[{\citenamefont{Migdal}(1958)}]{migdal58}
\bibinfo{author}{\bibnamefont{Migdal}, \bibfnamefont{A.~B.}},
  \bibinfo{year}{1958}, \bibinfo{journal}{Sov. Phys. JETP}
  \textbf{\bibinfo{volume}{7}}, \bibinfo{pages}{996}.

\bibitem[{\citenamefont{Migdal}(1967)}]{migdalbook67}
\bibinfo{author}{\bibnamefont{Migdal}, \bibfnamefont{A.~B.}},
  \bibinfo{year}{1967}, \emph{\bibinfo{title}{Theory of {F}inite {F}ermi
  {S}ystems and {A}pplications to {A}tomic {N}uclei}}
  (\bibinfo{publisher}{Pergamon}, \bibinfo{address}{London}).

\bibitem[{\citenamefont{Miller and Huse}(1993)}]{re:Miller93}
\bibinfo{author}{\bibnamefont{Miller}, \bibfnamefont{J.}}, and
  \bibinfo{author}{\bibfnamefont{D.~A.} \bibnamefont{Huse}},
  \bibinfo{year}{1993}, \bibinfo{journal}{Phys. Rev. Lett.}
  \textbf{\bibinfo{volume}{70}}, \bibinfo{pages}{3147}.

\bibitem[{\citenamefont{Millis}(1993)}]{millis}
\bibinfo{author}{\bibnamefont{Millis}, \bibfnamefont{A.~J.}},
  \bibinfo{year}{1993}, \bibinfo{journal}{Phys. Rev. B}
  \textbf{\bibinfo{volume}{48}}(\bibinfo{number}{10}), \bibinfo{pages}{7183}.

\bibitem[{\citenamefont{Millis}(1999)}]{millissces99}
\bibinfo{author}{\bibnamefont{Millis}, \bibfnamefont{A.~J.}},
  \bibinfo{year}{1999}, \bibinfo{journal}{Physica B}
  \textbf{\bibinfo{volume}{259-261}}, \bibinfo{pages}{1169}.

\bibitem[{\citenamefont{Millis} \emph{et~al.}(2001)\citenamefont{Millis, Morr,
  and Schmalian}}]{millisetal01}
\bibinfo{author}{\bibnamefont{Millis}, \bibfnamefont{A.~J.}},
  \bibinfo{author}{\bibfnamefont{D.~K.} \bibnamefont{Morr}}, and
  \bibinfo{author}{\bibfnamefont{J.}~\bibnamefont{Schmalian}},
  \bibinfo{year}{2001}, \bibinfo{journal}{Phys. Rev. Lett.}
  \textbf{\bibinfo{volume}{87}}, \bibinfo{pages}{167202}.

\bibitem[{\citenamefont{Millis} \emph{et~al.}(2002)\citenamefont{Millis, Morr,
  and Schmalian}}]{millisetal02}
\bibinfo{author}{\bibnamefont{Millis}, \bibfnamefont{A.~J.}},
  \bibinfo{author}{\bibfnamefont{D.~K.} \bibnamefont{Morr}}, and
  \bibinfo{author}{\bibfnamefont{J.}~\bibnamefont{Schmalian}},
  \bibinfo{year}{2002}, \bibinfo{journal}{Phys. Rev. B}
  \textbf{\bibinfo{volume}{66}}, \bibinfo{pages}{174433}.

\bibitem[{\citenamefont{Millis} \emph{et~al.}(2004)\citenamefont{Millis, Morr,
  and Schmalian}}]{millisetal04}
\bibinfo{author}{\bibnamefont{Millis}, \bibfnamefont{A.~J.}},
  \bibinfo{author}{\bibfnamefont{D.~K.} \bibnamefont{Morr}}, and
  \bibinfo{author}{\bibfnamefont{J.}~\bibnamefont{Schmalian}},
  \bibinfo{year}{2004}, \bibinfo{journal}{cond-mat/0411738} .

\bibitem[{\citenamefont{Milovanovi\'{c}}
  \emph{et~al.}(1989)\citenamefont{Milovanovi\'{c}, Sachdev, and
  Bhatt}}]{milovanovicetal89}
\bibinfo{author}{\bibnamefont{Milovanovi\'{c}}, \bibfnamefont{M.}},
  \bibinfo{author}{\bibfnamefont{S.}~\bibnamefont{Sachdev}}, and
  \bibinfo{author}{\bibfnamefont{R.~N.} \bibnamefont{Bhatt}},
  \bibinfo{year}{1989}, \bibinfo{journal}{Phys. Rev. Lett.}
  \textbf{\bibinfo{volume}{63}}, \bibinfo{pages}{82}.

\bibitem[{\citenamefont{Miranda and Dobrosavljevi\'{c}}(1999)}]{mirandavlad3}
\bibinfo{author}{\bibnamefont{Miranda}, \bibfnamefont{E.}}, and
  \bibinfo{author}{\bibfnamefont{V.}~\bibnamefont{Dobrosavljevi\'{c}}},
  \bibinfo{year}{1999}, \bibinfo{journal}{Physica B}
  \textbf{\bibinfo{volume}{259-261}}, \bibinfo{pages}{359}.

\bibitem[{\citenamefont{Miranda and
  Dobrosavljevi\'{c}}(2001{\natexlab{a}})}]{mirandavlad2}
\bibinfo{author}{\bibnamefont{Miranda}, \bibfnamefont{E.}}, and
  \bibinfo{author}{\bibfnamefont{V.}~\bibnamefont{Dobrosavljevi\'{c}}},
  \bibinfo{year}{2001}{\natexlab{a}}, \bibinfo{journal}{J. Magn. Magn. Mat.}
  \textbf{\bibinfo{volume}{226-230}}, \bibinfo{pages}{110}.

\bibitem[{\citenamefont{Miranda and
  Dobrosavljevi\'{c}}(2001{\natexlab{b}})}]{mirandavlad1}
\bibinfo{author}{\bibnamefont{Miranda}, \bibfnamefont{E.}}, and
  \bibinfo{author}{\bibfnamefont{V.}~\bibnamefont{Dobrosavljevi\'{c}}},
  \bibinfo{year}{2001}{\natexlab{b}}, \bibinfo{journal}{Phys. Rev. Lett.}
  \textbf{\bibinfo{volume}{86}}, \bibinfo{pages}{264}.

\bibitem[{\citenamefont{Miranda} \emph{et~al.}(1996)\citenamefont{Miranda,
  Dobrosavljevi\'{c}, and Kotliar}}]{mirandavladgabi1}
\bibinfo{author}{\bibnamefont{Miranda}, \bibfnamefont{E.}},
  \bibinfo{author}{\bibfnamefont{V.}~\bibnamefont{Dobrosavljevi\'{c}}}, and
  \bibinfo{author}{\bibfnamefont{G.}~\bibnamefont{Kotliar}},
  \bibinfo{year}{1996}, \bibinfo{journal}{J. Phys.: Condens. Matter}
  \textbf{\bibinfo{volume}{8}}, \bibinfo{pages}{9871}.

\bibitem[{\citenamefont{Miranda}
  \emph{et~al.}(1997{\natexlab{a}})\citenamefont{Miranda, Dobrosavljevi\'{c},
  and Kotliar}}]{mirandavladgabi2}
\bibinfo{author}{\bibnamefont{Miranda}, \bibfnamefont{E.}},
  \bibinfo{author}{\bibfnamefont{V.}~\bibnamefont{Dobrosavljevi\'{c}}}, and
  \bibinfo{author}{\bibfnamefont{G.}~\bibnamefont{Kotliar}},
  \bibinfo{year}{1997}{\natexlab{a}}, \bibinfo{journal}{Phys. Rev. Lett.}
  \textbf{\bibinfo{volume}{78}}, \bibinfo{pages}{290}.

\bibitem[{\citenamefont{Miranda}
  \emph{et~al.}(1997{\natexlab{b}})\citenamefont{Miranda, Dobrosavljevi\'{c},
  and Kotliar}}]{mirandavladgabi3}
\bibinfo{author}{\bibnamefont{Miranda}, \bibfnamefont{E.}},
  \bibinfo{author}{\bibfnamefont{V.}~\bibnamefont{Dobrosavljevi\'{c}}}, and
  \bibinfo{author}{\bibfnamefont{G.}~\bibnamefont{Kotliar}},
  \bibinfo{year}{1997}{\natexlab{b}}, \bibinfo{journal}{Physica B}
  \textbf{\bibinfo{volume}{230}}, \bibinfo{pages}{569}.

\bibitem[{\citenamefont{Monthus} \emph{et~al.}(1997)\citenamefont{Monthus,
  Golinelli, and Jolicoeur}}]{monthusetal97}
\bibinfo{author}{\bibnamefont{Monthus}, \bibfnamefont{C.}},
  \bibinfo{author}{\bibfnamefont{O.}~\bibnamefont{Golinelli}}, and
  \bibinfo{author}{\bibfnamefont{T.}~\bibnamefont{Jolicoeur}},
  \bibinfo{year}{1997}, \bibinfo{journal}{Phys. Rev. Lett.}
  \textbf{\bibinfo{volume}{79}}, \bibinfo{pages}{3254}.

\bibitem[{\citenamefont{Monthus} \emph{et~al.}(1998)\citenamefont{Monthus,
  Golinelli, and Jolicoeur}}]{monthusetal98}
\bibinfo{author}{\bibnamefont{Monthus}, \bibfnamefont{C.}},
  \bibinfo{author}{\bibfnamefont{O.}~\bibnamefont{Golinelli}}, and
  \bibinfo{author}{\bibfnamefont{T.}~\bibnamefont{Jolicoeur}},
  \bibinfo{year}{1998}, \bibinfo{journal}{Phys. Rev. B}
  \textbf{\bibinfo{volume}{58}}, \bibinfo{pages}{805}.

\bibitem[{\citenamefont{Moryia}(1985)}]{moryia}
\bibinfo{author}{\bibnamefont{Moryia}, \bibfnamefont{T.}},
  \bibinfo{year}{1985}, \emph{\bibinfo{title}{Spin {F}luctuations in
  {I}tinerant {E}lectron {M}agnetism}} (\bibinfo{publisher}{Springer-Verlag},
  \bibinfo{address}{Berlin}).

\bibitem[{\citenamefont{Motrunich} \emph{et~al.}(2001)\citenamefont{Motrunich,
  Mau, Huse, and Fisher}}]{motrunichetal01}
\bibinfo{author}{\bibnamefont{Motrunich}, \bibfnamefont{O.}},
  \bibinfo{author}{\bibfnamefont{S.-C.} \bibnamefont{Mau}},
  \bibinfo{author}{\bibfnamefont{D.~A.} \bibnamefont{Huse}}, and
  \bibinfo{author}{\bibfnamefont{D.~S.} \bibnamefont{Fisher}},
  \bibinfo{year}{2001}, \bibinfo{journal}{Phys. Rev. B}
  \textbf{\bibinfo{volume}{61}}, \bibinfo{pages}{1160}.

\bibitem[{\citenamefont{Mott}(1990)}]{mott-book90}
\bibinfo{author}{\bibnamefont{Mott}, \bibfnamefont{N.~F.}},
  \bibinfo{year}{1990}, \emph{\bibinfo{title}{Metal-Insulator Transition}}
  (\bibinfo{publisher}{Taylor \& Francis}, \bibinfo{address}{London}).

\bibitem[{\citenamefont{Muller and Ioffe}(2004)}]{muller04prl}
\bibinfo{author}{\bibnamefont{Muller}, \bibfnamefont{M.}}, and
  \bibinfo{author}{\bibfnamefont{L.~B.} \bibnamefont{Ioffe}},
  \bibinfo{year}{2004}, \bibinfo{journal}{Phys. Rev. Lett.}
  \textbf{\bibinfo{volume}{93}}, \bibinfo{pages}{256403}.

\bibitem[{\citenamefont{Naqib} \emph{et~al.}(2003)\citenamefont{Naqib, Cooper,
  Tallon, and Panagopoulos}}]{naqib-physicaC03}
\bibinfo{author}{\bibnamefont{Naqib}, \bibfnamefont{S.~H.}},
  \bibinfo{author}{\bibfnamefont{J.~R.} \bibnamefont{Cooper}},
  \bibinfo{author}{\bibfnamefont{J.~L.} \bibnamefont{Tallon}}, and
  \bibinfo{author}{\bibfnamefont{C.}~\bibnamefont{Panagopoulos}},
  \bibinfo{year}{2003}, \bibinfo{journal}{Physica C}
  \textbf{\bibinfo{volume}{387}}, \bibinfo{pages}{365}.

\bibitem[{\citenamefont{Narozhny} \emph{et~al.}(2001)\citenamefont{Narozhny,
  Aleiner, and Larkin}}]{narozhnyetal00}
\bibinfo{author}{\bibnamefont{Narozhny}, \bibfnamefont{B.~N.}},
  \bibinfo{author}{\bibfnamefont{I.~L.} \bibnamefont{Aleiner}}, and
  \bibinfo{author}{\bibfnamefont{A.~I.} \bibnamefont{Larkin}},
  \bibinfo{year}{2001}, \bibinfo{journal}{Phys. Rev. B}
  \textbf{\bibinfo{volume}{62}}, \bibinfo{pages}{14898}.

\bibitem[{\citenamefont{Nozi{\`e}res}(1974)}]{nozieres74}
\bibinfo{author}{\bibnamefont{Nozi{\`e}res}, \bibfnamefont{P.}},
  \bibinfo{year}{1974}, \bibinfo{journal}{J. Low Temp. Phys.}
  \textbf{\bibinfo{volume}{17}}, \bibinfo{pages}{31}.

\bibitem[{\citenamefont{Nozi{\`e}res and Blandin}(1980)}]{nozieresblandin80}
\bibinfo{author}{\bibnamefont{Nozi{\`e}res}, \bibfnamefont{P.}}, and
  \bibinfo{author}{\bibfnamefont{A.}~\bibnamefont{Blandin}},
  \bibinfo{year}{1980}, \bibinfo{journal}{J. Phys. (Paris)}
  \textbf{\bibinfo{volume}{41}}, \bibinfo{pages}{193}.

\bibitem[{\citenamefont{Ott} \emph{et~al.}(1993)\citenamefont{Ott, Felder, and
  Bernasconi}}]{ottetal93}
\bibinfo{author}{\bibnamefont{Ott}, \bibfnamefont{H.~R.}},
  \bibinfo{author}{\bibfnamefont{E.}~\bibnamefont{Felder}}, and
  \bibinfo{author}{\bibfnamefont{A.}~\bibnamefont{Bernasconi}},
  \bibinfo{year}{1993}, \bibinfo{journal}{Phys. B}
  \textbf{\bibinfo{volume}{186-188}}, \bibinfo{pages}{207}.

\bibitem[{\citenamefont{Ovadyahu and Pollak}(1997)}]{films23}
\bibinfo{author}{\bibnamefont{Ovadyahu}, \bibfnamefont{Z.}}, and
  \bibinfo{author}{\bibfnamefont{M.}~\bibnamefont{Pollak}},
  \bibinfo{year}{1997}, \bibinfo{journal}{Phys. Rev. Lett.}
  \textbf{\bibinfo{volume}{79}}, \bibinfo{pages}{459}.

\bibitem[{\citenamefont{Paalanen and Bhatt}(1991)}]{paalanen91}
\bibinfo{author}{\bibnamefont{Paalanen}, \bibfnamefont{M.~A.}}, and
  \bibinfo{author}{\bibfnamefont{R.~N.} \bibnamefont{Bhatt}},
  \bibinfo{year}{1991}, \bibinfo{journal}{Physica B}
  \textbf{\bibinfo{volume}{169}}, \bibinfo{pages}{231}.

\bibitem[{\citenamefont{Paalanen} \emph{et~al.}(1988)\citenamefont{Paalanen,
  Graebner, Bhatt, and Sachdev}}]{paalanen}
\bibinfo{author}{\bibnamefont{Paalanen}, \bibfnamefont{M.~A.}},
  \bibinfo{author}{\bibfnamefont{J.~E.} \bibnamefont{Graebner}},
  \bibinfo{author}{\bibfnamefont{R.~N.} \bibnamefont{Bhatt}}, and
  \bibinfo{author}{\bibfnamefont{S.}~\bibnamefont{Sachdev}},
  \bibinfo{year}{1988}, \bibinfo{journal}{Phys. Rev. Lett.}
  \textbf{\bibinfo{volume}{61}}, \bibinfo{pages}{597}.

\bibitem[{\citenamefont{Paalanen} \emph{et~al.}(1982)\citenamefont{Paalanen,
  Rosenbaum, Thomas, , and Bhatt}}]{paalanen82}
\bibinfo{author}{\bibnamefont{Paalanen}, \bibfnamefont{M.~A.}},
  \bibinfo{author}{\bibfnamefont{T.~F.} \bibnamefont{Rosenbaum}},
  \bibinfo{author}{\bibfnamefont{G.~A.} \bibnamefont{Thomas}}, , and
  \bibinfo{author}{\bibfnamefont{R.~N.} \bibnamefont{Bhatt}},
  \bibinfo{year}{1982}, \bibinfo{journal}{Phys. Rev. Lett.}
  \textbf{\bibinfo{volume}{48}}, \bibinfo{pages}{1284}.

\bibitem[{\citenamefont{Panagopoulos and
  Dobrosavljevi\'c}(2005)}]{christos-vlad04}
\bibinfo{author}{\bibnamefont{Panagopoulos}, \bibfnamefont{C.}}, and
  \bibinfo{author}{\bibfnamefont{V.}~\bibnamefont{Dobrosavljevi\'c}},
  \bibinfo{year}{2005}, \bibinfo{journal}{Phys. Rev. B}
  \textbf{\bibinfo{volume}{72}}, \bibinfo{pages}{014536}.

\bibitem[{\citenamefont{Panagopoulos}
  \emph{et~al.}(2003)\citenamefont{Panagopoulos, Tallon, Rainford, Cooper,
  Scott, and Xiang}}]{panagopoulos-ssc03}
\bibinfo{author}{\bibnamefont{Panagopoulos}, \bibfnamefont{C.}},
  \bibinfo{author}{\bibfnamefont{J.~L.} \bibnamefont{Tallon}},
  \bibinfo{author}{\bibfnamefont{B.~D.} \bibnamefont{Rainford}},
  \bibinfo{author}{\bibfnamefont{J.~R.} \bibnamefont{Cooper}},
  \bibinfo{author}{\bibfnamefont{C.~A.} \bibnamefont{Scott}}, and
  \bibinfo{author}{\bibfnamefont{T.}~\bibnamefont{Xiang}},
  \bibinfo{year}{2003}, \bibinfo{journal}{Solid State Comm.}
  \textbf{\bibinfo{volume}{126}}(\bibinfo{number}{47}), \bibinfo{pages}{47},
  \bibinfo{note}{(Special issue Eds. A.J. Millis and Y. Uemura); and references
  therein.}

\bibitem[{\citenamefont{Panagopoulos}
  \emph{et~al.}(2002)\citenamefont{Panagopoulos, Tallon, Rainford, Xiang,
  Cooper, and Scott}}]{panagopoulos-prb02}
\bibinfo{author}{\bibnamefont{Panagopoulos}, \bibfnamefont{C.}},
  \bibinfo{author}{\bibfnamefont{J.~L.} \bibnamefont{Tallon}},
  \bibinfo{author}{\bibfnamefont{B.~D.} \bibnamefont{Rainford}},
  \bibinfo{author}{\bibfnamefont{T.}~\bibnamefont{Xiang}},
  \bibinfo{author}{\bibfnamefont{J.~R.} \bibnamefont{Cooper}}, and
  \bibinfo{author}{\bibfnamefont{C.~A.} \bibnamefont{Scott}},
  \bibinfo{year}{2002}, \bibinfo{journal}{Phys. Rev. B}
  \textbf{\bibinfo{volume}{66}}, \bibinfo{pages}{064501}.

\bibitem[{\citenamefont{Pankov and Dobrosavljevi\'c}(2005)}]{pankov05prl}
\bibinfo{author}{\bibnamefont{Pankov}, \bibfnamefont{S.}}, and
  \bibinfo{author}{\bibfnamefont{V.}~\bibnamefont{Dobrosavljevi\'c}},
  \bibinfo{year}{2005}, \bibinfo{journal}{Phys. Rev. Lett.}
  \textbf{\bibinfo{volume}{94}}, \bibinfo{pages}{046402}.

\bibitem[{\citenamefont{Parcollet and Georges}(1999)}]{parcolletgeorges}
\bibinfo{author}{\bibnamefont{Parcollet}, \bibfnamefont{O.}}, and
  \bibinfo{author}{\bibfnamefont{A.}~\bibnamefont{Georges}},
  \bibinfo{year}{1999}, \bibinfo{journal}{Phys. Rev. B}
  \textbf{\bibinfo{volume}{59}}(\bibinfo{number}{8}), \bibinfo{pages}{5341}.

\bibitem[{\citenamefont{Pastor and Dobrosavljevi\'c}(1999)}]{pastor-prl99}
\bibinfo{author}{\bibnamefont{Pastor}, \bibfnamefont{A.~A.}}, and
  \bibinfo{author}{\bibfnamefont{V.}~\bibnamefont{Dobrosavljevi\'c}},
  \bibinfo{year}{1999}, \bibinfo{journal}{Phys. Rev. Lett.}
  \textbf{\bibinfo{volume}{83}}, \bibinfo{pages}{4642}.

\bibitem[{\citenamefont{Pastor} \emph{et~al.}(2002)\citenamefont{Pastor,
  Dobrosavreljevi\'c, and Horbach}}]{horbach02}
\bibinfo{author}{\bibnamefont{Pastor}, \bibfnamefont{A.~A.}},
  \bibinfo{author}{\bibfnamefont{V.}~\bibnamefont{Dobrosavreljevi\'c}}, and
  \bibinfo{author}{\bibfnamefont{M.~L.} \bibnamefont{Horbach}},
  \bibinfo{year}{2002}, \bibinfo{journal}{Phys. Rev. B}
  \textbf{\bibinfo{volume}{66}}, \bibinfo{pages}{014413}.

\bibitem[{\citenamefont{Pazmandi} \emph{et~al.}(1999)\citenamefont{Pazmandi,
  Zarand, and Zimanyi}}]{re:Pazmandi99}
\bibinfo{author}{\bibnamefont{Pazmandi}, \bibfnamefont{F.}},
  \bibinfo{author}{\bibfnamefont{G.}~\bibnamefont{Zarand}}, and
  \bibinfo{author}{\bibfnamefont{G.~T.} \bibnamefont{Zimanyi}},
  \bibinfo{year}{1999}, \bibinfo{journal}{Phys. Rev. Lett.}
  \textbf{\bibinfo{volume}{83}}, \bibinfo{pages}{1034}.

\bibitem[{\citenamefont{Pfeuty}(1970)}]{pfeuty70}
\bibinfo{author}{\bibnamefont{Pfeuty}, \bibfnamefont{P.}},
  \bibinfo{year}{1970}, \bibinfo{journal}{Ann. Phys. (N. Y.)}
  \textbf{\bibinfo{volume}{57}}, \bibinfo{pages}{79}.

\bibitem[{\citenamefont{Pich} \emph{et~al.}(1998)\citenamefont{Pich, Young,
  Rieger, and Kawashima}}]{pichetal98}
\bibinfo{author}{\bibnamefont{Pich}, \bibfnamefont{C.}},
  \bibinfo{author}{\bibfnamefont{A.~P.} \bibnamefont{Young}},
  \bibinfo{author}{\bibfnamefont{H.}~\bibnamefont{Rieger}}, and
  \bibinfo{author}{\bibfnamefont{N.}~\bibnamefont{Kawashima}},
  \bibinfo{year}{1998}, \bibinfo{journal}{Phys. Rev. Lett.}
  \textbf{\bibinfo{volume}{81}}, \bibinfo{pages}{5916}.

\bibitem[{\citenamefont{Pietri} \emph{et~al.}(1997)\citenamefont{Pietri,
  Andraka, Troc, and Tran}}]{pietrietal97}
\bibinfo{author}{\bibnamefont{Pietri}, \bibfnamefont{R.}},
  \bibinfo{author}{\bibfnamefont{B.}~\bibnamefont{Andraka}},
  \bibinfo{author}{\bibfnamefont{R.}~\bibnamefont{Troc}}, and
  \bibinfo{author}{\bibfnamefont{V.~H.} \bibnamefont{Tran}},
  \bibinfo{year}{1997}, \bibinfo{journal}{Phys. Rev. B}
  \textbf{\bibinfo{volume}{56}}, \bibinfo{pages}{14505}.

\bibitem[{\citenamefont{Pines and Nozi{\`e}res}(1965)}]{pinesnozieres}
\bibinfo{author}{\bibnamefont{Pines}, \bibfnamefont{D.}}, and
  \bibinfo{author}{\bibfnamefont{P.}~\bibnamefont{Nozi{\`e}res}},
  \bibinfo{year}{1965}, \emph{\bibinfo{title}{The {T}heory of {Q}uantum
  {L}iquids}} (\bibinfo{publisher}{Benjamin}, \bibinfo{address}{New York}).

\bibitem[{\citenamefont{Pollak}(1984)}]{re:Pollak84}
\bibinfo{author}{\bibnamefont{Pollak}, \bibfnamefont{M.}},
  \bibinfo{year}{1984}, \bibinfo{journal}{Philos. Mag. B}
  \textbf{\bibinfo{volume}{50}}, \bibinfo{pages}{265}.

\bibitem[{\citenamefont{Pruschke} \emph{et~al.}(1995)\citenamefont{Pruschke,
  Jarrell, and Freericks}}]{pruschkeetalrev95}
\bibinfo{author}{\bibnamefont{Pruschke}, \bibfnamefont{T.}},
  \bibinfo{author}{\bibfnamefont{M.}~\bibnamefont{Jarrell}}, and
  \bibinfo{author}{\bibfnamefont{J.~K.} \bibnamefont{Freericks}},
  \bibinfo{year}{1995}, \bibinfo{journal}{Adv. Phys.}
  \textbf{\bibinfo{volume}{44}}, \bibinfo{pages}{187}.

\bibitem[{\citenamefont{Quirt and Marko}(1971)}]{marko}
\bibinfo{author}{\bibnamefont{Quirt}, \bibfnamefont{J.~D.}}, and
  \bibinfo{author}{\bibfnamefont{J.~R.} \bibnamefont{Marko}},
  \bibinfo{year}{1971}, \bibinfo{journal}{Phys. Rev. Lett.}
  \textbf{\bibinfo{volume}{26}}, \bibinfo{pages}{318}.

\bibitem[{\citenamefont{Rappoport} \emph{et~al.}(2003)\citenamefont{Rappoport,
  Saguia, Boechat, and Continentino}}]{rappoportetal03}
\bibinfo{author}{\bibnamefont{Rappoport}, \bibfnamefont{T.~G.}},
  \bibinfo{author}{\bibfnamefont{A.}~\bibnamefont{Saguia}},
  \bibinfo{author}{\bibfnamefont{B.}~\bibnamefont{Boechat}}, and
  \bibinfo{author}{\bibfnamefont{M.~A.} \bibnamefont{Continentino}},
  \bibinfo{year}{2003}, \bibinfo{journal}{Europhys. Lett.}
  \textbf{\bibinfo{volume}{61}}, \bibinfo{pages}{831}.

\bibitem[{\citenamefont{Read and Newns}(1983)}]{readnewns2}
\bibinfo{author}{\bibnamefont{Read}, \bibfnamefont{N.}}, and
  \bibinfo{author}{\bibfnamefont{D.~M.} \bibnamefont{Newns}},
  \bibinfo{year}{1983}, \bibinfo{journal}{J. Phys. C}
  \textbf{\bibinfo{volume}{16}}, \bibinfo{pages}{L1055}.

\bibitem[{\citenamefont{Read} \emph{et~al.}(1995)\citenamefont{Read, Sachdev,
  and Ye}}]{re:Read95}
\bibinfo{author}{\bibnamefont{Read}, \bibfnamefont{N.}},
  \bibinfo{author}{\bibfnamefont{S.}~\bibnamefont{Sachdev}}, and
  \bibinfo{author}{\bibfnamefont{J.}~\bibnamefont{Ye}}, \bibinfo{year}{1995},
  \bibinfo{journal}{Phys. Rev. B} \textbf{\bibinfo{volume}{52}},
  \bibinfo{pages}{384}.

\bibitem[{\citenamefont{Refael} \emph{et~al.}(2002)\citenamefont{Refael,
  Kehrein, and Fisher}}]{refaeletal02}
\bibinfo{author}{\bibnamefont{Refael}, \bibfnamefont{G.}},
  \bibinfo{author}{\bibfnamefont{S.}~\bibnamefont{Kehrein}}, and
  \bibinfo{author}{\bibfnamefont{D.~S.} \bibnamefont{Fisher}},
  \bibinfo{year}{2002}, \bibinfo{journal}{Phys. Rev. B}
  \textbf{\bibinfo{volume}{66}}, \bibinfo{pages}{060402(R)}.

\bibitem[{\citenamefont{Rieger and Young}(1994)}]{riegeryoung94}
\bibinfo{author}{\bibnamefont{Rieger}, \bibfnamefont{H.}}, and
  \bibinfo{author}{\bibfnamefont{A.~P.} \bibnamefont{Young}},
  \bibinfo{year}{1994}, \bibinfo{journal}{Phys. Rev. Lett.}
  \textbf{\bibinfo{volume}{72}}, \bibinfo{pages}{4141}.

\bibitem[{\citenamefont{Rieger and Young}(1996)}]{riegeryoung96}
\bibinfo{author}{\bibnamefont{Rieger}, \bibfnamefont{H.}}, and
  \bibinfo{author}{\bibfnamefont{A.~P.} \bibnamefont{Young}},
  \bibinfo{year}{1996}, \bibinfo{journal}{Phys. Rev. B}
  \textbf{\bibinfo{volume}{54}}, \bibinfo{pages}{3328}.

\bibitem[{\citenamefont{Rosch}(1999)}]{rosch99}
\bibinfo{author}{\bibnamefont{Rosch}, \bibfnamefont{A.}}, \bibinfo{year}{1999},
  \bibinfo{journal}{Phys. Rev. Lett.} \textbf{\bibinfo{volume}{82}},
  \bibinfo{pages}{4280}.

\bibitem[{\citenamefont{Rosch}(2000)}]{rosch00}
\bibinfo{author}{\bibnamefont{Rosch}, \bibfnamefont{A.}}, \bibinfo{year}{2000},
  \bibinfo{journal}{Phys. Rev. B} \textbf{\bibinfo{volume}{62}},
  \bibinfo{pages}{4945}.

\bibitem[{\citenamefont{Rosch} \emph{et~al.}(1997)\citenamefont{Rosch,
  Schr{\"o}der, Stockert, and v.~L{\"o}hneysen}}]{roschetal97}
\bibinfo{author}{\bibnamefont{Rosch}, \bibfnamefont{A.}},
  \bibinfo{author}{\bibfnamefont{A.}~\bibnamefont{Schr{\"o}der}},
  \bibinfo{author}{\bibfnamefont{O.}~\bibnamefont{Stockert}}, and
  \bibinfo{author}{\bibfnamefont{H.}~\bibnamefont{v.~L{\"o}hneysen}},
  \bibinfo{year}{1997}, \bibinfo{journal}{Phys. Rev. Lett.}
  \textbf{\bibinfo{volume}{79}}, \bibinfo{pages}{159}.

\bibitem[{\citenamefont{Rosenbaum} \emph{et~al.}(1980)\citenamefont{Rosenbaum,
  Andres, Thomas, and Bhatt}}]{rosenbaum-prl80}
\bibinfo{author}{\bibnamefont{Rosenbaum}, \bibfnamefont{T.~F.}},
  \bibinfo{author}{\bibfnamefont{K.}~\bibnamefont{Andres}},
  \bibinfo{author}{\bibfnamefont{G.~A.} \bibnamefont{Thomas}}, and
  \bibinfo{author}{\bibfnamefont{R.~N.} \bibnamefont{Bhatt}},
  \bibinfo{year}{1980}, \bibinfo{journal}{Phys. Rev. Lett.}
  \textbf{\bibinfo{volume}{45}}, \bibinfo{pages}{1423}.

\bibitem[{\citenamefont{Rozenberg and Grempel}(1998)}]{rozenberg98prl}
\bibinfo{author}{\bibnamefont{Rozenberg}, \bibfnamefont{M.~J.}}, and
  \bibinfo{author}{\bibfnamefont{D.}~\bibnamefont{Grempel}},
  \bibinfo{year}{1998}, \bibinfo{journal}{Phys. Rev. Lett.}
  \textbf{\bibinfo{volume}{81}}, \bibinfo{pages}{2550}.

\bibitem[{\citenamefont{Rozenberg and Grempel}(1999)}]{rozenberg99prb}
\bibinfo{author}{\bibnamefont{Rozenberg}, \bibfnamefont{M.~J.}}, and
  \bibinfo{author}{\bibfnamefont{D.}~\bibnamefont{Grempel}},
  \bibinfo{year}{1999}, \bibinfo{journal}{Phys. Rev. B}
  \textbf{\bibinfo{volume}{60}}, \bibinfo{pages}{4702}.

\bibitem[{\citenamefont{Sachdev}(1999)}]{sachdevbook}
\bibinfo{author}{\bibnamefont{Sachdev}, \bibfnamefont{S.}},
  \bibinfo{year}{1999}, \emph{\bibinfo{title}{Quantum Phase Transitions}}
  (\bibinfo{publisher}{Cambridge University Press}, \bibinfo{address}{UK}).

\bibitem[{\citenamefont{Sachdev and Read}(1996)}]{re:Sachdev96}
\bibinfo{author}{\bibnamefont{Sachdev}, \bibfnamefont{S.}}, and
  \bibinfo{author}{\bibfnamefont{N.}~\bibnamefont{Read}}, \bibinfo{year}{1996},
  \bibinfo{journal}{J. Phys. Condens. Matter} \textbf{\bibinfo{volume}{8}},
  \bibinfo{pages}{9723}.

\bibitem[{\citenamefont{Sachdev} \emph{et~al.}(1995)\citenamefont{Sachdev,
  Read, and Oppermann}}]{sachdevreadopper}
\bibinfo{author}{\bibnamefont{Sachdev}, \bibfnamefont{S.}},
  \bibinfo{author}{\bibfnamefont{N.}~\bibnamefont{Read}}, and
  \bibinfo{author}{\bibfnamefont{R.}~\bibnamefont{Oppermann}},
  \bibinfo{year}{1995}, \bibinfo{journal}{Phys. Rev. B}
  \textbf{\bibinfo{volume}{52}}, \bibinfo{pages}{10286}.

\bibitem[{\citenamefont{Sachdev and Ye}(1993)}]{sachdevye}
\bibinfo{author}{\bibnamefont{Sachdev}, \bibfnamefont{S.}}, and
  \bibinfo{author}{\bibfnamefont{J.}~\bibnamefont{Ye}}, \bibinfo{year}{1993},
  \bibinfo{journal}{Phys. Rev. Lett.} \textbf{\bibinfo{volume}{70}},
  \bibinfo{pages}{3339}.

\bibitem[{\citenamefont{Saguia}
  \emph{et~al.}(2003{\natexlab{a}})\citenamefont{Saguia, Boechat, and
  Continentino}}]{saguiaetal}
\bibinfo{author}{\bibnamefont{Saguia}, \bibfnamefont{A.}},
  \bibinfo{author}{\bibfnamefont{B.}~\bibnamefont{Boechat}}, and
  \bibinfo{author}{\bibfnamefont{M.~A.} \bibnamefont{Continentino}},
  \bibinfo{year}{2003}{\natexlab{a}}, \bibinfo{journal}{Phys. Rev. Lett.}
  \textbf{\bibinfo{volume}{89}}, \bibinfo{pages}{117202}.

\bibitem[{\citenamefont{Saguia}
  \emph{et~al.}(2003{\natexlab{b}})\citenamefont{Saguia, Boechat, and
  Continentino}}]{saguiaetal03}
\bibinfo{author}{\bibnamefont{Saguia}, \bibfnamefont{A.}},
  \bibinfo{author}{\bibfnamefont{B.}~\bibnamefont{Boechat}}, and
  \bibinfo{author}{\bibfnamefont{M.~A.} \bibnamefont{Continentino}},
  \bibinfo{year}{2003}{\natexlab{b}}, \bibinfo{journal}{Phys. Rev. B}
  \textbf{\bibinfo{volume}{68}}, \bibinfo{pages}{020403(R)}.

\bibitem[{\citenamefont{Sanna} \emph{et~al.}(2004)\citenamefont{Sanna, Allodi,
  Concas, Hillier, and Renzi}}]{sanna-lanl04}
\bibinfo{author}{\bibnamefont{Sanna}, \bibfnamefont{S.}},
  \bibinfo{author}{\bibfnamefont{G.}~\bibnamefont{Allodi}},
  \bibinfo{author}{\bibfnamefont{G.}~\bibnamefont{Concas}},
  \bibinfo{author}{\bibfnamefont{A.~H.} \bibnamefont{Hillier}}, and
  \bibinfo{author}{\bibfnamefont{R.~D.} \bibnamefont{Renzi}},
  \bibinfo{year}{2004}, \bibinfo{journal}{preprint, cond-mat/0403608} .

\bibitem[{\citenamefont{Sarachik}(1995)}]{sarachik95}
\bibinfo{author}{\bibnamefont{Sarachik}, \bibfnamefont{M.~P.}},
  \bibinfo{year}{1995}, in \emph{\bibinfo{booktitle}{Metal-Insulator
  Transitions Revisited}}, edited by
  \bibinfo{editor}{\bibfnamefont{P.}~\bibnamefont{Edwards}} and
  \bibinfo{editor}{\bibfnamefont{C.~N.~R.} \bibnamefont{Rao}}
  (\bibinfo{publisher}{Taylor and Francis}).

\bibitem[{\citenamefont{Schaffer and Wegner}(1980)}]{wegner80}
\bibinfo{author}{\bibnamefont{Schaffer}, \bibfnamefont{L.}}, and
  \bibinfo{author}{\bibfnamefont{F.}~\bibnamefont{Wegner}},
  \bibinfo{year}{1980}, \bibinfo{journal}{Phys. Rev. B}
  \textbf{\bibinfo{volume}{38}}, \bibinfo{pages}{113}.

\bibitem[{\citenamefont{Schlager and Lohneysen}(1997)}]{schlager-epl97}
\bibinfo{author}{\bibnamefont{Schlager}, \bibfnamefont{H.~G.}}, and
  \bibinfo{author}{\bibfnamefont{H.~V.} \bibnamefont{Lohneysen}},
  \bibinfo{year}{1997}, \bibinfo{journal}{Europhys. Lett.}
  \textbf{\bibinfo{volume}{40}}, \bibinfo{pages}{661}.

\bibitem[{\citenamefont{Schmalian and Wolynes}(2000)}]{schmalian-prl00}
\bibinfo{author}{\bibnamefont{Schmalian}, \bibfnamefont{J.}}, and
  \bibinfo{author}{\bibfnamefont{P.~G.} \bibnamefont{Wolynes}},
  \bibinfo{year}{2000}, \bibinfo{journal}{Phys. Rev. Lett.}
  \textbf{\bibinfo{volume}{85}}, \bibinfo{pages}{836}.

\bibitem[{\citenamefont{Schr{\"o}der}
  \emph{et~al.}(1998)\citenamefont{Schr{\"o}der, Aeppli, Bucher, Ramazashvili,
  and Coleman}}]{schroederetal}
\bibinfo{author}{\bibnamefont{Schr{\"o}der}, \bibfnamefont{A.}},
  \bibinfo{author}{\bibfnamefont{G.}~\bibnamefont{Aeppli}},
  \bibinfo{author}{\bibfnamefont{E.}~\bibnamefont{Bucher}},
  \bibinfo{author}{\bibfnamefont{R.}~\bibnamefont{Ramazashvili}}, and
  \bibinfo{author}{\bibfnamefont{P.}~\bibnamefont{Coleman}},
  \bibinfo{year}{1998}, \bibinfo{journal}{Phys. Rev. Lett.}
  \textbf{\bibinfo{volume}{80}}, \bibinfo{pages}{5623}.

\bibitem[{\citenamefont{Schr{\"o}der}
  \emph{et~al.}(2000)\citenamefont{Schr{\"o}der, Aeppli, Coldea, Adams,
  Stockert, L{\"o}hneysen, Bucher, Ramazashvili, and
  Coleman}}]{schroedernature}
\bibinfo{author}{\bibnamefont{Schr{\"o}der}, \bibfnamefont{A.}},
  \bibinfo{author}{\bibfnamefont{G.}~\bibnamefont{Aeppli}},
  \bibinfo{author}{\bibfnamefont{R.}~\bibnamefont{Coldea}},
  \bibinfo{author}{\bibfnamefont{M.}~\bibnamefont{Adams}},
  \bibinfo{author}{\bibfnamefont{O.}~\bibnamefont{Stockert}},
  \bibinfo{author}{\bibfnamefont{H.}~\bibnamefont{L{\"o}hneysen}},
  \bibinfo{author}{\bibfnamefont{E.}~\bibnamefont{Bucher}},
  \bibinfo{author}{\bibfnamefont{R.}~\bibnamefont{Ramazashvili}}, and
  \bibinfo{author}{\bibfnamefont{P.}~\bibnamefont{Coleman}},
  \bibinfo{year}{2000}, \bibinfo{journal}{Nature}
  \textbf{\bibinfo{volume}{407}}, \bibinfo{pages}{351}.

\bibitem[{\citenamefont{Seaman} \emph{et~al.}(1991)\citenamefont{Seaman, Maple,
  Lee, Ghamaty, Torikachvili, Kang, Liu, Allen, and Cox}}]{seamanetal91}
\bibinfo{author}{\bibnamefont{Seaman}, \bibfnamefont{C.}},
  \bibinfo{author}{\bibfnamefont{M.~B.} \bibnamefont{Maple}},
  \bibinfo{author}{\bibfnamefont{B.~W.} \bibnamefont{Lee}},
  \bibinfo{author}{\bibfnamefont{S.}~\bibnamefont{Ghamaty}},
  \bibinfo{author}{\bibfnamefont{M.~S.} \bibnamefont{Torikachvili}},
  \bibinfo{author}{\bibfnamefont{J.~S.} \bibnamefont{Kang}},
  \bibinfo{author}{\bibfnamefont{L.~Z.} \bibnamefont{Liu}},
  \bibinfo{author}{\bibfnamefont{J.~W.} \bibnamefont{Allen}}, and
  \bibinfo{author}{\bibfnamefont{D.~L.} \bibnamefont{Cox}},
  \bibinfo{year}{1991}, \bibinfo{journal}{Phys. Rev. Lett.}
  \textbf{\bibinfo{volume}{67}}, \bibinfo{pages}{2882}.

\bibitem[{\citenamefont{Sengupta}(2000)}]{sengupta}
\bibinfo{author}{\bibnamefont{Sengupta}, \bibfnamefont{A.~M.}},
  \bibinfo{year}{2000}, \bibinfo{journal}{Phys. Rev. B}
  \textbf{\bibinfo{volume}{61}}, \bibinfo{pages}{4041}.

\bibitem[{\citenamefont{Sengupta and Georges}(1995)}]{senguptageorges}
\bibinfo{author}{\bibnamefont{Sengupta}, \bibfnamefont{A.~M.}}, and
  \bibinfo{author}{\bibfnamefont{A.}~\bibnamefont{Georges}},
  \bibinfo{year}{1995}, \bibinfo{journal}{Phys. Rev. B}
  \textbf{\bibinfo{volume}{52}}, \bibinfo{pages}{10295}.

\bibitem[{\citenamefont{Senthil and Sachdev}(1996)}]{senthilsachdev}
\bibinfo{author}{\bibnamefont{Senthil}, \bibfnamefont{T.}}, and
  \bibinfo{author}{\bibfnamefont{S.}~\bibnamefont{Sachdev}},
  \bibinfo{year}{1996}, \bibinfo{journal}{Phys. Rev. Lett.}
  \textbf{\bibinfo{volume}{77}}, \bibinfo{pages}{5292}.

\bibitem[{\citenamefont{Senthil} \emph{et~al.}(2003)\citenamefont{Senthil,
  Sachdev, and Vojta}}]{senthiletal}
\bibinfo{author}{\bibnamefont{Senthil}, \bibfnamefont{T.}},
  \bibinfo{author}{\bibfnamefont{S.}~\bibnamefont{Sachdev}}, and
  \bibinfo{author}{\bibfnamefont{M.}~\bibnamefont{Vojta}},
  \bibinfo{year}{2003}, \bibinfo{journal}{Phys. Rev. Lett.}
  \textbf{\bibinfo{volume}{90}}, \bibinfo{pages}{216403}.

\bibitem[{\citenamefont{Senthil} \emph{et~al.}(2004)\citenamefont{Senthil,
  Vojta, and Sachdev}}]{senthiletal2}
\bibinfo{author}{\bibnamefont{Senthil}, \bibfnamefont{T.}},
  \bibinfo{author}{\bibfnamefont{M.}~\bibnamefont{Vojta}}, and
  \bibinfo{author}{\bibfnamefont{S.}~\bibnamefont{Sachdev}},
  \bibinfo{year}{2004}, \bibinfo{journal}{Phys. Rev. B}
  \textbf{\bibinfo{volume}{69}}, \bibinfo{pages}{035111}.

\bibitem[{\citenamefont{Shah and Millis}(2003)}]{shahmillis03}
\bibinfo{author}{\bibnamefont{Shah}, \bibfnamefont{N.}}, and
  \bibinfo{author}{\bibfnamefont{A.~J.} \bibnamefont{Millis}},
  \bibinfo{year}{2003}, \bibinfo{journal}{Phys. Rev. Lett.}
  \textbf{\bibinfo{volume}{91}}, \bibinfo{pages}{147204}.

\bibitem[{\citenamefont{Shankar}(1994)}]{shankar94}
\bibinfo{author}{\bibnamefont{Shankar}, \bibfnamefont{R.}},
  \bibinfo{year}{1994}, \bibinfo{journal}{Rev. Mod. Phys.}
  \textbf{\bibinfo{volume}{66}}, \bibinfo{pages}{129}.

\bibitem[{\citenamefont{Shankar and Murthy}(1987)}]{shankarmurthy87}
\bibinfo{author}{\bibnamefont{Shankar}, \bibfnamefont{R.}}, and
  \bibinfo{author}{\bibfnamefont{G.}~\bibnamefont{Murthy}},
  \bibinfo{year}{1987}, \bibinfo{journal}{Phys. Rev. B}
  \textbf{\bibinfo{volume}{36}}, \bibinfo{pages}{536}.

\bibitem[{\citenamefont{Sherrington and Mihill}(1974)}]{sherrington74}
\bibinfo{author}{\bibnamefont{Sherrington}, \bibfnamefont{D.}}, and
  \bibinfo{author}{\bibfnamefont{K.}~\bibnamefont{Mihill}},
  \bibinfo{year}{1974}, \bibinfo{journal}{Journal de Physique Colloque}
  \textbf{\bibinfo{volume}{5}}, \bibinfo{pages}{199}.

\bibitem[{\citenamefont{Shklovskii and Efros}(1984)}]{doped-book}
\bibinfo{author}{\bibnamefont{Shklovskii}, \bibfnamefont{B.}}, and
  \bibinfo{author}{\bibfnamefont{A.}~\bibnamefont{Efros}},
  \bibinfo{year}{1984}, \emph{\bibinfo{title}{Electronic Properties of Doped
  Semiconductors}} (\bibinfo{publisher}{Springer-Verlag}).

\bibitem[{\citenamefont{Si} \emph{et~al.}(2001)\citenamefont{Si, Rabello,
  Ingersent, and Smith}}]{sietal}
\bibinfo{author}{\bibnamefont{Si}, \bibfnamefont{Q.}},
  \bibinfo{author}{\bibfnamefont{S.}~\bibnamefont{Rabello}},
  \bibinfo{author}{\bibfnamefont{K.}~\bibnamefont{Ingersent}}, and
  \bibinfo{author}{\bibfnamefont{J.~L.} \bibnamefont{Smith}},
  \bibinfo{year}{2001}, \bibinfo{journal}{Nature}
  \textbf{\bibinfo{volume}{413}}, \bibinfo{pages}{804}.

\bibitem[{\citenamefont{Si and Smith}(1996)}]{qimiao96prl}
\bibinfo{author}{\bibnamefont{Si}, \bibfnamefont{Q.}}, and
  \bibinfo{author}{\bibfnamefont{J.~L.} \bibnamefont{Smith}},
  \bibinfo{year}{1996}, \bibinfo{journal}{Phys. Rev. Lett.}
  \textbf{\bibinfo{volume}{77}}, \bibinfo{pages}{3391}.

\bibitem[{\citenamefont{Silin}(1958{\natexlab{a}})}]{silin58b}
\bibinfo{author}{\bibnamefont{Silin}, \bibfnamefont{V.~P.}},
  \bibinfo{year}{1958}{\natexlab{a}}, \bibinfo{journal}{Sov. Phys. JETP}
  \textbf{\bibinfo{volume}{6}}, \bibinfo{pages}{985}.

\bibitem[{\citenamefont{Silin}(1958{\natexlab{b}})}]{silin58a}
\bibinfo{author}{\bibnamefont{Silin}, \bibfnamefont{V.~P.}},
  \bibinfo{year}{1958}{\natexlab{b}}, \bibinfo{journal}{Sov. Phys. JETP}
  \textbf{\bibinfo{volume}{6}}, \bibinfo{pages}{387}.

\bibitem[{\citenamefont{Smith and Si}(2000)}]{qimiao00prb}
\bibinfo{author}{\bibnamefont{Smith}, \bibfnamefont{J.~L.}}, and
  \bibinfo{author}{\bibfnamefont{Q.}~\bibnamefont{Si}}, \bibinfo{year}{2000},
  \bibinfo{journal}{Phys. Rev. B} \textbf{\bibinfo{volume}{61}},
  \bibinfo{pages}{5184}.

\bibitem[{\citenamefont{Stewart}(1984)}]{stewartrev1}
\bibinfo{author}{\bibnamefont{Stewart}, \bibfnamefont{G.~R.}},
  \bibinfo{year}{1984}, \bibinfo{journal}{Rev. Mod. Phys.}
  \textbf{\bibinfo{volume}{56}}, \bibinfo{pages}{755}.

\bibitem[{\citenamefont{Stewart}(2001)}]{stewartNFL}
\bibinfo{author}{\bibnamefont{Stewart}, \bibfnamefont{G.~R.}},
  \bibinfo{year}{2001}, \bibinfo{journal}{Rev. Mod. Phys.}
  \textbf{\bibinfo{volume}{73}}, \bibinfo{pages}{797}.

\bibitem[{\citenamefont{Stockert} \emph{et~al.}(1998)\citenamefont{Stockert,
  L{\"o}hneysen, Rosch, Pyka, and Loewenhaupt}}]{stockertetal98}
\bibinfo{author}{\bibnamefont{Stockert}, \bibfnamefont{O.}},
  \bibinfo{author}{\bibfnamefont{H.~v.} \bibnamefont{L{\"o}hneysen}},
  \bibinfo{author}{\bibfnamefont{A.}~\bibnamefont{Rosch}},
  \bibinfo{author}{\bibfnamefont{N.}~\bibnamefont{Pyka}}, and
  \bibinfo{author}{\bibfnamefont{M.}~\bibnamefont{Loewenhaupt}},
  \bibinfo{year}{1998}, \bibinfo{journal}{Phys. Rev. Lett.}
  \textbf{\bibinfo{volume}{80}}, \bibinfo{pages}{5627}.

\bibitem[{\citenamefont{Stupp} \emph{et~al.}(1993)\citenamefont{Stupp, Hornung,
  Lakner, Madel, and L{\"o}hneysen}}]{stupp-prl93}
\bibinfo{author}{\bibnamefont{Stupp}, \bibfnamefont{H.}},
  \bibinfo{author}{\bibfnamefont{M.}~\bibnamefont{Hornung}},
  \bibinfo{author}{\bibfnamefont{M.}~\bibnamefont{Lakner}},
  \bibinfo{author}{\bibfnamefont{O.}~\bibnamefont{Madel}}, and
  \bibinfo{author}{\bibfnamefont{H.~V.} \bibnamefont{L{\"o}hneysen}},
  \bibinfo{year}{1993}, \bibinfo{journal}{Phys. Rev. Lett.}
  \textbf{\bibinfo{volume}{71}}, \bibinfo{pages}{2634}.

\bibitem[{\citenamefont{S{\"u}llow}
  \emph{et~al.}(2000)\citenamefont{S{\"u}llow, Mentink, Mason, Feyerherm,
  Nieuwenhuys, Menovsky, and Mydosh}}]{sullowetal00}
\bibinfo{author}{\bibnamefont{S{\"u}llow}, \bibfnamefont{S.}},
  \bibinfo{author}{\bibfnamefont{S.~A.~M.} \bibnamefont{Mentink}},
  \bibinfo{author}{\bibfnamefont{T.~E.} \bibnamefont{Mason}},
  \bibinfo{author}{\bibfnamefont{R.}~\bibnamefont{Feyerherm}},
  \bibinfo{author}{\bibfnamefont{G.~J.} \bibnamefont{Nieuwenhuys}},
  \bibinfo{author}{\bibfnamefont{A.~A.} \bibnamefont{Menovsky}}, and
  \bibinfo{author}{\bibfnamefont{J.~A.} \bibnamefont{Mydosh}},
  \bibinfo{year}{2000}, \bibinfo{journal}{Phys. Rev. B}
  \textbf{\bibinfo{volume}{61}}, \bibinfo{pages}{8878}.

\bibitem[{\citenamefont{Takagi} \emph{et~al.}(1992)\citenamefont{Takagi,
  Batlogg, Kao, Kwo, Cava, Krajewski, and W.~F.~Peck}}]{takagi-prl92}
\bibinfo{author}{\bibnamefont{Takagi}, \bibfnamefont{H.}},
  \bibinfo{author}{\bibfnamefont{B.}~\bibnamefont{Batlogg}},
  \bibinfo{author}{\bibfnamefont{H.~L.} \bibnamefont{Kao}},
  \bibinfo{author}{\bibfnamefont{J.}~\bibnamefont{Kwo}},
  \bibinfo{author}{\bibfnamefont{R.~J.} \bibnamefont{Cava}},
  \bibinfo{author}{\bibfnamefont{J.~J.} \bibnamefont{Krajewski}}, and
  \bibinfo{author}{\bibfnamefont{J.}~\bibnamefont{W.~F.~Peck}},
  \bibinfo{year}{1992}, \bibinfo{journal}{Phys. Rev. Lett}
  \textbf{\bibinfo{volume}{69}}, \bibinfo{pages}{2975}.

\bibitem[{\citenamefont{Tanaskovi\'{c}}
  \emph{et~al.}(2003)\citenamefont{Tanaskovi\'{c}, Dobrosavljevi\'{c},
  Abrahams, and Kotliar}}]{tanaskovicetal03}
\bibinfo{author}{\bibnamefont{Tanaskovi\'{c}}, \bibfnamefont{D.}},
  \bibinfo{author}{\bibfnamefont{V.}~\bibnamefont{Dobrosavljevi\'{c}}},
  \bibinfo{author}{\bibfnamefont{E.}~\bibnamefont{Abrahams}}, and
  \bibinfo{author}{\bibfnamefont{G.}~\bibnamefont{Kotliar}},
  \bibinfo{year}{2003}, \bibinfo{journal}{Phys. Rev. Lett.}
  \textbf{\bibinfo{volume}{91}}, \bibinfo{pages}{066603}.

\bibitem[{\citenamefont{Tanaskovi\'{c}}
  \emph{et~al.}(2005)\citenamefont{Tanaskovi\'{c},
  Dobrosavljevi\'{c}, and Miranda}}]{tanaskovicetal05}
\bibinfo{author}{\bibnamefont{Tanaskovi\'{c}}, \bibfnamefont{D.}},
  \bibinfo{author}{\bibfnamefont{V.}~\bibnamefont{Dobrosavljevi\'{c}}}, and
  \bibinfo{author}{\bibfnamefont{E.}~\bibnamefont{Miranda}},
  \bibinfo{year}{2005}, \bibinfo{journal}{Phys. Rev. Lett.
  (in press),} \bibinfo{note}{cond-mat/0412100}.

\bibitem[{\citenamefont{Tanaskovi\'{c}}
  \emph{et~al.}(2004)\citenamefont{Tanaskovi\'{c}, Miranda, and
  Dobrosavljevi\'{c}}}]{tanaskovicetal04}
\bibinfo{author}{\bibnamefont{Tanaskovi\'{c}}, \bibfnamefont{D.}},
  \bibinfo{author}{\bibfnamefont{E.}~\bibnamefont{Miranda}}, and
  \bibinfo{author}{\bibfnamefont{V.}~\bibnamefont{Dobrosavljevi\'{c}}},
  \bibinfo{year}{2004}, \bibinfo{journal}{Phys. Rev. B}
  \textbf{\bibinfo{volume}{70}}, \bibinfo{pages}{205108}.

\bibitem[{\citenamefont{Tesanovi\'c}(1986)}]{tesanovic-prb86}
\bibinfo{author}{\bibnamefont{Tesanovi\'c}, \bibfnamefont{Z.}},
  \bibinfo{year}{1986}, \bibinfo{journal}{Phys. Rev. B}
  \textbf{\bibinfo{volume}{34}}, \bibinfo{pages}{5212}.

\bibitem[{\citenamefont{Thill and Huse}(1995)}]{thillhuse95}
\bibinfo{author}{\bibnamefont{Thill}, \bibfnamefont{M.~J.}}, and
  \bibinfo{author}{\bibfnamefont{D.~A.} \bibnamefont{Huse}},
  \bibinfo{year}{1995}, \bibinfo{journal}{Physica A}
  \textbf{\bibinfo{volume}{214}}, \bibinfo{pages}{321}.

\bibitem[{\citenamefont{Thouless}(1969)}]{thouless69}
\bibinfo{author}{\bibnamefont{Thouless}, \bibfnamefont{D.~J.}},
  \bibinfo{year}{1969}, \bibinfo{journal}{Phys. Rev.}
  \textbf{\bibinfo{volume}{187}}, \bibinfo{pages}{732}.

\bibitem[{\citenamefont{de~la Torre} \emph{et~al.}(1998)\citenamefont{de~la
  Torre, Ellerby, Watmough, and McEwen}}]{delatorre98}
\bibinfo{author}{\bibnamefont{de~la Torre}, \bibfnamefont{M.~A.~L.}},
  \bibinfo{author}{\bibfnamefont{M.}~\bibnamefont{Ellerby}},
  \bibinfo{author}{\bibfnamefont{M.}~\bibnamefont{Watmough}}, and
  \bibinfo{author}{\bibfnamefont{K.~A.} \bibnamefont{McEwen}},
  \bibinfo{year}{1998}, \bibinfo{journal}{J. Magn. Magn. Mater.}
  \textbf{\bibinfo{volume}{177-181}}, \bibinfo{pages}{445}.

\bibitem[{\citenamefont{Tusch and Logan}(1993)}]{tuschlogan93}
\bibinfo{author}{\bibnamefont{Tusch}, \bibfnamefont{M.~A.}}, and
  \bibinfo{author}{\bibfnamefont{D.~E.} \bibnamefont{Logan}},
  \bibinfo{year}{1993}, \bibinfo{journal}{Phys. Rev. B}
  \textbf{\bibinfo{volume}{48}}, \bibinfo{pages}{14843}.

\bibitem[{\citenamefont{Ue and Maekawa}(1971)}]{maekawa}
\bibinfo{author}{\bibnamefont{Ue}, \bibfnamefont{H.}}, and
  \bibinfo{author}{\bibfnamefont{S.}~\bibnamefont{Maekawa}},
  \bibinfo{year}{1971}, \bibinfo{journal}{Phys. Rev.}
  \textbf{\bibinfo{volume}{12}}, \bibinfo{pages}{4232}.

\bibitem[{\citenamefont{Vaknin} \emph{et~al.}(1998)\citenamefont{Vaknin,
  Ovadyahu, and Pollak}}]{films24}
\bibinfo{author}{\bibnamefont{Vaknin}, \bibfnamefont{A.}},
  \bibinfo{author}{\bibfnamefont{Z.}~\bibnamefont{Ovadyahu}}, and
  \bibinfo{author}{\bibfnamefont{M.}~\bibnamefont{Pollak}},
  \bibinfo{year}{1998}, \bibinfo{journal}{Phys. Rev. Lett.}
  \textbf{\bibinfo{volume}{81}}, \bibinfo{pages}{669}.

\bibitem[{\citenamefont{Vaknin} \emph{et~al.}(2000)\citenamefont{Vaknin,
  Ovadyahu, and Pollak}}]{films25}
\bibinfo{author}{\bibnamefont{Vaknin}, \bibfnamefont{A.}},
  \bibinfo{author}{\bibfnamefont{Z.}~\bibnamefont{Ovadyahu}}, and
  \bibinfo{author}{\bibfnamefont{M.}~\bibnamefont{Pollak}},
  \bibinfo{year}{2000}, \bibinfo{journal}{Phys. Rev. Lett.}
  \textbf{\bibinfo{volume}{84}}, \bibinfo{pages}{3402}.

\bibitem[{\citenamefont{Varma} \emph{et~al.}(1989)\citenamefont{Varma,
  Littlewood, Schmitt-Rink, Abrahams, and Ruckenstein}}]{mfl}
\bibinfo{author}{\bibnamefont{Varma}, \bibfnamefont{C.~M.}},
  \bibinfo{author}{\bibfnamefont{P.~B.} \bibnamefont{Littlewood}},
  \bibinfo{author}{\bibfnamefont{S.}~\bibnamefont{Schmitt-Rink}},
  \bibinfo{author}{\bibfnamefont{E.}~\bibnamefont{Abrahams}}, and
  \bibinfo{author}{\bibfnamefont{A.~E.} \bibnamefont{Ruckenstein}},
  \bibinfo{year}{1989}, \bibinfo{journal}{Phys. Rev. Lett.}
  \textbf{\bibinfo{volume}{63}}, \bibinfo{pages}{1996}.

\bibitem[{\citenamefont{Vojta}(2003)}]{tvojta03}
\bibinfo{author}{\bibnamefont{Vojta}, \bibfnamefont{T.}}, \bibinfo{year}{2003},
  \bibinfo{journal}{Phys. Rev. Lett.} \textbf{\bibinfo{volume}{90}},
  \bibinfo{pages}{107202}.

\bibitem[{\citenamefont{Vojta and Schmalian}(2004)}]{vojtaschmalian04}
\bibinfo{author}{\bibnamefont{Vojta}, \bibfnamefont{T.}}, and
  \bibinfo{author}{\bibfnamefont{J.}~\bibnamefont{Schmalian}},
  \bibinfo{year}{2004}, \bibinfo{journal}{cond-mat/0405609} .

\bibitem[{\citenamefont{Vollmer} \emph{et~al.}(2000)\citenamefont{Vollmer,
  Pietrus, v.~L{\"o}hneysen, Chau, and Maple}}]{vollmeretal}
\bibinfo{author}{\bibnamefont{Vollmer}, \bibfnamefont{R.}},
  \bibinfo{author}{\bibfnamefont{T.}~\bibnamefont{Pietrus}},
  \bibinfo{author}{\bibfnamefont{H.}~\bibnamefont{v.~L{\"o}hneysen}},
  \bibinfo{author}{\bibfnamefont{R.}~\bibnamefont{Chau}}, and
  \bibinfo{author}{\bibfnamefont{M.~B.} \bibnamefont{Maple}},
  \bibinfo{year}{2000}, \bibinfo{journal}{Phys. Rev. B}
  \textbf{\bibinfo{volume}{61}}, \bibinfo{pages}{1218}.

\bibitem[{\citenamefont{Waffenschmidt}
  \emph{et~al.}(1999)\citenamefont{Waffenschmidt, Pfleiderer, and
  L{\"o}hneysen}}]{waffenschmit-prl99}
\bibinfo{author}{\bibnamefont{Waffenschmidt}, \bibfnamefont{S.}},
  \bibinfo{author}{\bibfnamefont{C.}~\bibnamefont{Pfleiderer}}, and
  \bibinfo{author}{\bibfnamefont{H.~V.} \bibnamefont{L{\"o}hneysen}},
  \bibinfo{year}{1999}, \bibinfo{journal}{Phys. Rev. Lett.}
  \textbf{\bibinfo{volume}{83}}, \bibinfo{pages}{3005}.

\bibitem[{\citenamefont{Weber} \emph{et~al.}(2001)\citenamefont{Weber,
  K{\"o}rner, Scheidt, Kehrein, and Stewart}}]{weberetal01}
\bibinfo{author}{\bibnamefont{Weber}, \bibfnamefont{A.}},
  \bibinfo{author}{\bibfnamefont{S.}~\bibnamefont{K{\"o}rner}},
  \bibinfo{author}{\bibfnamefont{E.-W.} \bibnamefont{Scheidt}},
  \bibinfo{author}{\bibfnamefont{S.}~\bibnamefont{Kehrein}}, and
  \bibinfo{author}{\bibfnamefont{G.~R.} \bibnamefont{Stewart}},
  \bibinfo{year}{2001}, \bibinfo{journal}{Phys. Rev. B}
  \textbf{\bibinfo{volume}{63}}, \bibinfo{pages}{205116}.

\bibitem[{\citenamefont{Wegner}(1976)}]{wegner76}
\bibinfo{author}{\bibnamefont{Wegner}, \bibfnamefont{F.}},
  \bibinfo{year}{1976}, \bibinfo{journal}{Z. Phys. B}
  \textbf{\bibinfo{volume}{25}}, \bibinfo{pages}{327}.

\bibitem[{\citenamefont{Wegner}(1979)}]{wegner79}
\bibinfo{author}{\bibnamefont{Wegner}, \bibfnamefont{F.}},
  \bibinfo{year}{1979}, \bibinfo{journal}{Phys. Rev. B}
  \textbf{\bibinfo{volume}{35}}, \bibinfo{pages}{783}.

\bibitem[{\citenamefont{Wegner}(1981)}]{dos0}
\bibinfo{author}{\bibnamefont{Wegner}, \bibfnamefont{F.}},
  \bibinfo{year}{1981}, \bibinfo{journal}{Z. Phys. B}
  \textbf{\bibinfo{volume}{44}}, \bibinfo{pages}{9}.

\bibitem[{\citenamefont{Westerberg}
  \emph{et~al.}(1995)\citenamefont{Westerberg, Furusaki, Sigrist, and
  Lee}}]{westerbergetal95}
\bibinfo{author}{\bibnamefont{Westerberg}, \bibfnamefont{E.}},
  \bibinfo{author}{\bibfnamefont{A.}~\bibnamefont{Furusaki}},
  \bibinfo{author}{\bibfnamefont{M.}~\bibnamefont{Sigrist}}, and
  \bibinfo{author}{\bibfnamefont{P.~A.} \bibnamefont{Lee}},
  \bibinfo{year}{1995}, \bibinfo{journal}{Phys. Rev. Lett.}
  \textbf{\bibinfo{volume}{75}}, \bibinfo{pages}{4302}.

\bibitem[{\citenamefont{Westerberg}
  \emph{et~al.}(1997)\citenamefont{Westerberg, Furusaki, Sigrist, and
  Lee}}]{westerbergetal}
\bibinfo{author}{\bibnamefont{Westerberg}, \bibfnamefont{E.}},
  \bibinfo{author}{\bibfnamefont{A.}~\bibnamefont{Furusaki}},
  \bibinfo{author}{\bibfnamefont{M.}~\bibnamefont{Sigrist}}, and
  \bibinfo{author}{\bibfnamefont{P.~A.} \bibnamefont{Lee}},
  \bibinfo{year}{1997}, \bibinfo{journal}{Phys. Rev. B}
  \textbf{\bibinfo{volume}{55}}, \bibinfo{pages}{12578}.

\bibitem[{\citenamefont{Wilson}(1975)}]{Wilson2}
\bibinfo{author}{\bibnamefont{Wilson}, \bibfnamefont{K.~G.}},
  \bibinfo{year}{1975}, \bibinfo{journal}{Rev. Mod. Phys.}
  \textbf{\bibinfo{volume}{47}}, \bibinfo{pages}{773}.

\bibitem[{\citenamefont{Yang} \emph{et~al.}(1996)\citenamefont{Yang, Hyman,
  Bhatt, and Girvin}}]{yangetal}
\bibinfo{author}{\bibnamefont{Yang}, \bibfnamefont{K.}},
  \bibinfo{author}{\bibfnamefont{R.~A.} \bibnamefont{Hyman}},
  \bibinfo{author}{\bibfnamefont{R.~N.} \bibnamefont{Bhatt}}, and
  \bibinfo{author}{\bibfnamefont{S.~M.} \bibnamefont{Girvin}},
  \bibinfo{year}{1996}, \bibinfo{journal}{J. Appl. Phys.}
  \textbf{\bibinfo{volume}{79}}, \bibinfo{pages}{5096}.

\bibitem[{\citenamefont{Ye} \emph{et~al.}(1993)\citenamefont{Ye, Sachdev, and
  Read}}]{ye93}
\bibinfo{author}{\bibnamefont{Ye}, \bibfnamefont{J.}},
  \bibinfo{author}{\bibfnamefont{S.}~\bibnamefont{Sachdev}}, and
  \bibinfo{author}{\bibfnamefont{N.}~\bibnamefont{Read}}, \bibinfo{year}{1993},
  \bibinfo{journal}{Phys. Rev. Lett.} \textbf{\bibinfo{volume}{70}},
  \bibinfo{pages}{4011}.

\bibitem[{\citenamefont{Young and Rieger}(1996)}]{youngrieger96}
\bibinfo{author}{\bibnamefont{Young}, \bibfnamefont{A.~P.}}, and
  \bibinfo{author}{\bibfnamefont{H.}~\bibnamefont{Rieger}},
  \bibinfo{year}{1996}, \bibinfo{journal}{Phys. Rev. B}
  \textbf{\bibinfo{volume}{53}}, \bibinfo{pages}{8486}.

\bibitem[{\citenamefont{Yusuf and Yang}(2002)}]{yusufyang1}
\bibinfo{author}{\bibnamefont{Yusuf}, \bibfnamefont{E.}}, and
  \bibinfo{author}{\bibfnamefont{K.}~\bibnamefont{Yang}}, \bibinfo{year}{2002},
  \bibinfo{journal}{Phys. Rev. B} \textbf{\bibinfo{volume}{65}},
  \bibinfo{pages}{224428}.

\bibitem[{\citenamefont{Yusuf and Yang}(2003{\natexlab{a}})}]{yusufyang03}
\bibinfo{author}{\bibnamefont{Yusuf}, \bibfnamefont{E.}}, and
  \bibinfo{author}{\bibfnamefont{K.}~\bibnamefont{Yang}},
  \bibinfo{year}{2003}{\natexlab{a}}, \bibinfo{journal}{Phys. Rev. B}
  \textbf{\bibinfo{volume}{67}}, \bibinfo{pages}{144409}.

\bibitem[{\citenamefont{Yusuf and Yang}(2003{\natexlab{b}})}]{yusufyang2}
\bibinfo{author}{\bibnamefont{Yusuf}, \bibfnamefont{E.}}, and
  \bibinfo{author}{\bibfnamefont{K.}~\bibnamefont{Yang}},
  \bibinfo{year}{2003}{\natexlab{b}}, \bibinfo{journal}{Phys. Rev. B}
  \textbf{\bibinfo{volume}{68}}, \bibinfo{pages}{024425}.

\bibitem[{\citenamefont{Yuval and Anderson}(1970)}]{Yuval-Anderson2}
\bibinfo{author}{\bibnamefont{Yuval}, \bibfnamefont{G.}}, and
  \bibinfo{author}{\bibfnamefont{P.~W.} \bibnamefont{Anderson}},
  \bibinfo{year}{1970}, \bibinfo{journal}{Phys. Rev. B}
  \textbf{\bibinfo{volume}{1}}, \bibinfo{pages}{1522}.

\bibitem[{\citenamefont{Zar{\'a}nd and Demler}(2002)}]{zaranddemler02}
\bibinfo{author}{\bibnamefont{Zar{\'a}nd}, \bibfnamefont{G.}}, and
  \bibinfo{author}{\bibfnamefont{E.}~\bibnamefont{Demler}},
  \bibinfo{year}{2002}, \bibinfo{journal}{Phys. Rev. B}
  \textbf{\bibinfo{volume}{66}}, \bibinfo{pages}{024427}.

\bibitem[{\citenamefont{Zhu and Si}(2002)}]{zhusi02}
\bibinfo{author}{\bibnamefont{Zhu}, \bibfnamefont{L.}}, and
  \bibinfo{author}{\bibfnamefont{Q.}~\bibnamefont{Si}}, \bibinfo{year}{2002},
  \bibinfo{journal}{Phys. Rev. B} \textbf{\bibinfo{volume}{66}},
  \bibinfo{pages}{024426}.

\end{thebibliography}

\end{document}